\documentclass[12pt]{amsart}

\usepackage{geometry}
\usepackage[utf8]{inputenc}
\usepackage[T1]{fontenc}
\usepackage{amsfonts}
\usepackage{graphicx}
\usepackage{amsmath}
\usepackage{color}
\usepackage{mathrsfs}
\usepackage{amssymb}
\usepackage{cite}

\def\br{\begin{remark}\rm\small}
\def\er{\end{remark}}
\newcommand{\dd}{\mathrm{d}}
\newcommand{\beq}{\begin{equation}}
\newcommand{\eeq}{\end{equation}}
\newcommand{\bea}{\begin{eqnarray}}
\newcommand{\eea}{\end{eqnarray}}
\newcommand{\Res}{\mathop{\,\rm Res\,}}
\newcommand{\bs}[1]{\ensuremath{\boldsymbol{#1}}}

\UseRawInputEncoding

\newtheorem{theorem}{Theorem}

\newtheorem{proposition}[theorem]{Proposition}
\newtheorem{lemma}[theorem]{Lemma}
\newtheorem{corollary}[theorem]{Corollary}

\theoremstyle{definition}
\newtheorem{definition}[theorem]{Definition}
\theoremstyle{remark}
\newtheorem{remark}[theorem]{Remark}

\numberwithin{equation}{section}
\numberwithin{theorem}{section}

\newcommand{\arxiv}[1]{\href{http://arxiv.org/abs/#1}{#1}}

\calclayout

\usepackage{hyperref}

\begin{document}

\hypersetup{pageanchor=false}
\title{Nesting statistics in the $O(n)$ loop model on random planar maps}
\date{\today}
\author{Ga\"etan Borot}
\address{Humboldt-Universit\"at zu Berlin, Institut f\"ur Mathematik und Institut f\"ur Physik, Unter den Linden 6, 10099 Berlin, Germany}
\email{gaetan.borot@hu-berlin.de}
\author{J\'er\'emie Bouttier}
\address{Universit\'e Paris-Saclay, CNRS, CEA, Institut de physique th\'eorique, 91191 Gif-sur-Yvette, France.}
\address{Universit\'e Lyon, \'Ecole Normale Sup\'erieure de Lyon, Universit\'e Claude Bernard, CNRS, Laboratoire de Physique, 69342 Lyon, France.}
\email{jeremie.bouttier@ipht.fr}
\author{Bertrand Duplantier}
\address{Universit\'e Paris-Saclay, CNRS, CEA, Institut de physique th\'eorique, 91191 Gif-sur-Yvette, France.} 
\email{bertrand.duplantier@ipht.fr}

\thanks{The authors wish to thank Timothy Budd, Philippe Di Francesco,
  Cl\'ement Hongler, Jason Miller, Scott Sheffield, Wendelin Werner,
  and an unknown referee, for insightful comments. We also thank Ivan
  Kostov for pointing references \cite{KostovS,KostovH}, and the
  organisers of the Journ\'ees Cartes at IH\'ES. 
  We acknowledge hospitality and support from the Isaac Newton
  Institute (INI) for Mathematical Sciences during the 2015 program
  ``Random Geometry'' supported by EPSRC Grant Number EP/K032208/
  (G.B.~and B.D.), and from the Max-Planck-Institut f\"ur Mathematik in Bonn
  Bonn (J.B.~and B.D.). %
  We acknowledge financial support from the following institutions:
  the Simons Foundation (G.B.~and B.D.); the Max-Planck Gesellschaft
  (G.B.); the Agence Nationale de la Recherche via the grants
  ANR-12-JS02-001-01 ``Cartaplus'' (J.B.) and ANR-14-CE25-0014
  ``GRAAL'' (J.B.~and B.D.); the City of Paris through the project
  ``Combinatoire \`a Paris'' (J.B.); the CNRS Projet international de
  coop\'eration scientifique (PICS) ``Conformal Liouville Quantum
  Gravity'' n$^{\mathrm o}$PICS06769 (B.D.). This work was started while G.B. was affiliated with MIT Mathematics Department, and mainly conducted during his affiliation with the Max-Planck-Institut f\"ur Mathematik in Bonn.}

\begin{abstract}
In the $O(n)$ loop model on random planar maps, we study the depth --- in terms of the number of levels of nesting --- of the loop configuration, by means of analytic combinatorics. We focus on the `refined' generating series of pointed disks  or cylinders, which keep track of the number of loops separating the marked point from the boundary (for disks), or the two boundaries (for cylinders). For the general $O(n)$ loop model, we show that these generating series satisfy functional relations obtained by a modification of those satisfied by the unrefined generating series. In a more specific $O(n)$ model where loops cross only triangles and have a bending energy, we explicitly compute the refined generating series. We analyse their non generic critical behavior in the dense and dilute phases, and obtain the large deviations function of the nesting distribution, which is expected to be universal. Using the framework of Liouville quantum gravity (LQG), we show that a rigorous functional KPZ relation can be applied to the multifractal spectrum of extreme nesting in the conformal loop ensemble (${\rm CLE}_{\kappa}$) in the Euclidean unit disk, as obtained by Miller, Watson and Wilson \cite{MWW}, or to its natural generalisation to the Riemann sphere. It allows us to recover the large deviations results obtained for the critical $O(n)$ random planar map models. This offers, at the refined level of large deviations theory, a rigorous check of the fundamental fact that the universal scaling limits of random planar map models as weighted by partition functions of critical statistical models are given by LQG random surfaces decorated by independent CLEs.
\end{abstract}

\maketitle

\hypersetup{pageanchor=true}

\newpage

\section{Introduction}\label{intro}

The enumeration of planar random maps, which are models for discretised surfaces, developed initially from the work of Tutte \cite{Tutte1,Tutte2,TutteQ}. The discovery of matrix model techniques \cite{BIPZ} and the development of bijective techniques based on coding by decorated trees \cite{CoriV,Schae} led in the past thirty years to a wealth of results. An important motivation comes from the physics conjecture that the geometry of large random maps is universal, \textit{i.e.}, there should exist ensembles of random metric spaces depending on a small set of data (like the central charge and a symmetry group attached to the problem) which describe the continuum limit of random maps. Two-dimensional quantum gravity aims at the description of these random continuum objects and physical processes on them, and the universal theory which should underlie it is \emph{Liouville quantum gravity}, possibly coupled to a conformal field theory \cite{MR623209,KPZ,Ginsparg-Moore,Ginsparg}. Understanding rigorously the emergent fractal geometry of such limit objects is nowadays a major problem in mathematical physics and in probability theory. Another important problem is to establish the convergence of discrete random planar maps towards such limit objects. Solving various problems of map enumeration is often instrumental in this program, as it provides useful probabilistic estimates.

As of now, the geometry of large random planar maps with faces of bounded degrees (\textit{e.g.}, quadrangulations) is fairly well understood, thanks to recent spectacular progress. In particular, their scaling limit is the so  called \emph{Brownian map} \cite{MarckertMokkadem,GallMap,MierMapP,Gall1,map_making}, with its convergence in the Gromov-Hausdorff sense  established by Le Gall and Miermont in Refs. \cite{Gall1,MierMapP}. Another major progress is the recent construction by Miller and Sheffield, via the so  called quantum Loewner evolution \cite{millersheffield2016}, of a \emph{metric structure} for Liouville  quantum gravity  (at Liouville parameter $\gamma=\sqrt{\frac{8}{3}}$), and the proof that it is indeed equivalent to that of the Brownian planar map \cite{qlebm,2016arXiv160503563M,2016arXiv160805391M,2017arXiv171201571M}.

This universality class is often called in physics that of  \emph{pure gravity}. Recent progress generalised part of this understanding to other universality classes, those of planar maps containing faces whose degrees are drawn from a heavy tail distribution. In particular, the limiting object is the so-called \emph{$\alpha$-stable map}, which can be coded in terms of stable processes whose parameter $\alpha$ is related to the power law decay of the degree distribution and to the Hausdorff dimension of the random map \cite{MierGall,BBG12a,BBG12b}.
 
The next class of interesting models concerns random maps equipped with a statistical physics model, like the Ising model \cite{1986PhLA..119..140K,BouKa}, percolation \cite{Kazakov}, the $O(n)$ model \cite{1988PhRvL..61.1433D,GaudinKostov,KOn,1990NuPhB.340..491D,KSOn,1992NuPhB.386..558E,BEThese,EKOn,EKOn2,BBG12a,BBG12b}, the $Q$-Potts model \cite{Daul,BonnetEynard,PZinn,BBG12c}, or non intersecting random walks \cite{PhysRevLett.61.2514,PhysRevLett.81.5489}. The $O(n)$ model admits a famous representation in terms of loops \cite{DoMuNiSc81,NienhuisCG} with $n$ being the fugacity per loop. 
It is also well  known, at least on fixed lattices \cite{FKcluster,BaKeWu,Truong,PerkWu,1983PhRvB..27.1674D,NienhuisCG,StatPhys1987}, that the critical $Q$-state Potts model, via its Fortuin--Kasteleyn (FK) cluster representation, can be reformulated as a fully packed loop model with a fugacity $\sqrt{Q}$ per loop; for planar random maps this equivalence is explained in detail in \cite{BBG12c,sheffield2016bis}. The interesting feature of the $O(n)$ or Potts models is that they give rise to universality classes which depend continuously on $n$ or $Q$, as can be detected at the level of critical exponents \cite{1983PhRvB..27.1674D,1982PhRvL..49.1062N,1984JSP....34..731N,NienhuisCG,0305-4470-20-2-031,PhysRevLett.57.941,0305-4470-19-13-009,PhysRevLett.57.3179,PhysRevLett.58.2325,PhysRevLett.83.1359,1988PhRvL..61.1433D,GaudinKostov,KOn,1990NuPhB.340..491D,KSOn,PagesjaunesOn}. The famous KPZ relations \cite{KPZ} (see also \cite{MR981529,MR1005268}) relate the critical exponents of these models on a fixed regular lattice, with the corresponding critical exponents on random planar maps, as was repeatedly checked for a series of models \cite{KPZ,1986PhLA..119..140K,1988PhRvL..61.1433D,1990NuPhB.340..491D,KOn,MR2112128}. In the framework of Liouville quantum gravity, the KPZ relations have now been mathematically proven for the Liouville measure defined  as the (renormalised) exponential of the Gaussian free field (GFF) times a parameter $\gamma\in [0,2]$ \cite{2008arXiv0808.1560D}, as well as in the context of Mandelbrot multiplicative cascades \cite{2009CMaPh.tmp...46B,Barral2014} and Gaussian multiplicative chaos  \cite{PSS:8474530,MR3215583,Barral2013}.

 It is widely believed that after a Riemann conformal map to a given planar domain, the correct conformal structure for the continuum limit of random planar maps  weighted by the partition function of a critical statistical model is described by the theory of Liouville quantum gravity (LQG), coupled to the conformal field theory (CFT) representing the conformally invariant model at its critical point (see, {\it e.g.}, the reviews \cite{Ginsparg-Moore,Ginsparg,MR2073993} and  \cite{DuplantierICM2014,LeGallICM2014}). In a more probabilistic setting, one expects  the continuum limit after conformal embedding to be some form of Liouville random surface decorated by Schramm--Loewner evolution (SLE) paths.
 
There are now several senses in which random planar maps with statistical models have been rigorously proved to converge to LQG surfaces, as path-decorated metric spaces in the self-avoiding walk and percolation models cases \cite{2016arXiv160800956G,2017arXiv170105175G,2017arXiv171201571M}, as mated pairs of trees \cite{sheffield2016bis,2015arXiv151104068K,2016arXiv160301194G,2016arXiv160309722G,LiSunWatson}, or as Tutte discrete embedding of so-called mated-CRT maps \cite{2017arXiv170511161G}, using results for the continuum mating of continuum random trees (CRT) \cite{LQGmating,quantum_spheres}. This approach was recently extended to graph distances \cite{2017arXiv171100723G} and random walk  \cite{2017arXiv171100836G} on  random planar maps.
  
The first instance was the proof by Sheffield \cite{sheffield2016bis} in the infinite volume case of the convergence of quadrangulations equipped with the FK clusters of a critical Potts model to LQG decorated by SLE, while the finite/sphere case was recently studied in   \cite{2015arXiv150200546G,gwynnesun2017,2015arXiv151006346G}. The convergence is here in the so-called {\it peanosphere topology}, obtained from the mating of trees approach \cite{LQGmating,quantum_spheres} (see also \cite{gwynne2017}).

In the case of the $O(n)$ model, the configuration of critical  loops after the Riemann conformal mapping is expected to be described in the continuous limit by the  \emph{conformal loop ensemble}  ${\rm CLE}_{\kappa}$ \cite{sheffield2009,zbMATH06121652}.  It depends on the continuous index $\kappa\in(\frac{8}{3},8)$ of the associated Schramm--Loewner evolution (SLE$_\kappa$),  with the correspondence
$$
n = 2\cos \pi\big(1 - \tfrac{4}{\kappa}\big)
$$
for $n\in (0,2]$  \cite{MR1964687,kg2004guide_to_sle,sheffield2016bis,MR2112128}. In Liouville quantum gravity, the ${\rm CLE}_{\kappa}$ is coupled to an independent GFF, which both govern the random measure with Liouville parameter $\gamma =\min\big(\sqrt{\kappa},\frac{4}{{\sqrt{\kappa}}}\big)$, and the  conformal welding of SLE$_\kappa$ curves according to the LQG-boundary measure \cite{sheffield2016,LQGmating,quantum_spheres,2011PhRvL.107m1305D,DuplantierICM2014}; see also \cite{Astala2011random}).

Yet, except for the pure gravity $n=0$, $\gamma^2=\frac{8}{3}$ case, little is known on the \emph{metric} properties of large random maps weighted by an $O(n)$ model, even from a physical point of view. In this work, we shall rigorously investigate the \emph{nesting} properties of loops in those maps. From the point of view of 2d quantum gravity, it is a necessary, albeit perhaps modest, step towards a more complete understanding of the geometry of these large random maps. For instance, one should first determine the typical `depth' ({\it i.e.}, the number of loops crossed) on a random map before trying to determine how deep geodesics are penetrating the nested loop configuration. While this last question seems at present to be out of reach, its answer is expected to be related to the value of the almost sure Hausdorff dimension of large random maps with an $O(n)$ model, a question which is under active debate (see, {\it e.g.}, Refs. \cite{AMBJORN2013328,AMBJORN2014676,2016arXiv161009998D,Dup1,2017arXiv171100723G}).

An early study of the depth via a transfer matrix approach can be found in the work by Kostov \cite{KostovS,KostovH}. Our approach is based on analytic combinatorics, and mainly relies on the substitution approach developed in \cite{BBG12a,BBG12b}, and uses transfer matrices as an intermediate step. For instance, we compute generating series of cylinders (planar maps with two boundary faces) weighted by $s^{P}$, where $P$ is the number of loops separating the two boundaries. This novel type of results  has a combinatorial interest \textit{per se}; we find that the new $s$ variable appears in a remarkably simple way in the generating series. While the present article is restricted to the case of planar maps, the tools that we present are applied in Ref. \cite{BGFhigher} to investigate the topology of nesting in maps of arbitrary genus, number of boundaries and marked points.
 
We also relate the asymptotics of our results in the critical scaling limit of large number of loops and large volume, to extreme nesting in  ${\rm CLE}_{\kappa}$ in a bounded planar domain in $\mathbb C$, as obtained by Miller, Watson and Wilson in Ref. \cite{MWW}, who built on earlier works \cite{JCR,Cardy2003,Dubedat,KW04,SSW}. The large deviations functions, obtained here for nesting on random planar maps, are rigorously shown to be  identical to some  transforms, in Liouville quantum gravity, of the Euclidean large deviations functions for  ${\rm CLE}_{\kappa}$ in the disk, as obtained in Ref. \cite{MWW}, which we also generalise to the Riemann sphere. These transforms represent subtle extensions of the KPZ relation. By matching continuous sets of critical exponents, {\it i.e., multifractal spectra}, our results strongly support the conjecture that CLE observed in Liouville quantum gravity describes the scaling limit of the loop ensemble on large maps carrying a critical $O(n)$ model.
\subsection*{Notations}
If $F$ and $G$ are non zero and depend on some parameter $\varepsilon \rightarrow 0$,
\begin{itemize}
\item $F \asymp G$ means that $\ln F \sim \ln G$;
\item $F \stackrel{.}{\asymp} G$ means that $F = e^{O(1)} G$; 
\item $F \stackrel{.}{\sim}  G$ means there exists $C > 0$ independent of $\varepsilon$ such that $F \sim CG$.
\end{itemize}
If $F$ is a formal series in some parameter $u$, $[u^{m}]\,F$ is the coefficient of $u^m$ in $F$.

\section{General definitions, reminders and main results}

\subsection{The \texorpdfstring{$O(n)$}{O(n)} loop model on random maps}

We start by reminding the definition of the model, following the
presentation of Refs. \cite{BBG12a,BBG12b}.

\subsubsection{Maps and loop configurations}

A \emph{map} is a finite connected graph (possibly with loops and
multiple edges) drawn on a closed orientable compact surface, in such
a way that the edges do not cross and that the connected components of
the complement of the graph (called \emph{faces}) are simply
connected. Maps differing by an orientation-preserving homeomorphism of their underlying
surfaces are identified, so that there are countably many maps. The map is \emph{planar} if the underlying surface is topologically a sphere.
The \emph{degree} of a vertex or a face is its number of incident edges (counted with multiplicity). To each map we may associate its \emph{dual map} which,
roughly speaking, is obtained by exchanging the roles of vertices and
faces. For $m \geq 1$, a \emph{map with $m$ boundaries} is a map with
$m$ distinguished faces, labeled from $1$ to $m$. By convention all the boundary faces are rooted,
that is to say for each boundary face $f$ we pick an oriented edge
(called a \emph{root}) having $f$ on its right.  The \emph{perimeter} of
a boundary is the degree of the corresponding face. Non boundary faces
are called \emph{inner} faces. A \emph{triangulation with $m$
  boundaries} (resp.\ a \emph{quadrangulation with $m$ boundaries}) is
a map with $m$ boundaries such that each inner face has degree $3$
(resp.\ $4$).

\begin{figure}[htpb]
  \centering
  \includegraphics[width=.7\textwidth]{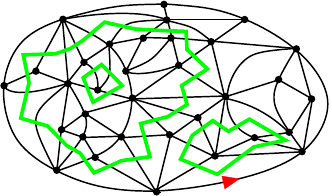}
  \caption{A planar triangulation with a boundary of perimeter $8$ (with
    root in red, the distinguished face being the outer face), endowed
    with a loop configuration (drawn in green).}
  \label{fig:myloopconfig_all}
\end{figure}

Given a map, a \emph{loop} is an undirected simple closed path on the
dual map (\textit{i.e.}, it covers edges and vertices of the dual map, and
hence visits faces and crosses edges of the original map). This is not
to be confused with the graph-theoretical notion of loop (an edge
incident twice to the same vertex), which plays no role here. A
\emph{loop configuration} is a collection of disjoint loops, and may
be viewed alternatively as a collection of \emph{crossed} edges such
that every face of the map is incident to either $0$ or $2$ crossed
edges. When considering maps with boundaries, we assume that the
boundary faces are not visited by loops. Finally, a
\emph{configuration} of the $O(n)$ loop model on random maps is a map
endowed with a loop configuration, see
Figure~\ref{fig:myloopconfig_all} for an example.

\begin{remark}
  In the original formulation in Refs. \cite{GaudinKostov,KOn,KSOn,EKOn,EKOn2},
  the loops cover vertices and edges the map itself. Our motivation
  for drawing them on the dual map is that it makes our combinatorial
  decompositions easier to visualise.
\end{remark}

\subsubsection{Statistical weights and partition functions}
\label{bendI}
Colloquially speaking, the $O(n)$ loop model is a statistical ensemble
of configurations in which $n$ plays the role of a fugacity per
loop. In addition to this ``nonlocal'' parameter, we need also some
``local'' parameters, controlling in particular the size of the maps
and of the loops. Precise instances of the model can be defined in
various ways.

The simplest instance is the \emph{$O(n)$ loop model on random
  triangulations} \cite{GaudinKostov,KOn,KSOn,EKOn,EKOn2}: here we require
the underlying map to be a triangulation, possibly with boundaries.
There are two local parameters $g$ and $h$, which are the weights per
inner face (triangle) which is, respectively, not visited and visited
by a loop. The Boltzmann weight attached to a configuration $C$ is
thus $w(C)=n^{\mathcal{L}} g^{T_1} h^{T_2}$, with $\mathcal{L}$ the number of loops of $C$,
$T_1$ its number of unvisited triangles and $T_2$ its number of
visited triangles.

\begin{figure}[htpb]
  \centering
  \includegraphics[width=.5\textwidth]{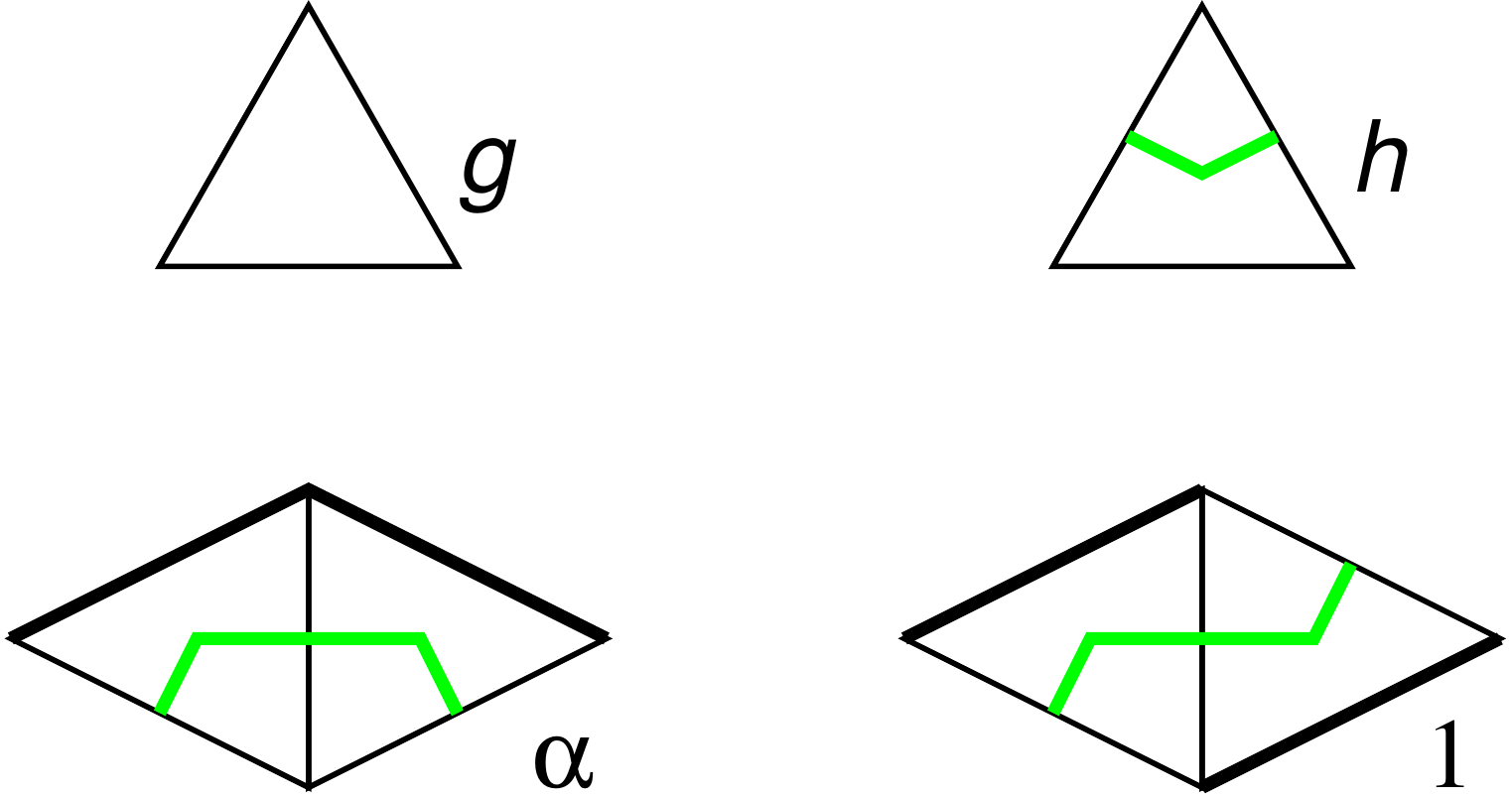}
  \caption{Top row: local weights for the $O(n)$ loop model on random
    triangulations. Bottom row: in the bending energy model, an extra
    weight $\alpha$ is attached to each segment of a loop between two
    successive turns in the same direction.}
  \label{fig:curv}
\end{figure}

A slight generalisation of this model is the \emph{bending energy
  model} \cite{BBG12b}, where we incorporate in the Boltzmann weight
$w(C)$ an extra factor $\alpha^B$, where $B$ is the number of pairs of
successive loop turns in the same direction, see
Figure~\ref{fig:curv}. Another variant is the $O(n)$ loop model on
random quadrangulations considered in \cite{BBG12a} (and its ``rigid''
specialisation). Finally, a fairly general model encompassing all the
above, and amenable to a combinatorial decomposition, is described in
\cite[Section 2.2]{BBG12b}.
We now define the partition function.  Fixing an integer $m \geq 1$, we consider the ensemble of allowed configurations where the underlying map is planar and has $m$ boundaries of respective perimeters $\ell_1,\ell_2,\ldots,\ell_m \geq 1$ (called perimeters). We will mainly be interested in $m = 1$ (\emph{disks}) and $m = 2$ (\emph{cylinders}). The corresponding partition function is then the sum of the Boltzmann weights $w(C)$ of all such
configurations. We find convenient to add an auxiliary weight $u$ per
vertex, and define the partition function as
\beq
  \label{eq:Fdef}
  F^{(m)}_{\ell_1,\ldots,\ell_m} = \delta_{m,1}\delta_{\ell_1,0}\,u + \sum_C u^{|V(C)|} w(C),
\eeq
where the sum runs over all desired configurations $C$, and $|V(C)|$
denotes the number of vertices of the underlying map of $C$, also called \emph{volume}.  By convention, the partition function for $m = 1$ includes an extra term $\delta_{\ell_1,0}\,u$, which means that we consider the map consisting of a single vertex on a sphere to be a planar map with one boundary of perimeter zero.
We also 
introduce the shorthand notation
\beq
  \label{eq:Fshort}
  F_\ell \equiv F^{(1)}_\ell.
\eeq

\subsection{Phase diagram and critical points}

When we choose the parameters to be real positive numbers such that the
sum \eqref{eq:Fdef} converges, we say that the model is
\emph{well defined} (it induces a probability distribution over the
set of configurations). Under mild assumptions on the model (\textit{e.g.}, the face degrees are bounded), this is the case for $u$ small, and there exists a
critical value $u_c$ above which the model ceases to be well defined:
\beq
  \label{eq:ucdef}
  u_c = \sup \{u \geq 0\,\,:\quad F^{(m)}_{\ell_1,\ldots,\ell_m} < \infty \}.
\eeq
It is not difficult to check that $u_c$ does not depend on $m$ and
$\ell_1,\ell_2,\ldots,\ell_m \geq 1$. If $u_c=1$ (resp.\ $u_c>1$, $u_c<1$), we say that the model is at a
\emph{critical} (resp.\ \emph{subcritical}, \emph{supercritical})
point.

\begin{figure}[htpb]
  \centering
  \includegraphics[width=.5\textwidth]{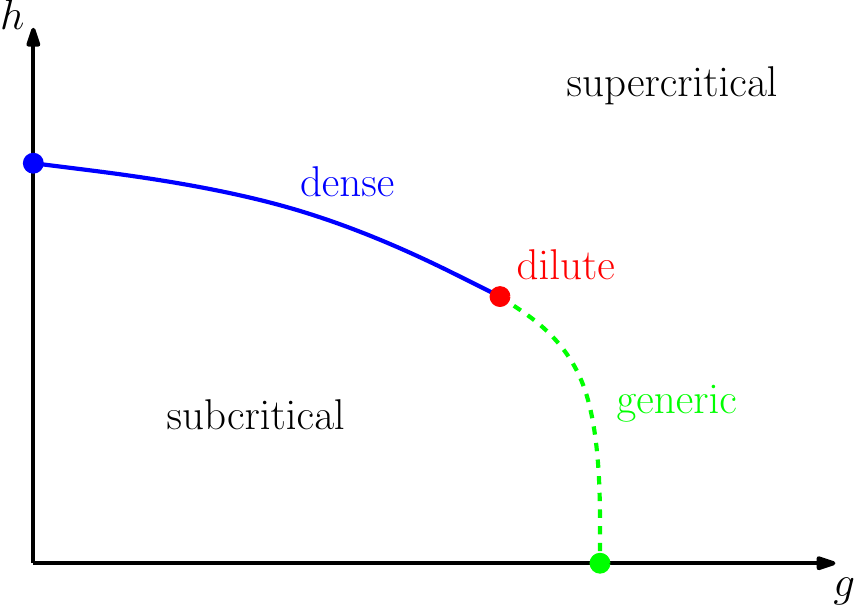}
  \caption{Qualitative phase diagram of $O(n)$ loop model on random
    maps: there is a line of critical points separating the
    subcritical and supercritical phase. Critical points may be in
    three different universality classes: generic, dilute and dense.}
  \label{fig:qualiphasediag}
\end{figure}

At a critical point, the partition function has a singularity at
$u=1$, and the nature (universality class) of this singularity is
characterised by some critical exponents, to be discussed below.  For
$n \in (0,2)$, three different universality classes of critical points
may be obtained in the $O(n)$ loop model on random triangulations \cite{KOn}, which we call \emph{generic}, \emph{dilute} and
\emph{dense}.

The generic universality class is that of ``pure gravity'', also obtained in models of maps without loops.
The location of these points in the $(g,h)$
plane forms the phase diagram of the model, displayed qualitatively on
Figure~\ref{fig:qualiphasediag}, and established in \cite{BBG12b} --- see also the earlier works \cite{KOn,GaudinKostov,KSOn,EKOn,EKOn2}. For the bending energy model, the phase diagram is similar for $\alpha$ not too large, but as $\alpha$ grows the
line of non generic critical points shrinks and vanishes eventually  \cite[Section 5.5]{BBG12c}. The same universality classes, and a similar phase diagram, is also obtained for the rigid $O(n)$ loop model on
quadrangulations \cite{BBG12a}, and is expected for more general loop models, where $g$ and $h$ should be thought as a
fugacity per empty and visited faces, respectively.

\subsection{Critical exponents}\label{Sec:ce}

We now discuss some exponents that characterise the different
universality classes of critical points of the $O(n)$ loop model. Some
of them are well known while others are introduced here for the
purposes of the study of nesting (for completeness all definitions are
given below). In the case of the dilute and dense universality
classes, the known exponents are rational functions of the parameter:
\beq
  \label{eq:bnot}
  b = \frac{1}{\pi} \arccos \left(\frac{n}{2}\right),
\eeq
which decreases from $\frac{1}{2}$ to $0$ as $n$ increases from $0$ to $2$. Let us mention that $b$ is closely related to the so-called coupling constant $\mathfrak{g}$ appearing in the Coulomb gas description of the $O(n)$ model on regular lattices, the relation being $\mathfrak{g}=1+b$ (dilute) or $\mathfrak{g}=1-b$ (dense).

\begin{table}[t]
  \bgroup
  
\begin{center}
  \def\arraystretch{1.5}
\begin{tabular}{|c||c|c|c|c|c|c|c|c|c|}
\hline
exponent & subcrit. & generic & dilute & dense   & $n = 0$ & Perc. & Ising & 3-Potts & KT  \\
\hline
 $b$        &                  &              &       $b$  &     $b$      & $\frac{1}{2}$    & $\frac{1}{3}$ & $\frac{1}{4}$ &  $\frac{1}{6}$  & 0   \\
 \hline
$\gamma_{{\rm str}}$ & & $-\frac{1}{2}$ & $-b$ & $-\frac{b}{1 - b}$  & $-1$ & $-\frac{1}{2}$ & $-\frac{1}{3}$ & $-\frac{1}{5}$ & $0$ \\
\hline
$\mathfrak{c}$ & & $0$ & $1 - \frac{6b^2}{1 + b}$ & $1 - \frac{6b^2}{1 - b}$  & $-2$ & $0$ & $\frac{1}{2}$ & $\frac{4}{5}$ & $1$ \\
         \hline
         $\kappa$ & & & $\frac{4}{1 + b}$ & $\frac{4}{1 - b}$  & $8$ & $6$ & $\frac{16}{3}$ & $\frac{24}{5}$ & $4$ \\
\hline
$c$ & & & $1$ & $\frac{1}{1 - b}$  & $2$ & $\frac{3}{2}$ & $\frac{4}{3}$ & $\frac{6}{5}$ & $1$ \\
\hline
$a$ & $\frac{3}{2}$ &  $\frac{5}{2}$ & $2 + b$ & $2 - b$  & $\frac{3}{2}$ & $\frac{5}{3}$ & $\frac{7}{4}$ & $\frac{11}{6}$ & $2$ \\
\hline
$d_{H}$ & $2$ & $4$ & $\bullet$ & $\bullet$  & $\bullet$ & $4$ & $\bullet$ & $\bullet$ & $\bullet$ \\ 
\hline 
$d^{{\rm gasket}}_{H}$ & $2$ & $4$ & $3 + 2b$ & $3 - 2b$  & $2$ & $\frac{7}{3}$ & $\frac{5}{2}$ & $\frac{8}{3}$ & $3$ \\
\hline
$\nu$ & $0$ & $0$ & $\frac{1}{2} - b$ & $\frac{1 - 2b}{2(1 - b)}$  & $0$ & $\frac{1}{4}$ & $\frac{1}{3}$ & $\frac{4}{10}$ & $\frac{1}{2}$ \\
\hline
 \end{tabular}
\end{center}
\egroup

\vspace{0.2cm}
\caption{\label{Fig:crit} Summary of the critical exponents for
    the $O(n)$ model on random maps as functions of
    $b = \frac{1}{\pi} \arccos \left(\frac{n}{2}\right)$. Pure gravity corresponds to the $n=0$ model in the dilute phase, critical percolation to the $n=1$ model in the dense phase, the critical Ising model and its interfaces to both the $n=1$ model in the dilute phase (for spin clusters) and the $n=\sqrt{2}$ model in the dense phase (for FK clusters). The Kosterlitz--Thouless transition is that of the $n=2$ model where the dilute and dense exponents are identical. More generally, the critical $Q$-Potts model and its FK cluster boundaries correspond to the $O(n=\sqrt{Q})$ model in the dense phase.}
\label{tab:critexp}
\end{table}

Before entering into definitions, we summarise the exponents on Table~\ref{tab:critexp}.
An entry $\bullet$ indicates that the exponent is unknown. At the time of writing, there is no consensus about the value of the Hausdorff dimension $d_{H}$ in the physics literature, although a so-called Watabiki formula has been proposed (see {\it e.g.}, \cite{Bowick,Dup1,AMBJORN2013328,AMBJORN2014676} and references therein) and critically analysed in view of recent mathematical results \cite{2016arXiv161009998D,2017arXiv171100723G}. All other exponents can be derived rigorously in the $O(n)$ model on triangulations, as well as the model with bending energy, and are expected to be universal. We actually reprove these results in the course of the article --- the only new statement concerns $\nu$ --- for the dense and dilute phases of the model with bending energy.

\subsubsection{Volume exponent}

The singularity of the partition
function in the vicinity of a critical point is captured in the so-called \emph{string
  susceptibility exponent} $\gamma_{\mathrm{str}}$:
\beq
  \label{eq:Fsing}
  \left. F_{\ell} \right|_{\mathrm{sing}} \stackrel{.}{\sim}  (1-u)^{1-\gamma_{\mathrm{str}}}, \qquad u \to 1,
\eeq
where $\ell$ is fixed, and $F_{\ell}|_{\mathrm{sing}}$ denotes the
leading singular part in the asymptotic expansion of
$F_{\ell}$ around $u=1$. As $u$ is coupled to the volume, the generating series of maps with fixed volume $V$ behaves as:
\begin{equation}
\label{gammastr} [u^V]\,\,F_{\ell} \stackrel{.}{\sim} V^{\gamma_{{\rm str}} - 2},\qquad V \rightarrow \infty.
\end{equation}
provided a delta-analyticity condition can be checked. In the context of the $O(n)$ loop model, $\gamma_{\mathrm{str}}$ may take the \emph{generic} value $-\frac{1}{2}$, already observed in models of maps without loops ($n=0$) ; the \emph{dilute} value $-b$ ; and  the \emph{dense} value $-\frac{b}{1-b}$. In all cases we consider, $\gamma_{{\rm str}}$ is comprised between $-1$ and $0$. Let us recall the celebrated KPZ relation \cite{KPZ} 
\begin{equation}
  \gamma_{{\rm str}} = \frac{\mathfrak{c}-1-\sqrt{(1-\mathfrak{c})(25-\mathfrak{c})}}{12},
\end{equation}
linking the string susceptibility exponent to the central charge $\mathfrak{c}$ of conformal field theory. For completeness, we also indicate in Table~\ref{tab:critexp} the value of the $\kappa$ parameter for the corresponding conformal loop ensemble (see Section~\ref{sec:clecompint}).

The parameter $c \in [1,2)$ defined by:
\begin{equation}
  \label{eq:cdef}
  c := -\frac{\gamma_{{\rm str}}}b
\end{equation}
will play an important role in this paper (note that it has nothing to do with $\mathfrak{c}$).

\subsubsection{Perimeter exponent}

Another exponent is obtained as we keep $u=1$ fixed but take
one boundary to be of large perimeter. Clearly, this requires
$F_{\ell}$ to be finite for all $\ell$, hence the model to be
either subcritical or critical, since $\gamma_{\mathrm{str}} \in (-1,0)$. We have the asymptotic behavior:
\beq
  \label{eq:Flasymp}
  F_\ell \stackrel{.}{\sim} \frac{\gamma_+^\ell}{\ell^a}, \qquad
  \ell \to \infty,
\eeq
where $\gamma_+$ is a non universal constant, and $a$ is a universal
exponent comprised between $\frac{3}{2}$ and $\frac{5}{2}$, which can take more
precisely four values for a given value of $n$: $\frac{3}{2}$ (subcritical
point), $\frac{5}{2}$ (generic critical point), $2+b$ (dilute critical point)
and $2-b$ (dense critical point).

\subsubsection{Gasket exponents}

\begin{figure}[h!]
\begin{center}
 \includegraphics[width=.6\textwidth]{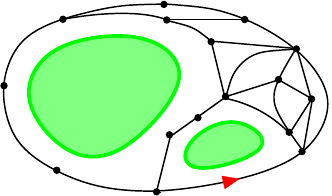}
\caption{\label{fig:myloopconfig_allgasket}The gasket of the map depicted in Figure~\ref{fig:myloopconfig_all}.}
\end{center}
\end{figure}

Consider a disk $\mathcal{D}$ with one boundary face and a loop
  configuration. The \emph{gasket} of $\mathcal{D}$ \cite{BBG12a} is
  the map formed by the vertices and edges which are exterior to all
  the loops, see Figure~\ref{fig:myloopconfig_allgasket}.

In Corollary~\ref{knowni}, we will combine known properties of the generating series of disks in the model with bending energy to show that the probability that a vertex chosen at random uniformly in a disk of volume $V$ and finite perimeter belongs to the gasket behaves as
\begin{equation}
\mathbb{P}\big[\bullet \in {\rm gasket}\,\big|\,V\big] \stackrel{.}{\sim} V^{-\nu},\qquad V \rightarrow \infty\,,
\end{equation}
with $\nu = c(\frac{1}{2} - b)$, modulo the check of a delta-analyticity condition.

Relying on the work of Le Gall and Miermont \cite{MierGall}, we showed in \cite{BBG12a} that the almost sure fractal dimension of the gasket when $V \rightarrow \infty$, denoted $d^{{\rm gasket}}_{H}$, is equal to $3 - 2b$ in the dense phase, $3 + 2b$ in the dilute phase. This exponent can also be extracted from Kostov \cite[Section 4.2]{KostovS} --- where $g$ is the Coulomb gas coupling constant $\mathfrak{g}$ mentioned above. This contrasts with the well known value $d_{H} = 4$ for the fractal dimension of disks at the generic critical point. We can only expect $d_{H} > d_{H}^{{\rm gasket}}$. Ref.~\cite{Dup1} relates it to the value of yet another critical exponent, which expresses how deep geodesics enter in the nested configuration of loops.

\subsection{Main results on random maps}
\label{sec:mainresults}
This paper is concerned with the statistical properties of nesting 
between loops. The situation is simpler in the planar case since every
loop is contractible, and divides the underlying surface into two
components. The nesting structure of large maps of arbitrary topology is analysed in the subsequent work \cite{BGFhigher}.

In the general $O(n)$ loop model, the generating series of disks and cylinders have been characterised in \cite{BBG12a,BBG12b,BEO13}, and explicitly computed in the model with bending energy in \cite{BBG12b}, building on the previous works \cite{EKOn,EKOn2,BEOn}. This characterisation is a linear functional relation which depends explicitly on $n$, accompanied by a nonlinear consistency relation depending implicitly on $n$. We remind the steps leading to this characterisation in Sections~\ref{FirstS}-\ref{Funcrel}. In particular, we review in Section~\ref{FirstS} the nested loop approach developed in \cite{BBG12a}, which allows enumerating maps with loop configurations in terms of generating series of usual maps. We then derive in Section~\ref{Funcrel} the functional relations for maps with loops as direct consequences of the well known functional relations for generating series of usual maps. The key to our results is the derivation in Section~\ref{Funcrelref} of an analogous characterisation for refined generating series of pointed disks (resp. cylinders), in which the loops which separate the origin (resp. the second boundary) and the (first) boundary face are counted with an extra weight $s$ each. We find that the characterisation of the generating series is only modified by replacing $n$ with $ns$ in the linear functional relation, while keeping $n$ in the consistency relation. Subsequently, in the model with bending energy, we can compute explicitly the refined generating series, in  Section~\ref{bend}. We analyse in Section~\ref{bendas} the behavior of those generating series at a non generic critical point which pertains to the $O(n)$ model. In the process, we rederive the phase diagram of the model with bending energy. More precisely, we perform an analysis of singularity in the canonical ensemble where the Boltzmann weight $u$ coupled to the volume tends to its critical value, which is equal to $1$ when suitably normalised. In order to convert it to large volume asymptotics, we establish in Appendices~\ref{sec:rigiddelta} and \ref{sec:anabend} a property of delta-analyticity of the generating series with respect to $u$, which partially relies on the explicit solution (see Theorem~\ref{theimdisk}) for the generating series of disks. One of our main result is then Theorem~\ref{mainT} in the text, restated below.

\begin{theorem}\label{theo:main}
\label{mainT0}Fix $(g,h,\alpha)$ and $n \in (0,2)$ such that the model with bending energy achieves a non generic critical point for the vertex weight $u = 1$. In the ensemble of random pointed disks of volume $V$ and perimeter $L$, the distribution of the number $P$ of separating loops between the marked point and the boundary face behaves when $V\rightarrow \infty$ as:
\begin{equation*}
\begin{split}
\mathbb{P}\Big[P = \big\lfloor \tfrac{c \ln V}{\pi}\,p\big\rfloor \,\Big|\,V\,,\,L = \ell\Big] &\,\stackrel{.}{\asymp}\,\, (\ln V)^{-\frac{1}{2}}\,V^{-\frac{c}{\pi}\,J(p)},  \\
\mathbb{P}\Big[P = \big\lfloor \tfrac{c \ln V}{2\pi}\,p \big\rfloor \Big|\,V\,,\,L = \lfloor V^{\frac{c}{2}}\ell\rfloor \Big] &\, \stackrel{.}{\asymp}\,\, (\ln V)^{-\frac{1}{2}}\,V^{-\frac{c}{2\pi}\,J(p)}, 
\end{split}
\end{equation*}
where
\[
J(p) = p\ln\left(\frac{2}{n}\,\frac{p}{\sqrt{1 + p^2}}\right) + {\rm arccot}(p) - {\rm arccos}\bigg(\frac{n}{2}\bigg).
\]
In the above estimates, $\ell$ and $p$ are bounded and bounded away from $0$ as  $V \rightarrow \infty$.
\end{theorem} 
We expect this result to be universal among all $O(n)$ loop models at
non generic critical points. The explicit, non universal finite
prefactors in those asymptotics are given in the more precise
Theorem~\ref{mainT}. We establish a similar result in
Section~\ref{CylS} and Theorem~\ref{mainTC} for the number of loops
separating the boundaries in cylinders.  Note that our derivation of
these theorems relies on the results of~\cite{BBG12b}, some of which
were justified using numerical evidence rather than formal
arguments. See Remark~\ref{rem:buddchen} below.

\begin{figure}
\begin{center}
\includegraphics[width=0.8\textwidth]{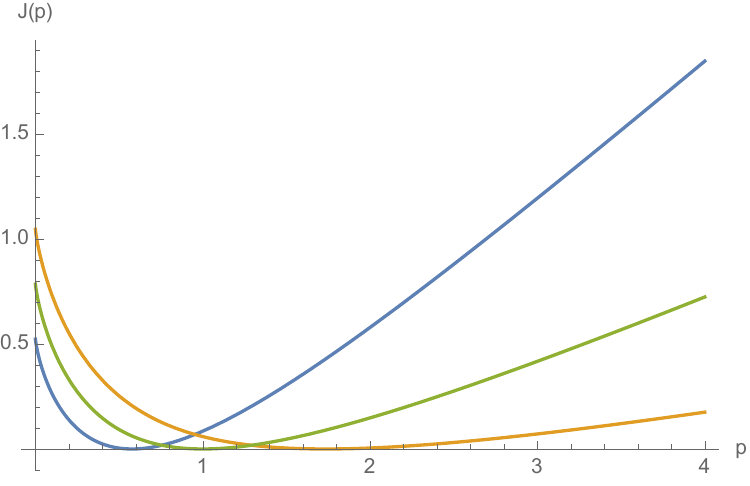}
\caption{\label{Jplotref}The function $J(p)$ for $n = 1$, $n = \sqrt{2}$ (Ising) and $n = \sqrt{3}$ (3-Potts). The larger is the value of $n$, the smaller is the slope when $p \rightarrow \infty$.}
\end{center}
\end{figure}

The large deviations function has the following properties (see Figure~\ref{Jplotref}):
\begin{itemize}
\item[$\bullet$] $J(p) \geq 0$ for positive $p$, and achieves its minimum value $0$ at $p_{{\rm opt}} = \frac{n}{\sqrt{4 - n^2}}$.
\item[$\bullet$] $J(p)$ is strictly convex, and $J''(p) = \frac{1}{p(p^2 + 1)}$.
\item[$\bullet$] $J(p)$ has a slope $\ln(2/n)$ when $p \rightarrow \infty$.
\item[$\bullet$] When $p \rightarrow 0$, we have $J(p) = {\rm arcsin}\big(\frac{n}{2}\big) + p\ln\big(\frac{2p}{n}\big) + O(p)$.
\end{itemize}
In Section \ref{DerivationCLT}, we prove a central limit theorem for fluctuations $P$ near its typical value. It is consistent with the Gaussian behavior of the large deviations function around its minimum $p_{{\rm opt}}$.
\begin{proposition}
\label{mainC} In pointed disks as above, the number of separating loops between the marked point and the boundary face behaves almost surely like $\frac{c p_{{\rm opt}}}{j\pi}\,\ln V$, with Gaussian fluctuations of order $O(\sqrt{\ln V})$:
\[
\frac{P - \frac{cp_{{\rm opt}}}{j\pi}\,\ln V}{\sqrt{\ln V}} \longrightarrow \mathcal{N}(0,\sigma^2),\qquad p_{{\rm opt}} = \frac{n}{\sqrt{4 - n^2}},\qquad \sigma^2 = \frac{2^{3 - j}nc}{\pi(4 - n^2)^{\frac{3}{2}}}.
\]
with $j = 1$ if we keep $L$ finite, or $j = 2$ if we scale $L = \lfloor \ell\,V^{\frac{c}{2}} \rfloor$ for a finite positive $\ell$.
\end{proposition}

Establishing the critical behavior of the generating series and the phase diagram requires analyzing special functions related to the Jacobi theta functions and elliptic functions in the trigonometric limit. The aformentioned variable $q$ is the elliptic nome. The lengthy computations with these special functions are postponed to Appendices to ease the reading. In Section~\ref{sec:weighting}, we generalise these results to a model where loops are weighted by independent, identically distributed random variables. Lastly, Section \ref{sec:CLE}, the content of which is briefly described below, uses a different perspective, and re-derives 
 the above results on random maps  from the Liouville quantum gravity approach. The latter is applied to similar earlier results obtained  in Ref. \cite{MWW} for a $\mathrm{CLE}_\kappa$ in the unit disk.

\subsection{Relation with other works}
\label{S2coment}

We now mention some closely related works, which appeared after the initial version of this paper was posted on the arXiv.

Chen, Curien and Maillard \cite{CCMtree} proposed an alternative study of the nesting, and proved the convergence of the nesting tree (see Section~\ref{Markm}) labeled by loop perimeters in rigid $O(n)$ loop model on random quadrangulations, to an explicit multiplicative cascade. This rigid $O(n)$ model is a variant of the one studied in the present article, for which an analogous explicit analysis can be carried out --- the seeds of the computation are in \cite{BBG12a} --- and lead to the same Theorem~\ref{theo:main} and Proposition~\ref{mainC}. Ref~\cite{CCMtree}  has proposed a heuristic argument confirming the result of Theorem~\ref{theo:main} from the properties of the offspring distribution of the cascade.

A detailed study of the rigid $O(n)$ loop model on random bipartite
maps was performed by Budd (with some input by Chen) in a series of
works. Budd's first observation~\cite{Buddtalk} was an unexpected
connection between planar maps and lattice walks on the slit plane. An
extension of his construction relates walks on $\mathbb{Z}^2$ with a
controlled winding angle around the origin to the rigid $O(n)$ loop
model. This led to new results~\cite{Buddwind} on the counting of
simple diagonal walks on $\mathbb{Z}^2$ with a prescribed winding
angle, hinging on the explicit diagonalisation of certain transfer
matrices acting on a $\ell^2$-space which are closely related to the
transfer matrices considered in the present article. Finally, the
paper~\cite{BuddOn} extends to loop-decorated maps the peeling process
of (undecorated) Boltzmann maps introduced in~\cite{Buddpeel}. This
approach brings many results:
\begin{itemize}
\item a formal justification of the phase diagram of the rigid model,
  see also~\cite[Chapter~II]{ChenThesis} and Remark~\ref{rem:buddchen}
  below,
\item a characterisation of the scaling limit of the perimeter
  process, which implies in turn the convergence of a certain rescaled
  first passage percolation distance,
\item exact and asymptotic results on the number of separating loops
  in a pointed rooted map, which are consistent with our own results
  (see Appendix~\ref{sec:rigid}), and also include the case $n=2$.
\end{itemize}

\subsection{Comparison with CLE properties} \label{sec:clecompint}
It is strongly believed that, if the random disks were embedded conformally to the unit disk $\mathbb D$, the loop configuration would be described in the thermodynamic $V \rightarrow \infty$ limit by the \emph{conformal loop ensemble} in presence of Liouville quantum gravity.   On a regular planar lattice, the critical $O(n)$-model is expected to converge in the continuum limit to the universality class of the $\mathrm{SLE}_\kappa$/$\mathrm{CLE}_\kappa$, for 
 \beq\label{nkappa0}
 n= 2\cos\big[\pi\big(1 - \tfrac{4}{\kappa}\big)\big]\qquad n\in (0,2] \qquad \begin{cases} \kappa\in (\tfrac{8}{3},4],\,\,\,{\rm dilute}\,\,{\rm phase}\\ \kappa\in [4,8),\,\,\,{\rm dense}\,\,{\rm phase} \end{cases} 
 \eeq
and the same is expected to hold at a non generic critical point in the dilute or dense phase on a random planar map. 
Although both conjectures are not yet mathematically fully established, we may try to relate the large deviations properties of nesting, as derived in the critical regime in the $O(n)$ loop model on a random planar map, to the corresponding nesting properties of ${\rm CLE}_{\kappa}$, in order to support both conjectures altogether.
 
Using the so-called Coulomb gas method for critical $O(n)$ and Potts models, Cardy and Ziff  provided the first prediction, numerically verified, for the expected number of loops surrounding a given point in a finite domain \cite{Cardy2003}. Elaborating on this work and on Refs. \cite{JCR,Dubedat,KW04,SSW}, Miller, Watson and Wilson  \cite{MWW} (see also \cite{MWW2}) were able to derive the almost sure multifractal dimension spectrum of \emph{extreme nesting} in the conformal loop ensemble. Let $\Gamma$ be a $\mathrm{CLE}_\kappa$ in $\mathbb D$. For each point $z\in \mathbb D$,  let $\mathcal N_z(\varepsilon)$ be the number of loops of $\Gamma$ which surround the ball $B(z,\varepsilon)$ centered at $z$ and of radius $\varepsilon >0$. For $\nu >0$, define 
 \begin{equation*} 
 \Phi_\nu=\Phi_\nu(\Gamma):=\left\{z\in D\,\,: \quad  \lim_{\varepsilon\to 0}\frac{\mathcal N_z(\varepsilon)}{\ln (1/ \varepsilon)}=\nu\right\}.
 \end{equation*} 
The almost-sure Hausdorff dimension of this set is given in terms of the distribution of  conformal radii of outermost loops in $\mathrm{CLE}_\kappa$. More precisely, let $\mathcal U$ be the connected component containing the origin in the complement $\mathbb D\setminus \mathcal L$  of the largest loop $\mathcal L$ of $\Gamma$ surrounding the origin in  $\mathbb D$, and ${\rm CR}(0,\mathcal{U})$ its conformal radius from $0$. The cumulant generating function of $T:=-\ln(\mathrm{CR}(0,\mathcal U))$ was computed independently in unpublished works \cite{JCR,Dubedat,KW04}, and rigorously confirmed in Ref. \cite{SSW}. It is given by
\beq\label{Lambda0}
\Lambda_\kappa(\lambda):=\ln {\mathbb E}\left[e^{\lambda T}\right]=\ln \left(\frac{-\cos(\tfrac{4\pi}{\kappa})}{\cos\left(\pi \sqrt{\left(1-\frac{4}{\kappa}\right)^2+\frac{8\lambda}{\kappa}}\right)}\right),
\eeq
for $\lambda \in (-\infty,1- \tfrac{2}{\kappa}- \tfrac{3\kappa}{32})$.  The symmetric Legendre--Fenchel transform, $\Lambda_\kappa^*:\mathbb R\rightarrow \mathbb R_+$ of $\Lambda_\kappa$ is defined by
\beq\label{Lambdastar0}
\Lambda^{\star}_\kappa(x):=\sup_{\lambda \in \mathbb R}\left(\lambda x-\Lambda_\kappa(\lambda)\right).
\eeq
The authors of \cite{MWW} then define 
\beq\label{gammanu0}
\gamma_\kappa(\nu):= \left\{\begin{array}{lll} \nu\Lambda^{\star}_\kappa(1/\nu), & \quad & \mathrm{if}\,\, \nu >0\\[3pt] 1-\frac{2}{\kappa}-\frac{3\kappa}{32} & \quad & \mathrm{if}\,\,\nu =0,
\end{array}\right.
\eeq 
which is right-continuous at $0$. Then, for $\kappa\in (\tfrac{8}{3},8)$, the Hausdorff dimension of the set $\Phi_\nu$ is almost surely \cite[Theorem 1.1]{MWW},
\begin{equation*}
\mathrm{dim}_{\mathcal H}\,\Phi_\nu =\max(0,2-\gamma_{\kappa}(\nu)).
\end{equation*}
As a Lemma for this result, the authors of Ref. \cite{MWW} estimate, for $\varepsilon \to 0$, the asymptotic nesting probability around point $z$,
\beq\label{Pnueps0} 
\mathbb P({\mathcal N}_z(\varepsilon) \approx \nu \ln(1/\varepsilon) \, |\,\varepsilon ) \asymp \varepsilon ^{\gamma_\kappa(\nu)},
\eeq
where the sign $\approx$ stands for a growth of the form $(\nu+o(1))  \ln(1/\varepsilon)$, and where $\asymp$ means an asymptotic equivalence of  logarithms.
In Section \ref{sec:CLE}, we consider the unit disk in Liouville quantum gravity (LQG), {\it i.e.}, we equip it with a random measure, formally written here as $\mu_\gamma=e^{\gamma h}\mathrm{d}^2z $, where $\gamma\in[0,2]$ and $h$ is an instance of a GFF on $\mathbb D$, $\mathrm{d}^2z $ being the Lebesgue measure. The random measure $\mu_{\gamma}$ is called the Liouville quantum gravity measure. We define accordingly  $\delta:=\int_{B(z,\varepsilon)}\mu_\gamma$ as the (random) quantum area of the ball $B(z,\varepsilon)$. In this setting, the KPZ formula, which relates a Euclidean conformal weight $x$ to its LQG counterpart $\Delta$ \cite{2008arXiv0808.1560D}, reads
\beq\label{KPZ0}
x=U_\gamma(\Delta):=\frac{\gamma^2}{4}\Delta^2+\left(1-\frac{\gamma^2}{4}\right)\Delta.
\eeq 

Studying extreme nesting in LQG then consists in looking for the distribution of loops of a $\mathrm{CLE}_\kappa$ around the same ball $B(z,\varepsilon)$, the latter being now \emph{conditioned} to have a given quantum measure $\delta$, and to measure this nesting in terms of the logarithmic variable $\ln(1/\delta)$, instead of $\ln(1/\varepsilon)$. We thus look for the probability,
\beq
\label{PnutPA0}
\mathbb P_{\mathcal Q}\left(\mathcal N_z\approx p \ln(1/\delta) \,|\,\delta\right),\,\,\,p\in \mathbb R_+,
\eeq 
which is the analogue of the left-hand side of \eqref{Pnueps0} in Liouville quantum gravity, and which we may call the \emph{quantum nesting probability}.

By taking into account the distribution of the Euclidean radius $\varepsilon$ for a given $\delta$ \cite{2009arXiv0901.0277D,2008arXiv0808.1560D}, we obtain two main results,  a first general one deriving via the KPZ relation the large deviations in nesting of a  ${\rm CLE}_{\kappa}$ in LQG from those in the Euclidean disk $\mathbb D$, as derived in Ref. \cite{MWW}, and a second one identifying these Liouville quantum gravity results to those obtained here for the critical $O(n)$ model on a random map. 
\begin{theorem}\label{theo:LambdaQ0}
In Liouville quantum gravity, 
the cumulant generating function $\Lambda_\kappa$ \eqref{Lambda0} with $\kappa\in(\tfrac{8}{3},8)$, is transformed into the quantum one, 
\beq
\label{LambdaG0}
\Lambda_\kappa^{\mathcal Q}:= \Lambda_\kappa\circ 2U_\gamma,
\eeq
where $\Lambda_\kappa$ is given by \eqref{Lambda0} and $U_\gamma$ is the KPZ function \eqref{KPZ0}, with $\gamma=\min\big(\sqrt{\kappa},\tfrac{4}{\sqrt{\kappa}}\big)$.    
The Legendre--Fenchel transform, $\Lambda^{{\mathcal Q} \star}_{\kappa}:\mathbb R\rightarrow \mathbb R_+$ of $\Lambda^{\mathcal Q}_\kappa$  is defined by
\begin{equation*} \Lambda^{{\mathcal Q} \star}_\kappa(x):=\sup_{\lambda \in \mathbb R}\left(\lambda x-\Lambda^{\mathcal Q}_\kappa(\lambda)\right).
\end{equation*} 
The quantum nesting distribution \eqref{PnutPA0} in the disk is then, when $\delta \rightarrow 0$, 
\begin{align*}
&\mathbb P_{\mathcal Q}\big(\mathcal N_z\approx p\ln(1/\delta)\,|\,\delta\big) \asymp \delta^{\Theta(p)},\\
&\Theta(p)= \left\{\begin{array}{lll} p\Lambda^{\mathcal Q\star}_\kappa\big(\tfrac{1}{p}\big), & \quad & \mathrm{if}\,\, p >0 \\[3pt] \tfrac{3}{4} - \tfrac{2}{\kappa} & & {\rm if}\,\,p = 0\,\,{\rm and}\,\,\kappa \in(\tfrac{8}{3},4] \\[3pt] \tfrac{1}{2} - \tfrac{\kappa}{16} & \quad & {\rm if}\,\,p = 0 \,\,{\rm and}\,\,\kappa \in[4,8).
\end{array}\right.
 \end{align*}
\end{theorem} 
\begin{corollary}\label{cor:LambdaQ}
The generating function associated with $\mathrm{CLE}_\kappa$ nesting in Liouville quantum gravity is explicitly given for $\kappa\in(\frac{8}{3},8)$ by  
\begin{align*}
&\Lambda^{\mathcal{Q}}_\kappa(\lambda)=\Lambda_\kappa\circ 2U_\gamma(\lambda)=\ln\left(\frac{\cos\big[\pi(1 - \frac{4}{\kappa})\big]}{\cos\Big[\pi \left(\frac{2\lambda}{c} +\left|1-\frac{4}{\kappa}\right|\right)\Big]}\right),\quad c=\max\{1,\tfrac{\kappa}{4}\},\\
\\
&\lambda\in \left[\tfrac{1}{2} - \tfrac{2}{\kappa},\tfrac{3}{4} - \tfrac{2}{\kappa}\right]\,\,\,\mathrm{for}\,\,\, \kappa \in\left(\tfrac{8}{3}, 4\right];\quad \,\,\, \lambda\in \left[\tfrac{1}{2} - \tfrac{\kappa}{8},\tfrac{1}{2} - \tfrac{\kappa}{16}\right]\,\,\,\mathrm{for}\,\,\, \kappa \in [4,8).
\end{align*}
\end{corollary}
\begin{remark}
$\Theta(p)$ is right-continuous at $p=0$. 
\end{remark}
\begin{remark}\label{rk:KPZ0}
Theorem \ref{theo:LambdaQ0} shows that the KPZ relation can directly act on an arbitrary continuum variable, here the conjugate variable in the cumulant generating function \eqref{Lambda0} for the $\mathrm{CLE}_\kappa$ log-conformal radius. This seems the first occurrence of such a role for the KPZ relation, which usually concerns scaling dimensions.
\end{remark}
\begin{remark}
As the derivation in Section~\ref{sec:CLE} will show, the map $\Lambda_\kappa \mapsto \Lambda^{\mathcal Q}_\kappa$ in \eqref{LambdaG0} to go from  Euclidean geometry to Liouville quantum gravity  is fairly general: the composition of $\Lambda$ by the KPZ function $U_\gamma$ would hold for any large deviations problem, the large deviations function being the Legendre--Fenchel transform of a certain generating function $\Lambda$. 
\end{remark}
 \begin{theorem} \label{theo:Jp'0} The quantum nesting probability of a $\mathrm{CLE}_\kappa$ in a proper simply connected domain $D \subsetneq \mathbb C$, for the number $\mathcal N_z$ of loops surrounding  a ball centered at $z$ and conditioned to have  a given Liouville quantum area $\delta$, has the large deviations form,
\begin{equation*}
\mathbb P_{\mathcal Q}\Big(\mathcal N_z\approx \frac{cp}{2\pi}\,\ln(1/\delta) \,\Big|\,\delta\Big)\asymp \delta^{\frac{c}{2\pi}J(p)},\qquad \delta \rightarrow 0,
\end{equation*}
where $c$ and $J$ are the same as in Theorem \ref{theo:main}.
\end{theorem}
A complementary result concerns the case of the Riemann sphere. The extreme nestings of CLE for this geometry is written in Theorem~\ref{ClassSphere} and seems to be new. After coupling to LQG, we obtain
\begin{theorem} \label{theo:sphere0} On the Riemann sphere $\widehat{\mathbb C}$, the large deviations function $\widehat \Theta$ which governs the quantum nesting probability,
\[\mathbb P^{\widehat {\mathbb C}}_{\mathcal Q}\big(\mathcal N \approx p \ln(1/\delta)\,|\,\delta\big)\asymp
\delta^{\widehat \Theta(p)},\qquad \delta \rightarrow 0,\]
is related to the similar function $\Theta$ for the disk topology, as obtained in Theorem \ref{theo:LambdaQ0}, by 
\[\widehat \Theta (2p)=2\Theta(p).\]
From Theorem \ref{theo:Jp'0}, we get explicitly,
\begin{equation*}
\mathbb P^{\widehat {\mathbb C}}_{\mathcal Q}\left(\mathcal N\approx \frac{cp}{\pi}\,\ln(1/\delta) \,\Big|\,\delta\right)\asymp \delta^{\frac{c}{\pi}J(p)},\qquad \delta\to 0,
\end{equation*}
where $c$ and $J$ are as in Theorem \eqref{theo:main}.
\end{theorem} 
\begin{remark}
The reader will have noticed the perfect matching of the LQG results for $\mathrm{CLE}_\kappa$ in Theorems \ref{theo:LambdaQ0}, \ref{theo:Jp'0} and \ref {theo:sphere0} with the main Theorem \ref{theo:main} for the $O(n)$ model on a random planar map, with the proviso that the first ones are local versions ({\it i.e.}, in the $\delta \to 0$ limit), while the latter one gives a global version ({\it i.e.}, in the $V \to \infty$ limit).
\end{remark}

\section{First combinatorial results on planar maps}
\label{FirstS}
\subsection{Reminder on the nested loop approach}

\label{sec:1} We remind that $F_{\ell}$ is the partition function for a loop model on a planar map with a boundary of perimeter $\ell$. The nested loop approach describes it in terms of the generating series $\mathcal{F}_{p} = \mathcal{F}_{p}(g_1,g_2,\ldots)$ of usual maps (\textit{i.e.}, without a loop configuration) which are planar, have a rooted boundary of perimeter $p$, and counted with a Boltzmann weight $g_k$ per inner face of degree $k$ ($k \geq 1$) and an auxiliary weight $u$ per vertex. To alleviate notations, the dependence on $u$ is left implicit in most expressions. By convention, we assume that boundaries are rooted. We then
have the fundamental relation \cite{BBG12b}
\beq
  \label{eq:Fnest}
  F_{\ell} = \mathcal{F}_{\ell}(G_1,G_2,\ldots),
\eeq
where the $G_k$'s satisfy the fixed point condition
\beq
  \label{eq:fixp}
  G_k = g_k + \sum_{\ell' \geq 0} A_{k,\ell'} \mathcal{F}_{\ell'}(G_1,G_2,\ldots) = g_k + \sum_{\ell' \geq 0} A_{k,\ell'}\,F_{\ell'},
\eeq
where $A_{k,\ell}$ is the generating series of sequences of faces visited by a loop, which are glued together so as to form an annulus, in which the outer boundary is rooted and has perimeter $k$, and the inner boundary is unrooted and has perimeter $\ell$. Compared to the notations of \cite{BBG12b}, we decide to include in $A_{k,\ell}$ the weight $n$ for the loop crossing all faces of the annulus. We call $G_k$ the renormalised face weights.

Throughout the text, unless it is specified in the paragraph headline that we are working with usual maps, the occurrence of $\mathcal{F}$ will always refer to $\mathcal{F}(G_1,G_2,\ldots)$.

\subsection{The nesting graphs}
\label{Markm}

In this paragraph, we introduce a notion of nesting graph attached to a configuration $C$ of the $O(n)$ model. Although this level of generality is not necessary for this article (see the discussion at the end of this paragraph), we include it to put our study in a broader context.

Let us cut the underlying surface
along every loop, which splits it into several connected components
$c_1,\ldots,c_N$. Let $T$ be the graph on the vertex set
$\{c_1,\ldots,c_N\}$ where there is an edge between $c_i$ and $c_j$ if
and only if they have a common boundary, \textit{i.e.}, they touch each other
along a loop (thus the edges of $T$ correspond to the loops of $C$).

\begin{figure}
\begin{center}
  \includegraphics[width=.8\textwidth]{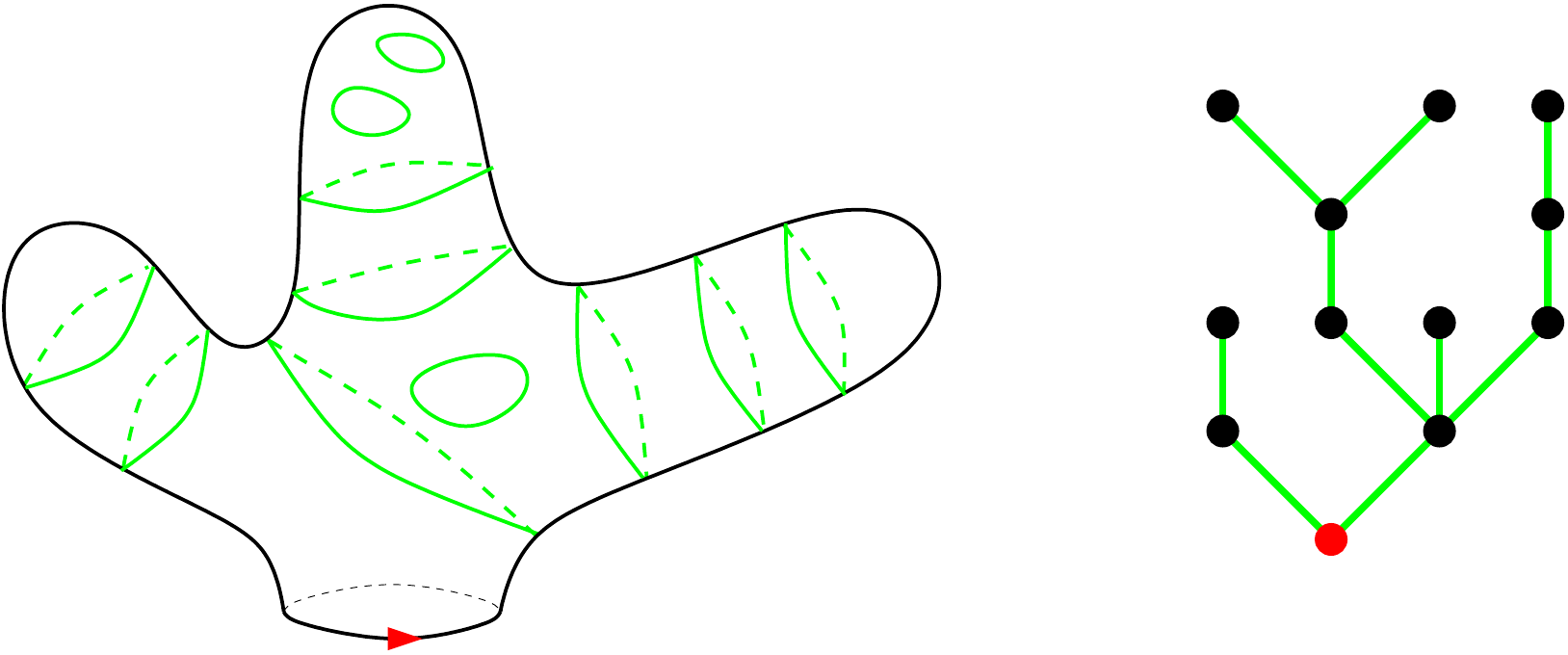}
  \caption{  \label{fig:nestingtree} Left: schematic representation of a loop configuration on a
    planar map with one boundary. Right: the associated nesting tree
    (the red vertex corresponds to the gasket).}
  \end{center}
\end{figure}

If the map is planar, $T$ is a tree called the \emph{nesting tree} of $C$, see
Figure~\ref{fig:nestingtree}. Each loop crosses a sequence of faces which form an annulus. This annulus has an outer and inner boundary, and we can record their perimeter on the half edges of $T$. As a result, $T$ is a rooted tree whose half edges carry non negative integers. If the map has a boundary face, we can root $T$ on the vertex corresponding to the connected component containing the boundary face. Then, for any vertex $v \in T$, there is a notion of parent vertex (the one incident to $v$ and closer to the root) and children vertices (all other incident vertices). We denote $\ell(v)$ the perimeter attached to the half-edge arriving to $v$ from the parent vertex. In this way, we can convert $T$ to a tree $T'$ where each vertex $v$ carries the non negative integer $\ell(v)$.

The nesting tree is closely related to the gasket decomposition introduced in \cite{BBG12a,BBG12b}. Consider the canonical ensemble of disks in the $O(n)$ model such that vertices receive a Boltzmann weight $u$, and the probability law it induces on the tree $T$'. The probability that a vertex $v$ with perimeter $\ell$ has $m$ children with perimeters $\{\ell_1,\ldots,\ell_m\}$ is:
\[
P_{\ell \rightarrow \ell_1,\ldots,\ell_m} = \frac{1}{m!}\,\frac{\sum_{k_1,\ldots,k_m \geq 0} \big[\prod_{i = 1}^m A_{k_i,\ell_i}F_{\ell_i}\big]\,\partial_{g_{k_1}}\cdots \partial_{g_{k_m}} \mathcal{F}_{\ell}(g_1,g_2,\ldots)}{F_{\ell}}.
\]
We see that $T'$ forms a Galton--Watson tree with infinitely many types. For the rigid $O(n)$ model on planar quadrangulation of a disk \cite{BBG12a}, the situation is a bit simpler as the inner and outer perimeters of the annuli carrying the loops coincide. We therefore obtain a random tree with one integer label for each vertex, whose convergence at criticality was studied in \cite{CCMtree} (see Section~\ref{S2coment}).

If one decides to consider a map $M$ with a given finite set of marked elements --- \textit{e.g.}, boundary faces or marked points ---, one can define the \emph{reduced nesting tree}  $(T_{{\rm red}},\mathbf{p})$ by:
\begin{itemize}
\item[$(i)$] For each mark in $M$, belonging to a connected component $c_i$, putting a mark on the  corresponding vertex of $T$ ;
\item[$(ii)$] erasing all vertices in $T$ which correspond to connected components which, in the complement of all loops and of the marked elements in $M$, are homeomorphic to disks ; this step should be iterated until all such vertices have disappeared ;
\item[$(iii)$] replacing any maximal simple path of the form $v_0 - v_1 - \cdots  - v_{p}$ with $p\geq 2$ where $(v_i)_{i = 1}^{p}$ represent connected components homeomorphic to cylinders, by a single edge
\[
v_0 \mathop{-}^{p} v_{p}
\]
carrying a length $p$. By convention, edges which are not obtained in this way carry a length $p = 1$.
\end{itemize}
The outcome is a tree, in which vertices may carry the marks that belonged to the corresponding connected components, and whose edges carry positive integers $\mathbf{p}$. By construction, given a finite set of marked elements, one can only obtain finitely many inequivalent $T_{{\rm red}}$.

In the subsequent article \cite{BGFhigher}, the first-named author and Garcia--Failde analyse the probability that a given topology of nesting tree is realised, conditioned on the lengths of the arms, as well as the generalisation to non simply connected maps. In the present article, we focus on the case of two marks: either a marked point and a boundary face, or two boundary faces. Then, the reduced nesting graph is either the graph with a single vertex (containing the two marked elements) and no edge, or the graph with two vertices (each of them containing a marked element) connected by an arm of length $P \geq 0$. Our goal consists in determining the distribution of $P$, which is the number of loops separating the two marked elements in the map (the pruning consisted in forgetting all information about the loops which were not separating). Yet, the tools we shall develop are important steps in the more general analysis of \cite{BGFhigher}.

\subsection{Maps with two boundaries}
\label{several}
We denote $F^{(2)}_{\ell_1,\ell_2}$ the partition function for a loop model on a random planar map with $2$ labeled boundaries of respective perimeters $\ell_1,\ell_2$, and similarly $\mathcal{F}^{(2)}_{\ell_1,\ell_2} \equiv \mathcal{F}^{(2)}_{\ell_1,\ell_2}(g_1,g_2,\ldots)$ for the partition function of usual maps. Such maps can be obtained from disks by marking an extra face and rooting it at an edge. At the level of partition functions, this amounts to: 
\beq
\label{eq:multi} \mathcal{F}^{(2)}_{\ell_1,\ell_2} =  \ell_2\frac{\partial}{\partial g_{\ell_2}} \mathcal{F}_{\ell_1}, \qquad F_{\ell_1,\ell_2}^{(2)} = \ell_2\frac{\partial}{\partial g_{\ell_2}}F_{\ell_1}.
\eeq
Differentiating the fixed point relation \eqref{eq:Fnest}, we can relate $F^{(2)}_{\ell_1,\ell_2}$ to partition functions of usual maps:
\beq
\label{eq:nestcyl} F^{(2)}_{\ell_1,\ell_2} = \mathcal{F}^{(2)}_{\ell_1,\ell_2} + \sum_{\substack{k \geq 1 \\ \ell \geq 0}} \mathcal{F}^{(2)}_{\ell_1,k}\,R_{k,\ell}\,F^{(2)}_{\ell,\ell_2},
\eeq
where we have introduced the generating series $R_{k,\ell} = A_{k,\ell}/k$, which now enumerate annuli whose outer and inner boundaries are both unrooted. In this equation, the evaluation of the generating series of usual maps at $G_k$ given by \eqref{eq:fixp} is implicit. 

\subsection{Separating loops and transfer matrix}
\label{sepa}
\begin{figure}
\begin{center}
\includegraphics[width=0.8\textwidth]{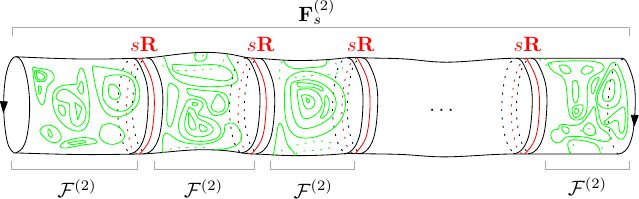}
\caption{\label{snake} Illustration of \eqref{eq:6}.}
\end{center}
\end{figure}

We say that a loop in a map $\mathcal{M}$ with $2$ boundaries is separating if after its removal, each connected component contains one boundary. The combinatorial interpretation of \eqref{eq:nestcyl} is transparent : the first term counts cylinders where no loop separates the two boundaries, while the second term counts cylinders with at least one separating loop (see Figure~\ref{snake}).

With this remark, we can address a refined enumeration problem. We denote by $F_{\ell_1,\ell_2}^{(2)}[s]$ the partition function of cylinders carrying a loop model, with an extra weight $s$ per loop separating the two boundaries. Obviously, the configurations without separating loops are enumerated by $\mathcal{F}^{(2)}_{\ell_1,\ell_2}$. If a configuration has at least one separating loop, let us cut along the first separating loop, and remove it. It decomposes the cylinder into : a cylinder without separating loops, that is adjacent to the first boundary ; the annulus that carried the first separating loop ; a cylinder with one separating loop less, which is adjacent to the second boundary. We therefore obtain the identity :
\beq
\label{eq:nestcyls} F^{(2)}_{\ell_1,\ell_2}[s] = \mathcal{F}_{\ell_1,\ell_2}^{(2)} + s\,\sum_{\substack{k \geq 1 \\ \ell \geq 0}} \mathcal{F}^{(2)}_{\ell_1,k}\,R_{k,\ell}\,F^{(2)}_{\ell,\ell_2}[s].
\eeq
We retrieve \eqref{eq:nestcyl} when $s = 1$, \textit{i.e.}, when separating and non separating loops have the same weight. We remind for the last time that $\mathcal{F}$'s should be evaluated at the renormalised face weights $G_k$.

Although it is not essential and will rarely be used in the body of this article, we point out that these relations can be rewritten concisely with matrix notations. Let $\mathbf{F}_{s}^{(2)}$ (resp. $\mathbf{R}$) be the semi-infinite matrices with entries $F^{(2)}_{\ell_1,\ell_2}[s]$ (resp. $R_{\ell_1,\ell_2}$) with row and column indices $\ell_1,\ell_2 \geq 0$, with the convention that $R_{0,\ell_2} = 0$. It allows the repackaging of \eqref{eq:nestcyls} as:
\beq
\label{eq:6}\mathbf{F}^{(2)}_{s} = \bs{\mathcal{F}}^{(2)} + s\,\bs{\mathcal{F}}^{(2)}\mathbf{R}\mathbf{F}^{(2)}_{s}.
\eeq
Therefore:
\beq
\label{eq:655}\mathbf{F}^{(2)}_{s} = \frac{1}{1 - s\,\bs{\mathcal{F}}^{(2)}\mathbf{R}}\,\bs{\mathcal{F}}^{(2)}.
\eeq
Then, $\bs{\Gamma}_{s} = (1 - s\,\bs{\mathcal{F}}^{(2)}\mathbf{R})^{-1}$ acts as a transfer matrix, where the inverse at least makes sense when $s$ is considered as a formal variable. Equations~\eqref{eq:6}-\eqref{eq:655} also appear in the early work of Kostov \cite{KostovS}.

\subsection{Pointed maps}
\label{pointed}
Remind that $u$ denotes the vertex weight. In general, a partition function $Z^\bullet$ of pointed maps can easily be obtained from the corresponding partition function $Z$ of maps :
\beq
Z^\bullet = u\frac{\partial}{\partial u}\,Z.
\eeq
We refer to the marked point as the \emph{origin} of the map. Let us apply this identity to disks with loops. We have to differentiate \eqref{eq:Fnest} and remember that the renormalised face weights depend implicitly on $u$:
\beq
F^\bullet_{\ell} = \mathcal{F}^\bullet_{\ell} + \sum_{\substack{k \geq 1 \\ \ell' \geq 1}} \mathcal{F}^{(2)}_{\ell,k}\,R_{k,\ell'}\,F^\bullet_{\ell'}.
\eeq
Obviously, the first term enumerates disks where the boundary and the origin are not separated by a loop.

Let us introduce a refined partition function $F^\bullet_{\ell}[s]$ that includes a Boltzmann weight $s$ per separating loop between the origin and the boundary. Cutting along the first (if any) separating loop starting from the boundary and repeating the argument of \S~\ref{sepa}, we find:
\beq
\label{eq:dotF}F^\bullet_{\ell}[s] = \mathcal{F}^\bullet_{\ell} + s\,\sum_{\substack{k \geq 1 \\ \ell' \geq 0}} \mathcal{F}^{(2)}_{\ell,k}\,R_{k,\ell'}\,F^\bullet_{\ell'}[s].
\eeq

If we introduce the semi-infinite line vectors $\mathbf{F}^\bullet_{s}$ (resp. $\bs{\mathcal{F}}^\bullet_{s}$) whose entries are $F^\bullet_{\ell}[s]$ (resp. $\mathcal{F}^\bullet_{\ell}[s]$) for $\ell \geq 0$, \eqref{eq:dotF} can be written in matrix form:
\beq
\label{recFs2}\mathbf{F}^\bullet_{s} = \bs{\mathcal{F}}^\bullet + s\,\bs{\mathcal{F}}^{(2)}\mathbf{R}\mathbf{F}^\bullet_{s}.
\eeq 
The solution reads:
\beq
\mathbf{F}^\bullet_{s} = \frac{1}{1 - s\,\bs{\mathcal{F}}^{(2)}\mathbf{R}}\,\bs{\mathcal{F}}^\bullet = \bs{\Gamma}_{s}\,\bs{\mathcal{F}}^\bullet,
\eeq
involving again the transfer matrix.

\section{Functional relations}
\label{Funcrel}
\subsection{More notations: boundary perimeters}

It is customary to introduce generating series for the perimeter of a boundary. Here, we will abandon the matrix notations of \S~\ref{sepa} unless explicitly mentioned, and rather introduce:
\beq
\label{eq:bfFdef}
\mathbf{F}(x) = \sum_{\ell \geq 0} \frac{F_{\ell}}{x^{\ell + 1}},\qquad \bs{\mathcal{F}}(x) = \sum_{\ell \geq 0} \frac{\mathcal{F}_{\ell}}{x^{\ell + 1}},
\eeq
which enumerate disks with loops (resp. usual disks) with a weight $x^{-(\ell + 1)}$ associated to a boundary of perimeter $\ell$, and similarly the generating series of pointed disks
\beq
\mathbf{F}^{\bullet}(x) = \sum_{\ell \geq 0} \frac{F_{\ell}^{\bullet}}{x^{\ell + 1}},\qquad \bs{\mathcal{F}}^{\bullet}(x) = \sum_{\ell \geq 0} \frac{\mathcal{F}_{\ell}^{\bullet}}{x^{\ell + 1}}
\eeq
and the generating series of pointed disks in which a weight $s^{P}$ is included when the boundary and the marked point are separated by $P$ loops:
\beq
\label{eq:bfFbsdef}
\mathbf{F}^{\bullet}_{s}(x) = \sum_{\ell \geq 0} \frac{F_{\ell}^{\bullet}[s]}{x^{\ell + 1}}
\eeq
Likewise, for the generating series of cylinders, we introduce:
\begin{equation}
\label{24}
\begin{split}
\mathbf{F}^{(2)}(x_1,x_2) & = \sum_{\ell_1,\ell_2 \geq 1} \frac{F^{(2)}_{\ell_1,\ell_2}}{x_1^{\ell_1 + 1}x_2^{\ell_2 + 1}}, \\
\mathbf{F}_{s}^{(2)}(x_1,x_2) & = \sum_{\ell_1,\ell_2 \geq 1} \frac{F_{\ell_1,\ell_2}^{(2)}[s]}{x_1^{\ell_1 + 1}x_2^{\ell_2 + 1}}, \\
\bs{\mathcal{F}}^{(2)}(x_1,x_2) & = \sum_{\ell_1,\ell_2 \geq 1} \frac{\mathcal{F}_{\ell_1,\ell_2}^{(2)}}{x_1^{\ell_1 + 1}x_2^{\ell_2 + 1}},
\end{split}
\end{equation}
etc. We will also find convenient to introduce generating series of annuli\footnote{Our definition for $\mathbf{A}$ differs by a factor of $n/x$ from the corresponding $A$ in \cite{BBG12b}.}:
\begin{equation}
\label{defR} 
\begin{split}
\mathbf{R}(x,z) & = \sum_{k + \ell \geq 1} R_{k,\ell}\,x^{k}\,z^{\ell}, \\
\mathbf{A}(x,z)  & = \sum_{\substack{k \geq 1 \\ \ell \geq 0}} A_{k,\ell}\,x^{k - 1}\,z^{\ell} = \partial_{x} \mathbf{R}(x,z).
\end{split}
\end{equation}

\subsection{Reminder on usual maps} \label{sec:remusualmaps}

The properties of the generating series of usual disks $\bs{\mathcal{F}}(x)$ have been extensively studied. We now review the results of \cite{BBG12b}. We say that a sequence of nonnegative weights $(u,g_1,g_2,\ldots)$ is admissible if for any $\ell \geq 0$, we have $\mathcal{F}^\bullet_{\ell} < \infty$ ; by extension, we say that a sequence of real-valued weights $(u,g_1,g_2,\ldots)_{k \geq 1}$ is admissible if $(|u|,|g_1|,|g_2|,\ldots)$ is admissible. Then, $\bs{\mathcal{F}}(x)$ satisfies the one-cut lemma and a functional relation coming from Tutte's combinatorial decomposition of rooted maps:
\begin{proposition}
\label{ponecut}If $(u,g_1,g_2,\ldots)$ is admissible, then the formal series $\bs{\mathcal{F}}(x)$ is the Laurent series expansion at $x = \infty$ of a holomorphic function in a maximal domain of the form $\mathbb{C}\setminus \gamma$, where $\gamma = [\gamma_-,\gamma_+]$ is a segment of the real line depending on the vertex and the face weights. Its endpoints are given by $\gamma_{\pm} = \mathfrak{s} \pm 2\sqrt{\mathfrak{r}}$ where $\mathfrak{r}$ and $\mathfrak{s}$ are the unique formal series in the variables $u$ and $(g_k)_{k \geq 1}$ such that:
\begin{equation}
\begin{split}
\label{ointg1}\oint_{\mathcal{C}(\gamma)} \frac{\dd x}{2{\rm i}\pi}\,\frac{\big(x - \sum_{k \geq 1} g_k\,x^{k - 1}\big)}{\sigma(x)} & = 0, \\
 -2u + \oint_{\mathcal{C}(\gamma)} \frac{\dd x}{2{\rm i}\pi}\,\frac{x\big(x - \sum_{k \geq 1} g_k\,x^{k - 1}\big)}{\sigma(x)} & = 0.
\end{split}
\end{equation}
where $\sigma(x) = \sqrt{x^2 - 2\mathfrak{s}x + \mathfrak{s}^2 - 4\mathfrak{r}}$ and $\mathcal{C}(\gamma)$ is a contour surrounding (and close enough to) $\gamma$ in the positive direction. Besides, the endpoints satisfy $|\gamma_-| \leq \gamma_+$, with equality iff $g_{k} = 0$ for all odd $k$'s.
\end{proposition}

\begin{remark} \label{rem:rs} The relations \eqref{ointg1} are equivalent to \cite[Equation 6.22]{BBG12b} after simple algebraic manipulations. In fact, $\mathfrak{r}$ and $\mathfrak{s}$ may be interpreted combinatorially as certain series of pointed rooted maps, see e.g.~\cite{BouGui}. In particular, we have
\begin{equation}
  \mathcal{F}_1^\bullet = u\,\mathfrak{s}, \qquad
  \mathcal{F}_2^\bullet = u\, (\mathfrak{s}^2 + 2\mathfrak{r}).
\end{equation}
\end{remark}

From now on, we shall use the same notation for the formal series and the holomorphic function.
\begin{proposition}
\label{prop2} $\bs{\mathcal{F}}(x)$ behaves like $\frac{u}{x} + O(\frac{1}{x^2})$ when $x \rightarrow \infty$, like $O\big(\sqrt{x - \gamma_{\pm}}\big)$ when $x \rightarrow \gamma_{\pm}$, and its boundary values on the cut satisfy the functional relation:
\beq
\label{eq:funcnu}\forall x \in \mathring{\gamma},\qquad \bs{\mathcal{F}}(x + {\rm i}0) + \bs{\mathcal{F}}(x - {\rm i}0) = x - \sum_{k \geq 1} g_k\,x^{k - 1}
\eeq
where $\mathring{\gamma} := (\gamma_-,\gamma_+)$. If $\gamma_-$ and $\gamma_+$ are given, there is a unique holomorphic function $\bs{\mathcal{F}}(x)$ on $\mathbb{C}\setminus\gamma$ satisfying these properties.
\end{proposition}

Although \eqref{eq:funcnu} arises as a consequence of Tutte's equation and analytical continuation, it has not received a direct combinatorial interpretation yet. 

With Proposition~\ref{ponecut} in hand, the analysis of Tutte's equation for generating series of maps with several boundaries, and their analytical continuation, has been performed in a more general setting in \cite{BEO13,Bstuff}. The outcome for usual cylinders (see also \cite{BEO13,Eynardbook}) is the following:

\begin{proposition}
\label{propF2nu} If $(g_k)_{k \geq 1}$ is admissible, the formal series $\bs{\mathcal{F}}^{(2)}(x,y)$ is the Laurent series expansion of a holomorphic function in $(\mathbb{C}\setminus\gamma)^2$ when $x,y \rightarrow \infty$, where $\gamma$ is as in Proposition~\ref{ponecut}. We have the functional relation, for $x \in \mathring{\gamma}$ and $y \in \mathbb{C}\setminus\gamma$:
\[
\bs{\mathcal{F}}^{(2)}(x + {\rm i}0,y) + \bs{\mathcal{F}}^{(2)}(x - {\rm i}0,y) = -\frac{1}{(x - y)^2}.
\]
It is subjected to the growth condition $\bs{\mathcal{F}}^{(2)}(x,y) \in O((x - \gamma_\pm)^{-\frac{1}{2}})$ when $x \rightarrow \gamma_+$ for fixed $y \in \mathbb{C}\setminus \gamma$, and a similar condition when $x$ and $y$ are exchanged.
\end{proposition}

\subsection{Reminder on maps with loops}

The relation \eqref{eq:Fnest} between disks with loops and usual disks allows carrying those results to the loop model.
We say that a sequence of face weights $(g_k)_{k \geq 1}$ and annuli weights $(A_{k,l})_{k,l \geq 0}$ is admissible if the sequence of renormalised face weights $(G_k)_{k \geq 1}$ given by \eqref{eq:fixp} is admissible as it is meant for usual maps. We say it is subcritical if the annuli generating series $\mathbf{A}(x,z)$ is holomorphic in a neighborhood of $\gamma \times \gamma$, where $\gamma$ is the segment determined by \eqref{ointg1} for the renormalised face weights. Being strictly admissible is equivalent to being admissible and not in the non generic critical phase in the terminology of \cite{BBG12a}. In the remaining of Section~\ref{Funcrel} and \ref{bend}, \emph{we always assume strict admissibility}.
 
In particular, $\mathbf{F}(x)$ satisfies the one-cut property (Proposition~\ref{ponecut}) on this segment $\gamma$, which now depends on face weights $(g_k)_{k}$ and annuli weights $(A_{k,l})_{k,l}$. And, its boundary values on the cut satisfy the functional relation: 
\begin{proposition}
\label{funcF} For any $x \in \mathring{\gamma}$, 
\beq
\label{eq:funcF}\mathbf{F}(x + {\rm i}0) + \mathbf{F}(x - {\rm i}0) + \oint_{\mathcal{C}(\gamma)} \frac{\dd z}{2{\rm i}\pi}\,\mathbf{A}(x,z)\,\mathbf{F}(z) = x - \sum_{k \geq 1} g_k\,x^{k - 1}.
\eeq
\end{proposition}

With Proposition~\ref{ponecut} in hand, the analysis of Tutte's equation for the partition functions of maps having several boundaries in the loop model, and their analytical continuation, has been performed in \cite{BEO13,Bstuff}. In particular, one can derive a functional relation for $\mathbf{F}^{(2)}(x_1,x_2)$, which matches  the one formally obtained by marking a face in Proposition~\ref{funcF} while considering the contour $\mathcal{C}(\gamma)$ independent of the face weights.

\begin{proposition}
\label{prop4} The formal series $\mathbf{F}^{(2)}(x,y)$ is the Laurent series expansion of a holomorphic function in $(\mathbb{C}\setminus\gamma)^2$ when $x,y \rightarrow \infty$, with $\gamma$ as in Proposition~\ref{funcF}. Besides, it satisfies the functional relation, for $x \in \mathring{\gamma}$ and $y \in \mathbb{C}\setminus\gamma$:
\beq
\label{26} \mathbf{F}^{(2)}(x + {\rm i}0,y) + \mathbf{F}^{(2)}(x - {\rm i}0,y) + \oint_{\mathcal{C}(\gamma)} \frac{\dd z}{2{\rm i}\pi}\,\mathbf{A}(x,z)\,\mathbf{F}^{(2)}(z,y) = -\frac{1}{(x - y)^2}.
\eeq
It is subjected to the growth condition $\mathbf{F}^{(2)}(x,y) \in O\big((x - \gamma_{\pm})^{-\frac{1}{2}}\big)$ when $x \rightarrow \gamma_{\pm}$ and $y \in \mathbb{C}\setminus\gamma$, and a similar condition when $x$ and $y$ are exchanged.
\end{proposition}

By similar arguments for the differentiation of \eqref{eq:funcF} with respect to the vertex weight $u$, one can derive for the generating series of pointed rooted disks a linear functional equation. This equation is in fact homogeneous because the right-hand side in \eqref{eq:funcF} does not depend on $u$, which leads to

\begin{proposition}
\label{prop9}For any $x \in \mathring{\gamma}$,
\beq
\mathbf{F}^\bullet(x + {\rm i}0) + \mathbf{F}^\bullet(x - {\rm i}0) + \oint_{\mathcal{C}(\gamma)} \frac{\dd z}{2{\rm i}\pi}\,\mathbf{A}(x,z)\,\mathbf{F}^{\bullet}(z) = 0.
\eeq
It is subjected to the growth conditions $\mathbf{F}^\bullet(x) = \frac{u}{x} + O(\frac{1}{x^2})$ when $x \rightarrow \infty$ and $\mathbf{F}^\bullet(x) \in O\big((x - \gamma_{\pm})^{-\frac{1}{2}}\big)$ when $x \rightarrow \gamma_{\pm}$.
\end{proposition}

\subsection{Separating loops}
\label{separ}
\label{Funcrelref}
The functional relations for the refined generating series (cylinders or pointed disks) including a weight $s$ per separating loop, are very similar to those of the unrefined case. 

\begin{proposition}
\label{prop76}At least for $|s| < 1$ and for $s = 1$, the formal series $\mathbf{F}_{s}^{(2)}(x,y)$ is the Laurent expansion of a holomorphic function in $(\mathbb{C}\setminus\gamma)^{2}$ when $x,y \rightarrow \infty$, and $\gamma$ is the segment already appearing in Proposition~\ref{funcF} and is independent of $s$. For any $x \in \mathring{\gamma}$ and $y \in \mathbb{C}\setminus\gamma$, we have:
\beq
\label{rrwt}\mathbf{F}_{s}^{(2)}(x + {\rm i}0,y) + \mathbf{F}^{(2)}_{s}(x - {\rm i}0,y) +  s \oint_{\mathcal{C}(\gamma)} \frac{\dd z}{2{\rm i}\pi}\,\mathbf{A}(x,z)\,\mathbf{F}^{(2)}_{s}(z,y) = - \frac{1}{(x - y)^2}.
\eeq
It is subjected to the growth condition $\mathbf{F}_s^{(2)}(x,y) \in O((x - \gamma_\pm)^{-\frac{1}{2}})$ when $x \rightarrow \gamma_+$ for fixed $y \in \mathbb{C}\setminus \gamma$, and a similar one when $x$ and $y$ are exchanged.
\end{proposition}

\begin{proposition}
\label{prop77}At least for $|s| < 1$ and for $s = 1$, the formal series $\mathbf{F}^\bullet_{s}(x)$ is the Laurent expansion of a holomorphic function in $(\mathbb{C}\setminus\gamma)$. It has the growth properties $\mathbf{F}^\bullet_{s}(x) = \frac{u}{x} + O(\frac{1}{x^2})$ when $x \rightarrow \infty$, and $\mathbf{F}^\bullet_{s}(x) \in O\big((x - \gamma_{\pm})^{-\frac{1}{2}}\big)$ when $x \rightarrow \gamma_{\pm}$. Besides, for any $x \in \mathring{\gamma}$, we have:
\beq
\mathbf{F}^\bullet_{s}(x + {\rm i}0) + \mathbf{F}^\bullet_{s}(x - {\rm i}0) + s \oint_{\mathcal{C}(\gamma)} \frac{\dd z}{2{\rm i}\pi}\,\mathbf{A}(x,z)\,\mathbf{F}^\bullet_{s}(z) = 0.
\eeq
\end{proposition}

\noindent\textbf{Proof.} Let us denote $\mathbf{F}^{(2)}_{[P]}$, the generating series of cylinders with exactly $P$ separating loops (discarding the power of $s$), and $\mathbf{F}^{(2)}_{[-1]} \equiv 0$ by convention. In particular
\beq
\label{seriesrepF2s}\mathbf{F}^{(2)}_{s} = \sum_{P \geq 0} \mathbf{F}^{(2)}_{[P]}\,s^{P}
\eeq
We first claim that for any $P \geq 0$, $\mathbf{F}^{(2)}_{[P]}(x,y)$ defines a holomorphic function in $(\mathbb{C}\setminus\gamma)^2$, and satisfies the functional relation: for any $x \in \mathring{\gamma}$ and $y \in \mathbb{C}\setminus\gamma$,
\begin{equation}
\label{theco}
\begin{split}
& \quad \mathbf{F}^{(2)}_{[P]}(x + {\rm i}0,y) + \mathbf{F}^{(2)}_{[P]}(x - {\rm i}0,y) \\
& = - \frac{\delta_{P,0}}{(x - y)^2} + \oint_{\mathcal{C}(\gamma)} \frac{\dd z_1}{2{\rm i}\pi}\,\bs{\mathcal{F}}^{(2)}(x,z_1)\,\oint_{\mathcal{C}(\gamma)} \frac{\dd z_2}{2{\rm i}\pi}\,\mathbf{A}(z_1,z_2)\,\mathbf{F}^{(2)}_{[P - 1]}(z_2,y) 
\end{split}
\end{equation}
The assumption of strict admissibility guarantees that $\mathbf{A}(\xi,\eta)$ --- and thus its $\xi$-anti\-de\-ri\-va\-ti\-ve $\mathbf{R}(\xi,\eta)$ --- is holomorphic in a neighborhood of $\gamma \times \gamma$, ensuring that the contour integrals in \eqref{theco} are well defined. Let us momentarily accept the claim.

Since $\mathbf{F}^{(2)}(x,y) = \mathbf{F}^{(2)}_{s = 1}(x,y)$, by dominated convergence we deduce that $\mathbf{F}^{(2)}_{s}(x,y)$ is an analytic function of $s$ --- uniformly for $x,y \in \mathbb{C}\setminus\gamma$ --- with radius of convergence at least $1$. Then, we can sum over $P \geq 0$ the functional relation \eqref{theco} multiplied by $s^{P}$ : the result is the announced \eqref{rrwt}, valid in the whole domain of analyticity of $\mathbf{F}^{(2)}_{s}$ as a function of $s$. Let $\mathbf{F}^{(2),||}(x,y)$ be the generating series of cylinders for face weights $(|g_k|)_{k}$ and annuli weight $(|A_{k,l}|)_{k,l}$. As the latter are strictly admissible by assumption, $\mathbf{F}^{(2),||}$ satisfies the growth condition in Proposition~\ref{prop4}. Since we have for $(s,x,y)$ in the aforementioned domain of analyticity the bound $|\mathbf{F}_{s}^{(2)}(x,y)| \leq \mathbf{F}^{(2),||}(|x|,|y|)$, we deduce that $\mathbf{F}_{s}^{(2)}(x,y)$ also satisfies the growth condition.

The claim is established by induction on $P$. Since $\mathbf{F}^{(2)}_{[0]} = \bs{\mathcal{F}}^{(2)}$, the claim follows by application of Proposition~\ref{propF2nu} for usual cylinders with renormalised face weights, \textit{i.e.}, vanishing annuli weights in the functional relation \eqref{26}. We however emphasise that the cut $\gamma$ is determined by Proposition~\ref{funcF}, thus depends on annuli weights via the renormalised face weights.

Assume the statement holds for some $P \geq 0$. We know from the combinatorial relation \eqref{eq:6} that:
\beq
\mathbf{F}^{(2)}_{[P + 1]} =  \bs{\mathcal{F}}^{(2)}\mathbf{R}\mathbf{F}^{(2)}_{[P]}
\eeq
with the matrix notations of \S~\ref{sepa}. The analytic properties of $\bs{\mathcal{F}}^{(2)}$ and of $\mathbf{F}^{(2)}_{[P]}$ --- as known from the induction hypothesis --- allows the rewriting:
\beq
\mathbf{F}^{(2)}_{[P + 1]}(x,y) = \oint_{\mathcal{C}(\gamma)} \frac{\dd z_1}{2{\rm i}\pi}\,\bs{\mathcal{F}}^{(2)}(x,z_1)\,\oint_{\mathcal{C}(\gamma)} \frac{\dd z_2}{2{\rm i}\pi}\,\mathbf{R}(z_1,z_2)\,\mathbf{F}^{(2)}_{[P]}(z_2,y).
\eeq
The expression on the right-hand side emphasises that the left-hand side, though initially defined as a formal Laurent series in $x$ and $y$, can actually be analytically continued to $(\mathbb{C}\setminus\gamma)^2$. Besides, for $x \in \mathring{\gamma}$ and $y \in \mathbb{C}\setminus\gamma$, we can compute the combination:
\begin{equation*}
\begin{split}
&\quad  \mathbf{F}^{(2)}_{[P + 1]}(x + {\rm i}0,y) + \mathbf{F}^{(2)}_{[P + 1]}(x - {\rm i}0,y)  \\
& =  \oint_{\mathcal{C}(\gamma)} \frac{\dd z_1}{2{\rm i}\pi}\big(\bs{\mathcal{F}}^{(2)}(x + {\rm i}0,z_1) +\bs{\mathcal{F}}^{(2)}(x - {\rm i}0,z_1)\big) \oint_{\mathcal{C}(\gamma)}\frac{\dd z_2}{2{\rm i}\pi}\,\mathbf{R}(z_1,z_2)\,\mathbf{F}^{(2)}_{[P]}(z_2,y) \\
& = -\oint_{\mathcal{C}(\gamma)} \frac{\dd z_1}{2{\rm i}\pi}\,\frac{1}{(x - z_1)^2}\,\oint_{\mathcal{C}(\gamma)} \frac{\dd z_2}{2{\rm i}\pi}\,\mathbf{R}(z_1,z_2)\,\mathbf{F}^{(2)}_{[P]}(z_2,y)  \\
& = -\oint_{\mathcal{C}(\gamma)} \frac{\dd z_2}{2{\rm i}\pi}\,\partial_{x} \mathbf{R}(x,z_2)\,\mathbf{F}^{(2)}_{[P]}(z_2,y).
\end{split}
\end{equation*}
and we recognise $\mathbf{A}(x,z_2) = \partial_{x}\mathbf{R}(x,z_2)$. Hence the statement is valid for $\mathbf{F}^{(2)}_{[P + 1]}$ and we conclude by induction. We thus have established the functional equation in Proposition~\ref{prop76}.

The proof of Proposition~\ref{prop77} is similar, except that we use $\mathbf{F}^\bullet_{[0]} = \bs{\mathcal{F}}^\bullet$ for initialisation, and later, the combinatorial relation \eqref{recFs2} instead of \eqref{eq:6}. \hfill $\Box$

\subsection{Depth of a vertex}

We now consider the depth $P$ of a vertex chosen at random in a disk configuration of the loop model. $P$ is by definition the number of loops that separate it from the boundary. This quantity gives an idea about how nested maps in the loop model are. Equivalently, $P$ is the depth of the origin in an ensemble of pointed disk configurations. We can study this ensemble in the microcanonical approach --- \textit{i.e.}, fixing the volume equal to $V$ and the perimeter equal to $L$ -- or in the canonical approach --- randomising the volume $V$ with a weight $u^{V}$ and the perimeter with a weight $x^{-(L + 1)}$.

In the canonical approach, the generating function of the depth distribution can be expressed in terms of the refined generating series of \S~\ref{pointed}:
\beq
\label{dodt}\mathbb{E}[s^{P}] = \frac{\mathbf{F}^\bullet_{s}(x)}{\mathbf{F}^\bullet(x)}.
\eeq

In the microcanonical approach,  the probability that, in an ensemble of pointed disks of volume $V$ and perimeter $L$, the depth takes the value $P$ reads:
\[
\mathbb{P}\big[P\,\big|\,V\,,\,L\big] = \frac{[u^{V}\cdot x^{-(L + 1)}\cdot s^{P}]\,\,\mathbf{F}_{s}^{\bullet}(x)}{[u^{V}\cdot x^{-(L + 1)}]\,\,\mathbf{F}^{\bullet}(x)}.
\]

\section{Computations in the loop model with bending energy}
\label{bend}
We shall focus on the class of loop models on triangulations with bending energy (see \S~\ref{bendI}) studied in \cite{BBG12b}, for which the computations can be explicitly carried out. The annuli generating series in this model are:
\beq
\label{rin}
\begin{split}
\mathbf{R}(x,z) & =  n\ln\Big(\frac{1}{1 - \alpha h(x + z) - (1 - \alpha^2)h^2xz}\Big), \\
\mathbf{A}(x,z) & = \frac{n}{\varsigma(z)-x} = n\Big(\frac{\varsigma'(x)}{z - \varsigma(x)} + \frac{\varsigma''(x)}{2\varsigma'(x)}\Big),
\end{split}
\eeq
where:
\beq
\label{varsigma}\varsigma(x) = \frac{1 - \alpha hx}{\alpha h + (1 - \alpha^2)h^2x}
\eeq
is a rational involution. In terms of the loop model, $h$ is the weight per triangle crossed by a loop, $\alpha$ is the bending energy, and we assume they are both non negative. Note that, for $\alpha = 1$, we have $\varsigma(x) = \frac{1}{h} - x$, so $\varsigma''(x) = 0$. In general:
\[
\frac{\varsigma''(x)}{2\varsigma'(x)} =  -\frac{1}{x + \frac{\alpha}{(1 - \alpha^2)h}} = -\frac{1}{x - \varsigma(\infty)}.
\]
If we assume $\varsigma(\gamma) \cap \gamma = \emptyset$ and $f$ is a holomorphic function in $\mathbb{C}\setminus\gamma$ such that $f(x) \sim c_{f}/x$ when $x \rightarrow \infty$, we can evaluate the contour integral:
\beq
\label{contueq}\oint_{\mathcal{C}(\gamma)} \frac{\dd z}{2{\rm i}\pi}\,\mathbf{A}(x,z)\,f(z) = -n\varsigma'(x)\,f(\varsigma(x)) + nc_{f}\,\frac{\varsigma''(x)}{2\varsigma'(x)}.
\eeq

\subsection{Preliminaries}
\label{elparam}

Technically, the fact that $\mathbf{A}(x,z)$ is a rational function with a single pole allows for an explicit solution of the model, and the loop model with bending energy provides a combinatorial realisation of such a situation. We review the solution of the functional equations for strictly admissible weights (see Section~\ref{Funcrel}), which amounts to requiring $\varsigma(\gamma) \cap \gamma = \emptyset$ or equivalently
\[
\gamma_+ < \gamma_+^* := \frac{1}{h(\alpha + 1)}
\]
The techniques to solve these functional equations have already been developed in \cite{BBG12b} slightly generalising \cite{EKOn2,TheseGB,BEOn}, and we refer to these works for more details. In the next Section~\ref{inequ}, we will study the non-generic critical weight by taking the limit $\gamma_+ \rightarrow \gamma_+^*$ in these solutions.

\begin{figure}
\begin{center}
\includegraphics[width=0.7\textwidth]{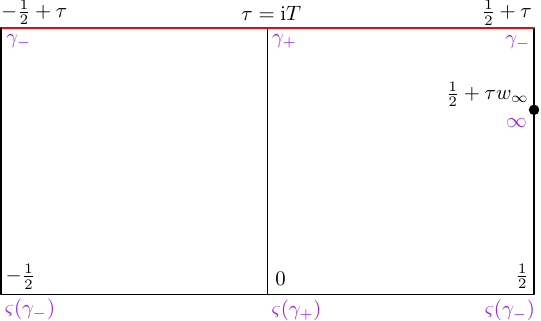}
\end{center}
\caption{\label{ParamF} The rectangle $\mathcal{R}$ in the $v$-plane. We indicate the image of special values of $x$ in purple, and the image of the cut $\gamma$ in red. The left (resp. right) panel is the image of ${\rm Im}\,x > 0$ (resp. ${\rm Im}\,x < 0$).}
\end{figure}

 The key to the solution is the use of an elliptic parametrisation $x = x(v)$. It depends on a parameter $\tau = {\rm i}T$ which is completely determined by the data of $\gamma_{\pm}$ and $\varsigma(\gamma_{\pm})$. The domain $\mathbb{C}\setminus\big(\gamma\cup\varsigma(\gamma)\big)$ will be the image via $v \mapsto x(v)$ of the rectangle (Figure~\ref{ParamF})
\beq
\label{rectan} \mathcal{R} := \big\{v \in \mathbb{C},\qquad  \mathrm{Re}\,v \in \big(-\tfrac{1}{2},\tfrac{1}{2}\big],\quad \mathrm{Im}\,v \in (0,T) \big\},
\eeq
with values:
\beq
\begin{array}{lcl}
x(\tau) = x(-\tau) = \gamma_{+}, & \qquad &  x\big(\tau + \tfrac{1}{2}\big) = x\big(-\tau + \tfrac{1}{2}\big) = \gamma_{-}, \\ 
x(0) = \varsigma(\gamma_{+}), & \qquad & x\big(\tfrac{1}{2}\big) = x\big(-\tfrac{1}{2}\big) = \varsigma(\gamma_{-}). \end{array} 
\eeq
We let
\beq
\hat{\mathcal{R}} = \big\{v \in \mathbb{C},\qquad {\rm Re}\,v \in \big(-\tfrac{1}{2},\tfrac{1}{2}\big],\quad {\rm Im}\,v \in [0,T] \big\}.
\eeq
and say that $x$ is in the physical sheet when $v(x) \in \overline{\mathcal{R}}$. For $x$ in the physical sheet, we have
\[
v(\varsigma(x)) = \tau - v(x).
\]
We call $v_{\infty}$ the point corresponding to $x = \infty$ in the physical sheet. With our assumptions, the involution $\varsigma$ is decreasing and we have $\gamma_+ < \gamma_+^* < \varsigma(\gamma_+)$. Therefore, the point $\infty$ can be to the right of $\varsigma(\gamma_+)$ and to the left of $\varsigma(\gamma_-)$, or to the right of $\varsigma(\gamma_-)$ and to the left of $\gamma_-$, that is
\[
v_{\infty} \in \big[0,\tfrac{1}{2}\big] \cup \big[\tfrac{1}{2},\tfrac{1}{2} + \tau\big]
\]
At least when we have $\varsigma^{-1}(\infty) \notin (-\gamma_+^*,\gamma_+^*)$, that is when $\alpha > \frac{1}{2}$, we must be in the second situation:
\beq
\label{vinfin} v_{\infty} = \tfrac{1}{2} + \tau w_{\infty},\qquad w_{\infty} \in (0,1).
\eeq
When $\alpha = 1$, by symmetry we must have $w_{\infty} = \frac{1}{2}$.

\begin{remark} \label{alphanottoosmall} For simplicity, \emph{we will assume in the remaining of the text that \eqref{vinfin} is satisfied} unless explicitly mentioned otherwise, i.e. that $\alpha$ is not too small; the main conclusions of our study are not affected when $v_{\infty}$ belongs to $\big[\tau,\frac{1}{2} + \tau\big]$, but some intermediate steps of analysis of the critical regime are a bit different.
\end{remark}

The function $v \mapsto x(v)$ is analytically continued for $v \in \mathbb{C}$ by the relations:
\beq
x(-v) = x(v + 1) = x(v + 2\tau) = x(v).
\eeq
This parametrisation allows the conversion \cite{EKOn2,BBG12b} of the functional equation:
\beq
\forall x \in \mathring{\gamma},\qquad f(x + {\rm i}0) + f(x - {\rm i}0) - n\,\varsigma'(x)\,f(\varsigma(x)) = 0
\eeq
for an analytic function $f(x)$ in $\mathbb{C}\setminus\gamma$, into the functional equation:
\beq
\label{reuh}\forall v \in \mathbb{C},\qquad \tilde{f}(v + 2\tau) + \tilde{f}(v) - n\,\tilde{f}(v + \tau) = 0,\qquad \tilde{f}(v) = \tilde{f}(v + 1) = -\tilde{f}(-v),
\eeq
for the analytic continuation of the function $\tilde{f}(v) = f(x(v))x'(v)$. The second condition in \eqref{reuh} enforces the continuity of $f(x)$ on $\mathbb{R}\setminus\gamma$. We set:
\beq \label{eq:b}
b = \frac{\mathrm{arccos}(n/2)}{\pi}.
\eeq
The new parameter $b$ ranges from $1$ to $0$ when $n$ ranges from $-2$ to $2$, and $b = \frac{1}{2}$ corresponds to $n = 0$. We emphasise the following uniqueness property which we will use repeatedly. It can be traced back to \cite{EKOn2} but we reproduce the argument for completeness.
\begin{lemma}
\label{uniquelemma} If $n \in (-2,2)$, there is at most one solution $f(v)$ to the equation
\[
\forall v \in \mathbb{C},\qquad f(v + 2\tau) + f(v) - nf(v + \tau) = 0,\qquad f(v) = f(v + 1)
\]
which is an entire function of $v$.
\end{lemma}
\noindent \textbf{Proof.} If $f$ is a solution, the functions
\[
f_{\pm}(v) = \frac{f(v) - e^{\pm {\rm i}\pi b}f(v + \tau)}{1 - e^{\pm {\rm i}\pi b}}
\]
satisfy $f_{\pm}(v + 1) = f_{\pm}(v) = e^{\mp {\rm i}\pi b} f(v + \tau)$ for $v \in \mathbb{C}$. Since $b$ is real-valued, $f_{+}$ and $f_{-}$ must be bounded entire functions, so must be constant by Liouville's theorem. The pseudo-periodicity condition in the $\tau$ direction then implies $f_{\pm} = 0$ hence $f = 0$. \hfill $\Box$

Solutions of the first two equations of \eqref{reuh} with prescribed divergent part at prescribed points in $\hat{\mathcal{R}}$ can be built from a fundamental solution $\Upsilon_b$, defined uniquely by the properties:
\beq
\label{propoup}\Upsilon_{b}(v + 1) = \Upsilon_{b}(v),\qquad \Upsilon_{b}(v + \tau) = e^{{\rm i}\pi b}\Upsilon_{b}(v),\qquad \Upsilon_{b}(v) \mathop{\sim}_{v \rightarrow 0} \frac{1}{v}.
\eeq
Its expression and main properties are reminded in Appendix~\ref{AppUpsilon}. In combination with Lemma~\ref{uniquelemma} this provides an effective way to solve the functional equations.

\vspace{0.2cm}

\subsubsection*{Remark} We will encounter the linear equation with non zero right-hand side given by a rational function $g(x)$:
\beq
\label{onir}f(x + {\rm i}0) + f(x - {\rm i}0) - n\,\varsigma'(x)\,f(\varsigma(x)) = g(x).
\eeq
It is enough to find a particular solution in the class of rational functions and subtract it from $f(x)$ to obtain a function $f^{{\rm hom}}(x)$ satisfying \eqref{onir} with vanishing right-hand side. This can be achieved for $n \neq \pm 2$ by:
\beq
f^{{\rm hom}}(x) = f(x) - \frac{1}{4 - n^2}\Big(2g(x) + n\varsigma'(x)g(\varsigma(x))\Big).
\eeq

\subsection{Disk and cylinder generating series}

We now review the results of \cite{BBG12b} for the generating series of disks $\mathbf{F}(x)$ for subcritical weights. Let $\mathbf{G}(v)$ be the analytic continuation of 
\beq
\label{gfggx}x'(v)\mathbf{F}(x(v)) - \frac{\partial}{\partial v}\Bigg(\frac{2\mathbf{V}(x(v)) + n\mathbf{V}(\varsigma(x(v)))}{4 - n^2} - \frac{nu \ln \big[\varsigma'(x(v))\big]}{2(2 + n)}\Bigg),
\eeq
where $\mathbf{V}(x) = \frac{x^2}{2} - \sum_{k \geq 1} g_k\,\frac{x^{k}}{k}$ collects the weights of empty faces. In the model we study, empty faces are triangles counted with weight $g$ each, so $\mathbf{V}(x) = \frac{x^2}{2} - g\,\frac{x^3}{3}$. However, there is no difficulty in including Boltzmann weights for empty faces of higher (bounded) degree as far as the solution of the linear equation is concerned, so we shall keep the notation $\mathbf{V}(x)$. Note that the last term in \eqref{gfggx} is absent if $\alpha = 1$. Let us introduce $(\tilde{g}_k)_{k \geq 1}$ as the coefficients of expansion:
\beq
\label{deftildeg} \frac{\partial}{\partial v}\Big(-\frac{2\mathbf{V}(x(v))}{4 - n^2} + \frac{2u \ln x(v)}{2 + n}\Big) = \sum_{k \geq 0} \frac{\tilde{g}_{k}}{(v - v_{\infty})^{k + 1}} + O(1),\qquad v \rightarrow v_{\infty}
\eeq
Their expressions for the model where all faces are triangles are recorded in Appendix~\ref{gdeter}.

\begin{theorem}[Disks \cite{BBG12b}]
\label{theimdisk}We have:
\[
\mathbf{G}(v) = \sum_{k \geq 0} \frac{1}{2}\,\frac{\tilde{g}_k}{k!}\,\frac{\partial^k}{\partial v_{\infty}^k}\Big[\Upsilon_{b}(v + v_{\infty}) + \Upsilon_{b}(v - v_{\infty}) - \Upsilon_{b}(-v + v_{\infty}) - \Upsilon_{b}(-v - v_{\infty})\Big]. 
\]
The endpoints $\gamma_{\pm}$ are determined by the two conditions:
\beq
\label{supportD}\mathbf{G}(\tau + \varepsilon) = 0,\qquad \varepsilon = 0,\tfrac{1}{2},
\eeq
which follow from the finiteness of the generating series $\mathbf{F}(x)$ at $x = \gamma_{\pm}$.
\end{theorem}
If $\alpha = 1$, the $4$ terms expression can be reduced to $2$ terms using $\tau - v_{\infty} = v_{\infty}\,\,{\rm mod}\,\,\mathbb{Z}$ and the pseudo-periodicity of the special function $\Upsilon_{b}$.

\begin{remark}
  \label{rem:buddchen}
  We refer to the original paper for the derivation of
  Theorem~\ref{theimdisk}. In all rigor, the
  conditions~\eqref{supportD} may yield several solutions for the cut
  endpoints $\gamma_\pm$, and the correct choice corresponds to the
  solution which leads to a series $\mathbf{F}$ with positive
  coefficients. The original paper used numerical evidence as a
  justification. For the rigid case~\cite{BBG12a}, a formal
  justification was later provided in~\cite{BuddOn} via two theorems,
  due to Timothy Budd and Linxiao Chen respectively, see
  also~\cite[Chapter~II]{ChenThesis}. Here we consider the bending
  energy model, to which these theorems do not apply directly. In
  Appendix~\ref{sec:equivadm}, we prove the analogue of Budd's theorem
  for the bending energy model, for $n \in (0,2)$. To keep a bound on the size of this
  paper, we do not prove the analogue of Chen's theorem, but we
  believe that there should be no unsurpassable obstacle in
  generalising his approach. Such an argument is also necessary to justify completely the phase diagram of the model.
\end{remark}

Remarkably, the generating series of pointed disks and of cylinders have very simple expressions. 
\begin{proposition}[Pointed disks]
\label{propPoint}Define $\mathbf{G}^\bullet(v)$ as the analytic continuation of:
\beq
\label{second}x'(v)\mathbf{F}^\bullet(x(v)) + \frac{\partial}{\partial v}\Bigg(\frac{nu\,\ln[\varsigma'(x(v))]}{2(2 + n)}\Bigg).
\eeq
(for $\alpha = 1$ the last term is absent). We have:
\beq
\label{2dsa}\mathbf{G}^\bullet(v) = \frac{u}{2 + n}\Big[-\Upsilon_{b}(v + v_{\infty}) - \Upsilon_{b}(v - v_{\infty}) + \Upsilon_{b}(-v + v_{\infty}) + \Upsilon_{b}(-v - v_{\infty})\Big].
\eeq
\end{proposition}
\noindent \textbf{Proof.} The strategy is similar to \cite{BBG12b}. In the functional equation of Proposition~\ref{prop9}, we can evaluate the contour integral using \eqref{contueq} and $\mathbf{F}^\bullet(x) \sim \frac{u}{x}$ when $x \rightarrow \infty$. Thus:
\beq
\label{jiuh}\forall x \in \mathring{\gamma},\qquad \mathbf{F}^\bullet(x + {\rm i}0) + \mathbf{F}^\bullet(x - {\rm i}0) - n\varsigma'(x)\,\mathbf{F}^\bullet(\varsigma(x)) = \frac{nu}{x - \varsigma(\infty)}.
\eeq
We can find a rational function of $x$ which is a particular solution to \eqref{jiuh}, and subtract it from $\mathbf{F}^\bullet(x)$ to obtain a solution of the linear equation with vanishing right-hand side. This is the origin of the second term in \eqref{second}. The construction reviewed in \S~\ref{elparam} then implies that $\mathbf{G}^\bullet(v)$ satisfies the functional relation:
\beq
\label{funq111}\mathbf{G}^\bullet(v + 2\tau) + \mathbf{G}^\bullet(v) - n\mathbf{G}^\bullet(v + \tau) = 0,\qquad \mathbf{G}^\bullet(v) = \mathbf{G}^\bullet(v + 1) = -\mathbf{G}^\bullet(-v).
\eeq
$\mathbf{G}^\bullet(v)$ inherits the singularities of \eqref{second}. If $\alpha \neq 1$, we have a simple pole in the fundamental domain at:
\beq
\label{res1}\Res_{v = v_{\infty}} \dd v\,\mathbf{G}^{\bullet}(v) = \frac{-2u}{2 + n},\qquad \Res_{v = (\tau - v_{\infty})} \dd v\,\mathbf{G}^{\bullet}(v) = \frac{-nu}{2 + n}.
\eeq
\eqref{2dsa} provides the (unique by Lemma~\ref{uniquelemma}) solution to this problem. When $\alpha  = 1$, we have $\varsigma(\infty) = \infty$, and $v_{\infty} = \frac{1 + \tau}{2}$, therefore $v_{\infty} = \tau - v_{\infty}\,\,{\rm mod}\,\,1$. Then, we have a unique simple pole in the fundamental domain:
\[
\Res_{v = v_{\infty}} \dd v\,\mathbf{G}^{\bullet}(v) = -u.
\]
In this case, we find:
\[
\mathbf{G}^{\bullet}(v) = \frac{u}{1 + e^{-{\rm i}\pi b}}\big[-\Upsilon_{b}(v - v_{\infty}) + \Upsilon_{b}(-v-v_{\infty})\big].
\]
Using the properties of $\Upsilon_{b}$ under translation, this is still equal to the right-hand side of \eqref{2dsa}. In other words, formula \eqref{2dsa} is well behaved when $v_{\infty} \rightarrow (\tau - v_{\infty})$.
\hfill $\Box$

\begin{proposition}[Cylinders] \label{p15} Define $\mathbf{G}^{(2)}(v_1,v_2)$ as the analytic continuation of:
\beq
\label{G2d2} x'(v_1)x'(v_2)\mathbf{F}^{(2)}(x(v_1),x(v_2)) + \frac{\partial}{\partial v_1}\frac{\partial}{\partial v_2}\Bigg(\frac{2\ln\big[x(v_1) - x(v_2)\big] + n\ln\big[\varsigma(x(v_1)) - x(v_2)\big]}{4 - n^2}\Bigg).
\eeq
We have:
\beq
\label{G2sol}\mathbf{G}^{(2)}(v_1,v_2) = \frac{1}{4 - n^2}\Big[\Upsilon_{b}'(v_1 + v_2) - \Upsilon_{b}'(v_1 - v_2) - \Upsilon_{b}'(-v_1 + v_2) + \Upsilon_{b}'(-v_1 - v_2)\Big].
\eeq
\end{proposition}
\noindent\textbf{Proof.} This result is proved in \cite[Section 3.4]{BEOn} for $\alpha = 1$, but its proof actually holds when $\varsigma$ is any rational involution. We include it for completeness. The fact that $\varsigma$ is an involution implies that $\mathbf{G}^{(2)}(v_1,v_2)$ is a symmetric function of $v_1$ and $v_2$, as:
\[
\frac{\dd x_1\dd x_2}{(x_1 - x_2)^2} = \frac{\dd \varsigma(x_1)\dd\varsigma(x_2)}{(\varsigma(x_1) - \varsigma(x_2))^2}.
\]
It must satisfy:
\begin{equation}
\begin{split}
\label{rel1}& \mathbf{G}^{(2)}(v_1,v_2) + \mathbf{G}^{(2)}(v_1 + 2\tau,v_2) - n\mathbf{G}^{(2)}(v_1 + \tau,v_2) = 0, \\
& \mathbf{G}^{(2)}(v_1,v_2) = \mathbf{G}^{(2)}(v_1 + 1,v_2) = - \mathbf{G}^{(2)}(-v_1,v_2).
\end{split}
\end{equation}
It has a double pole at $v_1 = v_2$ so that $\mathbf{G}^{(2)}(v_1,v_2) = \frac{2}{4 - n^2}\,\frac{1}{(v_1 - v_2)^2} + O(1)$, double poles at $v_1 = v_2 + (\mathbb{Z} \oplus \tau\mathbb{Z})$ ensuing from \eqref{rel1}, and no other singularities. Equation \eqref{G2sol} provides the (unique by Lemma~\ref{uniquelemma}) solution to this problem. \hfill $\Box$

\subsection{Refinement: separating loops}

We have explained in \S~\ref{separ} that the functional equation satisfied by refined generating series, with a weight $s$ per separating loop, only differs from the unrefined case by keeping the same cut $\gamma$, but replacing $n \rightarrow ns$ in the linear functional equations. Thus defining:
\beq
\label{eq:bsdef}
b(s) = \frac{\mathrm{arccos}(ns/2)}{\pi},
\eeq
we immediately find:
\begin{corollary}[Refined pointed disks] Let $\mathbf{G}^\bullet_{s}(v)$ be the analytic continuation of:
\label{corre}\beq
\label{cioj}x'(v)\mathbf{F}^\bullet_{s}(x(v)) + \frac{\partial}{\partial v}\Bigg(\frac{ns\,\ln[\varsigma'(x(v))]}{2(2 + ns)}\Bigg).
\eeq
We have:
\beq
\mathbf{G}^\bullet_{s}(v) = \frac{u}{2 + ns}\Big(-\Upsilon_{b(s)}(v + v_{\infty}) - \Upsilon_{b(s)}(v - v_{\infty}) + \Upsilon_{b(s)}(-v + v_{\infty}) + \Upsilon_{b(s)}(-v - v_{\infty})\Big).
\eeq
\hfill $\Box$
\end{corollary}

\begin{corollary}[Refined cylinders] \label{recycl}Let $\mathbf{G}_{s}^{(2)}(v_1,v_2)$ be the analytic continuation of:
\[
x'(v_1)x'(v_2)\mathbf{F}^{(2)}_{s}(x(v_1),x(v_2)) + \frac{\partial}{\partial v_1}\frac{\partial}{\partial v_2}\Bigg(\frac{2\ln\big[x(v_1) - x(v_2)\big] + ns\ln\big[\varsigma(x(v_1)) - x(v_2)\big]}{4 - n^2s^2}\Bigg).
\]
We have:
\begin{multline}
\label{G2solss}\mathbf{G}^{(2)}_{s}(v_1,v_2) = \\ \frac{1}{4 - n^2s^2}\Big[\Upsilon_{b(s)}'(v_1 + v_2) - \Upsilon_{b(s)}'(v_1 - v_2) - \Upsilon_{b(s)}'(-v_1 + v_2) + \Upsilon_{b(s)}'(-v_1 - v_2)\Big].
\end{multline}
\hfill $\Box$
\end{corollary}

\section{Depth of a vertex in disks}
\label{inequ}
\label{bendas}

We now study the asymptotic behavior of the distribution of the depth $P$ of the origin of a pointed disk, in loop model with bending energy. While the algebraic results that we have obtained in the previous sections are valid for nonpositive weights, we will in the rest of the paper assume that
\[
n \in (0,2),\qquad b \in \big(0,\tfrac{1}{2}\big),\qquad g,\alpha \geq 0,\qquad h > 0
\]
unless specified otherwise.

\subsection{Phase diagram and the volume exponent}
\label{prelem}

The phase diagram of the model with bending energy is Theorem~\ref{thphase} below, and was established in \cite{BBG12b}. We review its derivation, and push further the computations of \cite{BBG12b} to derive (Corollary~\ref{thvolume} below) the well known exponent $\gamma_{{\rm str}}$ appearing in the asymptotic number of pointed rooted disks of fixed, large volume $V$, and justify delta-analyticity statements that are used for the asymptotic analysis.  We remind that the model depends on the weight $g$ per empty triangle, $h$ per triangle crossed by a loop, and the bending energy $\alpha$, and the weight $u$ per vertex is set to $1$ unless mentioned otherwise. A non generic critical point occurs when $\gamma_+$ approaches the fixed point of the involution:
\begin{equation}
  \label{eq:gammastar}
  \gamma_+^* = \varsigma(\gamma_+^*) = \frac{1}{h(\alpha + 1)}.
\end{equation}
In this limit, the two cuts $\gamma$ and $\varsigma(\gamma)$ merge at $\gamma_+^*$, and one can justify on the basis of combinatorial arguments \cite[Section 6]{BBG12b} that $\gamma_- \rightarrow \gamma_-^*$ with:
\[
|\gamma_-^*| < |\gamma_+^*|\qquad {\rm and}\qquad \varsigma(\gamma_-^*) \neq \gamma_-^*.
\]
In terms of the parametrisation $x(v)$, it amounts to letting $T \rightarrow 0$, and this is conveniently measured in terms of the parameter:
\[
q = e^{-\frac{\pi}{T}} \rightarrow 0.
\]

To analyse the non generic critical regime, we first need to derive the asymptotic behavior of the parametrisation $x(v)$ and the special function $\Upsilon_{b}(v)$. This is performed respectively in Appendix~\ref{xbeh} and \ref{Upbeh}. The phase diagram and the volume exponent can then be obtained after a tedious algebra, which is summarised in Appendix~\ref{proofbeh}. Theorems~\ref{thphase}-\ref{thphase02} and a large part of the calculations done in Appendix appeared in \cite{BBG12b}. Here, we push these calculations further to present some consequences on generating series of pointed disks/gaskets  (Corollaries~\ref{corrrr1}-\ref{knowni} below), and we add a detailed description of the analytic properties with respect to $u$. It is then possible to apply transfer theorems, i.e. extracting asymptotic behavior of coefficients of the generating series from the analysis of their singularities.

\begin{theorem}\cite{BBG12b}
\label{thphase}
Assume $\alpha = 1$, and introduce the parameter:
\[
\rho := 1 - 2h\gamma_-^* = 1 - \frac{\gamma_-^*}{\gamma_+^*}.
\]
There is a non generic critical line, parametrised by $\rho \in (\rho_{\min},\rho_{\max}]$:
\begin{equation*}
\begin{split}
\frac{g}{h} & =  \frac{4(\rho b\sqrt{2 + n} - \sqrt{2 - n})}{-\rho^2(1 - b^2)\sqrt{2 - n} + 4\rho b\sqrt{2 + n} - 2\sqrt{2 - n}} \\
h^2 & =  \frac{\rho^2 b}{24\sqrt{4 - n^2}}\,\frac{\rho^2\,b(1 - b^2)\sqrt{2 + n}  - 4\rho\sqrt{2 - n} + 6b\sqrt{2 + n}}{-\rho^2(1 - b^2)\sqrt{2 - n} + 4\rho b\sqrt{2 + n} - 2\sqrt{2 - n}}.
\end{split}
\end{equation*}
It realises the dense phase of the model. The endpoint 
\[
\rho_{\max} = \frac{1}{b}\,\sqrt{\frac{2 - n}{2 + n}}
\]
corresponds to the fully packed model $g = 0$, with the critical value $h = \frac{1}{2\sqrt{2}\sqrt{2 + n}}$. The endpoint
\[
\rho_{\min} = \frac{\sqrt{6 + n} - \sqrt{2 - n}}{(1 - b)\sqrt{2 + n}}
\]
is a non  generic critical point realising the dilute phase, and it has coordinates:
\begin{equation*}
\begin{split}
\frac{g}{h} & =  1 + \sqrt{\frac{2 - n}{6 + n}}, \\
h^2 & =   \frac{b(2 - b)}{3(1-  b^2)(2 + n)}\bigg(1 - \frac{1}{4\sqrt{(2 - n)(6 + n)}}\bigg).
\end{split}
\end{equation*}
\end{theorem}
The fact that the non generic critical line ends at $\rho_{\max} < 2$ is in agreement with $|\gamma_-^*| < |\gamma_-^*|$.

\begin{remark} 
\label{thphase02} \label{alphanottoobig} In \cite{BBG12b}, it is proved that there exists $\alpha_c(n) > 1$ such that, in the model with bending energy $\alpha < \alpha_c(n)$, the qualitative conclusions of the previous theorem still hold, with a more complicated parametrisation of the critical line given in Appendix~\ref{proofbeh}. For $\alpha = \alpha_c(n)$, only a non generic critical point in the dilute phase exists, and for $\alpha > \alpha_c(n)$, non generic critical points do not exist.
\end{remark}

\begin{theorem}
\label{th38} Assume $(g,h)$ are chosen such that the model has a non  generic critical point for vertex weight $u = 1$. When $u < 1$ tends to $1$, we have:
\[
q \sim \Big(\frac{1 - u}{\Delta}\Big)^{c}.
\]
with the universal exponent:
\[
c = \left\{\begin{array}{lll} \frac{1}{1 - b} & & {\rm dense} \\ 1 & & {\rm dilute} \end{array}\right..
\]
The non universal constant reads, for $\alpha = 1$:
\[
\Delta = \left\{\begin{array}{lll} \frac{6(n + 2)}{b}\,\frac{\rho^2(1 - b)^2\sqrt{2 + n} + 2\rho(1 - b)\sqrt{2 - n} - 2\sqrt{2 + n}}{\rho^2b(1 - b^2)\sqrt{2 + n} - 4\rho(1 - b^2)\sqrt{2 - n} + 6b\sqrt{2 + n}} & & {\rm dense} \\ \frac{24}{b(1 - b)(2 - b)} & & {\rm dilute} \end{array}\right..
\]
For $\alpha \neq 1$, its expression is much more involved, but all the ingredients to obtain it are in Appendix~\ref{proofbeh}. 
\end{theorem}
We in fact obtain a stronger information in the Appendices.
\begin{lemma}
$u \mapsto q$ is delta-analytic.
\end{lemma}
This statement has two parts: delta-analyticity locally around $u = 1$, which is justified in Lemma~\ref{lemdeltanal}; and analytic continuation across the unit circle away from $u = 1$, which is justified in Theorem~\ref{thm:rigiddelta} for the rigid loop model, and more generally in Theorem~\ref{thm:rigiddeltaJ} for the bending energy model.

\subsection{Singular behavior of refined generating series}

We would like to study the asymptotic behavior of the weighted count of:
\begin{itemize}
\item[$(i)$] pointed disks with fixed volume $V$ and fixed depth $P$, in such a way that $V,P \rightarrow \infty$.
\item[$(ii)$] cylinders with fixed volume $V$, with two boundaries separated by $P$ loops, in such a way that $V,P \rightarrow \infty$.
\end{itemize}
This information can be extracted from the canonical ensemble where a map with a boundary of perimeter $L_i$ is weighted by $x^{-(L_i + 1)}$, each separating loop is counted with a weight $s$, and each vertex with a weight $u$. The generating series of interest are respectively $\mathbf{F}_{s}^{\bullet}(x)$ for $(i)$, and $\mathbf{F}_{s}^{(2)}(x_1,x_2)$ for $(ii)$. To retrieve the generating series of maps with fixed, large $V$ and $P$, we must  first obtain scaling asymptotics for these generating series when $u \rightarrow 1$.

As for fixing boundary perimeters, two regimes can be addressed. Either we want $L_i$ to diverge, in which case we should derive the previous asymptotics when $x$ approached the singularity $\gamma_+ \rightarrow \gamma_+^*$, since the other endpoint $|\gamma_-^*| < |\gamma_+^*|$ is subdominant. Or, we want to keep $L_i$ finite. In that case, we can work in the canonical ensemble by choosing $x$ away from $[\gamma_-,\gamma_+]$. We will actually consider the canonical ensemble with a control parameter $w_i$ such that $x_i = x\big(\frac{1}{2} + \tau w_i\big)$, and derive asymptotics for $w_i$ in some compact region containing $[0,1)$. The asymptotic count of maps with fixed, finite boundary perimeter $L_i$ can then be retrieved by a contour integration around $w_i = w_{\infty}^*$.

In a nutshell, we will set $x = x(v_i)$ with $v_i = \varepsilon_i + \tau w_i$ and $\varepsilon_i = 0$ to study a $i$-th boundary of large perimeter, and $\varepsilon_i = \frac{1}{2}$ to study finite boundaries.

The scaling behavior of $\mathbf{F}_{s}^{\bullet}(x)$ in the regime of large boundaries is established in Appendix~\ref{Sclpoin}.

\begin{theorem}
\label{th85} Let $(g,h)$ be a non generic critical point at $u = 1$. $F_{s}^{\bullet}(x)$ is an analytic family of meromorphic functions of $x$, parametrised by $u,s$ where $u$ belongs to a delta-domain centered at $1$ and $s$ to the strip $|{\rm Re}\,s| < \frac{2}{n}$. Besides, if ${\rm Re}\,b(s) \in \big(0,\tfrac{1}{2}\big)$, when $u \rightarrow 1$, in the two regimes $x \to \gamma_+$ and $x$ fixed away from
the cut, we have respectively\footnote{To be precise, we compute here the behavior of the singular part of $\mathbf{F}^{\bullet}_{s}(x)$, \textit{i.e.}, we did not include the shift in \eqref{cioj}, as it will always give zero when performing a contour integral against $x^{L}$ around the cut.}
\begin{equation}
\label{F1} 
\begin{split}
\mathbf{F}^{\bullet}_{s}(x)|_{{\rm sing}} & = \frac{q^{\frac{b(s) - 1}{2}}}{1 - q^{b(s)}}\,\Phi_{b(s)}\Big(\frac{x - \gamma_+}{q^{\frac{1}{2}}}\Big) + O(q^{\frac{b(s)}{2}}), \\ 
\mathbf{F}^{\bullet}_{s}(x)|_{{\rm sing}} & = \Psi_{b(s)}(x) + \frac{q^{b(s)}}{1 - q^{b(s)}}\,\tilde{\Psi}_{b(s)}(x) + O(q).
\end{split}
\end{equation}
The error in the first line of~\eqref{F1} is uniform for $\xi = q^{-\frac{1}{2}}(x - \gamma_+)$ in any fixed compact, and compatible\footnote{{\it I.e.}, it still yields a negligible term as compared to the previous ones.} with differentiation. For the expression of the scaling functions, we refer to \eqref{F5eq}-\eqref{F6eq} and \eqref{F7eq}-\eqref{F8eq} in the Appendix.
\end{theorem}

\begin{corollary}
\label{Covol}\label{thvolume} Assume $(g,h)$ are chosen such that the model has a non generic critical point for vertex weight $u = 1$. The number of pointed rooted planar maps of volume $V \rightarrow \infty$ behaves like:
\[
[u^{V}\cdot x^{-4}]\,\,\mathbf{F}^{\bullet}(x) \sim \frac{A}{\Delta^{bc}[-\Gamma(-bc)]\,V^{1 + bc}}.
\]
for some positive constant $A > 0$ given in Appendix, \eqref{AGasno}-\eqref{AGasno1}. Therefore, the critical exponent mentioned in \eqref{gammastr} is
\[
\gamma_{{\rm str}} = -bc.
\]
\end{corollary}

\begin{corollary}
\label{corrrr1} Assume $(g,h)$ are chosen such that the model has a non generic critical point for vertex weight $u = 1$. The number of rooted maps of volume $V \rightarrow \infty$ with a marked point in the gasket behaves as:
\[
[u^{V}\cdot x^{-4}]\,\,\mathbf{F}^{\bullet\,\,{\rm in}\,\,{\rm gasket}}(x) \sim \frac{A_{{\rm gasket}}}{\Delta^{\frac{c}{2}}[-\Gamma(-\frac{c}{2})]\,V^{1 + \frac{c}{2}}}.
\]
for a non universal constant $A_{{\rm gasket}} > 0$ given in Appendix, \eqref{AGas}-\eqref{AGas1}.
\end{corollary}

We can deduce the behavior when $V \rightarrow \infty$ of the probability that in a pointed rooted disk of volume $V$, the origin belongs to the gasket:
\begin{corollary}
\label{knowni}
Assume $(g,h)$ are chosen such that the model has a non generic critical point for vertex weight $u = 1$. When $V \rightarrow \infty$:
\[
\mathbb{P}\big[\bullet {\rm in}\,\,{\rm gasket}\,\big|\,V,L = 3\big] \sim \frac{A_{{\rm gasket}}}{A}\,\frac{\Gamma(-bc)}{\Gamma(-\frac{c}{2})}\,\frac{1}{\Delta^{c(\frac{1}{2} - b)}\,V^{c(\frac{1}{2} - b)}}.
\]
\end{corollary}

\subsection{Central limit theorem for the depth}
\label{DerivationCLT}
We are going to prove the following result.
\label{CLTsecder}
\begin{theorem}
\label{CLTthm} Let $(g,h)$ be a non generic critical point at $u = 1$. Consider an ensemble of refined pointed disks of volume $V$, boundary perimeter $L$. Let $P$ the random variable giving the depth, i.e. the number of loops separating the origin from the boundary. When $L$ is chosen independent of $V$, we have as $V \to \infty$ the convergence in law
\[
\frac{P - \frac{cp_{{\rm opt}}}{\pi}\,\ln V}{\sqrt{\ln V}} \rightarrow \mathscr{N}(0,\sigma^2),\qquad p_{{\rm opt}} = \frac{n}{\sqrt{4- n^2}},\qquad \sigma^2 = \frac{4nc}{\pi(4 - n^2)^{\frac{3}{2}}}.
\]
which is uniform for $L > 0$ bounded. When $L = \lfloor \ell V^{c/2} \rfloor$ and $V \rightarrow \infty$ while $\ell$ is bounded and bounded away from $0$, we have
\[ 
\frac{P - \frac{cp_{{\rm opt}}}{2\pi}\,\ln V}{\sqrt{\ln V}} \rightarrow \mathscr{N}(0,\sigma^2),\qquad \sigma^2 = \frac{2nc}{\pi(4 - n^2)^{\frac{3}{2}}}.
\] 
\end{theorem}
\begin{proof}
We first treat the case of $L$ being a fixed integer. By L\'evy's continuity, it is sufficient to prove that for $t \in \mathbb{R}$
\begin{equation}
\label{thevgyn}\lim_{V \rightarrow \infty} \phi_{V}(t) = \exp\bigg(\frac{\sigma^2 t^2}{2}\bigg)
,\qquad \phi_{V}(t) = \mathbb{E}\bigg[\exp\bigg({\rm i}t\,\frac{P - \frac{cp_{{\rm opt}}}{2\pi}}{\sqrt{\ln V}}\bigg) \,\,\bigg|\,\, V,L\bigg].  
\end{equation}
The characteristic function can be computed by
\[
\phi_{V}(t) = \exp\big(-{\rm i} t cp_{{\rm opt}}\sqrt{\ln V}\big)\,\frac{\oint \frac{x^{L}\dd x}{2{\rm i}\pi} \oint \frac{\dd u}{2{\rm i}\pi\,u^{V + 1}}\, \mathbf{F}_{\exp({\rm i}t/\sqrt{\ln V})}^{\bullet}(x)}{\oint \frac{x^{L}\,\dd x}{2{\rm i}\pi} \oint \frac{\dd u}{2{\rm i}\pi\,u^{V + 1}}\,\mathbf{F}^{\bullet}(x)},
\]
where the contours in $x$ surrounds $\infty$ and the contours in $u$ initially surrounds $u = 0$. We first look at the numerator. For fixed $x$ in a $u$-independent neighborhood of $\infty$, we first use Theorem~\ref{th85}, in particular the second line in \eqref{F1}, with a fixed $s$ in a small enough neighboorhood of $1$. The term $\Psi_{b(s)}(x)$ can be discarded as it does not contribute to the integral in $u$. The second term in $\mathbf{F}_{s}^{\bullet}(x)|_{{\rm sing}}$ is
\begin{equation}
\frac{q^{b(s)}}{1 -  q^{b(s)}}\,\Psi_{b(s)}(x) \sim \bigg(\frac{1 - u}{\Delta}\bigg)^{cb(s)}\,\Psi_{b(s)}(x)
\end{equation}
uniformly for $s$ and $x$ in their respective domains mentioned above. Computing the contour integral in $x$ therefore preserves the error, and by transfer theorem (here we rely on Lemma~\ref{lemdeltanal}), the $u \rightarrow 1$ asymptotics yields the $V \rightarrow \infty$ asymptotics
\[ 
\oint \frac{\dd x\,x^{L}}{2{\rm i}\pi} \oint \frac{\dd u}{2{\rm i}\pi\,u^{V + 1}}\,\mathbf{F}_{s}^{\bullet}(x) \sim \frac{\oint \frac{\dd x\,x^{L}}{2{\rm i}\pi}\,\tilde{\Psi}_{b(s)}(x)}{\Delta^{cb(s)}\,[-\Gamma(b(s)c)]\,V^{1 + cb(s)}},
\]  
again uniformly in $s$. We can therefore compute the integral over $x$ and substitute $s = s_V(t) := \frac{{\rm i}t}{\sqrt{\ln V}}$. Doing the same for the denominator --- this amounts to set $s = 1$ ---we get
\[
\phi_{V}(t) \sim \exp\big(-{\rm i} t cp_{{\rm opt}}\sqrt{\ln V}\big)\,\frac{\oint \frac{\dd x\,x^{L}}{2{\rm i}\pi}\,\tilde{\Psi}_{b(s_V(t))}(x)}{\oint \frac{\dd x\,x^{L}}{2{\rm i}\pi}\,\tilde{\Psi}_{b}(x)}\,\Delta^{c(b - b(s_V(t)))}\,\frac{-\Gamma(-bc)}{-\Gamma(-b(s_V(t))c)}\,V^{c(b - b(s_V(t)))}.
\]
Since $s_V(t) \rightarrow 1$, the prefactors disappear in the limit and expanding $V^{c(b - b(s_V(t)))}$ up to $o(1)$ we find
\[
\phi_V(t) \sim \exp\bigg\{{\rm i}t\sqrt{\ln V}\Big(\frac{cn}{\pi \sqrt{4 - n^2}} -cp_{{\rm opt}}\Big) + \frac{2cnt^2}{\pi(4 - n^2)^{\frac{3}{2}}}\bigg\}.
\] 
The value of $p_{{\rm opt}} = \frac{n}{\sqrt{4 - n^2}}$ exactly cancels the divergent term, and we obtain \eqref{thevgyn} with variance
\begin{equation}
\sigma^2 = \frac{4cn}{\pi(4 - n^2)^{\frac{3}{2}}}.
\end{equation}

When $L = \lfloor \ell V^{c/2} \rfloor$, we have $L = \ell V^{c/2} + o(V^{c/2})$. We now move the contour in $x$ to surround $[\gamma_-,\gamma_+]$ at distance $q^{\frac{1}{2}}$, hence depending on $u$, so that it can be converted into a $u$-independent contour $\mathcal{L}$ in the variable $w$ such that $q^{-\frac{1}{2}}(x - \gamma_+) = x_{0}^*(w) + O(q^{\frac{1}{2}})$. A difficulty is that now
\[
x^{L} = \bigg(\gamma_+ + q^{\frac{1}{2}}x_{\frac{1}{2}}^*(w) + O(q)\bigg)^{L}
\] 
with $q \rightarrow 0$ and $L \rightarrow \infty$. It is however possible to repeat the proof of the transfer theorem \cite[Theorem IV.3]{Flajolet} and show that we only need the asymptotic of the integrand when $u \rightarrow 1$ at scale $1 - u = O(V^{-1})$. In this case we have $q^{\frac{1}{2}}V^{\frac{c}{2}} = O(1)$ and thus we can use
\[
x^{L} \sim \gamma_+^{L} \exp\big(-\ell  x_{\frac{1}{2}}^*(w) [(1 - u)V/\Delta]^{\frac{c}{2}} + o(1)\big).
\]
The rest of the analysis is similar to the previous case, with factor $q^{b(s)}$ replaced by $q^{\frac{b(s) - 1}{2}}$. Omitting the details, we arrive to
\[
\phi_V(t) \sim \exp\bigg\{{\rm i}t\sqrt{\ln V}\Big(\frac{cn}{\pi \sqrt{4 - n^2}} -cp_{{\rm opt}}\Big) + \frac{2cnt^2}{\pi(4 - n^2)^{\frac{3}{2}}}\bigg\},
\] 
and this gives the central limit theorem with mean and variance divided by $2$ compared to the previous case.
\end{proof}

\subsection{Large deviations for the depth: main result}

The central limit theorem directly came from the analysis of the singularity $F_{s}^{\bullet}$. We now refine it to obtain large deviations for the depth.

\label{LargPPP}
\begin{theorem}
\label{mainT}Let $(g,h)$ be a non generic critical point at $u = 1$. Consider the random ensemble of refined disks of volume $V$, boundary perimeter $L$. When $V \rightarrow \infty$ and $\ell$ remains fixed positive, the probability that the origin is separated from the boundary by $P$ loops behaves like:
\bea
\label{mumff} \mathbb{P}\Big[P = \big\lfloor \tfrac{c\ln V}{\pi}\,p\big\rfloor\,\Big|\,V,L = \ell \Big] & \stackrel{.}{\asymp} & \frac{1}{\sqrt{\ln V}\,V^{\frac{c}{\pi}\,J(p)}}, \\
\label{mumf}\mathbb{P}\Big[P = \big\lfloor \tfrac{c\ln V}{2\pi}\,p\big\rfloor\,\Big|\,V,L = \lfloor \ell V^{\frac{c}{2}} \rfloor \Big] & \stackrel{.}{\asymp} & \frac{1}{\sqrt{\ln V}\,V^{\frac{c}{2\pi}\,J(p)}}.
\eea
These estimates are uniform for $p$ bounded and bounded away from $0$. The large deviations function reads:
\begin{equation}
\begin{split}
\label{larged}
J(p) & =  \sup_{s \in [0,2/n]}\big\{p\ln(s) + {\rm arccos}(ns/2) - {\rm arccos}(n/2)\big\}  \\
& =  p\ln\left(\frac{2}{n}\,\frac{p}{\sqrt{1 + p^2}}\right) + {\rm arccot}(p) - {\rm arccos}(n/2).
\end{split}
\end{equation}
\end{theorem}

From a macroscopic point of view, a pointed disk with a finite boundary looks like a sphere with two marked points, while a pointed disk with large boundary looks like a disk. We observe that in the regime where $P \stackrel{.}{\asymp} \ln V$ 
\[
\mathbb{P}\big[2P\,\big|\,V,L = \ell\big] \stackrel{.}{\asymp} \mathbb{P}\big[P\,\big|\,V,L = \ell V^{\frac{c}{2}}\big]^2
\]
Intuitively, this means that the nesting of loops in a sphere can be described by cutting the sphere in two independent halves (which are disks). In Section~\ref{subsec:sphere} and in particular Corollary~\ref{spher2}, we will find an analog property for CLE.

The remaining of this section is devoted to the proof of these results. The probability that the origin of a pointed disk is separated from the boundary by $P$ loops reads:
\[
\mathbb{P}[P\,|\,V,L] = \frac{\mathcal{P}(V,L;P)}{\tilde{\mathcal{P}}(V,L)}
\]
and we need to analyse, when $V \rightarrow \infty$, and $L$ and $P$ in various regimes, the behavior of the integrals:
\begin{equation}
\begin{split}
\label{PVL}
\mathcal{P}(V,L,P) & =  \oint\oint\oint \frac{\dd u}{2{\rm i}\pi\,u^{V + 1}}\,\frac{x^{L}\dd x}{2{\rm i}\pi}\,\frac{\dd s}{2{\rm i}\pi s^{P + 1}}\,\mathbf{F}_{s}^{\bullet}(x),   \\
 \tilde{\mathcal{P}}(V,L) & =  \oint \oint \frac{\dd u}{2{\rm i}\pi\,u^{V + 1}}\,\frac{x^{L}\dd x}{2{\rm i}\pi}\,\mathbf{F}^{\bullet}(x).
 \end{split}
 \end{equation}
The contours for $u$ and $s$ are initially small circles around $0$, and the contour for $x$ surrounds the union of the cuts $[\gamma_-,\gamma_+]$ for the corresponding $u$'s.

\subsection{Proof of Theorem~\ref{mainT} for finite perimeters}
\label{finiteA}
When $L$ is finite, we can keep the contour integral over $x$ away from the cut. So, we need to use \eqref{F1}. The first term disappears when integrating over $u$, and remains:
\beq
\label{till}\mathbf{F}_{s}^{\bullet}(x)|_{{\rm sing}} = - \frac{q^{b(s)}\,\tilde{\Psi}_{b(s)}(x)}{1 - q^{b(s)}} + O(q),
\eeq
where the error in \eqref{till} is uniform for $x$ in any compact away from the cut for $s$ in the strip $|{\rm Re}\, s| < \frac{2}{n}$ away from its boundaries. The first term does not depend on $u$, therefore it does not contribute to the contour integral and can be discarded. Since $q \sim \big(\frac{1 - u}{\Delta}\big)^{c}$ when $u \rightarrow 1$ and is delta-analytic, we find directly by transfer theorems:
\beq
\label{gof2} \tilde{\mathcal{P}}(V,L) \sim \bigg\{\oint_{\mathcal{C}([\gamma_-^*,\gamma_+^*])} \frac{x^{L}\,\dd x}{2{\rm i}\pi}\,\tilde{\Psi}_{b}(x)\bigg\}\,\frac{1}{[-\Gamma(-bc)]\Delta^{bc}}\,\frac{1}{V^{1 + bc}}.
\eeq
Due to the aforementioned uniformity of the estimates with respect to $x$ and $s$, we also have
\beq
\mathcal{P}(V,L,P) \sim \oint_{\mathcal{C}([\gamma_-^*,\gamma_+^*])} \oint \frac{\dd s}{2{\rm i}\pi\,s^{P + 1}}\,\frac{x^L\,\dd x}{2{\rm i}\pi}\,\frac{\tilde{\Psi}_{b(s)}(x)}{[-\Gamma(-b(s)c)]\Delta^{b(s)c}}\,\frac{1}{V^{1 + b(s)c}}. 
\eeq 
where the contour in $s$ initially surrounds $0$ and must remain away from the boundaries of the strip $|{\rm Re}\, s| < \frac{2}{n}$. Through the analysis the $x$-contour surrounding $[\gamma_-^*,\gamma_+^*]$ will be fixed independent of $s$. We are going to apply the saddle point method to analyse the behavior of the $s$-contour integral when $P \rightarrow \infty$. The integral to compute is 
$$
\hat{\mathcal{P}}(V,L,P) := \oint_{\mathcal{C}([\gamma_-^*,\gamma_+^*])} \frac{\dd x\,x^L}{2{\rm i}\pi}\,\oint \frac{\dd s}{2{\rm i}\pi s}\,\frac{\tilde{\Psi}_{b(s)}(x)\,e^{\mathscr{S}(s)}}{\Gamma(-b(s)c)\Delta^{b(s)c} V^{1 + bc}}\,
$$
where
\beq
\label{S1lim}\mathscr{S}(s) = - s\ln P  - cb(s)\ln V + cb
\eeq
This function has critical points at $s = \pm \,\mathfrak{s}\big(\frac{\pi P}{c \ln V}\big)$, where for $p > 0$ we have defined
\begin{equation}
\label{spspsfun} \mathfrak{s}(p) := \frac{2}{n}\,\frac{p}{\sqrt{1 + p^2}}
\end{equation}
and for the record we introduce
\begin{equation}
\label{spspsfun2}\mathfrak{b}(p) := b(\mathfrak{s}(p)) = \frac{{\rm arccot}(p)}{\pi}
\end{equation}
We also compute
\begin{equation}
\label{HessS} \partial_s^2 \mathscr{S}(\mathfrak{s}(p)) = \frac{c\ln V}{\pi}\,\frac{n^2(1 + p^2)^2}{4p} > 0
\end{equation}
and
$$  
\mathscr{S}(\mathfrak{s}(p)) = -\frac{cJ(p)}{\pi}\,\ln V,\qquad J(p) := p\ln\bigg(\frac{2}{n}\,\frac{p}{\sqrt{1 + p^2}}\bigg) + {\rm arccot}(p) - {\rm arccos}(n/2) 
$$ 
The location of the critical point suggests to take a fixed value of $p > 0$ and set
$$
P := \bigg\lfloor \frac{cp\ln V}{\pi} \bigg\rfloor
$$
We also define $\tilde{p}_V$ as the function of $(p,V)$ such that
\begin{equation}
\label{Pc1a}P = \frac{c\tilde{p}\ln V}{\pi},\qquad
\end{equation}
It is such that
\begin{equation}
\label{pptilde} \tilde{p} - p \in O\bigg(\frac{1}{\ln V}\bigg)
\end{equation}

\vspace{0.2cm}

\noindent \textit{Step 1.} Let $\epsilon > 0$ small so that $\rho(p) := \frac{n}{2}|\mathfrak{s}(p) + {\rm i}\epsilon| < 1$. Then for $V$ large enough, $|\mathfrak{s}(\tilde{p}) + {\rm i}\epsilon| < \frac{2}{n}$. We deform the $s$-contour to a contour $\mathscr{C}$ defined as follows. It is the union of the vertical segment $C_1^+$ from $\mathfrak{s}(\tilde{p}) - {\rm i}\epsilon$ to $\mathfrak{s}(\tilde{p}) + {\rm i}\epsilon$, followed by the counterclockwise arc of circle $C_2^+$ in the upper-half plane joining $\mathfrak{s}(\tilde{p}) + {\rm i}\epsilon$ to $-\mathfrak{s}(\tilde{p}) + {\rm i}\epsilon$, followed by the vertical segment  $C_1^-$ from $-\mathfrak{s}(\tilde{p}) + {\rm i}\epsilon$ to $-\mathfrak{s}(\tilde{p}) - {\rm i}\epsilon$, followed by the counterclockwise arc of circle $C_2^-$ in the lower-half plane joining $-\mathfrak{s}(\tilde{p}) - {\rm i}\epsilon$ to $\mathfrak{s}(\tilde{p}) - {\rm i}\epsilon$. We claim that there exists a choice of $\epsilon$ and of constant $\eta > 0$ depending on $p$ but independent of $V$, such that for any $s \in \mathscr{C}_2^{\pm}$
\begin{equation}
\label{masterRs}\mathscr{R}(s) := {\rm Re}\big(\mathscr{S}(s) - \mathscr{S}(\mathfrak{s}(\tilde{p}))\big) < -\eta\,\ln V
\end{equation}

Since $s \mapsto \mathscr{R}(s)$ is even and its first term is independent of $s \in C_2^{\pm}$ by definition of the contour, it is sufficient to prove the existence of $\eta > 0$ such that $\mathscr{R}(\mathfrak{s}(\tilde{p}) \pm {\rm i}\epsilon) < - \eta \ln V$ and that
$$
t \mapsto \mathscr{R}\big(\rho(\tilde{p})e^{\pm {\rm i}\pi t}\big)
$$
is a decreasing function of $t \in [0,\frac{1}{2}]$. The first point follows for $\epsilon$ small enough independently of $V$ from the computation of the second derivative in \eqref{HessS}, and we can choose $\eta$ depending on $p$ and not on $V$ because $\tilde{p} - p \in O\big(\frac{1}{\ln V}\big)$. To justify the second point, we compute
$$
\frac{\dd \mathscr{R}}{\dd t} = \mp \frac{c\ln V}{\pi}\,\frac{2\rho(\tilde{p})}{n}\,{\rm Im}\bigg(\frac{1}{\sqrt{e^{\mp 2{\rm i}\pi t} - \rho^2(\tilde{p})}}\bigg)
$$
where we use the standard determination of the square root. This quantity is nonnegative if and only if $\mp {\rm Im}\big(e^{\mp 2{\rm i}\pi t} - \rho^2(\tilde{p})\big) = \sin(2\pi t) \geq 0$, which indeed holds for $t \in [0,\frac{1}{2}]$.

We note that there exists a constant $M(p) > 0$ such that for $x \in \mathcal{C}([\gamma_-^*,\gamma_+^*])$ and $s \in C$, we have for $V$ large enough
$$
\bigg|\frac{\tilde{\Psi}_{b(s)}(x)}{\Gamma(-b(s)c)\Delta^{b(s)c}}\bigg| \leq M(p)
$$
Together with \eqref{masterRs} and \eqref{pptilde} we deduce the existence of a constant $M'(L,p) > 0$ such that for $V$ large enough
\begin{equation}
\label{nunusssnunun}\bigg|\oint_{\mathcal{C}([\gamma_-^*,\gamma_+^*])} \frac{\dd x\,x^L}{2{\rm i}\pi} \int_{C_2^+ \cup C_2^-} \frac{\dd s}{2{\rm i}\pi s}\,\frac{\tilde{\Psi}_{b(s)}(x)\,e^{\mathscr{S}(s)}}{\Gamma(-b(s)c)\Delta^{b(s)c} V}\,\bigg| \leq \frac{1}{V^{1 + bc + \frac{cJ(p)}{\pi}}}\,\frac{M'(L,p)}{V^{\eta}} 
\end{equation}

\vspace{0.2cm}

\noindent \textit{Step 2.} By parity in $s$, the contributions of $C_1^\pm$ to the $\hat{P}(V,L,P)$ are equal. To study the contribution of $C_1^+$, the order of magnitude $\ln V$ of the Hessian in \eqref{HessS} suggests to perform the change of variables
$$
\tilde{s} = \mathfrak{s}(\tilde{p}) + \frac{{\rm i}\tilde{s}}{\sqrt{\ln V}}
$$
Since $\mathfrak{s}(\tilde{p})$ corresponds to the critical point of $\mathscr{S}$, we obtain by Taylor approximation at order $2$
$$
\mathscr{S}(s) = -\frac{cJ(\tilde{p})}{\pi} - \frac{c}{\pi}\,\frac{n^2(1 + p^2)^2}{8p}\,\tilde{s}^2 + O\bigg(\frac{1}{\sqrt{\ln V}}\bigg)
$$
and the error is uniform when $s \in C_1^+$, that is $|\tilde{s}| \leq \epsilon \sqrt{\ln V}$. Besides, there exists a constant $M(\epsilon) > 0$ such that for any $x \in \mathcal{C}([\gamma_-^*,\gamma_+^*])$ and $s \in C_1^+$, 
$$
\frac{\tilde{\Psi}_{b(s)}(x)}{\Gamma(-b(s)c)\Delta^{b(s)c}} \leq M(\epsilon)
$$
and we have the convergence when $V \rightarrow \infty$, poinwise in $\tilde{s} \in \mathbb{R}$
$$
\frac{\tilde{\Psi}_{b(s)}(x)}{\Gamma(-b(s)c)\Delta^{b(s)c}}  \longrightarrow \frac{\tilde{\Psi}_{\mathfrak{b}(p)}(x)}{\Gamma(-\mathfrak{b}(p)c)\Delta^{\mathfrak{b}(p)c}}
$$
Dominated convergence then implies
\begin{equation*}
\begin{split}
& \quad \oint_{\mathcal{C}([\gamma_-^*,\gamma_+^*])} \frac{\dd x\,x^L}{2{\rm i}\pi} \int_{C_1^+}  \frac{\dd s}{2{\rm i}\pi s}\,\frac{\tilde{\Psi}_{b(s)}(x)\,e^{\mathscr{S}(s)}}{\Gamma(-b(s)c)\Delta^{b(s)c} V^{1 + bc}} \\  
& \dot{\sim} \bigg(\oint_{\mathcal{C}([\gamma_-^*,\gamma_+^*])} \frac{\dd x\,x^L}{2{\rm i}\pi}\,\frac{\tilde{\Psi}_{\mathfrak{b}(p)}(x)}{\Gamma(-\mathfrak{b}(p)c)\Delta^{\mathfrak{b}(p)c}}\bigg) \cdot \frac{1}{\sqrt{\ln V} V^{1 + bc + \frac{cJ(\tilde{p})}{\pi}}}\,\int_{\mathbb{R}} \dd \tilde{s}\,e^{- \frac{c}{\pi}\,\frac{n^2(1 + p^2)^2}{8p}\,\tilde{s}^2} 
\end{split}
\end{equation*}
The effect of replacing $\tilde{p}$ by $p$ in the argument of $J$ only results in changing the overall constant by a quantity that may now depend on $V$ (since $\tilde{p}$ depend on $V$) but remains bounded and bounded away from $0$. The prefactors bounded and bounded away from zero become irrelevant when we write
$$
\oint_{\mathcal{C}([\gamma_-^*,\gamma_+^*])} \frac{\dd x\,x^L}{2{\rm i}\pi} \int_{C_1^+ \cup C_1^-} \frac{\dd s}{2{\rm i}\pi s} \stackrel{.}{\asymp} \frac{1}{\sqrt{\ln V}}\,V^{1 + bc + J(p)}
$$
where we recall that $F \stackrel{.}{\asymp} G$ means that $\ln F = \ln G + O(1)$. In comparison to this, the contribution of $C_2^+ \cup C_2^-$ is negligible due to \eqref{nunusssnunun}, hence
$$ 
\hat{P}(V,L,P) \stackrel{.}{\asymp} \frac{1}{\sqrt{\ln V}}\,V^{1 + bc + \frac{cJ(p)}{\pi}}
$$
Taking the ratio with \eqref{gof2} cancels $V^{1 + bc}$ and leads to the desired estimate
$$
\mathbb{P}\Big[P = \big\lfloor \tfrac{cp\ln V}{\pi} \big\rfloor\,\big| \,L\Big] \stackrel{.}{\asymp} (\ln V)^{-\frac{1}{2}}\,V^{-\frac{cJ(p)}{\pi}}
$$

\subsection{Proof of Theorem~\ref{mainT} for large perimeters}
\label{infiniteA}
Now, we study (with less details) the case where the $(x^*,s^*)$-coordinates of the critical point are such that $\xi^* = \frac{x^* - \gamma_+^*}{(q^*)^{1/2}}$ has a limit, and $s^*$ has a limit away from $2/n$. We can then use \eqref{F1}:
\beq
\label{Fsfss}\mathbf{F}_{s}^{\bullet}(x)|_{{\rm sing}} \sim \frac{q^{\frac{b(s)}{2} - \frac{1}{2}}}{1 - q^{b(s)}}\,\Phi_{b(s)}\Big(\frac{x - \gamma^+}{q^{1/2}}\Big).
\eeq
We need to analyse the critical points of:
\[
\mathcal{S}_{2}(u,x,s) = - V\ln u - P\ln s + L\ln x + \frac{c}{2}(b(s) - 1)\,\ln\Big(\frac{1 - u}{\Delta}\Big) + \ln \Phi_{b(s)}\bigg(\frac{x - \gamma_+^*}{[(1 - u)/\Delta]^{\frac{c}{2}}}\bigg).
\]
Compared to \eqref{Fsfss}, we have replaced $\gamma_+$ by $\gamma_+^*$, as it only differ by $O(q)$. The equation $\partial_{u}\mathcal{S}_{2} = 0$ gives:
\[
V \sim \frac{-\frac{c}{2}}{1 - u^*}\Big(b(s^*) - 1 + \xi^*\,(\ln \Phi_{b(s^*)})'(\xi^*)\Big),
\]
while the equation $\partial_{x} \mathcal{S}_{2} = 0$ gives:
\[
\frac{L}{\gamma_+^*} \sim -\Big(\frac{\Delta}{1 - u^*}\Big)^{\frac{c}{2}}\,(\ln\Phi_{b(s^*)})'(\xi^*).
\]
It is then necessary that $L \stackrel{.}{\asymp} V^{\frac{c}{2}}$. The equation $\partial_{s}\mathcal{S}_{2} = 0$ gives:
\[
\frac{P}{s^*} \sim \frac{nc\,\ln(1 - u^*)}{2\pi\sqrt{4 - n^2(s^*)^2}}.
\]
If we set $P = \frac{c\ln V}{2\pi}\,\tilde{p}$, we obtain $s^* \sim \mathfrak{s}(\tilde{p})$ with the function introduced in \eqref{spspsfun}. Notice the factor of $\frac{1}{2}$ compared to \eqref{Pc1a} in the previous section, due to the occurrence of $q^{\frac{b(s)}{2}}$ here and $q^{b(s)}$ there in the scaling limits of $\mathbf{F}_{s}^{\bullet}(x)$. We also evaluate:
\[
\partial_{s}^2\mathcal{S}_{2}(u^*,x^*,\mathfrak{s}(\tilde{p})) \stackrel{.}{\asymp} \ln V.
\]

\begin{figure}
\begin{center}
\includegraphics[width=0.6\textwidth]{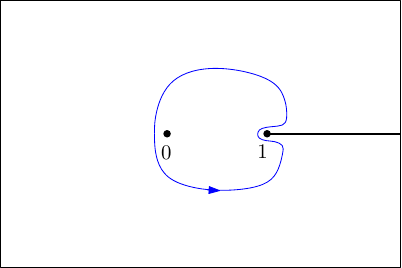}
\caption{\label{Cont1} The contour of integration for $u$.}
\end{center}
\end{figure}

\begin{figure}
\begin{center}
\includegraphics[width=0.6\textwidth]{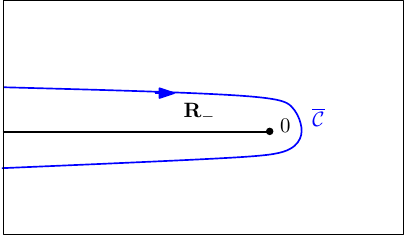}
\caption{\label{ContCbar} The contour $\overline{\mathcal{C}}$.}
\end{center}
\end{figure}

Therefore, let us now assume:
\[
L = \ell\,V^{\frac{c}{2}},\qquad P = \frac{c\tilde{p}\ln V}{2\pi}
\]
for a fixed positive $\ell$.  The previous discussion suggests the change of variable to compute $\tilde{\mathcal{P}}(V,L)$:
\[
u = 1 - \frac{\tilde{u}}{V},\qquad x = \gamma_+^* + \frac{\tilde{x}}{V^{\frac{c}{2}}}.
\]
We then find:
\[
\frac{\dd u}{u^{V + 1}}\,\dd x\,x^{L}\,\mathbf{F}_{s}^{\bullet}(x) \sim \frac{\dd\tilde{u}\,\dd\tilde{x}}{V^{1 + \frac{cb}{2}}}\,\Phi_{b}\Big(\frac{\tilde{x}}{(\tilde{u}/\Delta)^{\frac{c}{2}}}\Big)\,e^{\tilde{u} + \ell\tilde{x}/\gamma_+^*}\Big(\frac{u}{\Delta}\Big)^{\frac{c}{2}(b - 1)},
\]
where the convergence to the limit function in the right-hand side is uniform for $(\tilde{u},\tilde{x})$ in any compact away from $\mathbb{R}_{\leq 0}^2$. The contours can be deformed to steepest descent contours $\overline{\mathcal{C}}^2$ (see Figures~\ref{Cont1} and~\ref{ContCbar}), and we can conclude as before by dominated convergence:
\beq
\label{detg}\tilde{\mathcal{P}}(V,L) \sim \bigg\{ \oint_{\overline{\mathcal{C}}}\oint_{\overline{\mathcal{C}}} \frac{\dd\tilde{u}\,\dd\tilde{x}}{(2{\rm i}\pi)^2}\,e^{\tilde{u} + \ell\tilde{x}/\gamma_+^*}\Big(\frac{\tilde{u}}{\Delta}\Big)^{\frac{c}{2}(b - 1)}\,\Phi_{b}\Big(\frac{\tilde{x}}{(\tilde{u}/\Delta)^{\frac{c}{2}}}\Big)\bigg\}\,V^{-1 - \frac{cb}{2}}.
\eeq

Likewise, in order to compute $\mathcal{P}(V,L,P)$, we make the change of variable:
\[
u = 1 - \frac{\tilde{u}}{V},\qquad s = \mathfrak{s}(p) + \frac{\tilde{s}}{\sqrt{\ln V}},\qquad x = \gamma_+^* + \frac{\tilde{x}}{V^{\frac{c}{2}}}.
\]
We then find:
\small
\begin{equation*}
\begin{split}
& \quad \frac{\dd u}{u^{V + 1}}\,\frac{\dd s}{s^{P + 1}}\,\dd x\,x^{L}\,\mathbf{F}_{s}^{\bullet}(x)  \\
& \sim  \frac{\dd\tilde{u}\,\dd\tilde{x}\,\dd\tilde{s}\,(\gamma_+^*)^L}{\mathfrak{s}(\tilde{p})\,V^{1 + \frac{c}{2}\mathfrak{b}(\tilde{p}) + \frac{c\tilde{p}}{2\pi} \ln \mathfrak{s}(p)}\sqrt{\ln V}}\,\Phi_{\mathfrak{b}(s)}\Big(\frac{\tilde{x}}{(\tilde{u}/\Delta)^{\frac{c}{2}}}\Big)
\,e^{\tilde{u} + \ell\tilde{x}/\gamma_+^*} \Big(\frac{\tilde{u}}{\Delta}\Big)^{\frac{c}{2}(\mathfrak{b}(s) - 1)}\exp\Big\{\frac{cn^2(\tilde{p}^2 + 1)^2}{16\pi \tilde{p}}\,\tilde{s}^2\Big\}, \end{split}
\end{equation*}
\normalsize
where the convergence to the limit function in the right-hand side is uniform for $\tilde{s},\tilde{x},\tilde{u}$ in any compact with $\tilde{u}$ away from $0$. We deform the contours to steepest descent contours $\mathcal{C}^2$ in the variables $(x,u)$, and ${\rm i}\mathbb{R}$ in the variable $|\tilde{s}| \ll \sqrt{\ln V}$. By properties of steepest descent contours, we can apply dominated convergence and find:
\begin{equation}
\begin{split}
\label{cudg}
\mathcal{P}(V,L,P) & \sim \bigg\{\oint_{\overline{\mathcal{C}}}\oint_{\overline{\mathcal{C}}} \frac{\dd\tilde{x}\,\dd\tilde{u}}{(2{\rm i}\pi)^2}\,e^{\tilde{u} + \ell\tilde{x}/\gamma_+^*}\Big(\frac{\tilde{u}}{\Delta}\Big)^{\frac{c}{2}(\mathfrak{b}(s) - 1)}\,\Phi_{\mathfrak{b}(s)}\Big(\frac{\tilde{x}}{(\tilde{u}/\Delta)^{\frac{c}{2}}}\Big)\bigg\}  \\
& \quad \times \frac{(\ln V)^{-1/2}\,V^{-1 - \frac{c}{2}\mathfrak{b}(p) - \frac{c\tilde{p}}{2\pi}\ln \mathfrak{s}(p)}}{\sqrt{c\tilde{p}(\tilde{p}^2 + 1)}}.
\end{split}
\end{equation}
Taking the ratio of \eqref{cudg} and \eqref{detg} and replacing $\tilde{p}$ and $\tilde{\ell}$ with $p$ and $\ell$ such that $P = \big\lfloor \frac{cp\ln V}{2\pi} \big\rfloor$ and $L = \lfloor \ell V^{\frac{c}{2}} \rfloor$ gives the desired distribution \eqref{mumf}.

\section{Separating loops in cylinders}
\label{CylS}

Let us consider the probability that, in a random ensemble of planar maps of volume $V$, two boundaries of given perimeter $L_1$ and $L_2$ are separated by $P$ loops:
\[
\mathbb{P}[P\,|\,V,L_1,L_2] = \frac{\oint\oint\oint\oint \frac{\dd u}{2{\rm i}\pi\,u^{V + 1}}
\frac{x^{L_1}_1\dd x_1}{2{\rm i}\pi}\,\frac{x_2^{L_2}\,\dd x_2}{2{\rm i}\pi}\,\frac{\dd s}{2{\rm i}\pi\,s^{P + 1}}\,\mathbf{F}_{s}^{(2)}(x_1,x_2)}{\oint\oint\oint \frac{\dd u}{2{\rm i}\pi\,u^{V + 1}}
\frac{x^{L_1}_1\dd x_1}{2{\rm i}\pi}\,\frac{x_2^{L_2}\,\dd x_2}{2{\rm i}\pi}\,\mathbf{F}^{(2)}(x_1,x_2)}.
\]
The analog of Theorem~\ref{th85} for the behavior of $\mathbf{F}_{s}^{(2)}$ is derived in Appendix~\ref{apGg}, and it features singularities of the type\footnote{The fact that critical exponents for cylinders taking into account the number of separating loops are obtained by replacing $b$ with $b(s)$ can be observed indirectly in \cite[Section 4.2]{KostovS}, with the momentum $p$ playing the role of $b(s)$.} $q^{b(s)/j}$ with $j = 1$ for $x_1$ and $x_2$ both close to or both away from $\gamma_+^*$, and $j = 2$ for $x_1$ close to $\gamma_+^*$ and $x_2$ close to $\infty$. In that regard, the origin in pointed disks behaves as a boundary face whose perimeter is kept finite in a cylinder. As the type of singularities encountered in the asymptotic analysis is identical, the result can be directly derived from Sections \ref{finiteA}-\ref{infiniteA}:

\begin{theorem}\label{theo:cyl}
\label{mainTC} Let $(g,h)$ be a non generic critical point at $u = 1$. Let $\ell_1,\ell_2,p > 0$. When $V \rightarrow \infty$, we have
\begin{equation*}
\begin{split}
\mathbb{P}\Big[P = \big\lfloor \tfrac{c \ln V}{\pi}\,p \big\rfloor\,\Big|\,V,L_1 = \ell_1,L_2 = \ell_2\Big] & \stackrel{.}{\asymp} \frac{1}{\sqrt{\ln V}\,V^{\frac{c}{\pi}\,J(p)}},\\
\mathbb{P}\Big[P = \big\lfloor \tfrac{c\ln V}{2\pi}\,p \big\rfloor \,\Big|\,V,L_1 = \ell_1,L_2 = \lfloor \ell_2 V^{\frac{c}{2}} \rfloor \Big] & \stackrel{.}{\asymp} \frac{1}{\sqrt{\ln V}\,V^{\frac{c}{2\pi}\,J(p)}},
\end{split}
\end{equation*}
where the large deviations function $J(p)$ is the same as in \eqref{larged}.
\hfill $\Box$
\end{theorem}
In the regime where the two boundaries of the cylinder have perimeter of order $V^{\frac{c}{2}}$, the nesting distribution behaves differently and is not analysed here.

\section{Weighting loops by i.i.d. random variables}\label{sec:weighting}
\label{Weighsec}

\subsection{Definition and main result}
Following \cite{MWW}, we introduce a model of random maps with weighted loop configurations ; we describe it for pointed disks, but it will be clear that our reasoning extends to the cylinder topology. Let $\xi$ be a random variable, with distribution $\mu$, for which we assume that the cumulant function,
\[
\Lambda_{\mu}(\lambda) := \ln \mathbb E [e^{\lambda \xi}],
\]
exists for $\lambda$ in a neighborhood of $0$. Given a map with a self-avoiding loop configuration, let $(\xi_{l})_{l \in \mathcal{L}}$ be a sequence of i.i.d. random variables distributed like $\xi$, indexed by the set $\mathcal{L}$ of loops. Let $\mathcal{L}_{{\rm sep}} \subseteq \mathcal{L}$ be the set of loops separating the boundary from the marked point. We would like the describe the joint distribution of the depth $P = |\mathcal{L}_{{\rm sep}}|$ and of the sum $\Xi = \sum_{l \in \mathcal{L}_{{\rm sep}}} \xi_{l}$. 

Recall from the proof of Proposition~\ref{prop77} that $\mathbf{F}^{\bullet}_{[P]}(x)$ is the generating series of pointed disks with exactly $P$ separating loops. Our problem is solved by introducing the generating series $\mathbf{F}_{s,\lambda}^{\bullet}(x)$, as the $\mu$-expectation value of the generating series of pointed disks, whose usual weight in the loop model is multiplied by $\prod_{l \in \mathcal{L}_{{\rm sep}}} s\,e^{\lambda \xi_{l}}$. By construction, we have:
\[
\mathbf{F}_{s,\lambda}^{\bullet}(x) = \sum_{P \geq 0} (s e^{\Lambda_{\mu}(\lambda)})^{P}\,\mathbf{F}^{\bullet}_{[P]}(x) = \mathbf{F}_{s\exp(\Lambda_{\mu}(\lambda))}^{\bullet}(x).
\]
In the ensemble of pointed disks with volume $V$ and perimeter $L$, the joint distribution we look for reads:
\[
\mathbb{P}(P,\Xi|V,L) = \frac{\mathcal{P}(V,L ; P,\Xi)}{\tilde{\mathcal{P}}(V,L)}
\]
with a new numerator --- compare with \eqref{PVL}:
\[
\mathcal{P}(V,L;P,\Xi) = \oint\oint\oint\oint \frac{\dd u}{2{\rm i}\pi u^{V + 1}}\,\frac{\dd x\,x^{L}}{2{\rm i}\pi}\,\frac{\dd s}{2{\rm i}\pi\,s^{P + 1}}\,\frac{\dd \lambda\,e^{-\lambda\Xi}}{2{\rm i}\pi}\,\mathbf{F}_{s,\lambda}^{\bullet}(x).
\]

\begin{theorem}
\label{bivth}
Let $(g,h)$ be a non generic critical point at $u = 1$. Let\footnote{Note that $q$ here is a parameter with the same status as $p$, and does not refer to the elliptic nome controlling e.g. in Theorem~\ref{th38} the distance to criticality. As the context of their apparition is quite different, it should not lead to confusion.} $\ell,p,q > 0$. When $V \rightarrow \infty$, we have 
\begin{eqnarray}
\label{bivth1} \mathbb{P}\Big[P = \big\lfloor \tfrac{c\ln V}{\pi}\,p \big\rfloor\,\,{\rm and}\,\,\Xi =  \tfrac{c\ln V}{\pi}\,q\,\Big|\,V,L = \ell \Big]& \stackrel{.}{\asymp} & \frac{1}{(\ln V)\,V^{\frac{c}{\pi}\,J(p,q)}},  \\
\label{bivth2} \mathbb{P}\Big[P = \big\lfloor \tfrac{c\ln V}{2\pi}\,p \big\rfloor\,\,{\rm and}\,\,\Xi =  \tfrac{c\ln V}{2\pi}\,q\,\Big|\,V,L = \lfloor V^{\frac{c}{2}}\ell \rfloor \Big] & \stackrel{.}{\asymp} & \frac{1}{(\ln V)\,V^{\frac{c}{2\pi}\,J(p,q)}}.
\end{eqnarray}  
The bivariate large deviations function reads:
\[
J(p,q) = J(p) + q\lambda' - \Lambda_{\mu}(\lambda'),
\]
in terms of $J(p)$ defined in \eqref{larged}, and $\lambda'$ is the function of $(p,q)$ which is the unique solution to
\[
\frac{q}{p} = \frac{\partial \Lambda_{\mu}(\lambda')}{\partial \lambda'}.
\]
\end{theorem}
It is remarkable that the bivariate large deviations function is a sum of two terms, one being the usual $n$-dependent large deviations function for depth $J(p)$, the other being $\mu$-dependent but $n$-independent.

\subsubsection{Bernoulli weights}

For instance, if $\mu$ is the distribution of a signed Bernoulli random variable,
\[
\mu[\xi = -1] = \mu[\xi = 1] = \tfrac{1}{2},
\]
we have
\[
\Lambda_{\mu}(\lambda) = \ln{\rm cosh}(\lambda),\qquad \lambda' = {\rm arctanh}(q/p) = \frac{1}{2}\ln\Big(\frac{p + q}{p - q}\Big),
\]
and
\[
J(p,q) = J(p) + \frac{p + q}{2}\ln(p + q) + \frac{p - q}{2}\ln(p - q) - p\ln p.
\]
Note that, as $\xi \leq 1$, we have $\Xi = \sum_{l \in \mathcal{L}_{{\rm sep}}} \xi_{l} \leq P$, so we must have $q \leq p$.

\subsubsection{Gaussian weights}

If $\xi$ is a centered Gaussian variable with variance $\sigma^2$, we have:
\[
\Lambda_{\mu}(\lambda) = \frac{\sigma^2\lambda^2}{2},\qquad \lambda' = \frac{q}{p\sigma^2}
\]
and therefore:
\[
J(p,q) = J(p) + \frac{q^2}{2\sigma^2p^2}.
\]

\subsection{Proof of Theorem~\ref{bivth}}
We give some details of the proof in the case of finite perimeters, as the modifications necessary in the case of large perimeters, $L = \lfloor V^{\frac{c}{2}}\ell\rfloor$, are parallel to the changes of Section~\ref{finiteA} detailed in Section~\ref{infiniteA}. As the strategy is similar to Section~\ref{finiteA}, we leave the details of the analysis to the reader. To analyse $\mathcal{P}(V,L;P,\Xi)$, we should study the critical points of:
\begin{equation}
\begin{split}
& \quad \mathcal{S}_{1}(x,u,s,\lambda) \\
& = \mathcal{S}_{1}(x,u,se^{\Lambda_{\mu}(\lambda)}) - \lambda\Xi  \\
& = -V\ln u - P\ln s + cb(se^{\Lambda_{\mu}(\lambda)}) + \ln\Big(\frac{1 - u}{\Delta}\Big) - \ln\Big[1 - \Big(\frac{1 - u}{\Delta}\Big)^{cb(s\exp(\Lambda_{\mu}(\lambda)))}\Big] \\
& \quad + \ln \tilde{\Psi}_{b(s\exp(\Lambda_{\mu}(\lambda)))}(x) - \lambda\Xi
\end{split}
\end{equation}
Let $(s^*,\lambda^*)$ be the location of the critical point of $\tilde{\mathcal{S}}_{1}$, and assume that $s^*$ has a limit away from $\frac{2}{n}$, and $\lambda^*$ has a finite limit when $V \rightarrow \infty$. Using the scalings\footnote{Again, $q$ here is not the variable of Theorem~\ref{th38} controlling the distance to criticality.}
\[
P = \frac{c\ln V}{\pi}\,\tilde{p} ,\qquad \Xi = \frac{c\ln V}{\pi}\,q,
\]
we find that the equation $\partial_{s}\tilde{\mathcal{S}}_{1} = 0$ yields in the limit $V \rightarrow \infty$:
\beq
\label{muscol} \frac{ne^{\Lambda_{\mu}(\lambda^*)}}{\sqrt{4 - (ns^*e^{\Lambda_{\mu}(\lambda^*)})^2}} = \frac{\tilde{p}}{s^*},
\eeq
and the equation $\partial_{\lambda}\tilde{\mathcal{S}}_{1} = 0$ yields likewise:
\beq
\label{murcol} \frac{ne^{\Lambda_{\mu}(\lambda^*)}\,\Lambda_{\mu}'(\lambda^*)}{\sqrt{4 - (ns^*e^{\Lambda_{\mu}(\lambda^*)})^2}} = \frac{q}{s^*},
\eeq
while the equation $\partial_{u}\tilde{\mathcal{S}}_{1} = 0$ yields:
\[
V \sim -\frac{cb(s^*e^{\Lambda_{\mu}(\lambda^*)})}{1 - u^*}.
\]
Let us define $\lambda'$ as a function of $(p,q)$ in such a way that:
\beq
\frac{\partial \Lambda_{\mu}}{\partial \lambda'} = \frac{q}{p}.
\eeq  
As $\frac{\partial \Lambda}{\partial \lambda'}(0) = \mathbb{E}[\xi]$ and $\frac{\partial^2\Lambda_{\mu}}{\partial \lambda'^2}(0) = {\rm Var}[\xi] > 0$, $\lambda'$ is defined at least for $\frac{q}{p}$ in the neighborhood of the value $\mathbb{E}[\xi]$, corresponding to $\lambda'$ in a neighborhood of $0$. We assume that $\frac{q}{p}$ belongs to the (maximal) domain of definition of $\lambda'$. Combining \eqref{muscol} and \eqref{murcol}, we find that the saddle $\lambda^*$ is located at $\lambda'$, and:
\[
s^*e^{\Lambda_{\mu}(\lambda')} = \mathfrak{s}(p),\qquad b(s^*e^{\Lambda(\lambda')}) = \mathfrak{b}(p),
\]
in terms of the functions $\mathfrak{s}$ and $\mathfrak{b}$ defined in \eqref{spspsfun}-\eqref{spspsfun2}.

We compute the Hessian matrix of $\tilde{\mathcal{S}}_{1}$ with respect to the variables $(s,\lambda)$, and evaluated at the saddle point $(s^*,\lambda')$. At leading order in $V$,
\[
\tilde{\mathcal{S}}_{1} = \frac{c\ln V}{\pi}\,\Sigma(s,\lambda) + o(\ln V),\quad {\rm with}\,\,\,\, \Sigma(s,\lambda) = \pi b(se^{\Lambda_{\mu}(\lambda)}) - \tilde{p}\ln s,
\]
where the error $o(\ln V)$ is stable under differentiation. After a tedious, but straightforward computation, we find:
\begin{equation}
\begin{split}
\mathbf{H} & := \left(\begin{array}{cc} \partial_{s}^2 \Sigma & \partial_{\lambda}\partial_{s}\Sigma \\ \star & \partial_{\lambda}^2\Sigma\end{array}\right)\bigg|_{\substack{s = s^* \\ \lambda = \lambda'}} \\
& = \left(\begin{array}{cc} \frac{n^2(\tilde{p}^2 + 1)^2}{4}\,\exp(2\Lambda_{\mu}(\lambda')) & \frac{n(1 + \tilde{p}^2)^{3/2}}{2} \frac{\partial \Lambda_{\mu}}{\partial \lambda'}\,e^{\Lambda_{\mu}(\lambda')}\\ \star & \tilde{p}\big[\frac{\partial^2 \Lambda_{\mu}}{\partial \lambda'^2} + (\tilde{p}^2 + 1)(\frac{\partial \Lambda_{\mu}}{\partial \lambda'}\big)^2\big] \end{array}\right),
\end{split}
\end{equation}
where the lower corner of the matrix is deduced by symmetry. We also need to compute
\[
\det \mathbf{H} = \frac{n^2(\tilde{p}^2 + 1)^2}{4}\,\frac{\partial^2 \Lambda_{\mu}}{\partial \lambda'^2}\,e^{2\Lambda_{\mu}(\lambda')}.
\]

Now, if we introduce the change of variables:
\[
u = 1 - \frac{\tilde{u}}{V},\qquad s =e^{-\Lambda_{\mu}(\lambda')}\mathfrak{s}(\tilde{p}) + \frac{\tilde{s}}{\sqrt{\ln V}},\qquad \lambda = \lambda' + \frac{\tilde{\lambda}}{\sqrt{\ln V}},
\]
we obtain 
\begin{equation*}
\begin{split}
\frac{\dd u}{u^{V + 1}}\frac{\dd s}{s^{P + 1}} \dd \lambda\,e^{-\lambda\Xi} & \sim - \frac{\dd\tilde{u}\,\dd\tilde{\lambda}}{V\ln V}\,\frac{\tilde{\Psi}_{\mathfrak{b}(s\exp(\Lambda_{\mu}(\lambda'))}(x)}{\mathfrak{s}(\tilde{p})\exp(-\Lambda_{\mu}(\lambda'))}\,e^{\tilde{u}}\Big(\frac{\tilde{u}}{\Delta V}\Big)^{c\mathfrak{b}(\tilde{p})} \\ 
& \times V^{-\frac{c}{\pi}(\tilde{p}\ln \mathfrak{s}(\tilde{p}) -  \tilde{p}\Lambda_{\mu}(\lambda') + q\lambda')} \exp\Big\{\frac{c}{\pi}\,(\tilde{s},\tilde{\lambda})\cdot \mathbf{H} \cdot (\tilde{s},\tilde{\lambda})^{T}\Big\}.
\end{split}
\end{equation*}
We can perform the Gaussian integration in $\tilde{s}$ and $\tilde{\lambda}$, while the remaining integration on $\tilde{u}$ and $x$ result in a prefactor already appearing in Section~\ref{finiteA}. The result is:
\begin{equation*}
\begin{split}
\mathcal{P}\Big[V,L;P = \frac{c\tilde{p}\ln V}{\pi}\,;\,\Xi = \frac{cq\ln V}{\pi}\Big] & \sim \frac{\pi}{\Gamma(-c\mathfrak{b}(\tilde{p}))}\,\frac{n^2}{c\tilde{p}\sqrt{(\tilde{p}^2 + 1) \frac{\partial^2\Lambda_{\mu}}{\partial \lambda'^2}}}\bigg\{\oint_{\overline{\mathcal{C}}}\,\frac{x^{L}\dd x}{2{\rm i}\pi}\,\frac{\tilde{\Psi}_{\mathfrak{b}(\tilde{p})}(x)}{\Delta^{c\mathfrak{b}(\tilde{p})}}\bigg\} \\ 
& \times V^{-1 - c\mathfrak{b}(\tilde{p}) + \frac{c}{\pi}(- \ln \mathfrak{s}(\tilde{p}) + \Lambda_{\mu}(\lambda') - q\lambda')}\,(\ln V)^{-1}.  
\end{split}
\end{equation*}
We obtain the final result \eqref{bivth1} by dividing by $\tilde{\mathcal{P}}(V,L)$ given in \eqref{gof2}, and replacing $\tilde{p}$ with $p$ such that $P = \big\lfloor \frac{cp\ln V}{\pi} \rfloor$, which only affect a $O(1)$ term after taking logarithms. The proof of \eqref{bivth2} is similar. 
\hfill $\Box$

\section{Comparison with nesting in CLE via KPZ}\label{sec:CLE}
In this section, we compare the large deviations of loop nesting at criticality in the $O(n)$ model on a random planar map, as derived in the first sections of this work, with the large deviations of loop nesting in the so-called  \emph{conformal loop ensemble} in the plane. 

\label{CLEcomparison}
\subsection{Nesting in the conformal loop ensemble} The conformal loop ensemble $\mathrm{CLE}_\kappa$ for $\kappa\in (8/3,8)$ is the canonical conformally invariant measure on countably infinite collections of non crossing loops in a simply connected domain $D\subset \mathbb C$ \cite{sheffield2009,zbMATH06121652}. It is the analogue for loops of the celebrated \emph{Schramm--Loewner evolution} $\mathrm{SLE}_\kappa$, the canonical conformally invariant measure on non crossing paths \cite{schramm} in the plane, depending on the real positive parameter $\kappa$,  an  invention which is on par with Wiener's 1923 mathematical construction of continuous Brownian motion. It gives  the universal continuous scaling limit of 2d critical curves; of particular physical interest are the loop-erased random walk
($\kappa=2$) \cite{MR2044671}, the  self-avoiding walk ($\kappa= \frac{8}{3}$),
the Ising model interface ($\kappa=3$ or $\frac{16}{3}$) \cite{MR2680496,1257.82020}, the GFF contour lines ($\kappa=4$) \cite{MR2486487}, and the percolation interface ($\kappa=6$) \cite{MR1851632}. 
  Critical phenomena in the plane were earlier  well known to be related to conformal field theory \cite{MR757857}, a discovery 
anticipated in the so-called Coulomb gas approach to critical 2d statistical models (see, \textit{e.g.}, \cite{1982PhRvL..49.1062N,1983PhRvB..27.1674D}), and now including  SLE \cite{2003CMaPh.239..493B,2003CMaPh.243..105F,1280.81004}. 

In the same way as $\mathrm{SLE}_\kappa$ is proven or expected to be the scaling limit of a single interface in 2d critical discrete models, $\mathrm{CLE}_\kappa$ should be the limiting process of the collection of closed interfaces in such models. In particular, the critical $O(n)$-model on a regular planar lattice is expected to converge in the continuum limit to the universality class of the $\mathrm{SLE}_\kappa$/$\mathrm{CLE}_\kappa$, for 
 \beq\label{nkappa}
 n= 2\cos\big[\pi\big(1 - {4}/{\kappa}\big)\big], \qquad n\in [0,2],\,\,\,\begin{cases} \kappa\in (\frac{8}{3},4]\,\,\,{\rm in}\,\,{\rm dilute}\,\,{\rm phase} \\ \kappa\in (4,8)\,\,\,{\rm in}\,\,{\rm dense}\,\,{\rm phase}.\end{cases}
 \eeq
 
 In \cite{MWW} (see also \cite{MWW2}), Miller, Watson and Wilson were able to derive the almost sure multifractal dimension spectrum of \emph{extreme nesting} in the conformal loop ensemble. Fix a simply connected proper domain $D\subset \mathbb C$ and let $\Gamma$ be a configuration of $\mathrm{CLE}_\kappa$. For each point $z\in D$,  let $\mathcal N_z(\varepsilon)$ be the number of loops of $\Gamma$ which surround the ball $B(z,\varepsilon)$ centered at $z$ and of radius $\varepsilon >0$. For $\nu >0$, define the random set
 \beq\label{phinu}
\Phi_\nu= \Phi_\nu(\Gamma):=\left\{z\in D\,\,:\quad \lim_{\varepsilon \rightarrow 0} \frac{\mathcal{N}_z(\varepsilon)}{\ln(1/\varepsilon)} = \nu \right\}.
 \eeq 
This Hausdorff dimension of this set is almost surely equal to a constant, which is expressed in terms of the distribution of the conformal radius of the gasket of the origin in a $\mathrm{CLE}_\kappa$ in the unit disk $\mathbb D$. More precisely, the {\emph conformal radius} $\mathrm{CR}(z,\mathcal U)$ of a simply connected proper domain $\mathcal U\subset \mathbb C$ is defined to be $|\varphi'(0)|$, where $\varphi$ is the conformal map $\mathbb D \mapsto \mathcal U$ which sends $0$ to $z$. For a configuration $\Gamma$ of ${\rm CLE}_{\kappa}$ in $\mathbb{D}$, let then $\mathcal U_{\Gamma}$ be the connected component containing the origin in the complement $\mathbb D\setminus \mathcal L$  of the largest loop $\mathcal L$ of $\Gamma$ surrounding the origin in  $\mathbb D$, {\it i.e.} the interior of the outmost such loop. A formula for the cumulant generating function of $-\log(\mathrm{CR}(0,\mathcal U_{\Gamma}))$ was proposed in \cite{JCR,Cardy2003,Dubedat,KW04} and established in Ref. \cite{SSW} 
\beq\label{Lambda}
\Lambda_\kappa(\lambda):=\ln {\mathbb E}\Big[\big({\rm CR}(0,\mathcal{U}_{\Gamma})\big)^{-\lambda}\Big]=\ln\left(\frac{\cos\big[\pi(1 - \frac{4}{\kappa})\big]}{\cos\Big[\pi \sqrt{\left(1-\frac{4}{\kappa}\right)^2+\frac{8\lambda}{\kappa}}\Big]}\right),\quad \lambda \in \big(-\infty,1-\tfrac{2}{\kappa}-\tfrac{3\kappa}{32}\big).
\eeq 
The Legendre--Fenchel symmetric transform, $\Lambda_\kappa^{\star}:\mathbb R\rightarrow \mathbb R_+$ of $\Lambda_\kappa$ is defined by
\beq\label{Lambdastar}
\Lambda^{\star}_\kappa(x):=\sup_{\lambda \in \mathbb R}\left(\lambda x-\Lambda_\kappa(\lambda)\right).
\eeq 
It yields the continuous function on $\mathbb R_+$ \cite{MWW},
\beq\label{gammanu}
\gamma_\kappa(\nu):= \left\{\begin{array}{lll} \nu\Lambda^{\star}_\kappa(1/\nu) & \quad & {\rm if}\,\,\nu >0, \\ 1- \tfrac{2}{\kappa}-\tfrac{3\kappa}{32} & \quad & {\rm if}\,\,\nu = 0.\end{array}\right.
\eeq 
For $\kappa\in (8/3,8)$, the Hausdorff dimension of the set $\Phi_\nu$ is almost surely given by \cite[Theorem 1.1]{MWW},
\beq\label{dim}
\mathrm{dim}_{\mathcal H}\,\Phi_\nu = \max(0,2-\gamma_{\kappa}(\nu)),
\eeq
with $\Phi_\nu$ being a.s. empty if $\gamma_{\kappa}(\nu)>2$.
Note that the Legendre--Fenchel transform equations above can be recast for $\gamma_\kappa(\nu)$, $\nu >0$, as,
\beq
\label{nulambda}
\frac{\gamma_\kappa(\nu)}{\nu}=\frac{\lambda}{\nu} -\Lambda_\kappa(\lambda),\qquad  \frac{1}{\nu}= \frac{\partial \Lambda_\kappa(\lambda)}{\partial \lambda},
\eeq
from which we immediately get,
\beq\label{lambdanu}
\lambda =\frac{\partial}{\partial ({1}/{\nu})}\left( \frac{\gamma_\kappa(\nu)}{\nu}\right)
=\gamma_\kappa(\nu)-\nu\frac{\partial}{\partial \nu}\gamma_\kappa(\nu).
\eeq

\subsection{Liouville quantum gravity}

  Polyakov \cite{MR623209} suggested in 1981 that the summation
over random Riemannian metrics involved in a continuum theory of random surfaces 
 could be represented canonically by the now celebrated \textit{Liouville theory of quantum gravity} (see \cite{Ginsparg-Moore,Ginsparg,MR2073993,2008arXiv0808.1560D} and references therein). It is widely believed or proven in certain cases to provide, after a Riemann conformal map to a given planar domain, the correct conformal structure for the continuum limit of random planar maps, possibly weighted by the partition functions of various statistical models (see, \textit{e.g.}, the ICM reviews \cite{LeGallICM2014,DuplantierICM2014,2017arXiv171201571M}). In the case of usual random planar maps with faces of bounded degrees, the universal metric structure is that of the Brownian map \cite{Gall1,MierMapP}, which has been recently identified with that directly constructed from Liouville quantum gravity (LQG) \cite{map_making,qlebm,2016arXiv160503563M,2016arXiv160805391M}. Note also that different mathematical approaches to LQG exist \cite{2008arXiv0808.1560D,LQGmating,quantum_spheres,David2016}, whose  equivalence has been recently established \cite{Aru2017}.
 
As mentioned in the introduction, Section \ref{intro}, several models of random planar maps with critical statistical models have now been rigorously proven to converge to LQG surfaces, as path-decorated metric spaces  \cite{2016arXiv160800956G,2017arXiv170105175G}, as mated pairs of trees \cite{sheffield2016bis,2015arXiv150200546G,gwynnesun2017,2015arXiv151006346G,2015arXiv151104068K,2016arXiv160301194G,2016arXiv160309722G,LiSunWatson} in  the so-called peanosphere topology  of Refs. \cite{LQGmating,quantum_spheres}, or as Tutte embedding of mated-CRT maps \cite{2017arXiv170511161G}.

 Here, in order to compare the asymptotic findings of previous sections to a direct LQG approach, we focus on the \emph{measure}  aspects associated with Liouville quantum gravity.
 \subsubsection{Liouville quantum measure \cite{2008arXiv0808.1560D}}

Consider a  simply connected domain  $D\subset \mathbb C$ as the parameter domain of the random
 surface, and  $h$  an instance of the massless \textit{Gaussian free field} (GFF), a random distribution 
on $D$, associated with the  Dirichlet energy, 
 \begin{equation*}
(h,h)_{\nabla}:=\frac{1}{2\pi}
\int_D [\nabla h(z)]^2 \mathrm{d}^2z , 
\end{equation*} 
and whose two point correlations are given by the Green's function on $D$ with Dirichlet zero boundary conditions   \cite{MR2322706}. 
(Critical) Liouville quantum gravity consists in changing the Lebesgue area measure $\mathrm{d}^2z $ on $D$ to the \textit{quantum area measure}, formally written as 
 $\mu_\gamma(\mathrm{d}^2z ):=e^{\gamma h(z)} \mathrm{d}^2z $, where $\gamma$ is a real parameter. The GFF $h$ is a random distribution, not a function, but the random measure $\mu_\gamma$
can be constructed, for $\gamma \in [0,2]$, as the limit of regularised quantities, as follows. 
 
Given an instance $h$ of the GFF on $D$, for each $z \in D$, let 
$h_\varepsilon(z)$ denote the mean value of $h$ on the circle of
radius $\varepsilon$ centered at $z$  --- where $h(z)$ is defined to be
zero for $z \in \mathbb C \setminus D$ \cite{MR2322706}.  One then has 
\begin{equation*} 
\mathbb{E}\big[e^{\gamma h_\varepsilon(z)}\big] =e^{ \gamma^2{\mathrm{Var}}[h_\varepsilon(z)]/2}
= \left[\mathrm{CR}(z,D)/{\varepsilon} \right]^{\gamma^2/2},
\end{equation*} 
 where $\mathrm{CR}(z,D)$ the conformal radius of $D$
viewed from $z$.  

 This strongly suggests considering the limit, 
\beq
\label{liouvillemeasure}
\mu_{\gamma}(\mathrm{d}^2z ):=\lim_{\varepsilon\to 0}\varepsilon^{\gamma^2/2} e^{\gamma h_\varepsilon(z)}\mathrm{d}^2z,
\eeq
and one can indeed show that for $\gamma \in [0,2)$ this (weak) limit exists and is non degenerate, and is singular with respect to Lebesgue  measure \cite{2008arXiv0808.1560D}. This mathematically defines Liouville quantum gravity, 
in a way reminiscent of so-called Wick normal ordering in quantum field theory --- see also \cite{MR0292433} 
 for earlier work on the so-called H{\o}egh-Krohn model, and Kahane's general study of the so-called Gaussian multiplicative chaos \cite{MR829798}.

The critical case, $\gamma=2$, requires additional care, and it is shown in \cite{DRSV1,MR3215583} (see also \cite{2018arXiv180208433A})
that the weak limit, 
\beq
\label{liouvillemeasure2}
\mu_{\gamma=2}(\mathrm{d}^2z ):=\lim_{\varepsilon\to 0}\sqrt{\ln (1/\varepsilon)}\,\varepsilon^{2} e^{2 h_\varepsilon(z)}\mathrm{d}^2z ,
\eeq
exists and is almost surely non atomic. $\mu_{\gamma}(D)$ will be called the \emph{quantum area} of $D$.

\begin{remark} The Liouville quantum action is usually written as  
$S(h)=\frac{1}{2}(h,h)_{\nabla}+\mathfrak{b}\, \mathcal \mu_\gamma(D),$
where the  ``(bulk) cosmological constant", $\mathfrak{b} \geq 0$, weights the partition function according to the quantum area of the random surface. The corresponding Boltzmann 
statistical weight, $\exp[-S(h)]$,  should be integrated over with a ``flat'' uniform functional measure $\mathcal D h$ on $h$ --- which makes sense {\it a priori} for finite-dimensional approximations to $h$. The full Liouville quantum measure can then be constructed from the GFF one (see, \textit{e.g.}, \cite{David2016}), and for our purpose of studying the $\mathrm{CLE}_\kappa$ nesting properties, which are \emph{local} ones, it will suffice to consider this measure for  $\mathfrak b=0$, \textit{i.e.}, in the GFF case.\end{remark} 

\subsubsection{Canonical coupling of LQG to SLE}
 Various values of $\gamma$ are expected 
to describe weighting the random map by the 
partition function of a critical statistical physical model defined on that 
map (\textit{e.g.}, an Ising model, an $O(n)$ or a Potts model). 
The correspondence can be obtained by first considering \emph{conformal welding} in Liouville quantum gravity \cite{sheffield2016,sheffield2016bis,2011PhRvL.107m1305D,LQGmating} (see also \cite{Astala2011random}). It turns out that pieces of Liouville quantum gravity surfaces of parameter $\gamma \in [0,2)$ can be conformally welded together to produce as random seams $\mathrm{SLE}_\kappa$ curves, with the rigorous result,
\beq\label{gammakappa}
\gamma= \left\{\begin{array}{lll} \sqrt{\kappa} & \quad & {\rm if}\,\,\kappa < 4\\
\frac{4}{{\sqrt{\kappa}}} & \quad & {\rm if}\,\,\kappa>4. \end{array}\right.
\eeq
Together with \eqref{nkappa}, this provides us with the $(\gamma,\kappa,n)$ correspondence that we sought after for the  $O(n)$ model. 

\subsubsection{{\rm KPZ} formula} By the usual conformal invariance \textit{Ansatz} in phy\-sics, it is natural to expect
that if one conditions on the random map to be infinite, maps it into the plane, and then samples the loops or clusters in critical models, their law in the scaling limit will be {\em independent} of the random measure. This independence in turn leads  to the Knizhnik, Polyakov, and Zamolodchikov (KPZ) formula 
 \cite{KPZ} --- see also Refs. \cite{MR981529,MR1005268} --- which is a
relationship between (half) scaling dimensions (\textit{i.e.}, conformal weights $x$) of fields defined using Euclidean geometry and analogous dimensions ($\Delta$) 
defined via the Liouville quantum gravity measure $\mu_\gamma$, 
\beq
x =U_\gamma(\Delta):=\frac{\gamma^2}{4}\Delta^2+\left(1-\frac{\gamma^2}{4}\right)\Delta.
\label{KPZ}
\eeq
The inverse of the relation \eqref{KPZ} that is  positive is given by
\beq
\Delta=U^{-1}_{\gamma}(x):= \frac{1}{\gamma}\Big(\sqrt{4x +a^2_\gamma}-a_\gamma\Big),\qquad  a_\gamma:= \Big(\frac{2}{\gamma}-\frac{\gamma}{2}\Big) \geq 0.
\label{invKPZ}
\eeq
A mathematical proof of the KPZ relation, based on the stochastic properties of the GFF, first appeared in  \cite{2008arXiv0808.1560D}; it  was then also proved for multiplicative cascades \cite{2009CMaPh.tmp...46B} and in the framework of  Gaussian multiplicative chaos  \cite{PSS:8474530,MR3215583}. The KPZ formula holds for any
fractal structure sampled {\it independently} of the GFF,  and measured with the random measure $\mu_\gamma$, and  for any $0\leq
\gamma \leq 2$. 

\subsubsection{Quantum and Lebesgue measures} Define the (random) Liouville quantum measure of the Euclidean ball $B(z,\varepsilon)$,
\beq
\label{delta}
\delta:=\int_{B(z,\varepsilon)} \mu_{\gamma}(\mathrm{d}^2z ),
\eeq
and the logarithmic coordinates,
\beq
\label{tA}
t:= \ln(1/\varepsilon),\qquad A:= \gamma^{-1}\ln(1/\delta).
\eeq
For $z$ fixed, a given quantum area $\delta$, hence a given logarithmic coordinate $A$, corresponds through \eqref{delta} to a random Euclidean radius $\varepsilon$, and  the corresponding random value $T_A$ of $t$ in \eqref{tA}  can be seen as a stopping time of some Brownian process \cite{2009arXiv0901.0277D,2008arXiv0808.1560D}. The probability density of $T_A$, such that $\mathcal{P}(t\,|\,A)\dd t:= \mathbb{P}\big(T_A\in [t, t+\dd t]\big)$, is obtained as a by-product of the KPZ analysis in \cite{2009arXiv0901.0277D,2008arXiv0808.1560D}:
\beq \label{PA} \mathcal{P}(t\,|\,A)=\frac{A}{\sqrt{2\pi t^3}} \exp\bigg(-\frac{(A
-a_\gamma  t)^2}{2t}\bigg);
\eeq
it characterises, in logarithmic coordinates, the distribution of the Euclidean radius $\varepsilon$ of a ball of given quantum area $\delta$.  For a GFF in a domain $D$ with, say, Dirichlet boundary conditions, this local form is valid for a ball $B(z,\varepsilon)$ far away from $\partial D$, i.e., for $\varepsilon$ much smaller than the conformal radius $C(z,D)$, or, equivalently, than the Euclidean distance between $z$ and the boundary $\partial D$, since $\frac{1}{4}C(z,D)\leq \mathrm{dist}(z,D)\leq C(z,D)$ by Koebe $\frac{1}{4}$ theorem.

 Note that we can rewrite it as 
\beq \label{PA'} \mathcal{P}(t' A\,|\,A)=\frac{A^{-1/2}}{\sqrt{2\pi t'^3}} \exp\bigg(-\frac{A}{2t'}(1
-a_\gamma {t'})^2\bigg).
\eeq
In the regime  $\delta \to 0$, we have $A\to +\infty$, so the distribution  \eqref{PA'} becomes localised at  $a_\gamma t'=1$, thus $t=A/a_\gamma$. This gives the typical scaling of the quantum area of balls in $\gamma$-Liouville quantum gravity,  
$\ \delta \asymp \varepsilon^{\gamma a_\gamma}=\varepsilon^{2-\gamma^2/2}$ \cite{2009arXiv0902.3842H}.  The large deviations from this typical value, associated with \eqref{PA'}, will be the key in comparing the extreme nesting of $\mathrm{CLE}_\kappa$ in the plane, as seen with the Euclidean (Lebesgue) measure, or with the Liouville quantum measure $\mu_{\gamma}$.

\subsection{Nesting of \texorpdfstring{$\mathrm{CLE}_\kappa$}{CLEkappa} in Liouville quantum gravity}\label{CLE-LQG}\label{sec:NLQG}
\subsubsection{Definition}One ingredient in the proof of \eqref{dim} in Ref. \cite{MWW} is the following one-point estimate \cite[Lemma 3.2]{MWW}. For $z\in D$, define
\[
\widetilde {\mathcal N}_z(\varepsilon):=\frac{{\mathcal N}_z(\varepsilon)}{\ln (1/\varepsilon)}.
\]
Then  
\beq
\label{logPnu} 
\lim_{\varepsilon\to 0}\frac{\ln \mathbb P\big(\widetilde {\mathcal N}_z(\varepsilon) \in [\nu-\omega_-(\varepsilon),\,\nu+\omega_+(\varepsilon)]\big)}{\ln \varepsilon}=\gamma_\kappa(\nu) \quad \mathrm{for}\,\, \nu>0,
\eeq
for  $z\in D$ satisfying $a\leq \mathrm{CR}(z,D)\leq b$, with $0<a\leq b$, and for all $\omega_{\pm}(\varepsilon)$ decreasing to $0$ sufficiently slowly. A similar result holds for $\nu=0$. We shall rewrite the above result, for $\varepsilon \to 0$, in the more compact way,
\beq
\label{Pnueps} 
\mathbb P\big({\mathcal N}_z \approx \nu\ln (1/\varepsilon)\,\big|\,\varepsilon \big) \asymp \varepsilon^{\gamma_\kappa(\nu)},
\eeq
where the sign $\approx$ stands for a scaling of the form $(\nu+o(1))  \ln (1/\varepsilon)$. We also recall that the $\asymp$ sign means an asymptotic equivalence of logarithms, {\it i.e.}, a form $\varepsilon^{\gamma_\kappa(\nu)+o(1)}$ on the r.h.s. Recalling definition \eqref{tA}, this is also for $t\to +\infty$, 
\beq
\label{Pnut} 
\mathbb{P}\big({\mathcal N}_z \approx \nu t\, \big|\,t \big) \asymp e^{-\gamma_\kappa(\nu)t}.
\eeq

To define an analog of this nesting probability in LQG, 
instead of conditioning on the Euclidean radius $\varepsilon$ --- equivalently, on $t$ --- we condition on the quantum area $\delta$ \eqref{delta} of the ball $B(z,\varepsilon)$ --- equivalently, on $A$ \eqref{tA}. The number of loops $\mathcal N_z$ surrounding the ball $B(z,\varepsilon)$ stays the same. This conditional probability is then given by the convolution,
\beq
\label{PQ}
\mathbb P_{\mathcal{Q}}(\mathcal N_z\,|\,A) :=\int_0^\infty \dd t\, \mathbb P(\mathcal N_z\,|\,t)\,\mathcal{P}(t\,|\,A),
\eeq
where $\mathcal{P}(t\,|\,A)$ is as in \eqref{PA}-\eqref{PA'}. We call it the \emph{quantum nesting probability}.
 
\subsubsection{Saddle point computation}\label{Largedeviations}
For large $A$, if we let $\mathcal N_z$ scale as $\mathcal N_z \approx \gamma p A$, with $p\in \mathbb R_+$, we may also set $\mathcal N_z \approx \nu t$, where $\nu$ is now \emph{defined} as
\beq
 \label{Nnut} \nu=\nu(t)= \frac{\gamma p A}{t},
\eeq 
where $p$ and $A$ are considered as parameters. Note that expressions \eqref{Pnut} and \eqref{PA'} for the two probability distributions appearing in the integrand in  \eqref{PQ} are large deviations forms for $t$ and $A$ both large and in a finite ratio. Such is the case in  \eqref{Nnut}, which yields   
\beq
\label{PnutPA}
\mathbb P_{\mathcal{Q}}(\mathcal N_z \approx \gamma pA\,|\,A) \asymp \int_0^\infty \frac{\dd t\,A}{\sqrt{2\pi t^3}} \exp\bigg(-\frac{({A}
-a_\gamma {t})^2}{2t}-\gamma_\kappa(\nu)t\bigg),
\eeq
where for $A\to +\infty$ the integral over $t$ is consistently evaluated by a saddle point method. We thus look for the extremum of
\beq
{\mathcal E}(t):=\frac{1}{2t}\left({A}
-a_\gamma {t}\right)^2+\gamma_\kappa(\nu)t,
\eeq
along trajectories at \emph{constant} value of $\nu t=\gamma p A$, and for \emph{fixed} $p$ and $A$.  
We then have
\[
t \frac{\partial \gamma_\kappa}{\partial t} (\nu)=-\nu \frac{\partial \gamma_\kappa}{\partial \nu}(\nu),
\]
and using \eqref{lambdanu},
\begin{equation}\label{lambdanurec}
\frac{\partial}{\partial t}\big(\gamma_\kappa(\nu) t\big)=\gamma_\kappa(\nu)-\nu\frac{\partial \gamma_\kappa}{\partial \nu}(\nu) =\lambda.
\end{equation}
This in turn gives
\beq 
\label{eq:dEdt}
\frac{\partial \mathcal{E}}{\partial t} =\lambda -\frac{1}{2}\Big[\Big(\frac{A}{t}\Big)^2-a^2_\gamma\Big],
\eeq
and a saddle point value $t^*$ of $t$ at 
\beq
\label{Atstar}
\frac{A}{t^*} := u = u(\lambda) := \sqrt{2\lambda+a_\gamma^2}. 
\eeq
which is implicitly a function of $p$.
 
Note that from \eqref{lambdanu} again,
\[
\frac{\partial \lambda}{\partial t} =-\nu \frac{\partial}{\partial \nu}\Big(\gamma_\kappa(\nu)-\nu \frac{\partial \gamma_\kappa}{\partial \nu}\Big) =\nu^2\,\frac{\partial^2 \gamma_\kappa}{\partial \nu^2} > 0
\]
so that 
\[
\frac{\partial^2 {\mathcal E}}{\partial t^2} = \frac{\partial \lambda}{\partial t} + \frac{A^2}{t^3}>0.
\]
And the saddle point lies, as expected, at the minimum $\mathcal E^*$ of ${\mathcal E(t)}$, 
\beq
\label{Estar}
{\mathcal E}^*:={\mathcal E}(t^*)=A\bigg(\frac{(u -a_\gamma)^2}{2u}+\frac{\gamma_\kappa(\nu)}{u}\bigg),
\eeq
where, owing to  definition \eqref{Nnut} and to \eqref{Atstar}, $\nu$ is hereafter understood as the \emph{saddle point value},
\beq\label{nustar}
\nu =\nu(t^*)=\gamma p\,\frac{A}{t^*}=\gamma p \,u(\lambda).
\eeq
Owing to \eqref{nulambda}-\eqref{Atstar} and \eqref{nustar}, we have 
\beq
\label{gammak/u}
\frac{\gamma_\kappa(\nu)}{u} = \frac{\lambda -\nu\Lambda_\kappa(\lambda)}{u} = \frac{u^2-a_\gamma^2}{2u}-\gamma p\Lambda_\kappa(\lambda),
\eeq
 so that we finally get the simple form,
\beq
\label{Estarprefinal}
 \frac{\mathcal{E}^*}{A} =u(\lambda)-a_\gamma-\gamma p\,\Lambda_\kappa(\lambda).
\eeq
Notice that  \eqref{nulambda}, \eqref{Atstar} and \eqref{nustar} also imply 
\beq
\label{punu}
\frac{1}{\gamma p}=  \frac{u}{\nu}= u(\lambda)\,\frac{\partial \Lambda_\kappa(\lambda)}{\partial \lambda} = \frac{\partial \Lambda_\kappa(\lambda)}{\partial u(\lambda)}.
\eeq

\subsubsection{Role of the KPZ relation}
Let us define:
\[
\Theta(p):= \frac{\mathcal{E}^*}{\gamma A}.
\]
We have just computed:
\beq
\label{thetap}\Theta(p)=U^{-1}_\gamma(\tfrac{\lambda}{2})-p\,\Lambda_\kappa(\lambda),
\eeq
where $\lambda$ is the function of $p$ determined by \eqref{punu}, and where the inverse KPZ relation \eqref{invKPZ} precisely yields, 
\beq
\label{Uu}
 U^{-1}_\gamma\big(\tfrac{\lambda}{2}\big)=\frac{u(\lambda)-a_\gamma}{\gamma}.
\eeq
Note also that $\frac{1}{p}$ as in \eqref{punu} is the derivative of $\Lambda_\kappa$ with respect to \eqref{Uu}. 
Thus, setting $\lambda' :=U^{-1}_\gamma(\frac{\lambda}{2})$, we get the Legendre--Fenchel transform equations:
\beq
\label{Theta-p}
\Theta(p)= \lambda' -p\,(\Lambda_\kappa\circ 2 U_\gamma)(\lambda'),\qquad \frac{1}{p}= \frac{\partial (\Lambda_\kappa\circ 2 U_\gamma)(\lambda')}{\partial \lambda'}.
\eeq
 Comparing this result to the Legendre--Fenchel equations \eqref{nulambda} in the Euclidean case, we get 
\begin{theorem}\label{theo:LambdaQ}
 In presence of $\gamma$-Liouville quantum gravity, 
the generating function $\Lambda_\kappa$ \eqref{Lambda} is transformed into
\beq
\label{LambdaG}\Lambda_\kappa^{\mathcal{Q}} := \Lambda_\kappa\circ 2U_\gamma,
\eeq
where $U_\gamma$ is the KPZ function \eqref{KPZ}, with $\gamma$ given by \eqref{gammakappa}.  The nesting distribution around a ball of given quantum area $\delta$ \eqref{PnutPA} is then given asymptotically for $A= \gamma^{-1}\ln(1/\delta) \to +\infty$, by
\[
\mathbb P_{\mathcal{Q}}\big(\mathcal N_z \approx \gamma pA\,\big|\,A\big) \asymp e^{-\gamma \Theta(p) A} =\delta^{\Theta(p)}
\]
with $\Theta(p) = \lambda -p\Lambda^{\mathcal{Q}}_\kappa(\lambda)$ and $\lambda$ is determined as a function of $p$ by:
\[
\frac{1}{p} = \frac{\partial \Lambda^{\mathcal{Q}}_\kappa(\lambda)}{\partial \lambda}.
\]
\end{theorem} 
\begin{remark}
The occurrence of a factor 2 in the composition law \eqref{LambdaG} is simply due to a different choice of scale when measuring large deviations, {\it i.e.}, that of a \emph{quantum area} $\delta$ in the quantum case, as opposed to that of a \emph{radius} $\varepsilon$ in the Euclidean one. This is seen in particular in the $\kappa \to 0$ limit, where $U_\gamma $ simply becomes the identity function.  
\end{remark}
\begin{remark}\label{rk:KPZ}
Theorem \ref{theo:LambdaQ} shows that the KPZ relation, or its inverse as in \eqref{thetap}, can directly act on an arbitrary continuum variable, here the conjugate variable in the cumulant generating function \eqref{Lambda} for the $\mathrm{CLE}_\kappa$ log-conformal radius. To our knowledge, this is the first occurrence of such a role for the KPZ relation, which usually concerns scaling dimensions.
\end{remark}
\begin{remark}
The derivation above does not depend on the precise form of the large deviations function. Moreover, as shown in Refs. \cite{2009arXiv0901.0277D,2008arXiv0808.1560D}, the KPZ relation holds in full generality for any (fractal) random system in the plane and in Liouville quantum gravity, provided that the sampling of the random system is \emph{independent} of that of the Gaussian free field defining LQG. Therefore, the map $\Lambda \mapsto \Lambda^{\mathcal Q}=\Lambda\circ 2 U_\gamma$,  from  Euclidean geometry to Liouville quantum gravity,  holds for any large deviations problem, where the large deviations function is the Legendre--Fenchel transform of a certain  generating function $\Lambda$.
\end{remark}
\subsubsection{Comparison to Theorem \ref{theo:main}} 
Let us finally compute explicitly the Liouville large deviations function $\Theta$, in order to compare with the main results above  regarding  extreme nesting in the $O(n)$ model on a random planar map. 
The easiest way is to rewrite \eqref{Lambda} as 
\beq
\label{Lambdav}
\Lambda_\kappa(\lambda)=\ln\left(\frac{\cos\big[\pi\big(1 - \frac{4}{\kappa}\big)\big]}{\cos v}\right), \qquad 
 v = v(\lambda) := \frac{2\pi}{\sqrt{\kappa}}\sqrt{\left(\frac{\sqrt{\kappa}}{2} - \frac{2}{\sqrt{\kappa}}\right)^2 + 2\lambda}
 \eeq
for $\lambda \in (-\infty,1-\frac{2}{\kappa}-\frac{3\kappa}{32})$, and to notice that \eqref{Atstar} and \eqref{gammakappa} give 
\[
u(\lambda)=\frac{\sqrt{\kappa}}{2\pi}\,v(\lambda).
\]
Equations \eqref{punu} and \eqref{thetap} then take the compact form,
\beq
\label{Thetav}
\Theta(p)=\Theta=\frac{c}{2\pi}\,(v - a' - p'\Lambda_\kappa),\qquad \frac{1}{p'}=   \frac{\partial \Lambda_\kappa(v)}{\partial v},
\eeq
where we used the notations
\[
c:= \frac{\sqrt{\kappa}}{\gamma},\qquad p:= \frac{c}{2\pi}\,p',\qquad a_\gamma = \Big(\frac{2}{\gamma}-\frac{\gamma}{2}\Big) := \frac{\sqrt{\kappa}}{2\pi}\,a'.
\]
Because of \eqref{gammakappa}, we find as parameters, 
\beq
\label{ac}
c=\min\big(1,\tfrac{\kappa}{4}\big),\qquad a' =\pi b=\pi\big|1-\tfrac{4}{\kappa}\big|=\arccos\big(\tfrac{n}{2}\big),
\eeq
where $b$ and $c$ are the exponents defined in \eqref{eq:bnot} and Section \ref{Sec:ce}. 
The explicit form \eqref{Lambdav}  immediately yields the parametric solution to Legendre--Fenchel equations \eqref{Thetav},
\beq
\label{paramQ}
p' =\cot v,\qquad \Theta=\frac{c}{2\pi}\bigg[v- (\cot v) \ln\Big(\frac{n}{2\cos v}\Big)- \arccos(n/2)\bigg].
\eeq
One has $p' \in \mathbb R_+$ for $v\in [0,\frac{\pi}{2})$, so that
\[
\cos v= \frac{p'}{\sqrt{p'^2+1}} \geq 0,
\]  
which finally yields
\beq\label{Thetap'}
\Theta(p)=\frac{c}{2\pi}\,J(p'),\qquad J(p'):= \mathrm{arccot}(p') + p'\ln\bigg(\frac{2}{n}\frac{p'}{\sqrt{1+p'^2}}\bigg) -\arccos\Big(\frac{n}{2}\Big).
\eeq
Note that the $p = \frac{c}{2\pi}\,p'$ substitution above simply gives $\gamma p A = \frac{c}{2\pi}\,p'\ln(1/\delta)$. Theorem \ref{theo:LambdaQ} then yields
\begin{theorem} \label{theo:Jp'} The quantum nesting probability for $\mathrm{CLE}_\kappa$ loops,  with $\kappa\in(8/3,8)$ in a simply connected proper domain $D \subset \mathbb C$, surrounding a ball centered at $z$ with given quantum area $\delta$, behaves as
\begin{equation*}\label{PQp'}
\mathbb P_{\mathcal{Q}}\big(\mathcal N_z \approx \tfrac{c}{2\pi}p\,\ln(1/\delta) \,\big|\,\delta\big) \,\asymp\,\delta^{\frac{c}{2\pi}J(p)},\quad \delta \rightarrow 0,
\end{equation*} 
where the large deviations function $J$ is as in \eqref{Thetap'} and Theorem \ref{theo:main}, and where $c$ and $n$ are given in \eqref{ac} as functions of $\kappa$.
\end{theorem}
\begin{remark} We see that this result perfectly matches the second large deviations result in Theorem \ref{theo:main} for nesting in the $O(n)$ loop model on a random map with the topology of a pointed disk: one simply replaces $\frac{1}{{\delta}}$ here with the large volume $V$ of the map there. Indeed, one may assign elementary area $\frac{1}{V}$ to each face in the dual map, so that the dual map has in total unit area; then, the marked point corresponds in the dual to a face of elementary area $\frac{1}{V}$, and its depth $P = \frac{c}{2\pi} p\ln V$ is the number of loops separating this face from the boundary of the disk. \end{remark}
 
 It is interesting to compare the classical and quantum cases for nesting in $\mathrm{CLE}_\kappa$. In the classical case \cite{MWW}, the parametric equations of the Legendre-Fenchel transform \eqref{nulambda} are 
\begin{equation*}
\begin{split}
\nu & = \frac{\kappa}{(2\pi)^2}\,v \cot v \\
\gamma_\kappa(\nu) & = \frac{\kappa}{(2\pi)^2}\bigg[\frac{v^2}{2}-(v\cot v) \ln\Big(\frac{n}{2\cos v}\Big)-\frac{1}{2}(\arccos(n/2))^2\bigg]
\end{split}
\end{equation*}
for $v \in [0,\pi/2)$, and
\begin{equation*}
\begin{split}
\nu &= \frac{\kappa}{(2\pi)^2}\,w\,{\rm coth}\,w \\
\gamma_{\kappa}(\nu) & =  \frac{\kappa}{(2\pi)^2}\bigg[-\frac{w^2}{2} - (w\,{\rm coth}\,w)\ln\Big(\frac{n}{2\,{\rm cosh}\,w}\Big) - \frac{1}{2}(\arccos(n/2))^2\bigg]
\end{split}
\end{equation*}
for $v = {\rm i}w$ with $w \in \mathbb{R}_{> 0}$. These parametric equations
cannot be easily solved, whereas the quantum parametric equations \eqref{paramQ}, though similar, are simpler and explicitly solvable. Note also that in the classical case, the parameter  $\lambda$ is in the range $\lambda\in (-\infty,\lambda_{\max}]$, with the values $\lambda_{\max} =1-\frac{2}{\kappa}-\frac{3\kappa}{32}$ corresponding to $\nu \to 0$ or equivalently $v \rightarrow \frac{\pi}{2}$, while $\lambda \to -\infty$ corresponds to $\nu \to +\infty$ or $v = {\rm i}w$ with $w \rightarrow \infty$. We observe more precisely that
\beq
\label{gammagrowth} \gamma_{\kappa}(\nu) \sim \frac{(2\pi)^2}{\kappa}\,\frac{\nu^2}{2}, \qquad \nu \to +\infty.
\eeq
In the quantum case \eqref{paramQ}, $v$ is restricted to $v\in[0,\frac{\pi}{2})$, and $\lambda$ spans a finite interval only, $\lambda \in [\lambda_{\min},\lambda_{\max}]$, where $\lambda_{\min}=1-\frac{2}{\kappa}-\frac{\kappa}{8}$ is the point at which the square root in $v(\lambda)$ vanishes, corresponding to $p' \to +\infty$. And this results in $\lambda' = U_{\gamma}^{-1}\big(\frac{\lambda}{2}\big)$ spanning $[\frac{1}{2} - \frac{2}{\kappa},\frac{3}{4} - \frac{2}{\kappa}]$ if $\frac{8}{3}<\kappa \leq 4$, and $[\frac{1}{2} - \frac{\kappa}{8},\frac{1}{2} - \frac{\kappa}{16}]$ if $4\leq \kappa < 8$.

\subsection{Sphere topology} \label{subsec:sphere} Conformal loop ensembles can also be defined on the Riemann  sphere $\widehat{\mathbb{C}}$ \cite{MWW2,Kemppainen2015}. In particular, for any $\kappa \in ({8}/{3},4]$, the law of the simple nested $\mathrm{CLE}_\kappa$ in the full plane has been shown to 
be invariant under the inversion $z\mapsto \frac{1}{z}$ (and therefore under any M\"obius transformation of the Riemann sphere) \cite[Theorem 1]{Kemppainen2015}.  In this section, we connect the nesting statistics of ${\rm CLE}_{\kappa}$ in $\widehat{\mathbb{C}}$ with the nesting statistics in the $O(n)$ loop model on large random planar maps with the topology of the doubly punctured sphere. 

We first discuss the properties of ${\rm CLE}_{\kappa}(\widehat{\mathbb{C}})$. Let us pick two points (punctures), $z_1,z_2,$ on the sphere, which we may take to be $(z_1,z_2)=(0,\infty)$ using a suitable M\"obius transformation.
Consider the two balls $B(z_i,\varepsilon_i), i=1,2$, centered at these points. In stereographic projection, the connected domain $\widehat{\mathbb C}\setminus \overline{\left(B(z_1,\varepsilon_1) \cup B(z_2,\varepsilon_2)\right)}$ corresponds to the annulus $\mathbb A(\varepsilon_2^{-1},\varepsilon_1):=\varepsilon_2^{-1}\mathbb D\setminus \overline{B(z_1,\varepsilon_1)}$. 

Consider then in the whole ${\rm CLE}_{\kappa}(\widehat{\mathbb{C}})$ on the Riemann sphere, the loops which can be contracted to each one of the two punctures on $\hat{\mathbb C}$, {\it i.e.}, those loops which in projection belong to the above annulus. By scale invariance, their  number can depend only the \emph{product} $\varepsilon_1\varepsilon_2$, and we write it as $\mathcal N(\varepsilon_1\varepsilon_2)$. The nesting probability on the Riemann sphere is then defined as,  
\[
\mathbb P^{\widehat{\mathbb{C}}}\big[{\mathcal N}(\varepsilon_1\varepsilon_2) \approx \nu \ln(1/\varepsilon_1\varepsilon_2)\, |\,\varepsilon_1, \varepsilon_2\big],\] 
where we recall that $\approx$ is a short-hand notation for the event
\[
\frac{{\mathcal N}(\varepsilon_1\varepsilon_2)}{\ln(1/\varepsilon_1\varepsilon_2)} \in[\nu-\omega_-, \nu+\omega_+]
\]
for $\omega_{\pm}=\omega_{\pm}(\varepsilon_1\varepsilon_2)$ decreasing to $0$ sufficiently slowly with the $\varepsilon_i$'s (see Equation \eqref{logPnu}).

\subsubsection{Approximation to full-plane CLE and nesting estimates}\label{approxsphere}
\begin{figure}[h!]\label{fig:nestdisk}
\begin{center}
\includegraphics[width=0.45\textwidth]{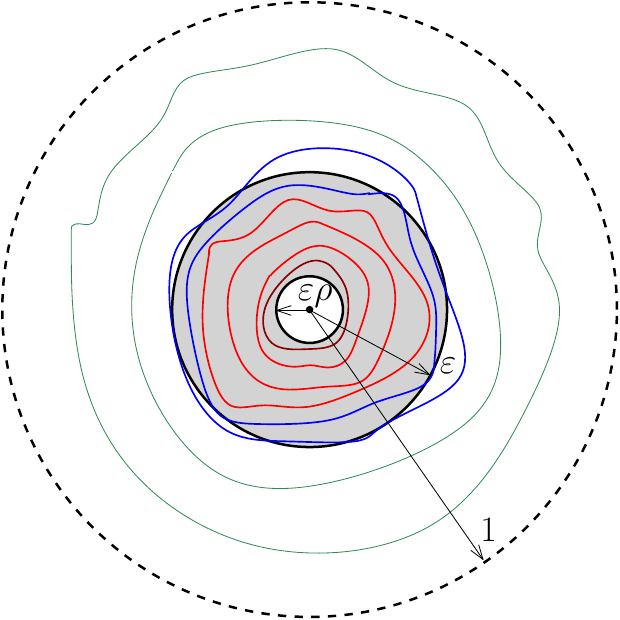}
\caption{\label{Nestdiskfig} The whole set of loops in the unit disk $\mathbb D$ is counted by $\mathcal{N}_0(\varepsilon\rho)$, with $\rho=\varepsilon_1\varepsilon_2$. The set of loops contained in the inner annulus, as counted by $\widehat{\mathcal{N}}(\varepsilon\rho)$ (resp. the set of intersecting loops, as counted by $\mathcal{N}_{\cap}(\varepsilon)$) appears in red (resp. blue).}
\end{center}
\end{figure}

Following Ref.  \cite[Appendix A]{MWW2}, about the rapid convergence of  CLE on a large disk to full-plane CLE, we can take as a large disk, $(\varepsilon \varepsilon_2)^{-1}\mathbb D$, with $0<\varepsilon<1$, which contains the annulus $\mathbb A(\varepsilon_2^{-1},\varepsilon_1)$ above.  Using scale invariance, we may simply consider events in $\mathbb D$ and in the annulus $\mathbb A(\varepsilon,\varepsilon\varepsilon_1\varepsilon_2)$ (see Figure~\ref{Nestdiskfig}), and by choosing $\varepsilon$ small enough,  approximate to any desired precision the probability of any event concerning a ball of radius $\varepsilon$ in the ensemble ${\rm CLE}_{\kappa}(\widehat{\mathbb{C}})$ (with probability law denoted by $\mathbb{P}^{\widehat{\mathbb{C}}}$) by the probability of the same event in the ensemble ${\rm CLE}_{\kappa}$ on the unit disk (with probability law simply denoted by $\mathbb{P})$. Ref.~\cite[Theorem A1]{MWW2} indeed states that with probability exponentially close to 1 in $\ln(1/\varepsilon)$ there exists a conformal map from whole-plane ${\rm CLE}_{\kappa}(\widehat{\mathbb{C}})$,  restricted to the interior of its smallest loop containing $B(0,\varepsilon)$, to ${\rm CLE}_{\kappa}(\mathbb D)$ similarly restricted to its smallest loop containing $B(0,\varepsilon)$, and  whose distorsion is exponentially close to 1.

As before, let  $\mathcal{N}_0(\varepsilon)$ be the number of loops surrounding the ball $B(0,\varepsilon)$  in $\mathbb D$, and let $\mathcal{N}_{\cap}(\varepsilon)$ be the number of loops surrounding the origin and intersecting $\partial B(0,\varepsilon)$. We seek for an estimation of the law of the number of loops in the annulus $B(0,\varepsilon)\setminus \overline{B(0,\varepsilon\rho)}$,
\beq
\label{Nestimsss}\widehat{\mathcal{N}}(\varepsilon\rho) := \mathcal{N}_0(\varepsilon\rho) - \mathcal{N}_0(\varepsilon) - \mathcal{N}_{\cap}(\varepsilon),\qquad \rho := \varepsilon_1\varepsilon_2,
\eeq
as illustrated in Figure~\ref{Nestdiskfig}. 

From Ref.~{\cite[Corollary 3.5]{MWW}}, we know that $\mathcal{N}_{\cap}(\varepsilon) < c_0\ln(1/\varepsilon)$ for some constant $c_0>0$, except	with	probability	exponentially small in $\ln (1/\varepsilon)$, since it is stochastically dominated by twice a geometric random variable with parameter $p(\kappa)>0$ which depends only on $\kappa$.

From Ref.~\cite[Lemma 3.2]{MWW}, and the convexity of $\gamma_\kappa(\nu)$, we know that there exists $\eta(\varepsilon) \rightarrow 0$, such that,
\beq
\label{smallt} \mathbb{P}\big(\mathcal{N}_0(\varepsilon) > \nu\ln(1/\varepsilon)\big) \leq \varepsilon^{\gamma_{\kappa}(\nu)-\eta(\varepsilon)},
\eeq
uniformly in $\nu \geq \nu_0$, with $\nu_0$ fixed but strictly larger than the point at which $\gamma_{\kappa}(\nu)$ reaches its minimum $0$.  We thus have in particular,
\beq
\label{lowbound1}\mathbb{P}\big(\mathcal{N}_0(\varepsilon) \leq \nu_0 \ln(1/\varepsilon)\big) \geq \frac{1}{2},
\eeq
for $\varepsilon$ small enough. Besides, we also know from \cite[Lemma 3.2]{MWW} that there exists $\omega(\varepsilon)$,  with $\omega(\varepsilon) \rightarrow 0$ as $\varepsilon\to 0$, such that,
\beq
\label{lowbound2}(\varepsilon\rho)^{\gamma_{\kappa}(\nu) + \eta(\varepsilon\rho)}\leq \mathbb{P}\Big(\big|\mathcal{N}_0(\varepsilon\rho) - \nu\ln(\varepsilon\rho)^{-1}\big| \leq \omega(\varepsilon\rho)\ln(\varepsilon\rho)^{-1}\Big) \leq (\varepsilon\rho)^{\gamma_{\kappa}(\nu) - \eta(\varepsilon\rho)}.
\eeq
Using these estimates will allow us shortly to show the existence of functions $\omega_{\pm}$ and $\eta$ of $\varepsilon$ and $\rho$, with $\omega_{\pm}(\varepsilon,\rho),\eta(\varepsilon,\rho) \rightarrow 0$ when $\rho \rightarrow 0$, such that,
\beq
\label{claim121} (\varepsilon\rho)^{\gamma_{\kappa}(\nu) + \eta(\varepsilon,\rho)} \leq \mathbb{P}\Big(\frac{\widehat{\mathcal{N}}(\varepsilon\rho)}{\ln(1/\varepsilon\rho)} \in [\nu-\omega_-,\nu+\omega_+]\Big) \leq (\varepsilon\rho)^{\gamma_{\kappa}(\nu) - \eta(\varepsilon,\rho)}. 
\eeq
\begin{proof}Let us denote by $\widehat{A}_{\nu}(\varepsilon,\rho)$ the event of interest in \eqref{claim121}, by $A_{\nu}(\varepsilon)$ the event $\{\mathcal{N}_0(\varepsilon) \leq \nu\ln(1/\varepsilon)\}$, and by $\bar{A}_{\nu}(\varepsilon)$ the latter's complement. Define also the logarithmic ratio,
\[
r= r(\varepsilon,\rho) := \frac{\ln(1/\varepsilon)}{\ln(1/\varepsilon\rho)};\quad r(\varepsilon,\rho)\to 0\,\,\textrm{as}\,\,\rho\to 0.
\]

For a lower bound, we write
\[
\mathbb{P}\big(\widehat{A}_{\nu}(\varepsilon,\rho)\big) \geq \mathbb{P}\big(\widehat{A}_{\nu}(\varepsilon,\rho)\cap A_{\nu_0}(\varepsilon)\big)=\mathbb{P}(A_{\nu_0}(\varepsilon))\,\mathbb{P}\big(\widehat{A}_{\nu}(\varepsilon,\rho)\,\big|\,A_{\nu_0}(\varepsilon)\big).
\]
Choosing $\nu_0$ as in \eqref{lowbound1},  the first factor is bounded from below by $\frac{1}{2}$, and using the lower bound \eqref{lowbound2} for the second factor, we get the desired lower bound in \eqref{claim121}, up to replacing $\omega(\varepsilon\rho)$ of \eqref{lowbound2} by $\omega_{\pm}(\varepsilon, \rho):= \omega(\varepsilon\rho) \mp (\nu_0+c_0)\,r(\varepsilon,\rho)$.

For the upper bound, we write
\beq
\label{lengg}\mathbb{P}\big(\widehat{A}_{\nu}(\varepsilon,\rho)\big) \leq \mathbb{P}(\widehat{A}_{\,\nu}(\varepsilon,\rho) \cap A_{\,\nu_1}(\varepsilon)\big) + \mathbb{P}(\bar{A}_{\nu_1}(\varepsilon)\big), 
\eeq
where, by using the estimate \eqref{smallt} for $\nu_1$ large enough, $\mathbb{P}(\bar A_{\nu_1}(\varepsilon)\big)\leq  \varepsilon^{\gamma_{\kappa}(\nu_1) - \eta(\varepsilon)}$. We will choose  $\nu_1= \nu_1(\varepsilon,\rho)$, such that $\nu_1(\varepsilon,\rho)\rightarrow \infty$ when $\rho \rightarrow 0$, as allowed by uniformity of \eqref{smallt}. As $\gamma_{\kappa}(\nu)$ grows quadratically in $\nu$ (see Equation \eqref{gammagrowth}) the latter estimate can be bounded, for large enough $\nu_1$, as
\beq
\label{lengg2}\varepsilon^{\gamma_{\kappa}(\nu_1) - \eta(\varepsilon)} \leq \varepsilon^{C\,\nu_1^2} = (\varepsilon\rho)^{C\, \nu_1^2\, r(\varepsilon,\rho)},
\eeq
for some constant $C > 0$. On the other hand, the first term in \eqref{lengg} can be estimated via the upper bound in \eqref{lowbound2} to yield an upper bound as in  \eqref{claim121}, provided that
\beq
\label{nu11bb} \nu_1(\varepsilon,\rho)\,r(\varepsilon,\rho)\rightarrow 0,\quad \rho\to 0,
\eeq
as this is the error term to be subtracted, together with $c_0r(\varepsilon,\rho)$, from $\omega(\varepsilon\rho)$ as the result of the restriction to event $A_{\nu_1}(\varepsilon)$. If we would like \eqref{lengg2} to be negligible in front of the first term in \eqref{lengg}, we would have to choose $\nu_1$ in such a way that
\beq
\label{nu112bb}\nu_1^2(\varepsilon,\rho)\,r(\varepsilon,\rho)  \rightarrow \infty,\quad \rho\to 0.
\eeq
To satisfy both \eqref{nu11bb} and \eqref{nu112bb},  choose for instance, 
$
\nu_1(\varepsilon,\rho) = r(\varepsilon,\rho)^{-3/4}.
$
Then, the second term in  \eqref{lengg} is bounded by $(\varepsilon\rho)^{C{r}^{-\frac{1}{2}}}$, which, since $r \rightarrow 0$ as $\rho\to 0$, is negligible as compared to the first term of order $(\varepsilon\rho)^{\gamma_{\kappa}(\nu)}$. This completes the proof of \eqref{claim121}.
\end{proof}

{As explained above, the exponentially fast convergence in $\ln(1/\varepsilon)$ when $\varepsilon \to 0$ (see \cite[Theorem A.1]{MWW2}), of the approximation of ${\rm CLE}_{\kappa}(\widehat{\mathbb C})$ by the restriction of ${\rm CLE}_{\kappa}(\mathbb D)$ to the ball $B(0,\varepsilon)$, allows us to translate result \eqref{claim121}, valid for $\rho=\varepsilon_1\varepsilon_2 \to 0$, into}
\begin{theorem}\label{ClassSphere}
The nesting probability in ${\rm CLE}_{\kappa}(\widehat{\mathbb C})$ between two balls of radius $\varepsilon_1$ and $\varepsilon_2$ and  centered at two distinct punctures, has the large deviations form,
\[
\mathbb{P}^{\widehat{\mathbb{C}}}\big({\mathcal{N}}(\varepsilon_1\varepsilon_2) \approx \nu\ln(1/\varepsilon_1\varepsilon_2)\big) \asymp  (\varepsilon_1\varepsilon_2)^{\gamma_{\kappa}(\nu)},\quad \nu\geq 0,\quad \varepsilon_1,\varepsilon_2\rightarrow 0,
\]
 where $\gamma_\kappa(\nu)$ is the same large deviations function \eqref{gammanu} as in the case of the disk topology, and where   notations are as in \eqref{logPnu}-\eqref{Pnueps}.
\end{theorem}
Even though the sphere and disk large deviations involve the same function $\gamma_\kappa$, the scalings involved actually differ by powers  of 2. Indeed, if we take the two balls on the Riemann sphere to have same radius $\varepsilon$, and measure nesting in $\ln\big(\frac{1}{\varepsilon}\big)$ units, we have from Theorem \ref{ClassSphere},
\begin{corollary}\label{spher2}
The nesting probability in ${\rm CLE}_{\kappa}(\widehat{\mathbb C})$ between two balls of same radius $\varepsilon$ and  centered at two distinct punctures, has the large deviations form,
\[
\mathbb{P}^{\widehat{\mathbb{C}}}\big({\mathcal{N}}(\varepsilon) \approx \nu\ln(1/\varepsilon)\big) \asymp  \varepsilon^{\widehat{\gamma}_{\kappa}(\nu)},\quad \nu\geq 0,\quad \varepsilon\rightarrow 0,
\]
 where $\widehat{\gamma}_\kappa(\nu)$ is related to the disk large deviations function \eqref{gammanu} by
 \[
 \widehat{\gamma}_\kappa(2\nu)=2\gamma_\kappa(\nu).
 \]
\end{corollary}

Using hereafter the variables $t_i := \ln\big(\frac{1}{\varepsilon_i}\big), i=1,2$, we have,
\beq \label{Pnut1t2}
\mathbb{P}^{\widehat{\mathbb{C}}}\big(\mathcal{N}(\varepsilon_1\varepsilon_2) \approx \nu(t_1 + t_2)\big) \asymp e^{-\gamma_{\kappa}(\nu)(t_1 + t_2)},\quad t_1,t_2\to +\infty.
\eeq
\subsubsection{Nesting on the quantum sphere}In Liouville quantum gravity, following the same steps as in Section \ref{CLE-LQG}, let us condition on each ball having the same quantum area $\delta=e^{-\gamma A}$. 
 The desired distribution should be given by the convolution
\beq\label{PQbis}
	\mathbb P^{\widehat{\mathbb C}}_{\mathcal{Q}}(\mathcal N\,|\,A):=\int_{0}^{\infty}\int_{0}^{\infty}  \dd t_1 \dd t_2\, \mathbb P^{\widehat{\mathbb C}}(\mathcal N\,|\,t_1,t_2)\,\mathcal{P}(t_1\,|\,A)\,\mathcal{P}(t_2\,|\,A),
\eeq
 {where $\mathcal{P}(t_i\,|\,A), i=1,2$ are as in \eqref{PA}-\eqref{PA'}. Note that this definition readily asserts a factorisation, hence independence, of these two distributions, because their two ball centers $(z_i, i=1,2)$ have been taken as $(0,\infty)$ on the Riemann sphere.}

For large $A$,  we let $\mathcal N$ scale as $\mathcal N \approx \gamma p A$, with $p \in \mathbb R_+$, and also set $\mathcal N \approx \nu (t_1+t_2)$, where $\nu$ is now defined such that,
\beq
\label{Nnutbis} \nu=\nu(t_1,t_2):=\gamma p A/(t_1+t_2),
\eeq 
where $p$ and $A$ are thought of as parameters. {By the same argument as in Section \ref{Largedeviations}}, the asymptotic result \eqref{Pnut1t2} then yields when $A\to +\infty$
\begin{equation}
\begin{split}
\mathbb P^{\widehat{\mathbb C}}_{\mathcal{Q}}(\mathcal N_z \approx \gamma pA\,|\,A) & \asymp \int_{0}^{\infty} \int_{0}^{\infty} \frac{A^2\,\dd t_1 \dd t_2}{2\pi\sqrt{t_1^3t_2^3}} \exp\left[-\mathcal E(t_1)-\mathcal E(t_2)\right]  \\
\label{PnutPAbis} \mathcal E(t_1)+\mathcal E(t_2) &= \frac{1}{2t_1}\left({A}
-a_\gamma {t_1}\right)^2+\frac{1}{2t_2}\left({A}
-a_\gamma {t_2}\right)^2+\gamma_\kappa(\nu)(t_1+t_2).
\end{split}
\end{equation}
The above integral is evaluated by a saddle point method, by looking for the minimum of $\mathcal E(t_1)+\mathcal E(t_2)$ at \emph{fixed} $\nu(t_1+t_2)=\gamma p A$.   
We then have for each $i=1,2$,
\[
(t_1+t_2) \partial_{t_i}\gamma_\kappa(\nu)=-\nu\frac{\partial \gamma_\kappa}{\partial \nu},
\]
and using \eqref{lambdanu},
\[
\frac{\partial}{\partial t_i}\big((t_1+t_2)\gamma_\kappa(\nu) \big)=\gamma_\kappa(\nu)-\nu \frac{\partial \gamma_\kappa}{\partial \nu} =\lambda.
\]
This in turn gives for each $i=1,2$,
 \[
\frac{\partial}{\partial t_i} \big({\mathcal E}(t_1)+{\mathcal E}(t_2)\big) =\lambda -\frac{1}{2}\Big[\Big(\frac{A}{t_i}\Big)^2-a^2_\gamma\Big],
 \]
so that both saddle points $t_1^*$ and $t_2^*$ for $t_1$ and $t_2$ are located at the same point $t^*$ as in \eqref{Atstar}  in the case of the disk topology. We thus have at this double saddle point
\begin{equation*}
\begin{cases}
2\nu t^* =\gamma p A\\
\mathcal E(t_1^*)+\mathcal E(t_2^*)=2\mathcal E(t^*).
\end{cases}
\end{equation*}
Setting:
\[
\widehat\Theta(p):= \frac{2\mathcal E(t^*)}{\gamma A}
\]
we deduce
\begin{theorem}
\label{theo:sphere}
The large deviations function $\widehat{\Theta}(p)$ which governs the quantum nesting probability of ${\rm CLE}_{\kappa}$ on $\widehat{\mathbb{C}}$,
\[
\mathbb P^{\widehat{\mathbb C}}_{\mathcal{Q}}(\mathcal N \approx p \ln(1/\delta)\,|\,\delta)\asymp 
\delta^{\widehat{\Theta}(p)},\quad \delta \rightarrow 0,
\]
is related to the large deviations function $\Theta(p)$ for the disk topology (Theorem~\ref{theo:LambdaQ}) by
\[
\widehat{\Theta}(2p)=2\Theta(p).
\]
\end{theorem} 
Using alternatively the explicit formulation, as in Theorem \ref{theo:Jp'}, we get
\begin{corollary}
In the same setting as in Theorem \ref{theo:sphere},
\[
\mathbb P^{\widehat{\mathbb C}}_{\mathcal{Q}}\Big(\mathcal N \approx \frac{cp}{\pi} \ln (1/\delta) \,\Big|\,\delta\Big) \,\asymp\, \delta^{\frac{c}{\pi}J(p)}, \quad \delta \rightarrow 0,
\]
where $J(p)$ is as in  \eqref{Thetap'}.
\end{corollary}
This is in complete agreement with:
\begin{itemize}
\item[$\bullet$] The first result in Theorem \ref{theo:main}, which describes the large deviations of the number of separating loops between a marked point and a microscopic boundary in a critical $O(n)$ model on a random map with a disk topology;
\item[$\bullet$] The first result in Theorem \ref{theo:cyl},  which describes the large deviations of the number of separating loops between two microscopic boundaries in a critical $O(n)$ model on a random map with a cylinder topology. 
\end{itemize}
These are indeed the sort of topologies considered in Section \ref{approxsphere} above.

\subsection{Weighted loops}\label{WL} 
\subsubsection{Weighting $\mathrm{CLE}_\kappa$} Our argument can be refined to include a model where loops receive independent random weights, in parallel to the results in Ref. \cite[Section 5]{MWW}. A motivation to introduce this model, beyond the fact it offers a natural generalisation of the counting of loops,
is that loops weighted with a Bernoulli random variable for $\kappa = 4$ are related to the extremes of the GFF \cite{2009arXiv0902.3842H}. 

Conditionally on a configuration $\Gamma$ of a $\mathrm{CLE}_\kappa$ in a proper simply connected domain $D$, let $(\xi_l)_{l \in \Gamma}$ be a collection of independent, identically distributed real random variables indexed by $\Gamma$. We denote by $\mu$ the law of each $\xi_{l}$. For $z\in D$ and $\varepsilon >0$, let $\Gamma_z(\varepsilon)$ be the set of loops which surround $B(z,\varepsilon)$, and define 
\[
\Xi_z(\varepsilon)=\sum_{l \in\Gamma_z(\varepsilon)}\xi_l,\qquad \qquad  \widetilde{\Xi}_z(\varepsilon)=\frac{\Xi_z(\varepsilon)}{ \ln(1/\varepsilon)}.
\] 
For a realisation of the ${\rm CLE}_{\kappa}$ and of the $(\xi_{l})_{l}$, and any fixed $(\nu,\alpha)\in \mathbb R_+ \times \mathbb R$, let 
\[
\Phi^\mu_{\nu,\alpha} = \bigg\{z\in D\,\,:\quad \lim_{\varepsilon\to 0} \widetilde{\mathcal N}_z(\varepsilon)=\nu\quad {\rm and}\quad \lim_{\varepsilon\to 0} \widetilde{\Xi}_z(\varepsilon)= \alpha\bigg\}.
\]
The cumulant generating function associated with the moments of $\mu$ is  
\beq
\label{Lambdamu}
\Lambda_\mu(\lambda):=\ln \mathbb E\big[e^{\lambda \xi}\big],
\eeq
where $\xi$ is a random variable whose distribution is $\mu$. Its symmetric Legendre-Fenchel transform, $\Lambda_\mu^{\star} :\mathbb R\rightarrow \mathbb R_+$, is defined as
\beq
\label{Lambdamustar}
\Lambda^{\star}_\mu(x):=\sup_{\lambda \in \mathbb R}\left(\lambda x-\Lambda_\mu(\lambda)\right).
\eeq
The Hausdorff dimension of the set $\Phi^\mu_{\nu,\alpha}$ is then almost surely constant, with value found in \cite[Theorem 5.1]{MWW}
\[
\mathrm{dim}_{\mathcal H}\,\Phi^\mu_{\nu,\alpha} = \max\{0,2-\gamma_{\kappa}(\nu,\alpha)\},
\]
as long as $\gamma_\kappa(\alpha,\nu) \leq 2$, with ${\Phi}^{\mu}_{\alpha,\nu}=\emptyset$ otherwise, 
and with the definition
\beq
\label{gammanualpha}
\gamma_{\kappa}(\nu,\alpha):= \left\{\begin{array}{lll} \nu\Lambda^{\star}_\mu(\alpha/\nu)+\nu\Lambda^{\star}_\kappa(1/\nu) & \quad & \mathrm{if}\, \nu >0 \\ 
\lim_{\nu'\to 0^+} \gamma_{\kappa}(\nu',\alpha) & \quad & {\rm if}\,\,\nu =0\,\,{\rm and}\,\,\alpha \neq 0\\
\lim_{\nu'\to 0^+} \gamma_{\kappa}(\nu')=1-\frac{2}{\kappa}-\frac{3\kappa}{32} & \quad & \mathrm{if}\,\,(\nu,\alpha) = (0,0), \end{array}\right.
\eeq
where the limits exist by convexity of  $\Lambda^{\star}_\kappa$ and $\Lambda^{\star}_\mu$. When $\nu\neq 0$, we thus have
\begin{equation*}
\begin{split}
\gamma_{\kappa}(\nu,\alpha) & = \gamma_\kappa(\nu)+\gamma_\mu(\nu,\alpha),  \\
 \label{gammanualphaexplicit} \gamma_\mu(\nu,\alpha) & := \nu\Lambda^{\star}_\mu\left({\alpha}/{\nu}\right)= \lambda'\alpha -\nu\Lambda_\mu(\lambda'),
 \end{split}
 \end{equation*}
 where $\lambda'$ is a function of $(\nu,\alpha)$ determined by:
\[
\frac{\alpha}{\nu}=  \frac{\partial \Lambda_\mu (\lambda')}{\partial \lambda'}.
\]
By homogeneity, we find the useful identity
\beq
\label{eq:nudnualdal}
\Big(\nu \frac{\partial}{\partial \nu} + \alpha\frac{\partial}{\partial \alpha}\Big)\gamma_\mu(\nu,\alpha) = \gamma_\mu(\nu,\alpha).
\eeq
Uniformly for a  point $z \in D$, we have the following joint probability scaling \cite{MWW}
\beq
\label{Pnualphat} 
\mathbb P({\mathcal N}_z \approx \nu t\,\,{\rm and}\,\,{\Xi}_z \approx \alpha t \,|\,t )\,\asymp\, e^{-\gamma_\kappa(\nu,\alpha)t}.
\eeq

\subsubsection{Weighted $\mathrm{CLE}_\kappa$ in Liouville Quantum Gravity} One follows exactly the same procedure as in Section \ref{sec:NLQG}.  We study the nesting around small  balls $B(z,\varepsilon)$ conditionally to a given quantum area $\delta$ \eqref{delta}, hence conditionally on $A$ \eqref{tA}, while the counts $\mathcal N_z$ and $\Xi_z$ are unchanged,
\beq
\label{PQmixed}
\mathbb{P}_{\mathcal{Q}}(\mathcal N_z,\,\Xi_z\,|\,A):= \int_0^{\infty} \dd t\,\mathbb P(\mathcal N_z,\,\Xi_z\,|\,t)\,\mathcal{P}(t\,|\,A),
\eeq
where $\mathcal{P}(t\,|\,A)$ is as in \eqref{PA}-\eqref{PA'}. 

For large $A$,  we let  $\mathcal N_z \approx \gamma p A$ and $\Xi_z \approx\gamma q A$, with $(p,q) \in \mathbb R_+ \times \mathbb{R}$, and also have $\mathcal{N}_z \approx \nu t,\,\Xi_z \approx \alpha t$, where $\nu$ and $\alpha$ are defined by:
\beq
\label{NnutAl} \gamma p A = \nu t,\qquad  \gamma q A= \alpha t,
\eeq
and $p,q,A$ are considered as parameters. {As in Section \ref{Largedeviations}}, the asymptotic result \eqref{Pnualphat} then yields, for $A\to +\infty$, 
\begin{equation}
\begin{split}
\label{PnualphatPA}
\mathbb P_{\mathcal{Q}}(&\mathcal N_z  \approx \gamma pA\,\,{\rm and}\,\,\Xi_z\approx \gamma q A\,|\,A) \asymp \int_0^\infty \frac{Ae^{-\mathcal{E}_{\mathrm{w}}(t)}\,\dd t}{\sqrt{2\pi t^3}}, \\
{\mathcal E}_{\mathrm{w}}(t) & := \frac{1}{2t}\left({A} -a_\gamma {t}\right)^2+\gamma_\kappa(\nu,\alpha)t.
\end{split}
\end{equation}
The above integral is evaluated by the saddle point method, looking for the minimum of 
${\mathcal E}_{\mathrm{w}}(t)$ along trajectories at constant values of $\nu t$ and $\alpha t$ according to \eqref{NnutAl}. We then have
\[
t \frac{\partial}{\partial t}  \gamma_\mu(\nu,\alpha)= -\Big(\nu\frac{\partial}{\partial \nu} + \alpha\frac{\partial}{\partial \alpha}\Big)\gamma_{\mu}(\nu,\alpha),
\]
and using \eqref{eq:nudnualdal},
\[
\frac{\partial}{\partial t}\big(t\,\gamma_\mu(\nu,\alpha)\big) = 0,
\]
so that,
\[
\frac{\partial}{\partial t}\big(t\,\gamma_{\kappa}(\nu,\alpha)\big) = \frac{\partial}{\partial t}\big(t\,\gamma_\kappa(\nu)\big)=\lambda,
\]
as in \eqref{lambdanurec}. This shows that $\frac{\partial \mathcal{E}_{\mathrm{w}}}{\partial t}$ is the same as in \eqref{eq:dEdt}, 
\[
\frac{\partial \mathcal{E}_{\mathrm{w}}}{\partial t}=\lambda -\frac{1}{2}\Big[\Big(\frac{A}{t}\Big)^2-a^2_\gamma\Big],
 \]
 with the same saddle point as in \eqref{Atstar},  
 \begin{equation*}\label{Atstarbis}
 \frac{A}{t^*}=u = u(\lambda) := \sqrt{2\lambda+a_\gamma^2}.
 \end{equation*}

The saddle point value resides at the minimum $\mathcal{E}_{\mathrm{w}}^*$ of $\mathcal{E}_{\mathrm{w}}(t)$,
\beq
\label{Estarbis}
\mathcal{E}_{\mathrm{w}}^* :=\mathcal{E}_{\mathrm{w}}(t^*)= A\Big[\frac{(u-a_\gamma)^2}{2u} + \frac{\gamma_{\kappa}(\nu,\alpha)}{u}\Big],
\eeq
where, because of  condition \eqref{NnutAl}, $\nu$ and $\alpha$ are now functions of $(p,q)$ determined by
 \beq\label{alphanustar}
 \nu =\gamma p\,\frac{A}{t^*} =\gamma p \,u(\lambda),\qquad \alpha =\gamma q\,\frac{A}{t^*}=\gamma q\,u(\lambda).
\eeq
It yields
\[
\frac{\gamma_\mu(\nu,\alpha)}{u} = \frac{\alpha\lambda' - \nu\Lambda_{\mu}(\lambda')}{u} = \gamma q \lambda' -\gamma p \Lambda_\mu(\lambda'),\qquad {\rm with}\quad 
\frac{\alpha}{\nu} = \frac{q}{p}= \frac{\partial \Lambda_\mu(\lambda')}{\partial \lambda'}.
\]
Recalling \eqref{gammak/u} and \eqref{Estarprefinal}, we get the simple form,
\[
\Theta(p,q):= \frac{{\mathcal E}_{\mathrm{w}}^*}{\gamma A} = \frac{u(\lambda)-a_\gamma}{\gamma} + q\lambda' - p\big(\Lambda_\kappa(\lambda)+\Lambda_\mu(\lambda')\big).
\]
Comparing to \eqref{thetap}-\eqref{Uu}, we get 
\begin{theorem}\label{theo:Thetapq} The joint distribution of the number of loops $\mathcal N_z$ surrounding a ball of given quantum area $\delta$ centered at $z$ in a simply connected domain $D\subset \mathbb C$, and of the sum of weights $\Xi_z$ on these loops in the ensemble of $\mu$-weighted loops in a $\mathrm{CLE}_\kappa$, satisfies the large deviations estimate,
\[
\mathbb P_{\mathcal{Q}}(\mathcal N_z \approx p\ln(1/\delta)\,\,{\rm and}\,\,\Xi_z\approx q\ln(1/\delta)\,|\,\delta) \, \asymp\, \delta^{\Theta(p,q)},\quad \delta \rightarrow 0,
\]
with
\[
\Theta(p,q)=\Theta(p)+ q\lambda' - p\Lambda_\mu(\lambda'),
\]
where $\Theta(p)$ is as in Theorem \ref{theo:LambdaQ}, and where the conjugate variable $\lambda'$ is the function of $(p,q)$ uniquely determined by 
\beq
\label{qnu}
\frac{q}{p} = \frac{\partial \Lambda_\mu(\lambda')}{\partial \lambda'}.
\eeq
\end{theorem}
We can also switch to parameters $(p',q')$ such that 
\beq
\label{tildepf} p=\frac{c}{2\pi}\,p', \quad q =\frac{c}{2\pi}\,q',
\eeq
where $c$ is the exponent defined in \eqref{ac}. Then, after writing $\Theta(p,q)=\frac{c}{2\pi}\,J(p',q')$, we get
\begin{corollary}\label{cor:biv}
In the same setting as in Theorem \ref{theo:Thetapq}, we have 
\[
\mathbb P_{\mathcal{Q}}\Big(\mathcal N_z \approx \tfrac{c}{2\pi}p\,\ln(1/\delta)\,\,{\rm and}\,\,\Xi_z \approx \tfrac{c}{2\pi}q\,\ln(1/\delta)\,\Big|\,\delta\Big) \,\asymp\, \delta^{\frac{c}{2\pi}\,J(p,q)}, \quad \delta \rightarrow 0,
\]
with the bivariate large deviations function
\[
J(p,q)= J(p) + q\lambda' - p\Lambda_\mu(\lambda'),
\]
where $J(p)$ is given by \eqref{Thetap'} and where $\lambda'$ is uniquely determined as a function of $(p,q)$ by
\[
\frac{q}{p}= \frac{\partial \Lambda_{\mu}(\lambda')}{\partial \lambda'}.
\]
 \end{corollary}
Corollary \ref{cor:biv} in LQG matches with the bivariate large deviations of nesting and sum of loop weights for critical $O(n)$ models on random maps with the topology of a pointed disk (first result of Theorem \ref{bivth}). The case of the bivariate distribution on the Riemann sphere can be analysed in exactly the same way as in Section \ref{subsec:sphere}, and we skip the details here.

\begin{theorem} \label{theo:spherebiv} On the Riemann sphere $\widehat{\mathbb{C}}$, the joint distribution of the nesting between two balls of given quantum area $\delta$ and the weight carried by the separating loops, behaves as
\[
\mathbb P^{\widehat {\mathbb C}}_{\mathcal{Q}}(\mathcal N \approx p \ln(1/\delta)\,\,{\rm and}\,\,\Xi \approx q\ln(1/\delta)\,|\,\delta)\,\asymp\, 
\delta^{\widehat{\Theta}(p,q)},\quad \delta \rightarrow 0,
\]
where the large deviations function $\widehat{\Theta}(p,q)$  is given in terms of the large deviations function $\Theta(p,q)$ for the quantum disk, as obtained in Theorem \ref{theo:Thetapq}, by
\[
\widehat{\Theta}(p,q)=2\,\Theta\big(\tfrac{p}{2},\tfrac{q}{2}\big).
\]
\end{theorem} 

Switching to variables  \eqref{tildepf}, we get
\begin{corollary}
In the same setting as in Theorem \ref{theo:spherebiv}, we have
\[
\mathbb P^{\widehat{\mathbb C}}_{\mathcal{Q}}\Big(\mathcal N \approx \frac{cp}{\pi}\,\ln(1/\delta)\,\,{\rm and}\,\,\Xi \approx \frac{cq}{\pi}\,\ln(1/\delta)\,\Big|\,\delta\Big)\,\asymp\, \delta^{\frac{c}{\pi}\,J(p,q)},\quad \delta \rightarrow 0,
\]
where $J(p,q)$ is the function as defined in Corollary~\ref{cor:biv}.
\end{corollary}
This last result is the exact analog, in Liouville quantum gravity, of the first large deviations result of Theorem~\ref{bivth} in the critical $O(n)$ model on random disks with $\mu$-weighted loops, for the topology of a pointed disk with a microscopic boundary.

\appendix
\section{Theta function}
\label{AppTheta}
 For properties of elliptic functions, we refer to \cite{GuoWang}. Let $\tau$ be a complex number in the upper-half plane. The Jacobi theta function is the entire function of $v \in \mathbb{C}$ defined by:
\beq
\label{A1theta}\vartheta_{1}(v|\tau) = -\sum_{m \in \mathbb{Z}} e^{{\rm i}\pi \tau (m + \frac{1}{2})^2 + {\rm i}\pi (v  + \frac{1}{2})(2m + 1)}.
\eeq
Its main properties are:
$$
\vartheta_{1}(-v|\tau) = \vartheta_{1}(v + 1|\tau) = -\vartheta_{1}(v|\tau),\qquad \vartheta_{1}(v + \tau|\tau) = -e^{-2{\rm i}\pi (v  + \frac{\tau}{2})}\,\vartheta_{1}(v|\tau),
$$
and the effect of the modular transformation:
\beq
\label{modularS}\vartheta_{1}(v|\tau) = {\rm i} \frac{e^{-\frac{{\rm i}\pi v^2}{\tau}}}{\sqrt{-{\rm i}\tau}}\,\vartheta_{1}\big(\tfrac{v}{\tau}\,\big|\,-\tfrac{1}{\tau}\big).
\eeq
We will also meet two other Jacobi theta functions
\begin{equation}
\begin{split}
\label{Jactheta}
\vartheta_2(v|\tau) & = \sum_{m \in \mathbb{Z}} e^{{\rm i}\pi \tau(m + \frac{1}{2})^2 + {\rm i}\pi v(2m + 1)}, \\
\vartheta_3(v|\tau) & = \sum_{m \in \mathbb{Z}} e^{{\rm i}\pi \tau m^2}\,e^{2{\rm i}\pi vm}.
\end{split}
\end{equation}

\section{Analytic properties of the parametrisation \texorpdfstring{$x \leftrightarrow v$}{x<->v}}

\label{App1}  
\label{xbeh}
In this section, we consider $\alpha,h$ fixed and $\gamma_-,\gamma_+$ as varying complex parameters. The purpose of this section is to introduce a real-analytic change of variable (Lemma~\ref{Lemachange}) which will turn handy in the study the critical limit; and, to show that it can be promoted to an analytic change of variables for complex parameters. As we are primarily interested in the neighboorhood of $\gamma_+ = \gamma_+^*$, we do not attempt to describe the largest possible domain of analyticity. The information gained will also serve in Appendix~\ref{sec:anabend}.

We define
\begin{equation}
\label{defM} \mathsf{m} := \frac{(\varsigma(\gamma_-) - \gamma_-)(\varsigma(\gamma_+) - \gamma_+)}{(\varsigma(\gamma_-) - \gamma_+)(\varsigma(\gamma_+) - \gamma_-)}
\end{equation}
and introduce the domains
\begin{equation}
\begin{split}
\mathcal{A} & := \big\{(\gamma_-,\gamma_+) \in \mathbb{C}^2 \quad \big| \quad |\mathsf{m}| < 1\big\} \\
\mathcal{A}^{\mathbb{R}} & := \big\{(\gamma_-,\gamma_+) \in \mathbb{R}^2 \quad |\quad |\gamma_-| < \gamma_+ < \varsigma(\gamma_+) < \varsigma(\gamma_-)\big\}
\end{split}
\end{equation}
We clearly have $\mathcal{A}^{\mathbb{R}} \subset \mathcal{A}$. We shall denote $B(z;r) \subset \mathbb{C}$ the open ball of radius $r$ centered at $z$.

\subsection{Formulas in \texorpdfstring{$\mathcal{A}^{\mathbb{R}}$}{A(R)}}

We first assume that $(\gamma_-,\gamma_+) \in \mathcal{A}^{\mathbb{R}}$, and set
\beq
\label{io} v = {\rm i}C\,\int^{x}_{\varsigma(\gamma_+)} \frac{\dd y}{\sqrt{(y - \varsigma(\gamma_-))(y - \varsigma(\gamma_+))(y - \gamma_+)(y - \gamma_-)}}.
\eeq
The normalising constant is chosen such that, for $x$ moving from the origin $\varsigma(\gamma_+)$ to $\varsigma(\gamma_-)$ with a small negative imaginary part, $v$ is moving from $0$ to $\tfrac{1}{2}$. When $x$ moves on the real axis from $\varsigma(\gamma_+)$ to $\gamma_+$, $v$ moves from $0$ to a purely imaginary value denoted $\tau = {\rm i}T$. Then, the function $v \mapsto x(v)$ has the properties:  
\[
x(v + 2\tau) = x(v + 1) = x(-v) = x(v),\qquad \varsigma(x(v)) = x(v - \tau),
\]
and is depicted in Figure~\ref{ParamF}. $x'(v)$ has zeroes when $v \in \tfrac{1}{2}\mathbb{Z} + \tau\mathbb{Z}$, and double poles at $v = v_{\infty} + \mathbb{Z} + 2\tau\mathbb{Z}$. From \eqref{io}, paying attention to the determination of the square root at infinity obtained by analytic continuation, we can read in particular:
\beq
x'(v) \sim \frac{{\rm i}C}{(v - v_{\infty})^2},\qquad v \rightarrow v_{\infty}.
\eeq

There is an alternative expression for \eqref{io} in terms of Jacobi functions:
\beq
\label{B4eq} v = \frac{2{\rm i}C\,{\rm arcsn}\Big[\sqrt{\frac{\varsigma(\gamma_-) - \gamma_+}{\varsigma(\gamma_+) - \gamma_+}\,\frac{x - \varsigma(\gamma_+)}{x - \varsigma(\gamma_-)}}\,;\,\mathsf{m}\Big]}{\sqrt{(\varsigma(\gamma_+) - \gamma_-)(\varsigma(\gamma_-) - \gamma_+)}},
\eeq
This formula can be easily checked by differentiating both sides with respect to $x$ and using
\[
\partial_{v}{\rm sn}[v;\mathsf{m}] = \sqrt{(1 - {\rm sn}^2[v;\mathsf{m}])(1 - \mathsf{m}\,{\rm sn}^2[v;\mathsf{m}])}.
\]
Note that $\mathsf{m}$ is for us the square of the elliptic modulus usually denoted $k$. For $(\gamma_-,\gamma_+) \in \mathcal{A}^{\mathbb{R}}$, we have $\mathsf{m} \in (0,1)$.

Recall the expression of the complete elliptic integral
\[
K(\mathsf{m}) = \int_{0}^{1} \frac{\dd t}{\sqrt{(1 - t^2)(1 - \mathsf{m}t^2)}} = {\rm arcsn}[1;\mathsf{m}].
\]
Matching $x = \gamma_+$ with $v = {\rm i}T$ we obtain
\begin{equation}
\label{CoverT} T = \frac{2CK(\mathsf{m})}{\sqrt{(\varsigma(\gamma_+) - \gamma_-)(\varsigma(\gamma_-) - \gamma_+)}}.
\end{equation}
Matching $x = \varsigma(\gamma_-)$ with $v = \frac{1}{2}$ we obtain
\begin{equation}
\label{XCeq} C = \frac{\sqrt{(\varsigma(\gamma_+) - \gamma_-)(\varsigma(\gamma_-) - \gamma_+)}}{4K(1 - \mathsf{m})}.
\end{equation}
Together this implies
\begin{equation}
\label{TCeq}
T = \frac{K(\mathsf{m})}{2K(1 - \mathsf{m})}.
\end{equation}

Later, we will need the following expansion.
\begin{lemma}
\label{Xinfexp} When $v \rightarrow v_{\infty}$, we have the expansion:
\begin{equation}
\label{xinfininit}
x(v) = \frac{-{\rm i}C}{v - v_{\infty}} + \frac{E_{1}}{4} + \frac{{\rm i}}{C}\,\frac{3E_{1}^2 - 8E_{2}}{48}\,(v - v_{\infty}) + O(v - v_{\infty})^2 \end{equation}
where we introduced the symmetric polynomials in the endpoints:
\begin{equation}
\label{DefE}
\begin{split}
 E_{1} & =  \gamma_- + \gamma_+ + \varsigma(\gamma_+) + \varsigma(\gamma_-), \\
 E_{2} & =  \gamma_-\big(\gamma_+ + \varsigma(\gamma_+) + \varsigma(\gamma_-)\big) + \gamma_+\big(\varsigma(\gamma_+) + \varsigma(\gamma_-)\big) + \varsigma(\gamma_+)\varsigma(\gamma_-),  
 \end{split}
 \end{equation}
More generally, the coefficient of $(v - v_{\infty})^k$ in this expansion is a homogeneous symmetric polynomial of degree $(k + 1)$ in the endpoints, with rational coefficients up to an overall factor $({\rm i}C)^{-k}$.
\end{lemma}
\begin{proof}
This can be easily derived from the integral representation \eqref{io}. 
\end{proof}

\subsection{Changes of parameters}

It is often convenient to trade $(\gamma_-,\gamma_+)$ for another set of parameters. The first change of variables we will consider is $\chi\,:\,(\gamma_-,\gamma_+) \mapsto (\mathsf{m},\mathsf{p})$ with
\begin{equation}
\label{pdef} \mathsf{p} = \frac{\varsigma(\gamma_-) - \gamma_-}{\varsigma(\gamma_+) - \gamma_-},
\end{equation}
and we assume here $\alpha \neq 1$. The case $\alpha = 1$ will be treated in Section~\ref{appa1}. $\chi$ is then an analytic map in the domain $\mathcal{A}$ which has degree $2$. It admits an inverse map $\chi^{-1}\,:\,(\mathsf{m},\mathsf{p}) \mapsto (\gamma_-,\gamma_+)$ given by the formula
\begin{equation}
\label{gmunngynon}\gamma_- = \frac{\alpha}{(\alpha^2 - 1)h} - \frac{\sqrt{1 - \mathsf{m}}}{(\alpha^2 - 1)(\mathsf{p} - 1)h},\qquad \gamma_+ = \frac{\alpha}{(\alpha^2 - 1)h} - \frac{\mathsf{p}\sqrt{1 - \mathsf{m}}}{(\alpha^2 - 1)(\mathsf{p} - \mathsf{m})h}.
\end{equation}
This map $\chi^{-1}$ is analytic in the domain
\[
\mathcal{A}_{1} := \big\{(\mathsf{m},\mathsf{p}) \in \mathbb{C}^2 \quad | \quad |\mathsf{m}| < 1\,\,{\rm and}\,\,\mathsf{p} \neq \mathsf{m}\big\},
\]
and in \eqref{gmunngynon} we use the square root with its standard determination and branchcut on the negative real axis. We have chosen the sign in front of the square root so that the $\mathsf{m} = 0$ specialisation of $\chi^{-1}$ gives $(\gamma_-^*,\gamma_+^*)$ where
\begin{equation}
\label{gplusstar}
\gamma_+^* = \frac{1}{h(\alpha + 1)}
\end{equation}
is the fixed point of the involution $\varsigma$. Note that $\gamma_-^*$ is still a function of $\mathsf{p}$ --- see later \eqref{pwinf} and \eqref{cosmim}.

When $(\gamma_-,\gamma_+) \in \mathcal{A}^{\mathbb{R}}$, we know that $\mathsf{m} \in (0,1)$ and $\mathsf{p} > 0$. We can rewrite \eqref{B4eq} using \eqref{XCeq} as
\begin{equation}
\label{vintermsx}v = \frac{{\rm i}\,{\rm arcsn}\big[\sqrt{\frac{\mathsf{p}}{\mathsf{m}}\,\frac{x - \varsigma(\gamma_+)}{x - \varsigma(\gamma_-)}}\,;\,\mathsf{m}\big]}{2K(1 - \mathsf{m})}.
\end{equation}
Specialising to $x = \infty$, we obtain
\begin{equation}
\label{vinfdsf} v_{\infty} = \frac{{\rm i}\,{\rm arcsn}\big[\sqrt{\frac{\mathsf{p}}{\mathsf{m}}}\,;\,\mathsf{m}\big]}{2K(1 - \mathsf{m})}.
\end{equation}
Decomposing $v_{\infty} = \frac{1}{2} + {\rm i}Tw_{\infty}$ with \eqref{TCeq}, we can rewrite using the transformation law of the Jacobi elliptic functions
\begin{equation}
\label{pksn} \sqrt{\mathsf{p}} = \sqrt{\mathsf{m}}\,{\rm sn}\big[-{\rm i}K(1 - \mathsf{m}) + K(\mathsf{m})w_{\infty}\,;\,\mathsf{m}\big] = \frac{1}{{\rm sn}\big[K(\mathsf{m})w_{\infty}\,;\,\mathsf{m}]}.
\end{equation}
With $\chi^{-1}$, we can see $w_{\infty}$ as a function of $(\mathsf{m},\mathsf{p})$ and we denote $w_{\infty}^*$ the specialisation of this function to $\mathsf{m} = 0$. As $K(\mathsf{m} = 0) = \frac{\pi}{2}$ and the Jacobi sn function degenerates to the sine function, we deduce
\begin{equation}
\label{pwinf}\mathsf{p} = \frac{1}{\sin^2\big(\frac{\pi w_{\infty}^*}{2}\big)}.
\end{equation}
Using \eqref{pdef} and trigonometric identities, we find equivalently
\begin{equation}
\label{cosmim}
\cos(\pi w_{\infty}^*) = \frac{1 - \alpha}{1 + \alpha}\,\frac{1 - h(1 + \alpha)\gamma_-^*}{1 + h(1 - \alpha)\gamma_-^*},
\end{equation}
where we recall that $\gamma_-^*$ is the function of $\mathsf{p}$ coming from the specialisation of $\chi^{-1}$ at $\mathsf{m} = 0$. Equation~\eqref{cosmim} uniquely determines $w_{\infty}^* \in (0,1)$. We can also see $w_{\infty}^*$ as a function of $(\gamma_-,\gamma_+) \in \mathcal{A}^{\mathbb{R}}$ by first applying the map $\chi\,:\,(\gamma_-,\gamma_+) \mapsto (\mathsf{m},\mathsf{p})$.

\begin{remark}
\label{rempossss} The condition $|\gamma_-| < \gamma_+ < \varsigma(\gamma_+)$ in $\mathcal{A}^{\mathbb{R}}$ also implies that $|\gamma_-| < \frac{1}{h(\alpha + 1)}$. Hence, \eqref{cosmim} implies
\[
\frac{\cos(\pi w_{\infty}^*)}{1 - \alpha^2} > 0
\]
when $(\gamma_-\gamma_+) \in \mathcal{A}^{\mathbb{R}}$. That is, $w_{\infty}^* \in (\tfrac{1}{2},1)$ (equivalently $\mathsf{p} < 2$) for $\alpha > 1$, and $w_{\infty}^* \in (0,\tfrac{1}{2})$ (equivalently $\mathsf{p} > 2$) for $\alpha < 1$.
\end{remark}

An important role in our analysis will be played by the elliptic nome
\begin{equation}
\label{q12de} q^{\frac{1}{2}} = e^{-\frac{{\rm i}\pi}{2\tau}} = e^{-\frac{\pi}{2T}} = e^{-\frac{\pi K(1 - \mathsf{m})}{K(\mathsf{m})}}.
\end{equation}

\begin{lemma}
\label{Lemachange} Let $\alpha \neq 1$. There exists domains $\mathcal{U},\mathcal{V} \subset \mathbb{C}^2$ (specified in the proof) so that the map $(\gamma_-,\gamma_+) \mapsto (q^{\frac{1}{2}},w_{\infty}^*)$ is analytic in $\mathcal{U}$, has image $\mathcal{V}$ and admits an analytic inverse on the domain $\mathcal{V}$. Besides, we have when $q \rightarrow 0$
\begin{equation}
\label{gmogu} 
\begin{split}
2h(\gamma_+^* - \gamma_+) & = \frac{16\,\cos(\pi w_{\infty}^*)}{1 - \alpha^2}\,q^{\frac{1}{2}} - \frac{64\cos(2\pi w_{\infty}^*)}{(1 - \alpha^2)}\,q + O(q^{\frac{3}{2}}), \\
2h(\gamma_-^* - \gamma_-) & = \frac{16\,\tan\big(\frac{\pi w_{\infty}^*}{2}\big)}{1 - \alpha^2}(q^{\frac{1}{2}} - 4q) + O(q^{\frac{3}{2}}), 
\end{split}
\end{equation}
\end{lemma}
\begin{proof}
The elliptic nome $q^{\frac{1}{2}} = e^{-\frac{\pi K(1- \mathsf{m})}{K(\mathsf{m})}}$ is an analytic function of $\mathsf{m} \in B(0;1) \setminus (-1,0]$. For later use, we record that
\begin{equation}
\label{qm16}
q^{\frac{1}{2}} \sim  \frac{\mathsf{m}}{16}.
\end{equation}
Since the analytic continuation of the elliptic integral has the property that for $\mathsf{m} \in (-1,0)$, ${\rm Im}(K(1 - \mathsf{m} \pm {\rm i}0)) = \mp K(m)$, there is in fact no discontinuity across $(-1,0)$. Together with \eqref{qm16}, this implies that $q^{\frac{1}{2}}$ is an analytic function of $\mathsf{m} \in B(0;1)$, and its image $\mathcal{Q}$ is a neighborhood of $0$. In fact, it is not hard to show that $\mathcal{Q} \subseteq B(0;1)$. A converse identity is known
\begin{equation}
\label{mq128} \mathsf{m} = \frac{\vartheta_2^4\big(0\,\big|\,-\frac{1}{2\tau}\big)}{\vartheta_3^4\big(0\,\big|\,-\frac{1}{2\tau}\big)} = 16q^{\frac{1}{2}}\,\left(\frac{1 + \sum_{l > 0} q^{\frac{l(l + 1)}{2}}}{1 + 2\sum_{l > 0} q^{\frac{l^2}{2}}}\right)^4,
\end{equation}
involving the Jacobi theta functions defined in \eqref{Jactheta}, showing that $\mathsf{m} \mapsto q^{\frac{1}{2}}$ admits an analytic inverse map $q^{\frac{1}{2}} \mapsto \mathsf{m}$ defined at least on the domain $q^{\frac{1}{2}} \in \mathcal{Q}$. 
From Equation~\eqref{pwinf}, we see that $\cos(\pi w_{\infty}^*) = 1 - \tfrac{2}{\mathsf{p}}$. As arccos is analytic in the open vertical strip of width $2$ centered on the imaginary axis, $w_{\infty}^*$ is an analytic function of $\mathsf{p}$ in the domain ${\rm Re}(\frac{1}{\mathsf{p}}) \in (0,1)$. So, if we define
\begin{equation}
\label{A2eps}
\mathcal{T} =  \big\{(\mathsf{m},\mathsf{p}) \in B(0;1) \times \mathbb{C} \quad \big| \quad |\mathsf{m}| < 1\,\,{\rm and}\,\,\mathsf{p} \neq \mathsf{m}\,\,{\rm and}\,\,{\rm Re}(\tfrac{1}{\mathsf{p}}) \in (0,1)\big\},
\end{equation}
the composed map
\begin{equation}
\label{thema1} (\gamma_-,\gamma_+) \mathop{\longmapsto}^{\chi} (\mathsf{m},\mathsf{p}) \longmapsto (q^{\frac{1}{2}},w_{\infty}^*)
\end{equation}
is analytic in the domain $\mathcal{U} = \mathcal{A} \cap \chi^{-1}(\mathcal{T})$. Then, the composed map
\begin{equation}
\label{thema2}
(q^{\frac{1}{2}},w_{\infty}^*) \longmapsto (\mathsf{m},\mathsf{p}) \mathop{\longmapsto}^{\chi^{-1}} (\gamma_-,\gamma_+)
\end{equation}
using \eqref{gmunngynon} is an analytic function in the domain
\begin{equation}
\label{A4eps} \mathcal{V} := B(0;1) \times \Big\{w_{\infty}^* \in \mathbb{C} \quad \Big| \quad {\rm Re}\,w_{\infty}^* \in (0,2)\,\,{\rm and}\,\,\big|\sin\big(\tfrac{\pi w_{\infty}^*}{2}\big)\big| < 1\Big\},
\end{equation}
where we imposed the last inequality to guarantee that $\mathsf{m} \neq \mathsf{p}$ in this domain. This function is by construction an inverse to $(\gamma_-,\gamma_+) \mapsto (q^{\frac{1}{2}},w_\infty^*)$.

Formula~\eqref{gmogu} comes from the expansion of \eqref{gmunngynon} when $\mathsf{m} \rightarrow 0$, using \eqref{pwinf}, \eqref{qm16} and trigonometric identities.
\end{proof}

\begin{remark}
\label{negarem} From \eqref{mq128} we see that that the preimage of $q \in (-1,0] \cap \mathcal{Q}$ via the map $\mathsf{m} \mapsto q$ is $\mathsf{m} \in (-1,0]$. 
\end{remark}

\subsection{Allowing complex parameters in \texorpdfstring{$x(v)$}{x(v)}}
\label{analyticon}
For $(\gamma_-,\gamma_+) \in \mathcal{A}^{\mathbb{R}}$ we recall that $\mathsf{m} \in (0,1)$ and $\mathsf{p} > 0$, and we can use the reciprocal of \eqref{vintermsx} to parametrise the complex $x$ plane:
\begin{equation}
\label{xintermv}
x(v) = \frac{\varsigma(\gamma_+) - \varsigma(\gamma_-)\,\frac{\mathsf{m}}{\mathsf{p}}\,{\rm sn}^2[2{\rm i}K(1 - \mathsf{m})v\,;\,\mathsf{m}]}{1 - \frac{\mathsf{m}}{\mathsf{p}}\,{\rm sn}^2[2{\rm i}K(1 - \mathsf{m})v\,;\,\mathsf{m}]}.
\end{equation} 
As ${\rm sn}[\cdot;\mathsf{m}]$ is an analytic family of meromorphic functions indexed by $\mathsf{m} \in B(0;1)$, we may continue \eqref{xintermv} analytically to complex values of $\gamma_-,\gamma_+$. In view of Lemma~\ref{Lemachange}, we can considering $x(v)$ as an analytic family of meromorphic function of $v$, parametrised by $(q^{\frac{1}{2}},w_{\infty}^*) \in \mathcal{V}$.

\subsection{Properties of \texorpdfstring{$x(v)$}{x(v)} when \texorpdfstring{$q \rightarrow 0$}{q tends to 0}}

Let us insert \eqref{gmunngynon} of $\gamma_{\pm}$ in terms of $(\mathsf{m},\mathsf{p})$ in the formula \eqref{XCeq} for $C$ to get
\[
C^2 = \bigg(\frac{\mathsf{p}^2 - 2\mathsf{p} + \mathsf{m}}{4h(1 - \alpha^2)K(1 - \mathsf{m})}\bigg)^2\,\frac{1}{\mathsf{p}(\mathsf{p} - 1)(1 - \mathsf{m})(\mathsf{p} -\mathsf{m})}.
\]
In view of Lemma~\ref{Lemachange}, the right-hand side is an analytic function of $(q^{\frac{1}{2}},w_{\infty}^*) \in \mathcal{V}$. We notice indeed that by definition of the domains, $(\gamma_-,\gamma_+)$ only take finite values, so $C^2$ has no singularities in $\mathcal{V}$. We now want to identify the sign of the square root to get $C$ itself. Up to restricting $\mathcal{V}$ to a domain
\[
\mathcal{V}'(\eta) := \big\{(q^{\frac{1}{2}},w_{\infty}^*) \in \mathcal{V} \quad \big| \quad {\rm Re}(\tfrac{1}{\mathsf{p}}) \in (\eta,1)\big\}
\]
where $\eta \in (0,1)$ is fixed, and as we have  $|\mathsf{p}| > 1$ due to the last condition in \eqref{A4eps}, we get
\[
{\rm Re}(\mathsf{p} - \mathsf{m}) \geq {\rm Re}(\tfrac{1}{\mathsf{p}}) - \eta' > 0.
\]
for some $\eta' > 0$ depending on $\eta$.
When $(\gamma_-,\gamma_+) \in \mathcal{A}^{\mathbb{R}}$, we know that $C > 0$ and taking into account Remark~\ref{rempossss} it follows that $\mathsf{p} > 2$ and therefore
\begin{equation}
\label{cppp} C = \frac{\mathsf{p}(\mathsf{p} - 2) + \mathsf{m}}{4h(1 - \alpha^2)K(1 - \mathsf{m})}\,\frac{1}{\sqrt{\mathsf{p}(\mathsf{p} - 1)(1 - \mathsf{m})(\mathsf{p} -\mathsf{m})}} 
\end{equation}
is the analytic continuation of $C$ to $(q^{\frac{1}{2}},w_{\infty}^*) \in\mathcal{V}'(\eta)$.

\begin{lemma}
\label{expict} Let $\eta \in (0,1)$. In view of Lemma~\ref{Lemachange}, $w_{\infty}$, $\frac{\pi C}{T}$ and $E_1,E_2$ can be considered as an analytic function of $(q^{\frac{1}{2}},w_{\infty}^*) \in \mathcal{V}'(\eta)$, and when $q \rightarrow 0$
\begin{equation}
\label{Eexp}
\begin{split}
w_{\infty} & = w_{\infty}^* - \frac{4\sin(\pi w_{\infty}^*)}{\pi}\,q^{\frac{1}{2}} + \frac{6\sin(2\pi w_{\infty}^*)}{\pi}\,q + O(q^{\frac{3}{2}}), \\
 \frac{\pi C}{T} & = \frac{2\,{\rm cot}(\pi w_{\infty}^*)}{h(1 - \alpha^2)} + \frac{8\,\sin(\pi w_{\infty}^*)}{h(1 - \alpha^2)}\,q^{\frac{1}{2}} + \frac{8\cos(\pi w_{\infty}^*)\big(2\cos^2(\pi w_{\infty}^*) + 1\big)}{h(1 - \alpha^2)\sin^2(\pi w_{\infty}^*)}\,q + O(q^{\frac{3}{2}}),
 \\
E_1 & = \frac{4\big(1 - \alpha\sin^2(\pi w_{\infty}^*)\big)}{h(1 - \alpha^2)\sin^2(\pi w_{\infty}^*)} + \frac{32 \cos(\pi w_{\infty}^*)}{h(1 - \alpha^2)\sin^2(\pi w_{\infty}^*)}\,q^{\frac{1}{2}} \\
& \quad + \frac{64\big(-\cos^4(\pi w_{\infty}^*) + 2\cos^2(\pi w_{\infty}^*) + 1\big)}{h(1 - \alpha^2)\sin^2(\pi w_{\infty}^*)}\,q  +  O(q^{\frac{3}{2}}), \\
E_2 & = \frac{2\big(3(1 - \alpha)^2 + (1 - 3\alpha^2)\cos^2(\pi w_{\infty}^*)\big)}{h^2(1 - \alpha^2)^2\sin^2(\pi w_{\infty}^*)} + \frac{32(2 - 3\alpha)\cos(\pi w_{\infty}^*)}{h^2(1 - \alpha^2)^2\sin^2(\pi w_{\infty}^*)} \\
& \quad + \frac{64\big((3\alpha + 2)\cos^4(\pi w_{\infty}^*) + (2 - 3\alpha)(2\cos^2(\pi w_{\infty}^*) + 1)\big)}{h^2(1 - \alpha^2)^2\sin^2(\pi w_{\infty}^*)}\,q + O(q^{\frac{3}{2}}),
\end{split} 
\end{equation}
\end{lemma}
\begin{proof}
Combining \eqref{pksn} and \eqref{pwinf} we have
\begin{equation}
\label{winfstar} w_{\infty}^* = \frac{2}{\pi}\,{\rm arcsin}\Big({\rm sn}\big[K(\mathsf{m})w_{\infty}\,;\,\mathsf{m}\big]\Big).
\end{equation}
Using the three first terms of the Fourier expansion
\begin{equation}
\label{splitsn} \begin{split}
{\rm sn}[K(\mathsf{m})w\,;\,\mathsf{m}] & =  \frac{2\pi}{\sqrt{\mathsf{m}}\,K(\mathsf{m})} \sum_{\ell \geq 0} \frac{q^{\frac{\ell}{2} + \frac{1}{4}}}{1 - q^{\ell + \frac{1}{2}}}\,\sin\big[(\ell + \tfrac{1}{2})\pi w\big]
\end{split}
\end{equation}
and solving for $w_{\infty}$ in terms of $(q^{\frac{1}{2}},w_{\infty}^*)$ in the limit  $q \rightarrow 0$, we find the desired expansion.
 
By combining \eqref{cppp} and \eqref{TCeq}, we have
\[ 
\frac{\pi C}{T} = \frac{\pi}{2K(\mathsf{m})}\,\frac{\mathsf{p}(\mathsf{p} - 2) + \mathsf{m}}{h(1 - \alpha^2)}\,\frac{1}{\sqrt{\mathsf{p}(\mathsf{p} - 1)(1 - \mathsf{m})(\mathsf{p} - \mathsf{m})}}.
\]
Using
\begin{equation}
\label{CKexp}
K(\mathsf{m}) = \frac{\pi}{2}\bigg(1 + \frac{\mathsf{m}}{4} + \frac{9\mathsf{m}^2}{64} + O(\mathsf{m}^3)\bigg)
\end{equation} 
and \eqref{mq128}, we obtain the desired expansion of $\frac{\pi C}{T}$. The expansion at $q \rightarrow 0$ of the $E_i$ comes from \eqref{gmogu}, the expression of $\varsigma$ and elementary identities between trigonometric functions.
\end{proof}

We shall need the asymptotic behavior of $x(v)$ near the vertical lines ${\rm Im}\,v = 0$ and ${\rm Im}\,v = \frac{1}{2}$.
\begin{lemma}
\label{LemB3} Let $v = \varepsilon + \tau w$ for $\varepsilon \in \{0,\frac{1}{2}\}$, and $\eta \in (0,1)$. We have when $q \rightarrow 0$ and $w_{\infty}^*$ such that $B(0;1) \times \{w_{\infty}^*\} \subset \mathcal{V}'(\eta)$
\begin{equation}
\label{xvnunv}x(v) - \gamma_+ = q^{\frac{1}{2} - \varepsilon}\,\big(x_{\varepsilon}^*(w) + O(q^{\frac{1}{2}})\big).
\end{equation}
The limit functions are:
\begin{equation*}
\begin{split}
x_{0}^*(w) & = \frac{16\cos(\pi w_{\infty}^*)}{h(1 - \alpha^2)}\,\cos^2\Big(\frac{\pi w}{2}\Big), \\
x_{\frac{1}{2}}^*(w) & = \frac{2\cos(\pi w_{\infty}^*)}{h(1 - \alpha^2)}\,\frac{1}{\cos(\pi w) - \cos (\pi w_{\infty}^*)},
\end{split}
\end{equation*}
where ${\rm cos}(\pi w_{\infty}^*)$ was given in \eqref{cosmim} and $\gamma_+^*$ in \eqref{gplusstar}. For $\varepsilon = 0$, the error in \eqref{xvnunv} is uniform for $w$ in any compact. For $\varepsilon = \frac{1}{2}$, the error is uniform for $w$ in any compact of $\mathbb{C} \setminus (w_{\infty}^* + \mathbb{Z})$. In both cases, the error is stable under differentiation with respect to $v$. 
\end{lemma}
\begin{proof}
Our starting point is the formula \eqref{xintermv} for $x(v)$. For $x = \tau w = {\rm i}T w$, we need to study
\[
{\rm sn}[2{\rm i}K(1 - \mathsf{m})\cdot {\rm i}Tw\,;\,\mathsf{m}] = -{\rm sn}[K(\mathsf{m})w\,;\,\mathsf{m}].
\]
For this we can use the Fourier expansion \eqref{splitsn}
\begin{equation}
\label{splitsn2} \begin{split}
{\rm sn}[K(\mathsf{m})w\,;\,\mathsf{m}] = 4q^{\frac{1}{4}}\sin\big(\tfrac{\pi w}{2}\big)\big(1 - 4q^{\frac{1}{2}} + O(q)\big),
\end{split}
\end{equation}
where the error term is uniform when $w$ belongs to any fixed compact and we have used \eqref{CKexp} to obtain the second line. Together with the expansion of $(\gamma_-,\gamma_+)$ from Lemma~\ref{Lemachange} and the formula $\mathsf{p} = \sin^{-2}(\frac{\pi w_{\infty}^*}{2})$, we deduce from \eqref{xintermv} after some elementary algebra
\[
x(\tau w) = \frac{16\cos(\pi w_{\infty}^*)}{h(1 - \alpha^2)}\,\cos^2\big(\tfrac{\pi w_{\infty}^*}{2}\big)\,q^{\frac{1}{2}} + O(q).
\]
For $x = \frac{1}{2} + \tau w = \frac{1}{2} + {\rm i}T w$, we rather need
\[
{\rm sn}\big[2{\rm i}K(1 - \mathsf{m})\cdot (\tfrac{1}{2} + {\rm i}Tw\big)\,;\,\mathsf{m}\big] = - \frac{1}{\sqrt{\mathsf{m}}\,{\rm sn}[K(\mathsf{m})w\,;\,\mathsf{m}]},
\]
which comes from the identity already used in \eqref{pksn}. Inserting this in \eqref{xintermv} and using again \eqref{splitsn2}, we obtain
\[
x\big(\tfrac{1}{2} + \tau w\big) = \frac{2\cos(\pi w_{\infty}^*)}{h(1 - \alpha^2)}\,\frac{1}{\cos(\pi w) - \cos(\pi w_{\infty}^*)} + O(q^{\frac{1}{2}}).
\]
\end{proof}

\subsection{The case \texorpdfstring{$\alpha = 1$}{alpha=1}}
\label{appa1}

There are some simplifications in absence of bending energy, \textit{i.e.}, $\alpha = 1$. We then have $\varsigma(x) = h^{-1} - x$ and the parameters $(\mathsf{p},\mathsf{m})$ defined in \eqref{defM}-\eqref{pdef} are not anymore independent as $\mathsf{p}(2 - \mathsf{p}) = \mathsf{m}$. This phenomenon also shows up as a singularity in the inverse change of variables \eqref{gmunngynon}, or equivalently, in the fact that since $\varsigma(\infty) = \infty$ we must have $w_{\infty} = \tfrac{1}{2}$ independently of $(\gamma_-,\gamma_+)$, hence $w_{\infty}^* = \tfrac{1}{2}$. We are therefore going to define a different change of variables $\chi_1\,:\,(\gamma_-,\gamma_+) \mapsto (\mathsf{m},\rho)$ with $\mathsf{m}$ still defined by \eqref{defM} and
\[
\rho = 1 - 2h\gamma_-.
\]

\begin{lemma}
\label{Lemachange1} The map $(\gamma_-,\gamma_+) \mapsto (q^{\frac{1}{2}},\rho)$ is analytic in the domain
\[
\mathcal{U}_1 = \mathcal{A}\cap \chi_1^{-1}(B(0;1) \times \mathbb{C})
\]
On its image $\mathcal{V}_1 \subset B(0;1) \times \mathbb{C}$, it admits an analytic inverse map $(q^{\frac{1}{2}},\rho) \mapsto (\gamma_-,\gamma_+)$. Besides, we have when $q \rightarrow 0$
\[
2h(\gamma_+^* - \gamma_+) = 4\rho q^{\frac{1}{2}} + O(q^{\frac{3}{2}}).
\]
where the error is uniform when $\rho$ is bounded.
\end{lemma}
\begin{proof}
The map $\chi_1$ admits an inverse $\chi_1^{-1}$ given by
\begin{equation}
\label{chi1inv} \gamma_- = \frac{1 - \rho}{2h},\qquad \gamma_+ = \frac{\rho + 1}{2h} - \frac{\rho}{h\mathsf{m}}\big(1 - \sqrt{1 - \mathsf{m}}\big).
\end{equation}
The map $\chi_1$ is analytic in the domain $\mathcal{A}$, while the inverse $\chi_1^{-1}$ is extends to an analytic function on $(\mathsf{m},\rho) \in B(0;1) \times \mathbb{C}$. We get the conclusion by further using the change of variable $\mathsf{m} \mapsto q^{\frac{1}{2}}$ already discussed in the proof of Lemma~\ref{Lemachange}.
\end{proof}

We then observe that $\mathsf{p}$ defined by \eqref{pdef} is equal to
\begin{equation}
\label{penwm}
\mathsf{p} = \frac{\mathsf{m}}{1 - \sqrt{1 - \mathsf{m}}}. 
\end{equation}
The discussion of Appendix~\ref{analyticon} can be specialised using this expression: it shows that $x(v)$ becomes an analytic family of meromorphic functions of $v$, parametrised by $(q^{\frac{1}{2}},\rho) \in \mathcal{V}_1$. We have the following analog of Lemma~\ref{expict}.

\begin{lemma}
\label{expict2} In view of Lemma~\ref{Lemachange1}, $\frac{\pi C}{T}$ and $E_1,E_2$ can be seen as analytic functions of $(q^{\frac{1}{2}},\rho) \in \mathcal{V}_1$. The function $E_1$ is constant
\[
E_1 = \frac{2}{h},
\] 
We also have the expansion when $q \rightarrow 0$
\begin{equation}
\begin{split}
\frac{\pi C}{T} & =  \frac{\rho}{2h} - \frac{2\rho}{h}\,q + O(q^2), \\
E_2 & = \frac{6 - \rho^2}{4h^2} - \frac{4\rho^2}{h^2}\,q + O(q^2),
\end{split}
\end{equation}
\end{lemma}
\begin{proof}
Inserting \eqref{chi1inv} in \eqref{CoverT} we find
\[
\frac{\pi C}{T} = \frac{\pi\rho(1 - \sqrt{1 - \mathsf{m}})}{2h\,\mathsf{m}K(\mathsf{m})}.
\]
Using \eqref{CKexp} and \eqref{mq128} then implies the desired expansion when $q \rightarrow 0$. Note that
\[
E_1 = \gamma_- + \gamma_+ + \bigg( \frac{1}{h} - \gamma_-\bigg) + \bigg(\frac{1}{h} - \gamma_+\bigg) = \frac{2}{h}.
\]
The $q \rightarrow 0$ expansions of $E_2$ and $E_3$ are obtained by inserting again \eqref{chi1inv}-\eqref{mq128} into the definitions \eqref{DefE}.
\end{proof}

The analog of Lemma~\ref{LemB3} for the expansion of $x(v)$ itself is
\begin{lemma}
Let $v = \varepsilon + \tau w$ for $\varepsilon \in \{0,\frac{1}{2}\}$. We have for $(q^{\frac{1}{2}},\rho) \in \mathcal{V}_1$, when $q \rightarrow 0$ and uniformly for $\rho$ bounded
\begin{equation}
\label{xv12nunun} x(v) = q^{\frac{1}{2} - \varepsilon}\big(x_{\varepsilon}^*(w) + O(q^{\frac{1}{2}})\big),
\end{equation}
where the limit functions are
\begin{equation}
\begin{split}
x_{0}^*(w) & = \frac{4\rho}{h}\,\cos\bigg(\frac{\pi w}{2}\bigg), \\
x_{\frac{1}{2}}^*(w) & = \frac{\rho}{2h\,\cos(\pi w)}.
\end{split}
\end{equation}
If $\varepsilon = 0$, the error in \eqref{xv12nunun} is uniform for $w$ in any compact. If $\varepsilon = \frac{1}{2}$, the error is uniform for $w$ in any compact of $\mathbb{C} \setminus (\frac{1}{2} + \mathbb{Z})$. In both cases, the error is stable under differentiation with respect to $v$.
\end{lemma}
\begin{proof}
The starting point is the expression \eqref{xintermv} for $x(v)$ with $\mathsf{p}$ replaced by \eqref{penwm} and using the change of variables \eqref{chi1inv}. The proof then becomes similar to that of Lemma~\ref{LemB3} and is omitted.
\end{proof}

Some of these results can be retrieved from those with $\alpha \neq 1$ by taking the limit $\alpha \rightarrow 1$ and $w_{\infty}^*$ in such a way that
\begin{equation}
\label{thelimimt}\bigg(\frac{1}{2} - w_{\infty}^*\bigg) \sim \frac{1 - \alpha}{2\pi}\,\rho.
\end{equation}
This asymptotic relation is compatible with identifying $\rho$ with $1 - 2h\gamma_-^*$ and using \eqref{cosmim}. However, as the change of variable $(\gamma_-,\gamma_+) \mapsto (q^{\frac{1}{2}},w_{\infty}^*)$ was singular at $\alpha \rightarrow 1$, this rule of thumb should not be used blindly. It gives correctly the constants in the above Lemmata that have a finite limit under this limiting procedure. But, for instance, it is meaningless for the second line of \eqref{gmogu}, due to the particular way we defined $\gamma_-^*$ for $\alpha \neq 1$.

\section{Coefficients \texorpdfstring{$(\tilde{g}_{k})_{k \geq 0}$}{(gk)k}}
\label{gdeter}

The coefficients $\tilde{g}_{k}$ have been defined in \eqref{deftildeg} and a priori depend on the parameters of the model: $g$ (resp. $h$) the weight per face not visited (resp. visited) by a loop, $\alpha$ the bending energy, $n$ the weight per loop, and the weight $u$ per vertex (when not set equal to $1$). We again recall that the latter determine $\gamma_-,\gamma_+$ in a way analysed later in Appendix~\ref{proofbeh}. For the moment, considering $\gamma_-,\gamma_+$ --- or equivalently $(q^{\frac{1}{2}},w_{\infty}^*)$ if $\alpha \neq 1$ and $(q^{\frac{1}{2}},\rho)$ if $\alpha = 1$ --- as variables, we can compute the $\tilde{g}_{k}$ using Lemma~\ref{Xinfexp}. Namely, if we introduce
\[
\tilde{g}_k = ({\rm i}C)^k\,\widehat{g}_{k},
\]
we find
\begin{equation}
\label{ghatvalues}
\widehat{g}_{3} =  \frac{2g}{4 - n^2}, \qquad
\widehat{g}_{2} =  \frac{2 - gE_1}{4 - n^2},  \qquad
\widehat{g}_{1} =  \frac{g(3E_{1}^2 - 4E_2) - 6E_1}{12(4 - n^2)}, \qquad
\widehat{g}_{0} =  -\frac{2u}{2 + n}.
\end{equation}
We remark that $\widehat{g}_{3}$ and $\widehat{g}_{0}$ are constants (depending on the parameters of the model) with respect to the variables $(\gamma_-,\gamma_+)$. We can deduce the analyticity properties and $q \rightarrow 0$ expansion of $\widehat{g}_1,\widehat{g}_2$ thanks to Lemma~\ref{expict} or \ref{expict2}.

\begin{corollary}
\label{Cohatg} Assume $\alpha \neq 1$. For $i = 0,1,2,3$, $(q^{\frac{1}{2}},w_{\infty}^*) \mapsto \widehat{g}_i$ is an analytic function in the domain $\mathcal{V}'(\eta)$ for $\eta$ small enough. We have the following expansions when $q \rightarrow 0$:
\begin{footnotesize}
\begin{equation*}
\begin{split}
\widehat{g}_{2} & =  \frac{2}{4 - n^2}\bigg[1 + \frac{2g}{h(1 - \alpha^2)}\bigg(\alpha - \frac{1}{\sin^2(\pi w_{\infty}^*)}\bigg)\bigg]  - \frac{32g}{h(1 - \alpha^2)(4 - n^2)}\,\frac{\cos(\pi w_{\infty}^*)}{\sin^2(\pi w_{\infty}^*)}\,q^{\frac{1}{2}} + O(q),  \\
\widehat{g}_{1} & =  \frac{2}{h(4 - n^2)(1 - \alpha^2)}\bigg[\frac{g}{h(1 - \alpha^2)}\bigg((1 + 3\alpha^2) - \frac{2(2 + 3\alpha)}{\sin^2(\pi w_{\infty}^*)} + \frac{6}{\sin^4(\pi w_{\infty}^*)}\bigg) + \alpha -\frac{1}{\sin^2(\pi w_{\infty}^*)}\bigg] \\ 
& \quad + \frac{16\cos(\pi w_{\infty}^*)}{h(4 - n^2)(1 - \alpha^2)\sin^2(\pi w_{\infty}^*)}\bigg(\frac{2g}{3h}\,\frac{(2 + 3\alpha)\cos^2(\pi w_{\infty}^*) + 4 - 3\alpha}{(1 - \alpha^2)\sin^2(\pi w_{\infty}^*)} - 1\bigg)q^{\frac{1}{2}} + O(q).
\end{split} 
\end{equation*}
\end{footnotesize}
\hfill $\Box$
\end{corollary}
We denote $\widehat{g}_i^*$ the value of this function $\widehat{g}_i$ at $q = 0$.

\vspace{0.2cm}

There are some simplifications for $\alpha = 1$. Owing to the exact relation $E_1 = \frac{2}{h}$, only $\widehat{g}_{1}$ has a non trivial dependence in $q$.
\begin{corollary}
\label{Cohatg2} Assume $\alpha = 1$. For $i = 0,1,2,3$, $(q^{\frac{1}{2}},\rho) \mapsto \widehat{g}_i$ is an analytic function in the domain $\mathcal{V}_1(\eta)$ for $\eta$ small enough, We have
\[
\widehat{g}_{2} =  \frac{2}{4 - n^2}\bigg(1 - \frac{g}{h}\bigg),
\]
and when $q \rightarrow 0$
\[
\widehat{g}_{1} =  \frac{1}{h(4 - n^2)}\bigg(- 1 + \frac{g}{12h}(\rho^2 + 6)\bigg) + \frac{4g\rho^2}{3h^2(4- n^2)}\,q + O(q^2).
\]
\hfill $\Box$
\end{corollary}

\section{The special function \texorpdfstring{$\Upsilon_{b}(v)$}{Ub(v)}}
\label{AppUpsilon}
\label{Upbeh}
$\Upsilon_{b}(v)$ is the unique meromorphic function with a simple pole at $v = 0$ with residue $1$, and the pseudo-periodicity properties:
\[
\Upsilon_{b}(v + 1) = \Upsilon_{b}(v),\qquad \Upsilon_{b}(v + \tau) = e^{{\rm i}\pi b}\Upsilon_{b}(v).
\]
We have several expressions:
\begin{equation}
\label{Upsfb}\begin{split}
\Upsilon_{b}(v) & = \sum_{m \in \mathbb{Z}} e^{-{\rm i}\pi b m}\,\pi\,\mathrm{cot}\big[\pi(v + m\tau)\big] \\
& = \frac{\vartheta_{1}'(0|\tau)}{\vartheta_{1}\big(-\frac{b}{2}\big|\tau\big)}\,\frac{\vartheta_{1}\big(v - \frac{b}{2}\big|\tau\big)}{\vartheta_{1}(v|\tau)} \\
& =  \frac{e^{\frac{{\rm i}\pi b v}{\tau}}}{{\rm i}T}\,\frac{\vartheta_1'\big(0\big|-\frac{1}{\tau}\big)}{\vartheta_1\big(-\frac{b}{2\tau}\big|-\frac{1}{\tau}\big)}\,\frac{\vartheta_{1}\big(\frac{v - b/2}{\tau}\big|-\frac{1}{\tau}\big)}{\vartheta_{1}\big(\frac{v}{\tau}|-\frac{1}{\tau}\big)}.
\end{split} 
\end{equation}
Curiously, this function also appears in the dynamical $R$-matrix of the elliptic Calogero system \cite{BBT}.

\begin{remark}\label{theremr} Due to the presence of $-\frac{1}{\tau}$ in the argument of $\vartheta_1$, $\Upsilon_b(v)$ is a family of meromorphic functions of $v \in \mathbb{C}$, only in the domain $q^{\frac{1}{2}} = e^{-\frac{{\rm i}\pi}{2\tau}} \in B(0;1)\setminus (-1,0]$. The values above and below the real negative axis in $q$ are different. The poles of $\Upsilon_b$ are located at $\mathbb{Z} \oplus \tau \mathbb{Z}$. 
\end{remark}

The last expression in \eqref{Upsfb} is convenient to study the regime $q^{\frac{1}{2}} \rightarrow 0$ in $B(0;1) \setminus (-1,0]$. 

\begin{lemma}
\label{lemUp}Let $v = \varepsilon + \tau w$ with $\varepsilon \in \{0,\frac{1}{2}\}$. We have, for $b \in (0,1)$:
\small
\[
\Upsilon_{b}(v) = \frac{2\pi q^{\varepsilon b}}{T(1 - q^b)} \cdot \left\{\begin{array}{lll} \Upsilon_{b,0}^*(w)  - q^b\Upsilon_{b + 2,0}^*(w) + O(q^{2 - b}) & & {\rm if}\,\,\varepsilon = 0 \\ \Upsilon_{b,\frac{1}{2}}^*(w) - (q^{1 - b} - q)\Upsilon_{b - 2,\frac{1}{2}}^*(w) + q\Upsilon_{b + 2,\frac{1}{2}}^*(w) + O(q^{1 + b}) & & {\rm if}\,\,\varepsilon = \tfrac{1}{2}\end{array}\right..
\]
\normalsize
The errors are uniform for $w$ in any compact independent of $\tau \rightarrow 0$, and the expressions for the limit functions are:
\begin{equation}
\label{theone}
\begin{split}
\Upsilon_{b,0}^*(w) & = \frac{e^{{\rm i}\pi (b -1)w}}{2{\rm i}\sin(\pi w)}, \\
\Upsilon_{b,\frac{1}{2}}^*(w) & = - e^{{\rm i}\pi b w}.
\end{split}
\end{equation}
\end{lemma}
\begin{proof} This is obtained by isolating carefully the terms dominating the $q$-series defining the theta functions that appear in the last line of \eqref{Upsfb}.
\end{proof}

\section{The phase diagram and volume exponent}
\label{proofbeh}

The parameters of the model $\alpha,g,h,n,u$ determine $\gamma_-,\gamma_+$ through the equations
\[
\mathbf{G}(\varepsilon + \tau) = 0\qquad {\rm for}\,\,\varepsilon \in \big\{0,\tfrac{1}{2}\big\}.
\]
With the previous notations $v_{\infty} = \frac{1}{2} + \tau w_{\infty}$ and $\tilde{g}_{k} = ({\rm i}C)^k\widehat{g}_{k}$ and according to Theorem~\ref{theimdisk}, these equations take the form:
\begin{multline}
\label{thetnirugngun}
\sum_{k \geq 0} \frac{\widehat{g}_{k}}{k!} \bigg(\frac{\pi C}{T}\bigg)^k\,\partial_{\pi w_{\infty}}^{k} \Big[\Upsilon_{b}(\overline{\varepsilon} + \tau (w_{\infty} + 1)) + \Upsilon_{b}(\overline{\varepsilon} + \tau(1 - w_{\infty}))   \\
 - \Upsilon_{b}(\overline{\varepsilon} + \tau(w_{\infty} - 1)) - \Upsilon_{b}(\overline{\varepsilon}  - \tau(1 + w_{\infty}))\Big]  = 0.
 \end{multline}
where $\overline{\varepsilon} = \frac{1}{2} - \varepsilon$.

\begin{lemma}
\label{predetla}Assume $n \in (0,2)$, that is $b \in (0,\frac{1}{2})$. In light of Lemma~\ref{Lemachange} for $\alpha \neq 1$, for $\eta$ small enough the equation \eqref{thetnirugngun} for $\varepsilon = 0$ determines a function $(q^{\frac{1}{2}},w_{\infty}^*) \mapsto u$ (resp. $(q^{\frac{1}{2}},\rho) \mapsto u$) which is analytic in the domain
\[
\mathcal{V}''(\eta) = \big(B(0;1)\setminus \mathbb{R}_{< 0}\big) \times \big\{w_{\infty}^* \in \mathbb{C} \quad \big| \quad {\rm Re}\,w_{\infty}^* \in (\eta,1)\,\,{\rm and}\,\,\big|\sin\big(\tfrac{\pi w_{\infty}^*}{2}\big)\big| < 1\big\},
\]
If $\alpha = 1$ and in view of Lemma~\ref{Lemachange1}, it determines likewise an analytic function $(q^{\frac{1}{2}},w_{\infty}^*) \mapsto u$ in the domain
\[
\mathcal{V}_1'' := \mathcal{V}_1 \setminus \big(\mathbb{R}_{\leq 0} \times \mathbb{C}\big).
\]
\end{lemma}
\begin{proof}
We first discuss $\alpha \neq 1$. From Lemmata \ref{Lemachange} and \ref{expict}, $w_{\infty}$, $(\widehat{g}_k)_{k = 1}^3$ and $\frac{\pi C}{T}$ are analytic functions of $(q^{\frac{1}{2}},w_{\infty}^*) \in \mathcal{V}'(\eta)$ for $\eta$ small enough. Taking into account Remark~\ref{theremr} and the fact that the argument of $\Upsilon_b$ always avoids the poles of $v \mapsto \Upsilon_b(v)$, we deduce that the left-hand side of \eqref{thetnirugngun} with $\overline{\varepsilon} = \frac{1}{2} - \varepsilon = \frac{1}{2}$ is an analytic function of $(q^{\frac{1}{2}},w_{\infty}^*) \in \mathcal{V}'(\eta) \setminus (\mathbb{R}_{\leq 0} \times \mathbb{C})$. $u$ only occurs in Equation~\eqref{thetnirugngun} through $\widehat{g}_0 = -\frac{2u}{2 + n}$ (cf. Equation \eqref{ghatvalues}). So, using Lemma~\ref{lemUp} and after dividing the equation by $\frac{2\pi q^{\frac{b}{2}}}{T(1 - q^{b})}$, the prefactor of $u$ in the left-hand side becomes in the $q \rightarrow 0$ limit
\begin{small}
\begin{equation}
\label{nigfudngiugn}\begin{split}
& \frac{-2}{2 + n}\bigg(\Upsilon_b\big(\tfrac{1}{2} + \tau(w_{\infty} + 1)\big) + \Upsilon_b\big(\tfrac{1}{2} + \tau(w_{\infty} - 1)\big) - \Upsilon_b\big(\tfrac{1}{2} + \tau(w_{\infty} - 1)\big) - \Upsilon_b\big(\tfrac{1}{2} - \tau(1 + w_{\infty})\big)\bigg) \\
& = -\frac{8{\rm i}\sin(\pi b)}{2 + n}\,\cos(\pi b w_{\infty}^*) + o(1).
\end{split}
\end{equation}
\end{small}
As we assumed $b \in (0,\tfrac{1}{2})$ and in the definition of the domain, restricting to the range of $w_{\infty}^*$ to ${\rm Re}\,w_{\infty}^* \in (\eta,1)$, i.e. restricting $(q^{\frac{1}{2}},w_{\infty}^*)$ further to the domain $\mathcal{V}''(\eta)$ shows that the leading term in Equation~\eqref{nigfudngiugn} does not vanish in a neighborhood of $q = 0$ in this domain. Therefore, Equation~\eqref{thetnirugngun} allows expressing $u$ as an analytic function of $(q^{\frac{1}{2}},w_{\infty}^*) \in \mathcal{V}''(\eta)$.

For $\alpha = 1$, the argument is similar but it is not necessary to restrict further the range of the extra variable $\rho$ because \eqref{nigfudngiugn} becomes $-\frac{8{\rm i}\sin(\pi b)\cos(\pi b/2)}{2 + n} \neq 0$ under the assumption $b \in (0,\tfrac{1}{2})$, so we can use the domain $\mathcal{V}''_1$.
\end{proof}

In the non generic critical regime, we have $\gamma_+ \rightarrow \gamma_+^* = \frac{1}{h(\alpha + 1)}$, thus $q \rightarrow 0$ in terms of the parametrisation of Lemma~\ref{Lemachange} or Lemma~\ref{Lemachange1}. Inserting the asymptotic expansions from Corollary~\ref{lemUp} yields:
\begin{equation}
\label{eq1} 
\begin{split}
& \quad \sum_{k = 0}^3 \frac{\widehat{g}_{k}}{k!}\,\bigg(\frac{\pi C}{T}\bigg)^{k}\,\Big[Y_{b,0}^{(k)}(\pi w_{\infty}) - q^{1 - b} Y_{b - 2,0}^{(k)}(\pi w_{\infty}) \\
& \quad\qquad\qquad\qquad\qquad  + q\big(Y_{b - 2,0}^{(k)}(\pi w_{\infty}) + Y_{b + 2,0}^{(k)}(\pi w_{\infty})\big) + O(q^{1 + b})\Big] = 0, \\
& \quad \sum_{k = 0}^3 \frac{\widehat{g}_{k}}{k!}\,\bigg(\frac{\pi C}{T}\bigg)^k\Big[Y_{b,\frac{1}{2}}^{(k)}(\pi w_{\infty}) - q^{b} Y_{b + 2,\frac{1}{2}}^{(k)}(\pi w_{\infty}) + O(q^{2 - b})\Big] = 0. 
\end{split}
\end{equation}
with coefficients:
\[
Y_{b,0}(w) = \cos(b w),\qquad Y_{b,\frac{1}{2}}(w) = \frac{\sin[(1 - b)w]}{\sin w}.
\]

\subsection{The non generic critical line}

At a non generic critical point, we must have $u = 1$ and $q = 0$, thus:
\[
-\frac{2}{2 + n} + \sum_{k = 1}^{3} \frac{\widehat{g}_k^*}{k!}\,\bigg(\frac{2\,{\rm cot}(\pi w_{\infty}^*)}{(1 - \alpha^2)h}\bigg)^k\,\frac{Y_{b,\varepsilon}^{(k)}(\pi w_{\infty}^*)}{Y_{b,\varepsilon}(\pi w_{\infty}^*)} = 0\qquad \varepsilon \in \{0,\tfrac{1}{2}\}.
\]
where we have used the expression for $\frac{\pi C}{T}$ at $q = 0$ given in Lemma~\ref{Cohatg} and we should insert the expression for $\widehat{g}_k^*$ given in Lemma~\ref{Cohatg2}.
We note that the critical values $\widehat{g}_k^*$ obtained in Section~\ref{gdeter} are such that \eqref{eq1} give a linear system of equations for $(\frac{g}{h},h^2)$, parametrised by $w_{\infty}^*$ for $\alpha \neq 1$, and by $\rho$ if $\alpha = 1$. These equations, as well as their explicit solution for $\alpha = 1$, already appeared in \cite[Sections 4.1 and 4.2]{BBG12b}.

For $\alpha = 1$, the solution is 
\begin{equation}
\label{gsurh}
\begin{split} 
\frac{g}{h} & = \frac{4(\rho b\sqrt{2 + n} - \sqrt{2 - n})}{-\rho^2(1 - b^2)\sqrt{2 - n} + 4\rho b\sqrt{2 + n} - 2\sqrt{2 - n}}, \\
h^2 & = \frac{\rho^2 b}{24\sqrt{4 - n^2}}\,\frac{\rho^2\,b(1 - b^2)\sqrt{2 + n}  - 4(1 - b^2)\rho\sqrt{2 - n} + 6b\sqrt{2 + n}}{-\rho^2(1 - b^2)\sqrt{2 - n} + 4\rho b\sqrt{2 + n} - 2\sqrt{2 - n}},
\end{split} 
\end{equation}
as claimed in Theorem~\ref{thphase} and in agreement with \cite[Equation 4.15]{BBG12b}. Since $\frac{g}{h}$ and $h^2$ must be non negative, we must have $\rho \in [\rho_{\min}',\rho_{\max}]$, with:
\begin{equation}
\begin{split}
\label{rhomin2}\rho_{\min}' & = \frac{2\sqrt{1 - b^2}\sqrt{2 - n} - \sqrt{2}\sqrt{(10 + n)b^2 - 4 + 2n}}{b\sqrt{1 - b^2}\sqrt{2 - n}}, \\
 \rho_{\max} & = \frac{1}{b}\sqrt{\frac{2 - n}{2 + n}}.
 \end{split}
\end{equation}
However, we will see later that the non generic critical line only exists until some value $ \rho_{\min} > \rho_{\min}'$, so \eqref{rhomin2} will become irrelevant.

In the general case $\alpha \neq 1$, the expression is more complicated.

\begin{equation}
\begin{split} 
\label{gsurh2}\frac{g}{h} & = 6(1 - \alpha^2)\sin^2(\pi w_{\infty}^*)\,\frac{\sum_{k = 0}^1 b^k\tilde{P}_{k}(\pi w_{\infty}^*)}{\sum_{k = 0}^2 b^k P_{k}(\pi w_{\infty}^*)},  \\
 h^2 & = \frac{2b\cos^2(\pi w_{\infty}^*)}{(1 - \alpha^2)^2(2 - n)\sin^4(\pi w_{\infty}^*)}\,\frac{\sum_{k = 0}^3 b^k \tilde{Q}_k(\pi w_{\infty}^*)}{\sum_{k = 0}^2 b^kP_{k}(\pi w_{\infty}^*)},
\end{split}
\end{equation}
with: \small
\begin{equation*}
\begin{split}
\tilde{P}_1(w) & = \sin(2w)\big(3 - 2\sin^2(bw) - \alpha \sin^2(w)\big), \\
\tilde{P}_0(w) & = \sin(2bw)\big(-3 + (2 + \alpha)\sin^2(w)\big), \\
P_3(w) & = -\sin(2bw), \\
P_2(w) & = -3\sin(2w)^2\sin(2bw), \\
P_1(w) & = 2\sin(2w)\Big((3\alpha^2 + 1)\cos^4(w) + \big(12(\alpha + 1)\cos^2(bw) - 6\alpha^2 + 6\alpha + 2\big)\cos^2(w) \\
& \quad + 12(1 - \alpha)\cos^2(bw) + 3(1 - \alpha)^2\Big), \\
P_0(w) & = 6\sin(2bw)\big(- (\alpha + 1)(\alpha + 3)\sin^4(w) + 6(\alpha + 2)\sin^2(w) - 10\big),
\end{split}
\end{equation*}
\normalsize and \small
\begin{equation*}
\begin{split}
\tilde{Q}_4(w) & = \sin^2(w)\sin^3(2w), \\
\tilde{Q}_3(w) & = -\sin^2(w)\sin(2w)\sin(2bw), \\
\tilde{Q}_2(w) & = 2\sin^2(w)\sin(2w)\Big((3\alpha^2 - 1)\cos^4(w) - 2\big((\alpha + 2)\cos^2(bw) + 3\alpha^2 - 5\alpha + 3\big)\cos^2(w), \\ 
& \quad + 4(1 - \alpha)\cos^2(bw) - 3\alpha^2 + 10\alpha - 7\Big), \\
\tilde{Q}_1(w) & = 2\sin^2(w)\sin(2bw)\Big((3\alpha^2 + 12\alpha + 7)\cos^4(w) + 2(-3\alpha^2 - 3\alpha + 7)\cos^2(w) + 3(1 - \alpha)^2\Big), \\
\tilde{Q}_0(w) & = 2\sin(2w)\Big((1 - 4\alpha - 3\alpha^2)\cos^4(w) + 2(3\alpha^2 - 2\alpha - 1)\cos^2(w) - 3\alpha^2 + 8\alpha - 5\Big), \\
Q_3(w) & = -\sin^2(2w), \\
Q_2(w) & = 3\sin^2(2w)\sin(2bw), \\
Q_1(w) & = 2\sin(2w)\Big((3\alpha^2 + 1)\cos^4(w) + 2\big(6(\alpha + 1)\cos^2(bw) - 3\alpha^2 + 3\alpha + 1\big)\cos^2(w), \\
& \quad + 12(1 - \alpha)\cos^2(bw) + 3(1 - \alpha)^2\Big), \\
Q_0(w) & = -3\sin(2bw)\Big((\alpha + 1)(\alpha + 3)\cos^4(w) + 2(-\alpha^2 - \alpha + 3)\cos^2(w) + (1 - \alpha)^2\Big).
\end{split}
\end{equation*}
\normalsize

We have checked that, in the limit $\alpha \rightarrow 1$ such that $(\frac{1}{2} - w_{\infty}^*) \sim \frac{1 - \alpha}{2}\,\rho$, these expressions retrieve \eqref{gsurh}. This completes the proof of Theorem~\ref{thphase}.

\subsection{Near criticality} Let us fix $(g,h)$ on the non generic critical line for $u = 1$. We now study the behavior when $u \neq 1$ but $u \rightarrow 1$ of the endpoints $\gamma_{\pm}$. In view of the change of variables in Lemma~\ref{Lemachange} for $\alpha \neq 1$ (Lemma~\ref{Lemachange1} for $\alpha = 1$), it amounts to determining the dependence of $u$ in the variable $q$ while the second parameter $w_{\infty}^*$ (resp. $\rho$) is specified by the position of $(g,h)$ on the non generic critical line via Equation~\ref{gsurh2} (resp. Equation~\ref{gsurh}).

For this purpose, we look at \eqref{eq1}, and note that $u$ only appears in $\widehat{g}_{0}$. There could be a term of order $q^{\frac{1}{2}}$ stemming from near-criticality corrections to $w_{\infty}$, $\widehat{g}_{k}$ and $\frac{\pi C}{T}$, but the computation reveals that it is absent. Therefore, we obtain:
\begin{equation}
1 - u  = \frac{n + 2}{2}\bigg(\sum_{k = 0}^3 \frac{\widehat{g}_{k}^*}{k!}\Big(\frac{2\,{\rm cot}(\pi w_{\infty}^*)}{(1 - \alpha^2)h}\Big)^k\,\frac{Y_{b - 2,0}^{(k)}(\pi w_{\infty}^*)}{Y_{b,0}(\pi w_{\infty}^*)}\bigg) q^{1 - b} + O(q),
\end{equation}
where $\widehat{g}^*_{0} = -\frac{2}{2 + n}$ and $(\widehat{g}^*_k)_{k \geq 1}$ should be replaced by their values in terms of $(g,h,w_{\infty}^*)$ from Corollary~\ref{Cohatg} or Corollary~\ref{Cohatg2}.

\subsubsection{Case $\alpha = 1$} Here, we rather use the parametrisation \eqref{gsurh}, and the resulting formula is relatively simple:
\beq
\label{Deltastar}1 - u = \Delta\,q^{1 - b} + O(q),
\eeq
with:
\[
\Delta = \frac{12}{b}\,\frac{\rho^2(1 - b)^2\sqrt{2 + n} + 2\rho(1 - b)\sqrt{2 - n} - 2\sqrt{2 + n}}{-\rho^2 b(1 - b^2)\sqrt{2 + n} + 4\rho(1 - b^2)\sqrt{2 - n} - 6b\sqrt{2 + n}}.
\]
When we restrict to real values of  $\gamma_-,\gamma_+$ --- hence real values of $\rho$ --- we have $\Delta \geq 0$ iff $\rho \in [\rho_{\min},\rho_{\max}]$ with
\beq
\label{rhominf}\rho_{\min} = \frac{\sqrt{6 + n} - \sqrt{2 - n}}{(1 - b)\sqrt{2 + n}}.
\eeq
and we note by comparing with \eqref{rhomin2} that $\rho_{\min} > \rho_{\min}'$ for $n \in [0,2]$.

At $\rho = \rho_{\min}$, we have $\Delta = 0$, and we need to go further in the $q \rightarrow 0$ expansion. The next term is of order $q$. To compute it, we need to take into account in Equation~\ref{thetnirugngun} the term of order $q$ arising from $\frac{\pi C}{T}$ and $\widehat{g}_k$ (Lemma~\ref{expict2}) and from the expansion of $\Upsilon_b$ (Lemma~\ref{lemUp}). The result for a general value of $\rho$ is
\[
1 - u = \Delta q^{1 - b} + \Delta_1\,q + O(q^{1 + b}),
\]
with
\[
\Delta_1  =  \frac{24}{b}\,\frac{-\rho^2(3b^2 + 1)\sqrt{2 + n} + 4\rho b\sqrt{2 - n} + 2\sqrt{2 + n}}{-\rho^2b(1 - b^2)\sqrt{2 + n} + 4\rho(1 - b^2)\sqrt{2 - n} - 6b\sqrt{2 + n}}.
\]
and for $\rho = \rho_{\min}$ it specialises to
\[
\Delta_1 = \frac{24(1 + b)}{b(1 - b)(2 - b)}.
\]

If we are at a dense critical point, we must have $\Delta > 0$, hence $\rho \in (\rho_{\min},\rho_{\max}]$. If we are at a dilute critical point, we must have $\Delta = 0$ and $\Delta_1 > 0$, hence $\rho = \rho_{\min}$. These necessary conditions were already obtained in \cite{BBG12b} --- where the lower bound arose from the constraint of positivity of the spectral density associated with the generating series of disks $\mathbf{F}(x)$. Modulo Remark~\ref{rem:buddchen} about the justification that this condition is also sufficient, this establishes the phase diagram of the model for $\alpha = 1$.

In particular, we see that at a dense critical point
\[
q \stackrel{.}{\sim} (1 - u)^{\frac{1}{1 - b}},
\]
while at a dilute critical point we rather have
\[
q \stackrel{.}{\sim} (1 - u).
\]

\subsubsection{General $\alpha$} 

The method is similar but the explicit results are cumbersome and we will not reproduce them here. We start from \eqref{eq1} but now have to take into account that $w_{\infty}$ now depends on $(q^{\frac{1}{2}},w_{\infty}^*)$, that it contains a $q^{\frac{1}{2}}$ term in its $q \rightarrow 0$, and that so do $\frac{\pi C}{T}$, $\widehat{g}_1$ and $\widehat{g}_2$. We have nevertheless checked that the $q^{\frac{1}{2}}$ term is absent in the $q \rightarrow 0$ expansion of \eqref{eq1}, which gives an expansion of the form
\[
1 - u = \Delta\,q^{1 - b} + \Delta_1 q + O(q^{1 + b})
\]
for some $\Delta$ and $\Delta_1$ which are complicated functions of $w_{\infty}^*$. If $\Delta > 0$ we have a dense critical point, if $\Delta = 0$ and $\Delta_1 > 0$ we have a dilute critical point. This reasoning is still valid although describing explicitly the dense and dilute critical locus requires the explicit expressions of $\Delta$ and $\Delta_1$.

\subsection{Delta-analyticity}
\label{gstrapp}

Let us recall an important notion of singularity analysis~\cite{Flajolet}.

\begin{definition}
  \label{defdeltadom}
  A \emph{delta-domain} at $z_0 \in \mathbb{C} \setminus \{0\} $ is an open subset of the complex plane of the form
  \[
    \big\{z \in \mathbb{C} \setminus \{z_0\}  \quad | \quad |\tfrac{z}{z_0} - 1| < R,\,\,|{\rm arg}(\tfrac{z}{z_0} - 1)| > \phi\big\}
  \]
  with $R>1$ and $\phi \in (0,\tfrac{\pi}{2})$. A function is said
  \emph{delta-analytic} if it is analytic in some delta-domain.
\end{definition}

Let us introduce a weaker notion: we say that a function is
\emph{delta-analytic locally at $z_0$} if it is analytic in the
intersection of a delta-domain at $z_0$ and of a neighborhood of
$z_0$. It is not difficult to check that a function is analytic in a
delta-domain at $z_0$ if and only if:
\begin{itemize}
\item it is analytic in the open disk of radius $|z_0|$ centered at $0$,
\item it is analytic at every point of modulus $|z_0|$ other than $z_0$,
\item it is delta-analytic locally at $z_0$.
\end{itemize}

\begin{lemma}
\label{lemdeltanal} If $(g,n,\alpha,h)$ is a non generic critical point (dense or dilute) and $n \in (0,2)$, the maps described in Lemma~\ref{predetla} have an inverse: specialising the inverse to the value of $w_{\infty}^*$ if $\alpha \neq 1$ (resp. $\rho$ if $\alpha = 1$), we obtain a map $u \mapsto q$ which is delta-analytic locally at $1$, and behaves like
\[
q \sim \bigg(\frac{1 - u}{\Delta}\bigg)^{c},\qquad c = \left\{\begin{array}{lll} \frac{1}{1 - b} & & {\rm dense}\,\,{\rm phase} \\ \,\,\,1 & & {\rm dilute}\,\,{\rm phase} \end{array}\right.
\]
for some constant $\Delta > 0$.
\end{lemma}
\begin{proof}
Let us assume first $\alpha = 1$. We have justified in the previous paragraphs that the maps $(q^{\frac{1}{2}},\rho) \mapsto u$ from Lemma~\ref{predetla} satisfy
\[
1 - u = \Delta\,q^{1 - b}  + O(q),
\]
where $\Delta$ is a function of $\rho$. At a dense critical point, the value of $\rho$ is such that $\Delta > 0$.  As $1 - b \in (\frac{1}{2},1)$. the image of $\mathbb{C}\setminus \mathbb{R}_{\leq 0}$ by the map $q \mapsto q^{1 - b}$ is a delta-domain. Therefore, for $\eta > 0$ small enough the image of $B(0;\eta)$ via the map $q \mapsto u$ of Lemma~\ref{predetla} contains a local delta-domain centered at $1$ and on the latter we can find an inverse map $u \mapsto q$ which is analytic, hence delta-analytic locally at $u = 1$, by definition. This map behaves like
\[
q \,\,\sim\,\, \bigg(\frac{1 - u}{\Delta}\bigg)^{\frac{1}{1 - b}}.
 \]
At a dilute critical point, we have $\Delta = 0$ and pushing further the expansion we have
\[
1 - u = \Delta_1\,q + O(q^{1 + b}).
\]
The image of $\mathbb{C}\setminus \mathbb{R}_{\leq 0}$ by the identity map being a delta-domain, we can repeat the same argument and conclude that we have a map $u \mapsto q$ which is delta-analytic locally at $u = 1$.

For positive $\alpha \neq 1$, we should distinguish (see Remark~\ref{alphanottoosmall}) between the case $w_\infty \in \big[\tau,\tau + \frac{1}{2}\big]$ which can occur when $\alpha$ is small enough, and $w_\infty \in \big[\frac{1}{2}, \frac{1}{2} + \tau\big]$ on which we have focused since Appendix~\ref{App1}. We only discuss the second case, as the first case can be obtained similarly, after an adaptation without further difficulty of the previous appendices. The condition $\Delta > 0$ (resp. $\Delta = 0$ and $\Delta_1 > 0$) is necessary to be at a dense (resp. dilute) critical point, so the discussion for $\alpha = 1$ extends here using as second parameter $w_{\infty}^*$ instead of $\rho$:  the explicit expressions for $\Delta,\Delta_1$ are not needed in this argument.
\end{proof}

\section{Scaling limits for pointed disks}
\label{Sclpoin}
We are going to prove Theorem~\ref{th85} and Corollaries~\ref{Covol}-\ref{corrrr1}, i.e. analyse the generating series of pointed disks
\begin{equation}
\label{Fbulbus}\mathbf{F}^{\bullet}(x) = v'(x)\,\mathbf{G}^{\bullet}(v(x)) - \partial_{x}\bigg(\frac{nu\,\ln[\varsigma'(x)]}{2(2 + n)}\bigg).
\end{equation}
Here $x$ is in the physical sheet, that is
\[
v(x) \in \hat{\mathcal{R}} = \big\{v \in \mathbb{C},\qquad {\rm Re}\,v \in \big(-\tfrac{1}{2},\tfrac{1}{2}\big],\quad {\rm Im}\,v \in [0,T] \big\}.
\]
and we also assume that $(g,h)$ are chosen at a non generic critical point when $u = 1$. According to the previous sections, it amounts to fixing a value of $w_{\infty}^*$ (if $\alpha \neq 1$) or $\rho$ (if $\alpha = 1$) and take $(g,h)$ given by \eqref{gsurh2} or \eqref{gsurh}.

Let us first fix the value of $b$. As the second term in \eqref{Fbulbus} is linear in $u$, it will not contribute to the singularity analysis when $u \rightarrow 1$. We therefore focus on the first term, as given by Proposition~\ref{propPoint}
\[
\mathbf{G}^{\bullet}(v) = \frac{u}{2 + n}\bigg(-\Upsilon_{b}(v + v_{\infty}) - \Upsilon_{b}(v - v_{\infty}) + \Upsilon_{b}(-v + v_{\infty}) + \Upsilon_{b}(-v-v_{\infty})\bigg).
\]
In view of Remark~\ref{theremr} and Lemma~\ref{lemdeltanal}, $\mathbf{G}^{\bullet}(v)$ can be considered as an analytic family of meromorphic functions of $v$, parametrised by $u$ in a delta-domain centered at $1$. On the other hand $x \mapsto v(x)$ given by \eqref{vintermsx} is an analytic family of analytic functions of $x$ in the physical sheet, parametrised by $\mathsf{m}$ in a small neighborhood of $0$. Changing the variable to $q$ (see Lemma~\ref{Lemachange} or Lemma~\ref{Lemachange1}) and then to $u$ (Lemma~\ref{lemdeltanal}), we deduce that $x \mapsto v'(x)\mathbf{G}^{\bullet}(v(x))$ is an analytic family of meromorphic functions of $x$ in the physical sheet, parametrised by $u$ in a delta-domain centered at $1$. The map $s \mapsto b(s) = \frac{1}{\pi}\,{\rm arccos}\big(\frac{ns}{2}\big)$ is an analytic function of $s$ in the strip $|{\rm Re}\,s| < \frac{2}{n}$. So, if we set $b = b(s)$, one can extend the previous arguments to prove the analyticity of the family of functions $\mathbf{F}_s^{\bullet}(x)$ with respect to $s$ in this strip.

We first analyse the regime $x = x(v)$ with $v = \tfrac{1}{2} + \tau w$ and $w$ in a compact region of the complex plane containing a $u$-independent neighborhood of $w_{\infty}^*$. This means that $x$ remains in a $u$-independent region away from $[\gamma_-,\gamma_+]$. We first need  $v'(x(v)) = \frac{1}{x'(v)}$. By Lemma~\ref{LemB3} we have $x\big(\tfrac{1}{2} + \tau w\big) = x_0^*(w) + O(q^{\frac{1}{2}})$ with
\begin{equation}
\label{PsisGx0} x_0^*(w) :=  \frac{2\cos(\pi w_{\infty}^*)}{h(1 - \alpha^2)}\,\frac{1}{\cos(\pi w) - \cos(\pi w_{\infty}^*)}.
\end{equation}
Differentiating with respect to $v$ we get
\[
x'\big(\tfrac{1}{2} + \tau w\big) = \frac{\pi}{{\rm i}T}\,\frac{2\cos(\pi w_{\infty}^*)}{h(1 - \alpha^2)}\,\frac{\sin(\pi w)}{(\cos(\pi w) - \cos(\pi w_{\infty}^*))^2} + O(q^{\frac{1}{2}}).
\]
Owing to Lemma~\ref{lemUp}, we have
\begin{equation}
\begin{split}
& \quad \mathbf{G}^{\bullet}\big(\tfrac{1}{2} + \tau w\big) \\
& = \frac{2\pi}{T}\,\frac{u}{2 + n}\bigg\{\Big[-\Upsilon_{b,0}^*(\tau(w + w_{\infty})) - \Upsilon_{b,0}^*(\tau(w - w_{\infty})) + \Upsilon_{b,0}^*(\tau(-w + w_{\infty})) + \Upsilon_{b,0}^*(\tau(-w-w_{\infty}))\Big] \\
& \quad + \frac{q^{b}}{1 - q^{b}}\Big[(\Upsilon_{b + 2,0}^* - \Upsilon_{b,0}^*)(w + w_{\infty}) + (\Upsilon_{b + 2,0}^*- \Upsilon_{b,0}^*)(w - w_{\infty}) \\
& \quad  - (\Upsilon_{b + 2,0}^* - \Upsilon_{b,0}^*)(-w + w_{\infty}) - (\Upsilon_{b + 2,0}^* - \Upsilon_{b,0}^*)(-w-w_{\infty})\Big] + O(q)\bigg\}.
\end{split}
\end{equation}
where we have used $\frac{1}{1 - q^{b}} = 1 + \frac{q^{b}}{1 - q^{b}}$ to write separately the leading term. The quantity in brackets can be computed from \eqref{theone}. The prefactor $u$ can be replaced by $1$ up to $O(q^{c}) = O(q^{\frac{1}{2}})$ since $c > \tfrac{1}{2}$ when $b \in \big(0,\tfrac{1}{2}\big)$. All calculations done we obtain
\begin{equation}
\label{PsisG} v'(x(v))\mathbf{G}^{\bullet}(v) = \Psi_{b}(x) - \frac{q^{b}}{1 - q^{b}}\tilde{\Psi}_{b}(x) + O(q^{\frac{1}{2}}),
\end{equation}
with
\begin{equation}
\label{F5eq} \begin{split}
\Psi_b(x_0^*(w)) & = \frac{-1}{2 + n}\,\frac{h(1 - \alpha^2)\big(\cos(\pi w) - \cos(\pi w_{\infty}^*)\big)^2}{\cos(\pi w_{\infty}^*)\sin(\pi w)} \\
& \quad \times \bigg(\frac{\cos[\pi(1 - b)(w + w_{\infty}^*)]}{\sin[\pi(w + w_{\infty}^*)]} + \frac{\cos[\pi(1 - b)(w - w_{\infty}^*)]}{\sin[\pi(w - w_{\infty}^*)]}\bigg) \\
\tilde{\Psi}_{b}(x_0^*(w)) & = \Psi_{b + 2}(x_0^*(w)) - \Psi_{b}(x_0^*(w)) \\
& = -\frac{4}{2 + n}\,\frac{h(1 - \alpha^2)\cos(\pi b w_{\infty}^*)\big(\cos(\pi w) - \cos(\pi w_{\infty}^*)\big)^2}{\cos(\pi w_{\infty}^*)}\,\frac{\sin(\pi b w)}{\sin(\pi w)}.
\end{split}
\end{equation}
For $\alpha = 1$, this simplifies to
\begin{equation}
\label{F6eq} \begin{split}
x_0^*(w) & = \frac{\rho}{2h\cos(\pi w)}, \\
\Psi_{b}(x_0^*(w)) & = \frac{2h\,{\rm cot}^2(\pi w)}{\rho\sqrt{2 + n}}\,\sin\big(\pi(1 - b)w), \\
\tilde{\Psi}_{b}(x_0^*(w)) & = -\frac{8h}{\rho\sqrt{2 + n}}\,\frac{\cos^2(\pi w)\sin(\pi b w)}{\sin(\pi w)}.
\end{split}
\end{equation}

In the regime $v = \tau w$ with $w$ in a compact, we have according to Lemma~\ref{LemB3}
\[
q^{-\frac{1}{2}}(x(v) - \gamma_+) = x_{\frac{1}{2}}^*(w) + O(q^{\frac{1}{2}}),
\]   
meaning that $q^{-\frac{1}{2}}(x - \gamma_+)$ remains in a $u$-independent compact. A similar analysis can be carried out and we only give the result
\[
v'(x(v))\mathbf{G}^{\bullet}(v) = \frac{q^{\frac{b - 1}{2}}}{1 - q^{b}}\,\Phi_{b}\bigg(\frac{x - \gamma_+}{q^{\frac{1}{2}}}\bigg) + O(q^{\frac{b}{2}}),
\]
with
\begin{equation}
\label{F7eq} \begin{split}
x_{\frac{1}{2}}^*(w) & = \frac{16\cos(\pi w_{\infty}^*)}{(1 - \alpha^2)h}\,\cos^2\bigg(\frac{\pi w}{2}\bigg), \\
\Phi_{b}\big(x_{\frac{1}{2}}^*(w)\big) & = \frac{h(1 - \alpha^2)}{2 + n}\,\frac{\cos(\pi b w_{\infty}^*)}{\cos(\pi w_{\infty}^*)}\,\frac{\sin(\pi b w)}{\sin(\pi w)}.
\end{split}
\end{equation}
For $\alpha = 1$, it simplifies to
\begin{equation}
\label{F8eq} \begin{split}
x_{\frac{1}{2}}^*(w) & = \frac{4\rho}{h}\,\cos^2\bigg(\frac{\pi w}{2}\bigg), \\
\Phi_{b}\big(x_{\frac{1}{2}}^*(w)\big) & = \frac{4h}{\rho\sqrt{2 + n}}\,\frac{\sin(\pi b w)}{\sin(\pi w)}.
\end{split}
\end{equation}

One can check that this analysis remains valid if $b = b(s)$ and $s$ is complex-valued but such that ${\rm Re}\,b(s) \in \big(0,\tfrac{1}{2}\big)$ --- noting this inequality also implies ${\rm Re}\big(\frac{1}{1 - b(s)}\big) > \frac{1}{2}$. This completes the proof of Theorem~\ref{th85}.

We move to the proof of Corollary~\ref{Covol}. Planar pointed rooted maps are pointed disks whose boundary face is a triangle. Therefore, their generating series with vertex weight $u$ is:
\begin{equation}
\label{XFB}
\begin{split}
[x^{-4}]\mathbf{F}^{\bullet} & = - \Res_{x = \infty} \dd x\,x^3\,\mathbf{F}^{\bullet}(x) \\
& = -\Res_{v = v_{\infty}} \dd v\,(x(v))^3\,\mathbf{G}^{\bullet}(v) \\
& = - {\rm i}T\,\Res_{w = w_{\infty}} \dd w\,\Big(x\big(\tfrac{1}{2} + \tau w\big)\Big)^3\,\mathbf{G}^{\bullet}\big(\tfrac{1}{2} + \tau w\big).
\end{split}
\end{equation}
After taking the residue and taking into account Remark~\ref{theremr} and the analyticity properties established in Appendix~\ref{App1}, we find that $[x^{-4}] \mathbf{F}^{\bullet}$ is an analytic function of $q \in B(0;\eta)\setminus \mathbb{R}_{\leq 0}$. We will see in Theorem~\ref{thm:rigiddeltaJ} that $q$ is a delta-analytic function of $u$, so this implies that $[x^{-4}] \mathbf{F}^{\bullet}$ is a delta-analytic function of $u$.

To compute \eqref{XFB}, we use \eqref{PsisGx0} and \eqref{PsisG}. We first treat $\alpha \neq 1$.  The leading term $u \cdot q^0$ does not contribute to the singularity as it is an entire function of $u$, so the relevant part for the singularity analysis therefore comes from $\tilde{\Psi}_{b}$. We then use the Laurent expansion of $x(v)$ when $v \rightarrow v_{\infty}$ from Lemma~\ref{Xinfexp} to get
\[
[x^{-4}] \mathbf{F}^{\bullet}|_{{\rm sing}} = \frac{16u\pi \sin(\pi b w_{\infty}^*)}{2 + n} \cdot \frac{3b\pi C^2 E_1}{4T^2}\,q^{b} + o(q^{b}).
\]
Taking the leading term when $q \rightarrow 0$ of $E_1$ and $\frac{\pi C}{T}$  from Lemma~\ref{expict}, and replacing $u$ by $1$ up to negligible terms, we obtain
\begin{equation}
\label{x4flu2} [x^{-4}] \mathbf{F}^{\bullet}|_{{\rm sing}}  = A\,q^{b} + o(q^{b}),
\end{equation}
with
\begin{equation}
\label{AGasno}
A = \frac{96 b\sin(\pi b w_{\infty}^*)\cos(\pi w_{\infty}^*)\big(1 - \alpha\sin^2(\pi w_{\infty}^*)\big)}{(2 + n)h^2(1 - \alpha^2)^2\sin^3(\pi w_{\infty}^*)},
\end{equation}
where it remains to substitute the value of $h^2$ in terms of $w_{\infty}^*$ from \eqref{gsurh2}. We can also do the computation for $\alpha = 1$ in terms of the parameter: the result is correctly reproduced by taking the limit \eqref{thelimimt} and yields
\begin{equation}
\label{AGasno1}
A = \frac{6b\rho\sqrt{2 - n}}{h^2(2 + n)},
\end{equation}
where one substitute the value of $h^2$ on the critical line given in \eqref{gsurh}. Then using $q \sim \big(\frac{1 - u}{\Delta}\big)^{c}$ and transfer theorems, we deduce from \eqref{x4flu2} the large volume asymptotics stated in Corollary~\ref{Covol}.

The parameter $b$ appears in two ways in the previous computation: firstly via the value of $c$ (coming from the critical behavior of $\gamma$ according to Theorem~\ref{th38}), and secondly via the value $n = 2\cos(\pi b)$ in the functional relation. If we substitute in this second dependence $b \rightarrow \frac{1}{2}$ but keep the first dependence unchanged, we obtain the generating series of disks with a marked point in the gasket:
\begin{equation}
\label{x4fIu}[x^{-4}] \mathbf{F}^{\bullet\,\,{\rm in}\,\,{\rm gasket}}|_{{\rm sing}} = A_{{\rm gasket}}\,q^{b} + o(q^{b}),
\end{equation}
with
\begin{equation}
\label{AGas} A_{{\rm gasket}} = \frac{24\sin(\frac{\pi w_{\infty}^*}{2})\cos(\pi w_{\infty}^*)\big(1 - \alpha\sin^2(\pi w_{\infty}^*)\big)}{h^2(1 - \alpha^2)^2\sin^3(\pi w_{\infty}^*)},
\end{equation}
where one should still use \eqref{gsurh2} for $h^2$. For $\alpha = 1$ this is
\begin{equation}
\label{AGas1} A_{{\rm gasket}} = \frac{3\sqrt{2}\rho}{4h^2}
\end{equation}
where one should still use \eqref{gsurh} for $h^2$. By transfer theorems, \eqref{x4fIu}  implies the large volume asymptotics stated in Corollary~\ref{corrrr1}.

\section{Scaling limits for cylinder generating series}
\label{apGg}
We distinguish whether the variable $x_i$ coupled to the perimeter of the $i$-th boundary is away from $\gamma_+^*$  --- in which case the perimeter remains typically finite --- or close to $\gamma_+^*$ at scale $O(q^{\frac{1}{2}})$ --- in which case the perimeter typically diverges.

\subsection{Refined cylinders: finite/finite}

This is governed by the regime $x_i = x\big(\frac{1}{2} + \tau w_i)$, and leads to:
\begin{equation*}
\begin{split}
\mathbf{F}^{(2)}_{s}(x_1,x_2) & = \frac{(1 - \alpha^2)^2h^2}{4\pi\cos^2(\pi w_{\infty}^*)}\Big[\prod_{i = 1}^2 \frac{(\cos(\pi w_i) - \cos(\pi w_{\infty}^*))^2}{\sin(\pi w_i)}\Big]  \\
& \quad  \times \bigg\{\frac{R_{b(s)}(w_1,w_2) - q^{b(s)} R_{b(s) + 2}(w_1,w_2)}{(4 - n^2s^2)(1 - q^{b(s)})} + O(q)\bigg\}  \\ 
& =  \frac{(1 - \alpha^2)^2h^2}{4\pi(4 - n^2s^2)\cos^2(\pi w_{\infty}^*)}\bigg[\prod_{i = 1}^2 \frac{\big(\cos(\pi w_i) - \cos(\pi w_{\infty}^*)}{\sin(\pi w_i)\big)^2}\bigg]  \\
& \quad \times \bigg\{R_{b(s)}(w_1,w_2) + \frac{q^{b(s)}}{1 - q^{b(s)}}(R_{b(s)}(w_1,w_2) - R_{b(s) + 2}(w_1,w_2)) + O(q)\bigg\},
\end{split}
\end{equation*}
where:
\[
R_{b}(w_1,w_2) = 2{\rm i}\partial_{w_1}\big[\Upsilon_{b,0}^*(w_1 + w_2) - \Upsilon_{b,0}^*(w_1 - w_2) + \Upsilon_{b,0}^*(-w_1 + w_2) - \Upsilon_{b,0}^*(-w_1 - w_2)\big].
\]
The first term $R_{b(s)}$ does not feature a singularity when $u \rightarrow 1$, and thus will not contribute to large volume asymptotics. We compute using the expression of $\Upsilon_{b,0}^*$ in \eqref{theone}:
\beq
\label{fsugfs}R_{b(s)}(w_1,w_2) - R_{b(s) + 2}(w_1,w_2) = -8\pi b(s)\,\sin(\pi b(s)w_1)\sin(\pi b(s)w_2).
\eeq
Therefore:
\[
\mathbf{F}^{(2)}_{s}(x_1,x_2)|_{{\rm sing}} = -\frac{b(s)\,q^{b(s)}}{1 - q^{b(s)}}\,\Xi_{b(s),3}(x_{\frac{1}{2}}^*(w_1),x_{\frac{1}{2}}^*(w_2)) + O(q),
\]
with:
\[
\Xi_{b(s),3} = \frac{2h^2(1 - \alpha^2)^2}{4 - n^2s^2}\bigg[\prod_{i = 1}^2 \frac{(\cos(\pi w_i) - \cos(\pi w_{\infty}^*))^2 \sin(\pi b(s)w_i)}{\cos(\pi w_{\infty}^*)\sin(\pi w_i)}\bigg].
\]
For $\alpha = 1$, it specialises to:
\[
\Xi_{b(s),3} = \frac{-32 h^2}{\rho^2(4 - n^2s^2)} \Big[\prod_{i = 1}^2 \frac{\cos^2(\pi w_i)\,\sin(\pi b(s)w_i)}{\sin(\pi w_i)}\Big].
\]

\subsection{Refined cylinders: finite/large}

This is governed by $x_1 = x\big(\frac{1}{2} + \tau w_1\big)$ and $x_2 = x(\tau w_2)$, and leads to:
\begin{equation*}
\begin{split} 
\mathbf{F}^{(2)}_{s}(x_1,x_2)|_{{\rm sing}} & = \frac{q^{(b(s) -1)/2}}{1 - q^{b(s)}}\,\frac{(1 - \alpha^2)^2h^2(\cos(\pi w_1) - \cos(\pi w_{\infty}^*))^2}{16\pi(4 - n^2s^2)\cos^2(\pi w_{\infty}^*)\sin(\pi w_1)\sin(\pi w_2)}  \\
& \quad  \times \big\{\tilde{R}_{b(s)}(w_1,w_2) + O(q^{(1 - b(s))/2})\big\},
\end{split}
\end{equation*}
with:
\begin{equation*}
\begin{split}
\tilde{R}_{b(s)}(w_1,w_2) & =  -2{\rm i}\big\{(\Upsilon^*_{b(s),\frac{1}{2}})'(w_1 + w_2) - (\Upsilon^*_{b(s),\frac{1}{2}})'(w_1 - w_2)  \\
& \quad - (\Upsilon^*_{b(s),\frac{1}{2}})'(w_2 - w_1) + (\Upsilon^*_{b(s),\frac{1}{2}})'(-w_1 - w_2)\big\}   \\
& =  8\pi\,b(s)\,\sin(\pi b(s)w_1) \sin(\pi b(s) w_2).
\end{split}
\end{equation*}
Therefore:
\[
\mathbf{F}^{(2)}_{s}(x_1,x_2)|_{{\rm sing}} = \frac{b(s)\,q^{(b(s) - 1)/2}}{1 - q^{b(s)}}\,\Xi_{b(s),4}(x_{\frac{1}{2}}^*(w_1),x_{0}^*(w_2)) + O(q^{(1 - b(s))/2}),
\]
with:
\[
\Xi_{b(s),4} = \frac{(1 - \alpha^2)^2h^2[\cos(\pi w_1) - \cos(\pi w_{\infty}^*)]^2}{2(4 - n^2s^2)\cos^2(\pi w_{\infty}^*)}\,\bigg[\prod_{i = 1}^2 \frac{\sin(\pi b(s) w_i)}{\sin(\pi w_i)}\bigg].
\]
In particular for $\alpha = 1$, we find:
\[
\Xi_{b(s),4} = \frac{8h^2}{\rho^2(4 - n^2s^2)}\,\cos^2(\pi w_{1})\bigg[\prod_{i = 1}^2 \frac{\sin(\pi b(s)w_i)}{\sin(\pi w_i)}\bigg].
\]

\section{Admissibility vs. finiteness in the bending energy model}
\label{sec:equivadm}

In this appendix, we consider
the bending energy model, as defined in Section~\ref{bendI}. It
depends on the parameters $n$, $g$, $h$ and $\alpha$, see again
Figure~\ref{fig:curv} (the vertex weight $u$ is set to $1$). The
corresponding generating series of annuli $A_{k,\ell}(n,h,\alpha)$ are
given by \eqref{defR}, \eqref{rin} and \eqref{varsigma}.  Recall also
the definition of admissibility for a sequence of face weights
$(g_k)_{k\geq 1}$ given in Section~\ref{sec:remusualmaps}. Then, we
have the following analogue of~\cite[Theorem~1]{BuddOn}:

\begin{theorem}
  \label{thm:equivadm}
  For $n \in (0,2)$ and $g,h,\alpha \geq 0$, the bending energy model
  with parameters $(n,g,h,\alpha)$ is well defined (i.e., the
  partition function $F_\ell(n,g,h,\alpha)$ is finite for all $\ell$)
  if and only if there exists an admissible weight sequence
  $(G_k)_{k \geq 1}$ such that
  \begin{equation}
    \label{eq:equivadmcond}
    G_k- \sum_{k' \geq 0} A_{k,k'}(n,h,\alpha) \mathcal{F}_{k'}(G_1,G_2,\ldots) = g_k, \qquad
    g_k :=
    \begin{cases}
      g & \text{if $k=3$,}\\
      0 & \text{otherwise.}
    \end{cases}
  \end{equation}
  In that case, we have
  $F_\ell(n,g,h,\alpha)=\mathcal{F}_{\ell}(G_1,G_2,\ldots)$, and the
  expected number of vertices in a disk of perimeter $\ell$ is finite.
\end{theorem}

The case $\alpha=1$ is established in~\cite{Korzhenkova2022}.  We now
explain how to prove the result for general $\alpha$ by adapting the
strategy given in~\cite{BuddOn}. We will heavily use the notions of
this paper, as well as those of~\cite{Buddpeel}. The reader is invited
to consult these references for more details. The proof does not
depend on the specific form of $g_k$, the only important assumption
being that it is nonnegative for all $k$. Therefore,
Theorem~\ref{thm:equivadm} remains valid for a more general model in
which we allow for unvisited faces of arbitrary degrees. Our proof
however depends on the specific form of $A_{k,k'}$ in the bending
energy model, precisely within the proof of Lemma~\ref{lem:ricochet}
below. We also assume that $n<2$, but we believe the case $n=2$ could
be included as well by a slight adaptation.

The core idea is to use the peeling exploration of loop-decorated
maps, as defined in~\cite[Sections~2 and 3]{BuddOn}. There are
actually two types of peeling explorations: untargeted peeling, which
applies to disks, and targeted peeling, which applies to pointed
disks. Their definitions are adapted straightforwardly to the bending
energy model, with the following modifications:
\begin{itemize}
\item since the maps are no longer assumed bipartite, we should keep
  track of perimeters rather than half-perimeters,
\item since the model is not rigid, annuli (rings) may have different
  outer and inner perimeters.
\end{itemize}
For these reasons, we now encounter the following possible events in
untargeted peeling:
\begin{itemize}
\item $\mathrm{C}_k$: discovering a new (unvisited) face of degree
  $k \geq 1$ (when considering triangulations, only $\mathrm{C}_3$ may
  occur, but the arguments allow for unvisited faces of arbitrary
  degrees),
\item $\mathrm{G}_{k_1,k_2}$: splitting a hole of degree $k_1+k_2+2$
  into two holes of degrees $k_1$ and $k_2$, by identifying two active
  edges incident to it,
\item $\mathrm{L}_{k,k'}$: discovering an annulus with outer perimeter
  $k \geq 1$ and inner perimeter $k'\geq 0$.
\end{itemize}
The analogous events in targeted peeling are easy to deduce, but we do
not enter into details here since we only allude to targeted peeling
in Remark~\ref{rem:targeted}.

Assuming that the model is well-defined, the probabilities of these
events when peeling an active edge incident to a hole of degree
$\ell \geq 1$ are given by
\begin{equation}
  \begin{split}
    P_\ell(\mathrm{C}_k) &= \frac{g_k F_{\ell+k-2}}{F_\ell}, \\
    P_\ell(\mathrm{G}_{k_1,k_2}) &= \frac{F_{k_1} F_{k_2}}{F_\ell}, \qquad (\ell=k_1+k_2+2) \\
    P_\ell(\mathrm{L}_{k,k'}) &= \frac{A_{k,k'} F_{\ell+k-2}
      F_{k'}}{F_\ell}.
  \end{split}
\end{equation}

If we instead assume that~\eqref{eq:equivadmcond} is satisfied for an
admissible weight sequence $(G_k)_{k \geq 1}$, then we follow the
strategy of~\cite[Section~5]{BuddOn} and \emph{define} the
probabilities of these events by the putative expressions
\begin{equation}
  \label{eq:probput}
  \begin{split}
    P_\ell(\mathrm{C}_k) &= \left( \nu(k-2) - \frac{\sum_{k'\geq 0} A_{k,k'} \gamma_+^{k+k'} \nu(-k'-2)}{2} \right) \frac{\nu(-\ell-k)}{\nu(-\ell-2)},\\
    P_\ell(\mathrm{G}_{k_1,k_2}) &= \frac{\nu(-k_1-2)\nu(-k_2-2)}{2 \nu(-\ell-2)}, \qquad (\ell=k_1+k_2+2) \\
    P_\ell(\mathrm{L}_{k,k'}) &= \frac{A_{k,k'} \gamma_+^{k+k'} \nu(-k'-2)}{2} \cdot
    \frac{\nu(-\ell-k)}{\nu(-\ell-2)},
  \end{split}
\end{equation}
where $\nu$ is the probability measure\footnote{Note that
  $\sum_{k \in \mathbb{Z}}
  \nu(k)=\frac{1}{\gamma_+}(2\mathbf{\mathcal{F}}(\gamma_+)+\sum_{k
    \geq 1} G_k \gamma_+^{k-1})$ is indeed equal to $1$, by taking
  $x \to \gamma_+$ in~\eqref{eq:funcnu}.} on $\mathbb{Z}$ corresponding
to the admissible weight sequence $(G_k)_{k \geq 1}$
via~\cite[Proposition~3]{Buddpeel}, namely
\begin{equation}
  \label{eq:nudef}
  \nu(k) =
  \begin{cases}
    G_{k+2} \gamma_+^k & \text{for $k \geq -1$,} \\
    2 \mathcal{F}_{-k-2}(G_1,G_2,\ldots) \gamma_+^k & \text{for $k \leq -2$,}
  \end{cases}
\end{equation}
and $\gamma_+=\lim_{k \to \infty}(\mathcal{F}_{k})^{\frac{1}{k}}$ (the limit
exists for any admissible weight sequence). Following the same
reasoning as in~\cite[p.~29]{BuddOn}, we see that the quantities
appearing in~\eqref{eq:probput} are nonnegative and add up to one,
which allows to interpret them as probabilities. Then, we have the
analogue of \cite[Proposition~8 and Lemma~5]{BuddOn}:

\begin{proposition}
  \label{prop:termination}
  Consider an admissible weight sequence
  satisfying~\eqref{eq:equivadmcond}.  If we construct a random
  loop-decorated map via the untargeted peeling algorithm with the
  event probabilities~\eqref{eq:probput}, then the algorithm
  terminates almost surely, and produces a sample of the bending
  energy model with parameters $(n,g,h,\alpha)$.
\end{proposition}

The proof of this proposition again follows Budd's steps. The most
delicate point is the termination of the algorithm, which comes from a
martingale argument. To a map $\mathcal{M}$ with holes of degrees
$k_1,k_2,\ldots$ and with $N$ vertices not incident to a hole, we
assign the quantity
\begin{equation}
  \mathcal{V}(\mathcal{M}) = N + f^\downarrow(k_1) + f^\downarrow(k_2) + \cdots, \qquad
  f^\downarrow(k):=\frac{\nu(-2) h^\downarrow(k)}{\nu(-k-2)}
\end{equation}
where we sum over all holes, and where the function $h^\downarrow$ is
yet unspecified. Mimicking the proof of~\cite[Lemma~4]{BuddOn}, we see
that $\mathcal{V}$ defines a martingale for the untargeted peeling
algorithm if $h^\downarrow(0)=1$ and
\begin{equation}
  \label{eq:hdown}
  h^\downarrow(p) = \sum_{k=0}^\infty \nu(k-p) h^\downarrow(k) + \sum_{k=1}^\infty \sum_{k'=0}^\infty
  \frac{A_{k,k'} \gamma_+^{k+k'}}{2} \nu(-k-p) h^\downarrow(k'), \qquad p \geq 1.
\end{equation}
We claim that such a function $h^\downarrow$ exists and can be
interpreted as the probability that a certain generalisation of the
ricocheted random walk of~\cite[Section~4]{BuddOn} gets trapped at
$0$. This generalisation is in fact rather straightforward: we
consider a random walk with step distribution $\nu$ which gets trapped
when touching $0$ and, when it would jump to some position $-k < 0$,
gets ricocheted to position $k' \geq 0$ with probability
$\frac{1}{2}A_{k,k'} \gamma_+^{k+k'}$, and otherwise remains trapped at $-k$
with probability $1-\frac{1}{2}\sum_{k'\geq 0} A_{k,k'} \gamma_+^{k+k'}$. The
fact that these are indeed probabilities is ensured by the following
lemma, whose proof is postponed to the end of the section.

\begin{lemma}
  \label{lem:ricochet}
  For any admissible weight sequence satisfying~\eqref{eq:equivadmcond},
  we have
  \begin{equation}
    \label{eq:gambound}
    \gamma_+ \leq \gamma_+^*
  \end{equation}
  where $\gamma^*=\frac{1}{h(\alpha+1)}$ is as
  in~\eqref{eq:gammastar}, and we have
   \begin{equation}
    \label{eq:Aidentgam}
    \sum_{k' \geq 0} A_{k,k'} (\gamma_+^*)^{k+k'} = n, \qquad k \geq 1.
  \end{equation}
\end{lemma}

Note that, when $\gamma_+=\gamma_+^*$, the probability to remain
trapped when jumping to position $-k<0$ is equal to $1-\frac{n}{2}>0$
independently of $k$. It is higher and dependent on $k$ for
$\gamma_+<\gamma_+^*$. It follows that the ricocheted walk gets
trapped almost surely, since the walk cannot drift to $+\infty$
by~\cite[Proposition~4]{Buddpeel}, and therefore there indeed exists a
nonnegative function $h^\downarrow$ satisfying~\eqref{eq:hdown}. Note
that the case $n=2$ and $\gamma_+=\gamma_+^*$ is slightly subtler
since the walk always ricochets when jumping to a negative
position. We do not consider this case here, even though we expect the
same conclusion to hold.

Straightforward adaptations of the other arguments given
in~\cite[Section~5]{BuddOn} lead to
Proposition~\ref{prop:termination}, and furthermore imply that the
expected number of vertices of a loop-decorated map with perimeter $p$
is \emph{equal} to $f^\downarrow(p)$, hence is finite. This concludes
the proof of Theorem~\ref{thm:equivadm}.

\begin{remark}
  Here we are considering a martingale, whereas Budd considers a
  supermartingale, since he only defines the ricocheted random walk
  corresponding to the non generic critical case.
\end{remark}

\begin{remark}
  \label{rem:targeted}
  Using targeted peeling and the uniqueness of the solution
  of~\eqref{eq:hdown} for given boundary conditions, we may identify
  \begin{equation}
    f^\downarrow(p)=\frac{F_p^\bullet}{F_p}, \qquad
    h^\downarrow(p)=\frac{F_p^\bullet}{\gamma_+^p},
  \end{equation}
  and we may interpret $h^\downarrow$ as the harmonic function used to
  condition the ricocheted walk to get trapped at $0$ via a Doob
  $h$-transform. Changing the boundary condition of $h^\downarrow$ to
  condition the walk to get trapped at $-k<0$ should give a relation
  with the generating series of cylinders.
\end{remark}

\begin{proof}[Proof of Lemma~\ref{lem:ricochet}]
  Let us prove that $\gamma_+ \leq \gamma_+^*$.  We first establish a
  lower bound for $G_k$, namely that $G_k \geq \varsigma(\beta)^{-k}$
  for any $\beta\in (0,\gamma_+)$ and $k$ large enough. Here,
  $\varsigma$ is the rational involution defined by~\eqref{varsigma}.

  Fix some $\beta\in (0,\gamma_+)$ and consider $m$ such that
  $\mathcal{F}_{k'} \geq \beta^{k'}$ for $k' > m$. Then, we have
  \begin{equation}
    \label{eq:Amin}
    G_k \geq \sum_{k' \geq 0} A_{k,k'} \mathcal{F}_{k'} \geq  \sum_{k'=0}^{\infty} A_{k,k'} \beta^{k'} - \sum_{k'=0}^{m} A_{k,k'} \beta^{k'}.
  \end{equation}
  We may estimate both sums in the right-hand side using the explicit
  form~\eqref{rin} of the generating $\mathbf{A}(x,z)$, or
  alternatively the combinatorial arguments
  of~\cite[Section~3.2]{BBG12b}.  We find
  \begin{equation}
    \label{eq:Acomput}
    \sum_{k'=0}^{\infty} A_{k,k'} \beta^{k'} =
    \begin{cases}
      n \varsigma(\beta)^{-k} & \text{if $\beta<\frac{1}{\alpha h}$,}\\
      \infty & \text{if $\beta \geq \frac{1}{\alpha h}$,}
    \end{cases} \qquad
    \sum_{k'=0}^{m} A_{k,k'} \beta^{k'}=O\left( \varsigma(0)^{-k} k^{m}\right)
  \end{equation}
  where the second estimate follows from the expression
  \begin{equation}
    [z^{k'}] \mathbf{A}(x,z)=
    \begin{cases}
      \frac{\alpha h n}{1-\alpha h x} & \text{if $k'=0$,}\\
      -\frac{n \varsigma'(x)}{\varsigma(x)^{k+1}} & \text{if $k'\geq 1$,}
    \end{cases}
  \end{equation}
  which, by~\eqref{varsigma}, shows that $[z^{k'}] \mathbf{A}(x,z)$ is
  a rational series in $x$ with a pole of order $k'+1$ at
  $\varsigma(0)=\frac{1}{h\alpha}$.  This establishes the wanted
  lower bound.

  On the other hand, the fact that~\eqref{eq:nudef} defines a
  probability distribution entails that
  $\lim_{k \to \infty} G_k \gamma_+^k=0$. We conclude that
  $\varsigma(\gamma_+) \geq \gamma_+$, hence $\gamma_+$ is smaller
  than or equal to the positive fixed point
  $\gamma_+^*=\frac{1}{h(\alpha+1)}$ of the decreasing involution
  $\varsigma$.  Finally, the identity~\eqref{eq:Aidentgam} is another
  consequence of~\eqref{eq:Acomput}, which was in fact valid for all
  $\beta \geq 0$ hence for $\beta=\gamma_+^*$.
\end{proof}

\section{On the rigid \texorpdfstring{$O(n)$}{O(n)} loop model on
  bipartite maps}
\label{sec:rigid}

In this appendix, we consider
the rigid $O(n)$ loop model, introduced originally in~\cite{BBG12a} in
the case where the underlying map is a quadrangulation, and
generalised in~\cite{BuddOn} by allowing the unvisited faces to have
arbitrary even degrees. In all cases, the faces visited by loops are
always quadrangles, and the rigidity constraint entails that, in each
visited quadrangle, the loop enters and exits through opposite
edges. This greatly simplifies the nested loop approach for the two
following reasons.
\begin{itemize}
\item The annuli always have equal inner and outer perimeters hence
  the generating series $A_{k,k'}$ vanishes unless $k=k'$. Therefore,
  the fixed point equation~\eqref{eq:fixp} is simpler, as well as many
  related arguments.
\item It is well-known that working with bipartite maps makes life
  easier.
\end{itemize}
The rigid loop model may actually be viewed as a limit of the bending
energy model: contemplating again Figure~\ref{fig:curv}, we see that
for $\alpha=0$ the visited triangles always come top to tail by pairs,
and may therefore be merged in quadrangles satisfying the rigidity
constraint. Note however that, in our computations, we assume that the
unvisited faces are triangles, hence the underlying map is not
necessarily bipartite. As pointed out in Appendix~\ref{sec:equivadm},
we could generalise the model by allowing the unvisited faces to have
arbitrary degrees, which does not affect the validity of the nested
loop approach, and then we could constrain the degrees to be even to
recover precisely the model of~\cite{BuddOn} for $\alpha=0$. In this
identification, the weight per visited quadrangle, denoted $g$ in
Budd's paper and $h_1$ in~\cite{BBG12a}, is equal to $h^2$ in our
present notations.

\subsection{An explicit expression for $\mathbf{F}^\bullet_s(x)$ at a non generic critical point.}
\label{sec:rigidexpli}

In this section we set the vertex weight $u$ to $1$.
Consider the generating series $\mathbf{F}(x)$ as defined
in~\eqref{eq:bfFdef}. Noting that the sum is restricted to even
$\ell$'s, we see that the cut of $\mathbf{F}$ is necessarily symmetric
and we denote it by $[-\gamma,\gamma]$, dropping the $\pm$ subscripts.
The condition for non generic criticality reads
\begin{equation}
  \gamma = \gamma^* := h^{-1}
\end{equation}
which is nothing but the case $\alpha=0$ of \eqref{eq:gammastar}.  It
was shown in~\cite{BBG12a} that, at a non generic point of the model
on quadrangulations with a weight $g_4$ per unvisited quadrangle and
for $n \in (0,2)$, we have the explicit expression
\begin{equation}
  \label{eq:rigidF}
  \mathbf{F}(x) = \mathbf{F}_{\mathrm{part}}(x)+\mathbf{F}_{\mathrm{hom}}(x)
\end{equation}
where
\begin{equation}
  \label{eq:rigidFaux}
  \begin{split}
    \mathbf{F}_{\mathrm{part}}(x) &:= \frac{2(x-g_4 x^3)-n\big(\frac{\gamma^4}{x^3}-\frac{g_4 \gamma^8}{x^5}\big)}{4-n^2}+\frac{n}{(2+n)x} \\
    \mathbf{F}_{\mathrm{hom}}(x) &:= \left( \mathfrak{f}(x) - {\textstyle
        \frac{\gamma^2}{x^2} \mathfrak{f}\big(\frac{\gamma^2}{x} \big)
      }\right) \left(\frac{x - \gamma}{x + \gamma}\right)^b -
    \left( \mathfrak{f}(-x) - {\textstyle \frac{\gamma^2}{x^2} \mathfrak{f}\big(-
          \frac{\gamma^2}{x} \big) } \right)
    \left(\frac{x + \gamma}{x - \gamma}\right)^b \\
    \mathfrak{f}(x) &:= \frac{g_4}{4-n^2} \left( x^3 + 2 b \gamma x^2 + 2 b^2
      \gamma^2 x + \frac{2(b+2 b^3)\gamma^3}{3}\right) -
    \frac{x + 2 b \gamma}{4-n^2} 
  \end{split}
\end{equation}
and we recall $b = \tfrac{1}{\pi} \arccos \big(\tfrac{n}{2}\big)$. A
nice feature of this explicit expression is that it bypasses the use
of the elliptic parametrisation, which makes the singularity analysis
somewhat simpler. In particular, we immediately see why $b$ appears in
the singular expansions for $x \to \pm \gamma$. The method that was
used to obtain the expressions~\eqref{eq:rigidFaux} may be generalised
to the case where the unvisited faces have arbitrary even degrees.

It is natural to ask whether similar explicit expression exist for the
other generating series that we consider in this paper. We will not
consider the case of cylinders here but, in the case of pointed disks,
the question admits an affirmative answer.

\begin{proposition}[see also~{\cite[Section~3.4]{BouHDR}}]
  \label{prop:universalFs}
  At a non generic critical point of the rigid model with
  $n \in (0,2)$, possibly with unvisited faces of arbitrary even
  degrees, the refined generating function $\mathbf{F}^\bullet_s(x)$
  of pointed disks, defined as in~\eqref{eq:bfFbsdef}, admits the
  universal expression
  \begin{equation}
    \label{eq:universalFs}
    \mathbf{F}^\bullet_s(x) = \frac{1}{(2 + n s) x} \left( ns  + 
    \left(\frac{x + \gamma}{x - \gamma}\right)^{b(s)-1} +
      \left(\frac{x - \gamma}{x + \gamma}\right)^{b(s)-1} \right)
  \end{equation}
  where $b(s)=\frac{1}{\pi}\mathrm{arccos}\big(\tfrac{ns}{2}\big)>0$
  is defined as in~\eqref{eq:bsdef}. Notice (again, considering
  $\gamma$ fixed) that this expression depends on $n$ and $s$ only via
  the combination $ns$.
\end{proposition}

\begin{proof}
  We specialise the functional equation~\eqref{jiuh} with
  $\varsigma(x)=\frac{\gamma^2}x$ (since $\alpha = 0$,
  $h=\gamma^{-1}$, $\gamma$ denoting now the positive endpoint of the
  cut) and $u = 1$.  This yields
   \begin{equation}
    \forall x \in (-\gamma,\gamma),\qquad \mathbf{F}^\bullet(x + {\rm i}0) + \mathbf{F}^\bullet(x - {\rm i}0) + \frac{n \gamma^2}{x^2} \,\mathbf{F}^\bullet\Big(\frac{\gamma^2}{x}\Big) = \frac{n}{x}.
  \end{equation}
  We may solve this equation using the method described
  in~\cite[Section~6.2]{BBG12a}, the function $\mathfrak{f}(x)$ being here
  proportional to $(x-\gamma)^{-1}$. This
  yields~\eqref{eq:universalFs} for $s=1$. Then, we recall from
  \S~\ref{separ} that adding the refinement parameter $s$ amounts to
  changing $n$ into $ns$.
\end{proof}

Interestingly, the universal form~\eqref{eq:universalFs} also appears
implicitly in~\cite{BuddOn}. To make the connection precise, we
combine Equations~(23), (25) and (42) of that paper, which yield the relation
\begin{equation}
  \sum_{\ell=0}^\infty h^\downarrow_{n/2}(\ell) x^{2\ell} = \frac{\gamma}{x}
  \mathbf{F}^\bullet\left(\frac{\gamma}{x}\right) 
\end{equation}
for $s=1$. Here, $h^\downarrow_{\mathfrak{p}}(\ell)$ denotes the
probability that Budd's $\mathfrak{p}$-ricocheted random walk gets
trapped at $0$.

It is very easy to analyse the singular behaviour
of~\eqref{eq:universalFs}. We find
\begin{equation}
  \mathbf{F}^\bullet_s(x)  \mathop{\sim}_{x \to \pm \gamma} \pm 
  \frac{2^{1-b(s)}}{(2+n s) \gamma}
  \cdot (1 \mp \gamma/x)^{b(s)-1},
\end{equation}
which implies for the refined generating series of pointed disks of perimeter $2\ell$
\begin{equation}
  \label{eq:rigFpsasy}
F_{2\ell}^\bullet[s] \mathop{\sim}_{\ell \rightarrow \infty} 2 \Gamma(b(s)) \frac{4^{-b(s)}}{\pi} \sqrt{\frac{2-ns}{2+ns}} \cdot \frac{\gamma^{2\ell}}{\ell^{b(s)}},
\end{equation}
consistently with~\cite[Equation~(24)]{BuddOn}.  This yields the asymptotic behaviour
\begin{equation}
  \label{eq:rigdepthpgfasy}
  \mathbb{E}\left(s^{P}\middle\vert L=2\ell\right) = \frac{F_{2\ell}^{\bullet}[s]}{F_{2\ell}^{\bullet}[1]} \mathop{\sim}_{\ell \rightarrow \infty} C(s) \ell^{b-b(s)}.
\end{equation}
for the probability generating function of the number $P$ of
separating loops in a pointed disk of prescribed perimeter $2\ell$ and
fluctuating volume. The varying exponent in~\eqref{eq:rigdepthpgfasy}
is a smoking gun evidence that $P$ grows logarithmically with
$\ell$. In fact, a straightforward adaptation of the method of the proof --- which crucially on Hwang's quasi-powers theorems~\cite[Theorems~IX.8 and
IX.15]{Flajolet}  --- of Proposition~\ref{CLTthm} and Theorem~\ref{mainT} in the main text, we obtain the following variant of
Theorem~\ref{theo:main} and Proposition~\ref{mainC}.

\begin{theorem}[see also~{\cite[Theorem~5]{BuddOn}} and {\cite[Theorem~3.3]{BouHDR}}]
  \label{thm:rigidper}
  At a non generic critical point of the rigid model with $n \in (0,2)$, possibly with unvisited faces of arbitrary even
  degrees, the distribution of the number $P$ of separating loops in
  the ensemble of random pointed disks with prescribed perimeter
  $L=2\ell$ and fluctuating volume behaves when $\ell \to \infty$ as
  \begin{equation}
    \label{eq:rigidper}
    \mathbb{P}\Big[P = \big\lfloor \tfrac{\ln \ell}{\pi}\,p \big\rfloor \Big|L = 2\ell\Big]\,\,\stackrel{.}{\asymp}\,\,(\ln \ell)^{-\frac{1}{2}}\,\ell^{-\frac{1}{\pi}\,J(p)}
  \end{equation}
 where $J$ is the same rate function as in
  Theorem~\ref{theo:main}, and $p$ bounded, bounded away from $0$ and chosen so that $P$ is an integer. We have the
  convergence in law
  \begin{equation}
    \frac{P - \frac{p_{{\rm opt}}}{\pi}\,\ln \ell}{\sqrt{\ln \ell}} \longrightarrow
    \mathcal{N}(0,\sigma^2)
  \end{equation}
  with the same $p_{{\rm opt}} = \frac{n}{\sqrt{4 - n^2}}$ as in Proposition~\ref{mainC} and with
  $\sigma^2=\frac{4n}{\pi(4 - n^2)^{\frac{3}{2}}}$.
\end{theorem}

Two remarks are now in order. First, we note that Budd's
theorem~\cite[Theorem~5]{BuddOn} also treats the case $n=2$. Nothing
would prevent from rederiving his result by taking $n \to 2$ in
Proposition~\ref{prop:universalFs}.

Second, we observe that Theorem~\ref{thm:rigidper} makes no
distinction between dilute or dense critical points. In contrast,
Theorem~\ref{theo:main} and Proposition~\ref{mainC} involved the
critical exponent $c$, which depends on the nature of the non generic critical
point ($c=1$ in the dilute phase and $c=\frac{1}{1-b}$ in the
dense phase). In fact, the large perimeter versions of
Theorem~\ref{theo:main} and Proposition~\ref{mainC} are totally
consistent with Theorem~\ref{thm:rigidper}, when we replace
$\ln V$ with $\frac{c}{2} \ln \ell$. This means that, if we
fix a large perimeter $L$, then further conditioning on a large volume
$V \dot{\sim} L^{\frac{2}{c}}$ does not affect the asymptotics of $P$ under consideration.

\subsection{Delta-analyticity with respect to the vertex weight $u$}
\label{sec:rigiddelta}

Recall Definition~\ref{defdeltadom} of delta-domain and
delta-analyticity. In this section, we consider the rigid loop model
on bipartite maps, with the parameters $n$ (weight per loop), $h^2$
(weight per visited square) and $g_{2k}$ (weight per unvisited face of
degree $2k \geq 2$) fixed to nonnegative real values. We assume that
$n,h>0$ (to have a nontrivial loop model), that $g_2<1$ (to avoid the
proliferation of digons) and that $g_{2k}=0$ for $k$ large enough
(i.e., the face degrees are bounded). We will vary the vertex weight
$u$, hence we will keep the dependence on $u$ explicit.

The partition function $F_{2\ell}(u)$ is a series in $u$ with
nonnegative coefficients and, for $\ell$ positive, has a finite radius
of convergence $u_c$ which does not depend on $\ell$. Recall from
Proposition~\ref{ponecut} the definition of
$\mathfrak{r}(u)=\frac{1}{2}\gamma^2_+(u)$ (in the bipartite case, we
have $\mathfrak{s}(u)=0$ since the cut is symmetric). By
Remark~\ref{rem:rs}, it is essentially the same as the series
$\mathcal{F}_2^\bullet(u)=2u\mathfrak{r}(u)$ of rooted maps with a
marked point in the gasket, thus has positive coefficients and radius
of convergence $u_c$.  The purpose of this section is to establish the
following:

\begin{theorem}
  \label{thm:rigiddelta}
  If there exists an integer $k$ such that $g_{4k}>0$, then
  $\mathfrak{r}(u)$ and $F_{2\ell}(u)$ are analytic functions of $u$
  in a delta-domain at $u_c$, for all $\ell > 0$. When $g_{4k}$
  vanishes for all integer $k$, they are analytic in the intersection
  of a delta-domain at $u_c$ and a delta-domain at $-u_c$, since we
  have the parity relations $\mathfrak{r}(u)=-\mathfrak{r}(-u)$ and
  $F_{2\ell}(u) = (-1)^{\ell+1} F_{2\ell}(-u)$. $q$, seen as a function of $u$, has the same properties.
\end{theorem}

The delta-analyticity of $q$ is a direct corollary of the delta-analyticity of $\mathfrak{r}(u)$ and the fact that $q$ is an analytic function of $\mathfrak{r}(u)$, which we observe in two steps: $\mathsf{m}$ is an analytic function of $\mathfrak{r}(u)$ as we see later in \eqref{eq:mrigid} and $q$ is an analytic function of $\mathsf{m}$ by the definition \eqref{q12de}.

The reason why the case $g_{4k}=0$ is special is the following. In a
bipartite map, an elementary counting argument based on Euler's
formula shows that the number of vertices and the number of faces with
degree divisible by $4$ have the same parity. When the map carries a
rigid loop configuration, the number of visited squares is necessarily
even, since the loops form annuli whose contours must have even
length. Therefore, if we consider a map contributing to $F_{2\ell}(u)$
with no unvisited inner face of degree divisible by $4$, then the
parity of the number of vertices is opposite to that of $\ell$. For a
map contributing to $\mathfrak{r}(u)$ with the same constraint, the
number of vertices is even, but we assign no weight to the marked
vertex.

We will make use of the equation~\eqref{ointg1} determining
$\mathfrak{r}(u)$ which, in the rigid bipartite case, may be rewritten
in the form
\begin{equation}
  \label{eq:req}
  \mathfrak{r}(u) = u + \sum_{k \geq 1} \binom{2k-1}{k} \left( g_{2k} + n h^{4k} F_{2k}(u)\right) \mathfrak{r}(u)^k
\end{equation}
making sense whenever the series $\mathfrak{r}(u)$ is absolutely
convergent. We have the following elementary lemma, relying on the
nonnegativity of the coefficients of $\mathfrak{r}(u)$.

\begin{lemma}
  \label{lem:rbd}
  We have $\mathfrak{r}(u_c) \leq \frac1{4h^2}$ so that
  $\mathfrak{r}(u)$ converges normally on its closed disk of
  convergence.  For every $u$ in this disk, different from $u_c$, and
  different from $-u_c$ if $g_{4k}$ vanishes for all integer $k$, we
  have
  \begin{equation}
    \label{eq:daff}
    \lvert \mathfrak{r}(u) \rvert < \mathfrak{r}(u_c).
  \end{equation}
\end{lemma}

\begin{proof}
  By monotone convergence, we have
  $\mathfrak{r}(u_c) = \lim_{u \to u_c^-} \mathfrak{r}(u)$, so we need
  to prove that $\mathfrak{r}(u) \leq \frac1{4h^2}$ for every
  $u \in (0,u_c)$. For such $u$, we know by Proposition~\ref{ponecut}
  that
  $$
  \lim_{k \rightarrow \infty} F_{2k}(u)^{\frac{1}{k}} \to 4\mathfrak{r}(u).
  $$
  Since the sum appearing in~\eqref{eq:req} must converge, we
  conclude that $16h^4\mathfrak{r}(u)^2 \leq 1$ as wanted. To
  obtain~\eqref{eq:daff}, we apply the Daffodil
  Lemma~\cite[Lemma~VI]{Flajolet}, which requires checking that
  $\mathfrak{r}(u)$ is aperiodic if one $g_{4k}$ is nonzero, and has
  period $2$ otherwise. For this, we observe that~\eqref{eq:req}
  implies that $[u^{2k+1}]\mathfrak{r}(u) > 0$ for all $k$. Indeed, 
  $[u^{k+1}]F_{2k}(u)>0$ as trees with $k+1$ vertices contribute, and when $g_{4k} > 0$ we also have
  $[u^{2k}]\mathfrak{r}(u) > 0$.
  \end{proof}

We will also make use of the exact solution of the model, discussed in
Section~\ref{bend} and in the appendices, via the following:

\begin{proposition}
  \label{prop:meropar}
  There exists functions $\varphi$ and $\psi_\ell$ ($\ell \geq 0$)
  which are meromorphic in the domain
  $\mathbb{C} \setminus ( (-\infty,-\frac{1}{4h^2}] \cup
  [\frac{1}{4h^2},\infty) )$ and such that
  \begin{equation}
    u = \varphi(\mathfrak{r}(u)), \qquad F_{2\ell}(u) = \psi_\ell(\mathfrak{r}(u)).
  \end{equation}
  The set of poles of $\psi_\ell$ is included in that of $\varphi$.
\end{proposition}

\begin{proof}
  For $u \in (0,u_c)$ and given the values of
  $\gamma_\pm(u)=\pm 2 \sqrt{\mathfrak{r}(u)}$,
  Theorem~\ref{theimdisk} (specialised to the rigid bipartite case)
  gives an explicit expression of $\mathbf{G}(v)$, which is related to
  $\mathbf{F}(x)$ via~\eqref{gfggx}. Notice that the expression of
  $\mathbf{G}(v)$ depends linearly on $u$ via the $\tilde{g}_k$
  defined in~\eqref{deftildeg}. Thus, the conditions~\eqref{supportD}
  determining $\gamma_\pm(u)$, which turn out to be equivalent to each other
 in the rigid case, yield a linear condition on $u$, which is
  nontrivial by the discussion of Appendix~\ref{proofbeh}. Solving
  this linear condition, we express $u$ as a rational function of
  derivatives of the function $\Upsilon_{b}(v)$ at $v=\tau \pm \frac{1}{4}$
  (we have $v_\infty=\frac{1}{4}$ in the rigid case). By~\eqref{Upsfb},
  $\Upsilon_{b}$ is itself a rational function of the Jacobi theta
  functions, which are analytic in the elliptic parameter $\mathsf{m}$
  except at the singularities $0,1,\infty$. Specialising~\eqref{defM}
  to the rigid bipartite case, the elliptic parameter is given by
  \begin{equation}
    \label{eq:mrigid}
    \mathsf{m} = \left( \frac{1 - 4 h^2 \mathfrak{r}(u)}{1 + 4 h^2 \mathfrak{r}(u)} \right)^2.
  \end{equation}
  Grouping all the steps together, we get a meromorphic function
  $\varphi$ in
  $\mathbb{C}^* \setminus ((-\infty,-\frac{1}{4h^2}] \cup
  [\frac{1}{4h^2},\infty) )$ such that $u =
  \varphi(\mathfrak{r}(u))$. Since $\mathfrak{r}(u)$ is analytic at
  $u=0$ with $\mathfrak{r}(u)=\frac{u}{1-g_2} + o(u)$, $\varphi$ has
  no singularity at $0$.

  Now, the $F_{2\ell}$ are given by the expansion of $\mathbf{F}(x)$
  around $x=\infty$, or equivalently of $\mathbf{G}(v)$ around
  $v=\frac{1}{4}$.  This expansion involves those of $\Upsilon_{b}(v)$ around
  $v=0,\frac{1}{2}$ and of $x(v)$ around $v=\frac{1}{4}$. The functions
  $\Upsilon_{b}(v)$ and $x(v)$ are precisely constructed to have
  simple poles at respectively $v=0$ and $v=\frac{1}{4}$ for all $\mathsf{m}$,
  and the coefficients in their expansions are analytic in
  $\mathsf{m}$. Finally, we substitute $u=\varphi(\mathfrak{r}(u))$,
  to find that $F_{2\ell}(u)=\psi_\ell(\mathfrak{r}(u))$ for some
  meromorphic function $\psi_\ell$ whose set of poles in included in
  that of $\varphi$.
\end{proof}

By the discussion at the beginning of Appendix~\ref{gstrapp}, the proof
of Theorem~\ref{thm:rigiddelta} requires to check two things: local
delta-analyticity at $u_c$, and analyticity at every other point of
modulus $u_c$ (except $-u_c$ when $g_{4k}=0$, which we can be treated by
symmetry).

In the case of a non generic critical point
($\mathfrak{r}(u_c)= \frac1{4h^2}$), local delta-analyticity at $u_c$
is found in Lemma~\ref{lemdeltanal}, whose proof was detailed for the loop model triangulations with bending energy $\alpha$ not too small, but can be adapted to all nonnegative $\alpha$ including the rigid loop model at $\alpha = 0$. Thus, what we still have to prove is local delta-analyticity at $u_c$
for generic critical points ($\mathfrak{r}(u_c) < \frac1{4h^2}$), and
analyticity at other points of modulus $u_c$. In view of
Lemma~\ref{lem:rbd} and Proposition~\ref{prop:meropar}, it is
sufficient to establish these for $\mathfrak{r}(u)$. We start with the
analyticity statement.

\begin{proposition}
  \label{eq:globalana}
  Let $u^*$ be such that $|u^*|=u_c$ and
  $\lvert \mathfrak{r}(u^*) \rvert < \mathfrak{r}(u_c)$. Then,
  $\mathfrak{r}(u)$ admits an analytic continuation in a neighborhood
  of $u^*$. 
\end{proposition}

\begin{proof}
  Let $r^* = \mathfrak{r}(u^*)$, which lies in the domain of
  analyticity of $\varphi$. We want to show that
  $\varphi'(r^*) \neq 0$, since we may then apply the analytic inverse
  function theorem to show that $\varphi$ admits an inverse in a
  neighborhood of $u^*$, providing the analytic continuation we are
  looking for.  Instead of trying to compute $\varphi'(r^*)$
  directly, we will rather use~\eqref{eq:req}, which amounts to the
  relation
  \begin{equation}
    \label{eq:phieq}
    \varphi(r) = r - \sum_{k \geq 1} \binom{2k-1}{k} \left( g_{2k} + n h^{4k} \psi_k(r)\right) r^k
  \end{equation}
  valid for $r$ in the closure of the domain
  \begin{equation}
    \mathcal{D} = \{ \mathfrak{r}(u) \,\,|\,\, |u|<u_c \}.
  \end{equation}
  The convergence of the sum in the right-hand side
  of~\eqref{eq:phieq} is normal, and therefore (by Cauchy's integral
  formula) we may differentiate the relation at every point inside
  $\mathcal{D}$. We may rearrange the result in the form
  \begin{equation}
    \label{eq:ABphi}
    1-A(r) = B(r) \varphi'(r), \qquad
    \begin{cases}
      A(r) := \sum_{k \geq 1} k \binom{2k-1}{k} \left( g_{2k} + n h^{4k} \psi_k(r)\right) r^{k-1}, \\
      B(r) := 1 + n \sum_{k \geq 1} \binom{2k-1}{k} h^{4k}
      \frac{\psi'_k(r)}{\varphi'(r)} r^k.
    \end{cases}
  \end{equation}
  We will show that, for $r \to r^*$, the quantity $1-A(r)$ tends to a nonzero
  limit and $B(r)$ to a finite limit, which implies that
  $\varphi'(r^*) \neq 0$ since $\varphi'$ is continuous at $r^*$.

  For this, we observe that
  \begin{equation}
    \begin{split}
      A(\mathfrak{r}(u)) &= \sum_{k \geq 1} k \binom{2k-1}{k} \left( g_{2k} + n h^{4k} F_{2k}(u)\right) \mathfrak{r}(u)^{k-1}, \\
      B(\mathfrak{r}(u)) &= 1 + n \sum_{k \geq 1} \binom{2k-1}{k} h^{4k}
      F_{2k}'(u) \mathfrak{r}(u)^k
    \end{split}
  \end{equation}
  are both series in $u$ with nonnegative coefficients.

  Let $r_c=\mathfrak{r}(u_c)$. Since $\varphi$, $A$, $B$ are all
  strictly increasing functions on the real interval $(0,r_c)$, we see that
  $1-A$ cannot vanish, and therefore $A(r_c) \leq 1$. Thus, for
  $r \to r^*$, $A(r)$ tends to a limit $A(r^*)$ of modulus strictly
  smaller than $1$, and $1-A(r^*)$ is nonzero.

  On the other hand, by Proposition~\ref{prop9} --- or directly
  \eqref{eq:rigFpsasy} in the non generic critical case --- we know that
  $F_{2k}'(u_c)$ is finite and grows as $(4r_c)^k$ for $k \to
  \infty$. Thus, as $r$ tends to $r^*$ in $\mathcal{D}$, $B(r)$ tends
  to a finite limit $B(r^*)$ since $16h^4(r_cr^*)^2<1$. Thus,
  $\varphi'(r^*) \neq 0$ as wanted.
\end{proof}

We finish with the analysis of generic singularities.

\begin{proposition}
  \label{prop:genana}
  If $\mathfrak{r}(u_c)< \frac1{4h^2}$, then $\mathfrak{r}(u)$ is
  delta-analytic locally at $u_c$. More precisely, there exists
  $\delta>0$ such that $\mathfrak{r}(u)$ admits in the slit disk
  $\{|u-u_c|<\delta,\arg(u-u_c)\neq 0\}$ an analytic continuation of
  the form $\mathfrak{r}(u_c)-\tilde{\mathfrak{r}}(\sqrt{u_c-u})$, with
  $\tilde{\mathfrak{r}}$ analytic.
  
\end{proposition} 
In other words, $\mathfrak{r}(u)$ has a square root type singularity, and the volume exponent is that of pure
  gravity, as expected.
  
\begin{proof}
  We will reuse ideas and notations from the proof of
  Proposition~\ref{eq:globalana}. Since $\varphi$ is analytic at
  $r_c$, we must have $\varphi'(r_c)=0$, as otherwise
  $\mathfrak{r}(u)$ could be analytically continued at $u_c$,
  contradicting Pringsheim's theorem~\cite[Theorem~IV.6]{Flajolet}.
  By differentiating~\eqref{eq:ABphi}, we get
  \begin{equation}
    \varphi''(r) = - \frac{A'(r) + B'(r) \varphi'(r)}{B(r)}.
  \end{equation}
  By the same arguments as before, since $16h^4r_c^4<1$, $B(r)$ tends
  for $r \to r_c$ to a finite limit, which is now manifestly
  positive. We also have $A'(r_c)$ and $B'(r_c)$ positive (the latter
  possibly infinite), thus $\varphi''(r_c)<0$ (this quantity being
  necessarily finite by analyticity of $\varphi$ at $r_c$).

  Let us define
  $\tilde{\varphi}(x)=\frac{\varphi(r_c)-\varphi(r_c-x)}{(r_c-x)^2}$. It
  is analytic at $x=0$ and
  $\tilde{\varphi}(0)=-\frac{\varphi''(r_c)}{2}$ is positive. Thus,
  $\sqrt{\tilde{\varphi}(x)}$ is analytic and nonzero at $x=0$ (we
  pick the principal branch of the square root). By the inverse
  function theorem, $x \mapsto x \sqrt{\tilde{\varphi}(x)}$ admits an
  inverse $\tilde{\mathfrak{r}}$ locally around zero.  The function
  $u \mapsto r_c-\tilde{\mathfrak{r}}\big(\sqrt{u_c-u}\big)$ is the
  analytic continuation we are looking for, since it coincides with
  $\mathfrak{r}(u)$ on an interval $(u_c-\delta,u_c)$.
\end{proof}

\section{Delta-analyticity in the bending energy model}
\label{sec:anabend}

The purpose of this appendix is to establish an analogue of
Proposition~\ref{eq:globalana} in the more general case of the bending
energy model. Our proof works in the even more general setting where
we allow for unvisited faces of arbitrary degrees. The parameters are
then the weight per loop $n$, the weight per visited triangle $h$, the
bending energy factor $\alpha$, the weight $g_k$ per unvisited face of
degree $k \geq 1$, and the weight $u$ per vertex. We fix $n$, $h$,
$\alpha$ and $(g_k)_{k \geq 2}$ to nonnegative real values, with
$\alpha,h>0$, $n \in (0,2)$, $g_2<1$ and $g_k=0$ for $k$ large enough. We will however let
$u$ and $g_1$ vary in the complex plane, thus we will keep the
dependence on them explicit (we may think of $u$ as being $g_0$). We aim at establishing the following result.

\begin{theorem}
 \label{thm:rigiddeltaJ}
Assume $u_c(0)$ is a non generic critical point, i.e. $\mathfrak{s}(u_c(0),0) + 2\mathfrak{r}(u_c(0),0) = \gamma_+^*$. Then $\mathfrak{r}(u,g_1 = 0)$, $\mathfrak{s}(u,g_1 = 0)$, $q(u,g_1 = 0)$, $F_{\ell}(u,g_1 = 0)$ ($\ell > 0$) are delta-analytic functions of $u$.
 \end{theorem}
 
In this theorem, the delta-analyticity of $q$ is directly implied by the delta-analyticity of $\mathfrak{r}$ and $\mathfrak{s}$, as it is an analytic function of $\gamma_{\pm} = \mathfrak{s} \pm 2\sqrt{\mathfrak{r}}$ (consider together  \eqref{defM}, \eqref{TCeq}, and \eqref{q12de}). As the delta-analyticity statement locally around $u = u_c(0)$ is a consequence of Lemma~\ref{lemdeltanal} and the preceding discussion on analyticity in $q^{\frac{1}{2}}$ of all the quantities of concern near $q^{\frac{1}{2}} \rightarrow 0$, we focus on the justification of the existence of an analytic continuation at the other points on the closure of the disk of convergence in the $u$-plane. Our proof will in fact give a similar property when $g_1$ is set to small enough (perhaps non-zero) value. Treating $g_1$ as a parameter is the trick making possible to adapt the strategy of Appendix~\ref{sec:rigiddelta} to this bivariate situation.

Consider the partition functions $F_\ell(u,g_1)$ ($\ell > 0$) and
the series $\mathfrak{r}(u,g_1)$ and $\mathfrak{s}(u,g_1)$ of
Proposition~\ref{ponecut} and Remark~\ref{rem:rs}. All these are
series in $u$ and $g_1$ with nonnegative coefficients, which are
absolutely converging in an ($\ell$-independent) open subset $\mathcal{B} \subset \mathbb{C}^2$ which is a neighborhood of $(0,0)$. This domain of convergence is such that for any $g_1$ there exists some $u_c(g_1) \geq 0$ such that $\mathcal{B} \cap (\mathbb{C} \times \{g_1\}) = B(0;u_c(g_1)) \times \{g_1\}$, and $u_c(g_1) > 0$ for $g_1$ small enough.

Following the notations from~\cite[Section~6]{BBG12b}, let us define
\begin{equation}
  P_k(r,s) := \sum_{i=0}^{\lfloor \frac{k}{2} \rfloor} \frac{k!}{(i!)^2(k-2i)!} \,r^i s^{k-2i},
\end{equation}
and define similarly
\begin{equation}
  Q_k(r,s) := \sum_{i=0}^{\lfloor \frac{k-1}{2} \rfloor} \frac{k!}{i!(i+1)!(k-2i-1)!}\, r^{i+1} s^{k-2i-1}.
\end{equation}
These are polynomials with nonnegative coefficients, counting certain
lattice paths. For $r,s>0$, we have
$$
\lim_{k \to \infty} P_k(r,s)^{\frac{1}{k}}= \lim_{k \to \infty}
Q_k(r,s)^{\frac{1}{k}}=s+2\sqrt{r}.
$$
Then, the relations~\eqref{ointg1}
determining $(\mathfrak{r},\mathfrak{s})$ may be rewritten
\begin{equation}
  \label{eq:rseq}
  \begin{split}
    \mathfrak{r}(u,g_1) &= u + \sum_{k \geq 2} G_k(u,g_1)\, Q_{k-1}(\mathfrak{r}(u,g_1),\mathfrak{s}(u,g_1)),\\
    \mathfrak{s}(u,g_1) &= \sum_{k \geq 1} G_k(u,g_1)\, P_{k-1}(\mathfrak{r}(u,g_1),\mathfrak{s}(u,g_1)),
  \end{split}
\end{equation}
where $G_k(u,g_1)$ are the renormalised face weights given
of~\eqref{eq:fixp}. To connect with~\cite[Equation~(6.7)]{BBG12b},
note that $Q_{k-1}(r,s)=\frac{1}{2}(P_{k}(r,s)-sP_{k-1}(r,s))$. We give the following analog of Proposition~\ref{eq:globalana}, which due to the previous remarks proves Theorem~\ref{thm:rigiddeltaJ}.

\begin{proposition}
  \label{prop:anabend}
  Let $(u^*,g_1^*)$ be a point on the boundary of $\mathcal{B}$ such that
  $\lvert \mathfrak{r}(u^*,g_1^*) \rvert < \mathfrak{r}(\lvert
  u^*\rvert,\lvert g_1^*\rvert)$. Then, $\mathfrak{r}$, $\mathfrak{s}$
  and $F_\ell$ ($\ell > 0$) all admit an analytic continuation in a
  neighborhood of $(u^*,g_1^*)$.
\end{proposition}

The only way for the condition
$\lvert \mathfrak{r}(u^*,g_1^*) \rvert < \mathfrak{r}(\lvert
u^*\rvert,\lvert g_1^*\rvert)$ not to be satisfied is to have $u^*$
and $g_1^*$ both real and positive, or some periodicity phenomenon to
occur: it may be seen that this only happens for the situation already
discussed in Appendix~\ref{sec:rigiddelta}, namely $u^*$ negative,
$\alpha=0$, $g_1^*=0$ and $g_k=0$ for $k \neq 2 \mod 4$. As here $\alpha > 0$, there are no non-vanishing conditions on $g_k$ in Theorem~\ref{thm:rigiddeltaJ}. Before proving Proposition~\ref{prop:anabend}, we first state the analog of
Proposition~\ref{prop:meropar} --- recall that
$\gamma^*_+=\frac{1}{h(\alpha+1)}$.

\begin{proposition}
  \label{prop:meroparbis}
  There exists functions $\varphi(r,s)$ and $\psi_\ell(r,s)$
  ($\ell > 0$) which are meromorphic in a domain containing
  $\mathcal{Y} = \big\{(r,s)\in \mathbb{C}^2 \quad | \quad |s|+2\sqrt{|r|} < \gamma^*_+ \big\}$ and such
  that, for $(u,g_1) \in \mathcal{B}$,
  \begin{equation}
    (u,g_1) = \varphi(\mathfrak{r}(u,g_1),\mathfrak{s}(u,g_1)),
    \qquad F_{\ell}(u,g_1) = \psi_\ell(\mathfrak{r}(u,g_1),\mathfrak{s}(u,g_1)).
  \end{equation}
  The set of poles of $\psi_\ell$ is included in that of $\varphi$.
\end{proposition}

\begin{proof}[Proof of Proposition~\ref{prop:meroparbis}]


We shall use the exact solution of the model, in particular  Theorem~\ref{theimdisk}. The vanishing
  conditions $\mathbf{G}(\tau) = \mathbf{G}(\tau+\frac12)=0$ form
  a $2 \times 2$ linear system for $u$ and $g_1$, namely
\begin{equation}
\label{thesystemsdun}  \forall \varepsilon \in \{0,\tfrac{1}{2}\},\qquad {\rm i}\sqrt{4 - n^2} \bigg(- \frac{u}{2 + n} f(v_\infty + \varepsilon)  + \frac{2{\rm i}C g_1}{4 - n^2} f'(v_\infty + \varepsilon) + \cdots\bigg) = 0, 
\end{equation}
  where we set $f(v) := \Upsilon_b(v) + \Upsilon_b(-v)$, we have used Lemma~\ref{Xinfexp}, and $\cdots$ are functions of the fixed parameters $(g_k)_{k \geq 2},n,h,\alpha$ and $\gamma_-,\gamma_+$. More precisely, according to Lemma~\ref{predetla} the coefficients in these equations are analytic functions of  $\gamma_-,\gamma_+$ (hence of $r,s$) such that $(q^{\frac{1}{2}},w_\infty)$ belongs to $\mathcal{V}''(0)$. We explain in the next paragraph that they can in fact analytically continue to a larger domain containing $\mathcal{Y}$. The matrix of this $2 \times 2$ system reads, up to multiplication of each column by a non-vanishing prefactor:
$$
\mathbf{D} = \left(\begin{array}{cc} f(v_\infty) & f\big(v_{\infty} + \frac{1}{2}\big) \\[1ex] f'(v_{\infty}) & f'\big(v_\infty + \frac{1}{2}\big)\end{array}\right),
$$ 
Due to the previous discussion, $\det(\mathbf{D})$ is an analytic function of $(r,s) \in \mathcal{Y}$. We will justify at the end of the proof that it is not identically zero. Therefore, the system for $u,g_1$ has a solution which is a meromorphic function of $\gamma_-,\gamma_+$ (zeroes of $\det(\mathbf{D})$ may create poles on a divisor). The rest of the proof is then similar to that of Proposition~\ref{prop:meropar} and thus omitted.

To show analytic continuation to $\mathcal{Y}$, we first discuss the case $\alpha \neq 1$. It is not hard to check that there exists analytic continuations for $\Upsilon_b^{(l)}(v_\infty + \varepsilon)$ for $\varepsilon \in \{0,\frac{1}{2}\}$ and $l \in \mathbb{N}$, as functions of $(q^{\frac{1}{2}},w_\infty)$ across the $q^{\frac{1}{2}}$-negative real axis. Therefore, it is enough to discuss the analyticity and analytic continuation of $(q^{\frac{1}{2}},w_\infty)$ seen as a function of $(\gamma_-,\gamma_+)$. For the latter, can perform analytic continuation across the branch cuts on the loci $\mathsf{m} \in (1,+\infty)$ and $\mathsf{p} \in (-\infty,0)$, hence the only possible singularities occur on the locus $\mathsf{m} \in \{1,\infty\}$ or $\mathsf{p} \in \{0,\infty\}$. When $(r,s) \in \mathcal{Y}$, the points $\gamma_{\pm} = s \pm 2\sqrt{r}$ are contained $B(0;\frac{1}{h(\alpha +1)})^2$. Recall that $\gamma_+^*$ is a fixed point of $\varsigma$, and the other fixed point is $\frac{1}{h(\alpha - 1)}$ which is outside $B(0;\frac{1}{h(\alpha - 1)})$. We then observe that the image of $B(0;\frac{1}{h(\alpha +1)})$ via the involution $\varsigma$ is disjoint from $B(0;\frac{1}{h(\alpha +1)})$. By comparing with the definition of $\mathsf{m},\mathsf{p}$ in \eqref{defM}-\eqref{pdef}, it means that $\mathsf{m}$ and $\mathsf{p}$ as functions on $\mathcal{Y}$ avoid the values $0$ and $\infty$. As a result, the only problem that could be met in continuing analytically $(q^{\frac{1}{2}},w_\infty)$ occurs on the intersection of the locus $\mathsf{m} = 1$ with $\mathcal{Y}$. In fact, substituting $\gamma_\pm = s + 2\sqrt{r}$ into the definition \eqref{defM} of $\mathsf{m}$, we have
$$
\mathsf{m} = 1 -  \frac{16h^2 r}{\big(1 - 2h\alpha s + h^2(4r - s^2)(1 - \alpha^2)\big)^2}.
$$
Therefore, $\mathsf{m} = 1$ corresponds to $r = 0$ and the complement of this locus in $\mathcal{Y}$ is not simply-connected (its fundamental group is $\mathbb{Z}$). We are going to show, using modular transformations for the theta functions and Jacobi elliptic functions, that there are in fact no singularities at $\mathsf{m} = 1$. Since $\mathcal{Y}$ is simply-connected, this will justify that the coefficients of the system \eqref{thesystemsdun} can be continued (uniquely) to analytic functions on the whole $\mathcal{Y}$. The important variable after modular transformation is
 $$
 \tilde{q} = e^{2{\rm i}\pi\tau} = e^{-\frac{\pi K(\mathsf{m})}{K(1-\mathsf{m})}} \mathop{\sim}_{\mathsf{m} \rightarrow 1} \frac{1 - \mathsf{m}}{16}.
 $$
This is an analytic function of $\mathsf{m}$ in a neighborhood of $1$. The expression \eqref{A1theta} of $\vartheta_1(v|\tau)$ is a series of terms involving $\tilde{q}^{\frac{1}{8} + \frac{m(m + 1)}{2}}$. Up to the $\tilde{q}^{\frac{1}{8}}$ prefactor this contains only integer powers of $\tilde{q}$. So, the ratio $\Upsilon_b(v)$ in \eqref{Upsfb} involves only integer powers of $\tilde{q}$, hence is an analytic family (over $\tilde{q}$ in a neighborhood of $0$) of meromorphic functions of $v$. With Jacobi imaginary transformations we also convert \eqref{vinfdsf} into
$$
v_\infty = \frac{{\rm arcsc}\big[{\rm i}\sqrt{\frac{\mathsf{p}}{\mathsf{m}}}\,;\,1 - \mathsf{m}\big]}{2K(1-\mathsf{m})} \mathop{\sim}_{\mathsf{m} \rightarrow 1} \frac{{\rm arctan}\big({\rm i}\sqrt{\mathsf{p}})}{\pi},
$$
where ${\rm arcsc}$ is (an analytic continuation of) the reciprocal function of ${\rm sc} = \frac{{\rm sn}}{{\rm cn}}$. The singularity at $\mathsf{m} = 1$ is now absent as these quantities are analytic near the locus $\mathsf{m} = 1$, and $\Upsilon_b^{(l)}(v_\infty + \varepsilon)$ for $\varepsilon \in \{0,\frac{1}{2}\}$ and $l \in \mathbb{N}$ are analytic functions near the locus $\mathsf{m} = 1$, as desired. The case $\alpha = 1$ is in fact simpler since $w_\infty = \frac{1}{2}$ is independent of $\gamma_\pm$ and the previous discussion of analytic continuation for $q^{\frac{1}{2}}$ was sufficient to conclude.
 
It remains to justify that $\det(\mathbf{D})$  is not identically zero. For this, we evaluate it when $\gamma_+$ approaches $\gamma_+^*$. Using the variables $(\mathsf{m},\mathsf{p})$ of Section~\ref{App1}, this corresponds to $\mathsf{m} \rightarrow 0$, i.e. $q \rightarrow 0$. According to \eqref{winfstar}, we have $v = \frac{1}{2} + \tau w_\infty$ with
$$
w_\infty \sim w_\infty^* = \tfrac{2}{\pi}\,{\rm arcsin}\bigg(\frac{1}{\sqrt{\mathsf{p}}}\bigg).
$$
Using Lemma~\ref{lemUp}, we obtain
\begin{equation}
\begin{split}
f(v_\infty) & \sim - \frac{4\pi}{T}\cos(\pi b w_\infty^*), \\
f'(v_\infty) & \sim \frac{4\pi}{T}\,\frac{\pi b}{{\rm i}T}\,\sin(\pi b w_\infty^*), \\
f\big(v_\infty + \tfrac{1}{2}\big) & \sim \frac{2\pi}{T}\,\frac{\sin\big(\pi(b - 1)w_\infty^*)}{\sin(\pi b w_\infty^*)}, \\
f'\big(v_\infty + \tfrac{1}{2}\big) & \sim \frac{2\pi}{T}\,\frac{\pi}{{\rm i}T}\bigg(\frac{(b - 1)\cos\big(\pi(b - 1)w_\infty^*)}{\sin(\pi w_\infty^*)} - \frac{\cos(\pi w_\infty^*)\sin\big(\pi(b - 1)w_\infty^*)}{\sin^2(\pi w_\infty^*)}\bigg).
\end{split}
\end{equation}
After trigonometric simplifications, this yields
$$
{\rm det}(\mathbf{D}) \sim \frac{8\pi^3}{{\rm i}T^3} \bigg(-b\,{\rm cot}(\pi w_\infty^*) + \frac{1}{2}\,\frac{\sin(2\pi b w_\infty^*)}{\sin^2(\pi w_\infty^*)}\bigg),
$$
which does not vanish identically.
\end{proof}

\begin{proof}[Proof of Proposition~\ref{prop:anabend}]

We will again invoke the
analytic implicit function theorem, now in its bivariate form. Note
that, by Lemma~\ref{lem:ricochet}, for any $u>0$ and $g_1 \geq 0$ such
that $(u,g_1) \in \mathcal{B}$, we have
$\gamma_+(u,g_1) \leq \gamma_+^*$, and by monotone convergence the
same property holds for $(u,g_1) \in \partial \mathcal{B}$. Thus, for
any $(u^*,g_1^*)$ satisfying the hypotheses of
Proposition~\ref{prop:anabend},
$(r^*,s^*):=(\mathfrak{r}(u^*,g_1^*),\mathfrak{s}(u^*,g_1^*))$ belongs
to the domain of analyticity of $\varphi$. What remains to check is
that the differential of $\varphi$ is nondegenerate at $(r^*,s^*)$.

We now substitute $(u,g_1)=\varphi(r,s)$ in~\eqref{eq:rseq}, and
differentiate, which is possible whenever $(r,s)$ is inside
$\mathcal{D}=\varphi^{-1}(\mathcal{B})$. The result may be written
compactly in matrix form
\begin{equation}
  {\rm Id}- \mathbf{M} = ({\rm Id} + \mathbf{N}) J_\varphi,
\end{equation}
where $J_\varphi$ is the Jacobian matrix of $\varphi$:
\begin{equation}
  J_\varphi =
  \begin{pmatrix}
    \partial_r u & \partial_s u \\[0.7ex]
    \partial_r g_1 & \partial_s g_1 \\
  \end{pmatrix},
\end{equation}
and
\begin{equation}
  \begin{split}
    \mathbf{M} = \sum_{k \geq 1} (k-1) G_k
    \begin{pmatrix}
      P_{k-2} & Q_{k-2} \\[0.7ex]
      \frac{1}{r}Q_{k-2} & P_{k-2}
    \end{pmatrix}, \qquad
   \mathbf{N} = \sum_{k \geq 1} \sum_{\ell \geq 0} A_{k,\ell}
    \begin{pmatrix}
     Q_{k-1}\,\partial_u F_\ell\, & Q_{k-1}\,\partial_{g_1} F_\ell \\[0.7ex]
   P_{k-1}\, \partial_u F_\ell  &  P_{k-1}\,\partial_{g_1} F_\ell
    \end{pmatrix}
  \end{split}.
\end{equation}
Here, we are using the relation
$\partial_r Q_{k-1}= (k-1) P_{k-1}$ and its variants. Note that $\frac{1}{r}Q_{k-2}$ is still a polynomial in $r,s$.

Notice that, when expressing $\mathbf{M}$ and $\mathbf{N}$ as power series in $u$ and $g_1$,
all their coefficients are nonnegative, and $\mathbf{M}$ has no constant
coefficient. The eigenvalues of $\mathbf{M}$ are of the form
$X \pm Y \sqrt{r}$, with $X,Y$ also power series with nonnegative
coefficients and no constant term. When restricting to $u$ and $g_1$
real nonnegative in $\mathcal{B}$, the nonvanishing of
$\det J_\varphi$ (since $\varphi$ is a local diffeomorphism at $(r,s) \in \varphi^{-1}(\mathcal{B})$) implies that $X+Y\sqrt{r}$ must remain strictly smaller than
$1$. Thus, going back to complex values, the eigenvalues of $\mathbf{M}$ have
modulus strictly smaller than $1$ in all $\mathcal{B}$, and the same holds for
their limits at $(r^*,s^*)$, since it is assumed that
$|r^*|<\mathfrak{r}(|u^*|,|g_1^*|)$. We conclude that $\det({\rm Id}-\mathbf{M})$
tends to a nonzero limit at $(r^*,s^*)$.

On the other hand, in $\mathbf{N}$ we may bound the derivatives
$\partial_u F_\ell$ and
$\partial_{g_1} F_\ell$ by their values at
$(|u^*|,|g_1^*|)$ which, by Propositions~\ref{prop4} and \ref{prop9},
are finite and grow as $O\big((\gamma_+^*)^\ell\big)$ for $\ell \to
\infty$. By~\eqref{eq:Aidentgam}, we deduce that
$\sum_{\ell \geq 0} A_{k,\ell} \,\partial_u F_\ell$ and
$\sum_{\ell \geq 0} A_{k,\ell} \,\partial_{g_1} F_\ell$
are $O\big((\gamma_+^*)^{-k}\big)$. As $Q_{k-1}$ and $P_{k-1}$ are $O(\gamma^k)$,
with $\gamma=|s^*|+2\sqrt{|r^*|}<\gamma_+^*$, we find that $N$ has a finite
limit at $(r^*,s^*)$, and we conclude by continuity that
$\det J_\varphi(r^*,s^*) \neq 0$, as wanted.
\end{proof}

We expect that Proposition~\ref{prop:genana} can be extended to treat generic critical points of the bending energy model, but we shall not attempt to do so here.

\end{document}